%% file: sisa_2021.tex
\documentclass[sigconf]{acmart}

\newif\ifQA
\QAfalse

\copyrightyear{2021}
\acmYear{2021}
\setcopyright{licensedothergov}\acmConference[MICRO '21]{MICRO-54: 54th Annual IEEE/ACM International Symposium on Microarchitecture}{October 18--22, 2021}{Virtual Event, Greece}
\acmBooktitle{MICRO-54: 54th Annual IEEE/ACM International Symposium on Microarchitecture (MICRO '21), October 18--22, 2021, Virtual Event, Greece}
\acmPrice{15.00}
\acmDOI{10.1145/3466752.3480133}
\acmISBN{978-1-4503-8557-2/21/10}

\include{mac-includes}
\usepackage{comment}

\conftrue
\allfalse
\trfalse
\todosfalse

\conffalse
\allfalse
\trtrue
\todosfalse

\newif\iftrLOW
\trLOWfalse

\newif\iftrNEW
\trNEWfalse

\newif\iftrDYN
\trDYNfalse

\usepackage{colortbl}

\newcommand{\VfaBatteryFull}{{\tiny\begin{turn}{90}\faBatteryFull\end{turn}}}
\newcommand{\VfaBatteryHalf}{{\tiny\begin{turn}{90}\faBatteryHalf\end{turn}}}

\newcommand{\marginparN}[1]{\marginpar{}}
\newcommand{\marginparX}[1]{}

\renewcommand{\marginpar}[1]{}
\renewcommand{\hl}[1]{#1}

\newcommand{\faBatteryFullT}{{\tiny\faBatteryFull}}
\newcommand{\faBatteryHalfT}{{\tiny\faBatteryHalf}}
\newcommand{\faTimesT}{{\tiny\faTimes}}

\ifconf
\fi

\begin{document}

%
%

\iftr
\title{SISA: Set-Centric Instruction Set Architecture\\ for Graph Mining on Processing-in-Memory Systems} 
\fi

\author{Maciej Besta$^1$, Raghavendra Kanakagiri$^2$, Grzegorz Kwasniewski$^1$, Rachata Ausavarungnirun$^3$,
Jakub Beránek$^4$, Konstantinos Kanellopoulos$^1$, Kacper Janda$^5$, Zur Vonarburg-Shmaria$^1$,
Lukas Gianinazzi$^1$, Ioana Stefan$^1$, Juan Gómez-Luna$^1$, Marcin Copik$^1$, Lukas Kapp-Schwoerer$^1$,
Salvatore Di Girolamo$^1$, Nils Blach$^1$, Marek Konieczny$^5$, Onur Mutlu$^1$, Torsten Hoefler$^1$}
       \affiliation{\vspace{0.3em}$^1$ETH Zurich, Switzerland\quad\quad
       {$^2$}IIT Tirupati, India\quad\quad
       {$^3$}King Mongkut's University of Technology North Bangkok, Thailand\quad\quad
       {$^4$}Technical University of Ostrava, Czech Republic\quad\quad
       {$^5$}AGH-UST, Poland\vspace{0.3em}\country{}\\
}




\begin{abstract}
%
%
Simple graph algorithms such as PageRank have been the target of
numerous hardware accelerators. Yet, there also exist much more complex graph
\emph{mining} algorithms for problems such as clustering or maximal clique
listing. 
These algorithms are memory-bound and thus could be accelerated by hardware
techniques such as Processing-in-Memory (PIM). However, they also come with
non-straightforward parallelism and complicated memory access patterns.
%
%
In this work, we address this problem with a simple yet surprisingly powerful
observation: operations on sets of vertices, such as intersection or union,
form a large part of many complex graph mining algorithms, and can offer rich
and simple parallelism at multiple levels. 
%
%
This observation drives our cross-layer design, in which we (1) expose set
operations using a novel programming paradigm, (2) express and execute these operations
efficiently with carefully designed \emph{set-centric} ISA extensions called
SISA, and (3) use PIM to accelerate SISA instructions.
The key design idea is to alleviate the bandwidth needs of SISA instructions by
mapping set operations to two types of PIM: in-DRAM bulk bitwise computing for
bitvectors representing high-degree vertices, and near-memory logic layers for
integer arrays representing low-degree vertices. 
\all{We enhance SISA with a simple
hardware scheme that automatically selects the form of PIM.}
Set-centric SISA-enhanced algorithms are efficient and outperform
hand-tuned baselines, offering more than 10$\times$ speedup over the
established Bron-Kerbosch algorithm for listing maximal cliques.
%
%
%
We deliver more than 10 SISA set-centric algorithm formulations,
illustrating SISA's wide applicability. 
%
%
%
%
\all{The resulting SISA design enables programmable memory-accelerated graph
mining: large parts of graph algorithms are expressed with a small group of
well-defined set-centric operations.}
\all{We show that SISA's set-centric algorithms are theoretically efficient,
matching the work and depth complexity of best-known baselines.
SISA outperforms hand-tuned comparison targets, for example offering more than
10$\times$ speedups over the established Bron-Kerbosch algorithm for listing
maximal cliques, and more than 100$\times$ over a state-of-the-art graph
pattern matching framework Peregrine.}
%
%
\all{SISA can become a standard tool for harnessing the capabilities of
state-of-the-art hardware acceleration such as in-memory execution in the broad
context of graph mining analytics.}

\end{abstract}



\maketitle

\pagestyle{plain}


\input{intro.tex}

\input{background.tex}

\input{overview.tex}

\input{abstractions.tex}

\input{formulations.tex}
\input{structure.tex}
\input{theory.tex}

\input{isa-implementation.tex}

\input{eval.tex}

\input{related.tex}
\input{conclusion.tex}


%


\begin{acks}

We thank Mark Klein, Hussein Harake, Colin McMurtrie, Angelo Mangili, and the
whole CSCS team granting access to the Ault and Daint machines, and for their
excellent technical support. We thank Timo Schneider for his immense help with
computing infrastructure at SPCL. 
This project received funding from the European Research Council
(ERC) under the European Union's Horizon 2020 programme (grant
agreement DAPP, No.~678880). 

\end{acks}

{
  \bibliographystyle{ACM-Reference-Format}
  \bibliography{refs}
}

\end{document}

%% file: mac-includes.tex
\newif\iftr     
\newif\ifall    
\newif\ifconf   
\newif\ifsq     
\newif\ifnonb   
\newif\iftodos  

\newif\ifsqCAP
\newif\ifsqVS
\newif\ifsqEN
\newif\ifsqTIT

\newcommand{\tr}[1]{\iftr #1 \fi}
\newcommand{\trNEW}[1]{}
\newcommand{\all}[1]{}
\newcommand{\cnf}[1]{}

\newcommand{\ignore}[1]{}

\sqtrue
\sqCAPtrue
\sqVStrue

\sqfalse
\sqCAPfalse
\sqENfalse
\sqVSfalse
\sqTITfalse

\usepackage{etex}
\usepackage{balance}
\usepackage{epstopdf}
\usepackage{placeins}

\usepackage{graphicx}
\usepackage{float}
\usepackage{dblfloatfix}
\usepackage{multirow}
\usepackage{rotating}
\usepackage[switch]{lineno} 
\usepackage{makecell}
\usepackage{parcolumns}
\usepackage{tikz}
\usetikzlibrary{tikzmark}

\usepackage{xpatch}
\expandafter\xpatchcmd
\csname pgfk@/tikz/every picture/.@cmd\endcsname
{\thepage}{\arabic{page}}{}{}

\tikzstyle{comment} = [draw, fill=blue!70, text=white, text width=3cm, minimum height=1cm, rounded corners, align=left, font=\scriptsize]
\tikzstyle{background_alg} = [draw, fill=blue!20, opacity=0.4, inner sep=4pt, rounded corners=2pt]

\usetikzlibrary{shapes}
\usetikzlibrary{plotmarks}
\usetikzlibrary{calc, fit}

\usepackage{enumitem}

\usepackage{mathtools}

\newtheorem{theorem}{Theorem}[section]

\newtheorem{observation}[theorem]{Observation}

\usepackage{soul}
\usepackage{fontawesome}
\usepackage{pifont}
\usepackage{textcomp}
\usepackage{booktabs}
\usepackage{url}
\usepackage{pbox}
\usepackage[10pt]{moresize}

\usepackage{cleveref}
\usepackage[utf8]{inputenc}
\crefname{section}{§}{§§}
\Crefname{section}{§}{§§}

\newcommand{\macb}[1]{\textbf{{#1}}}

\usepackage[font={normalfont, scriptsize}]{subfig}

\ifsq
\usepackage[font={normalfont, scriptsize}]{caption}
\else
\usepackage[font={normalfont}]{caption}
\fi

\newcommand{\vspaceSQ}[1]{\ifsqVS\vspace{#1}\fi}
\newcommand{\enlargeSQ}[1]{\ifsqEN\enlargethispage{\baselineskip}\fi}

\ifsqTIT
\usepackage[compact]{titlesec}
\titlespacing*{\section}{0pt}{6pt}{2pt}
\titlespacing*{\subsection}{0pt}{5pt}{1pt}
\titlespacing*{\subsubsection}{0pt}{5pt}{1pt}
\fi

\usepackage{xcolor}
\definecolor{darkgrey}{RGB}{70,70,70}
\definecolor{lightgrey}{RGB}{200,200,200}
\definecolor{lyellow}{RGB}{255,255,100}
\definecolor{llyellow}{RGB}{250,250,180}
\definecolor{lgreen}{RGB}{144,238,144}

\usepackage[customcolors]{hf-tikz}
\hfsetbordercolor{white}
\hfsetfillcolor{vlgray}

\definecolor{vlgray}{rgb}{0.77 0.77 0.77}
\definecolor{ablack}{rgb}{0.2 0.2 0.2}
\definecolor{vllgray}{rgb}{0.9 0.9 0.9}
\definecolor{bblue}{rgb}{0.7 0.7 0.99}

\usepackage{colortbl}

\usepackage{listings}

\ifsq
\else
\fi

\ifsq

\else

\fi

\newcommand{\maciej}[1]{\textcolor{blue}{[Maciej: #1]}}
\newcommand{\m}[1]{\textcolor{blue}{[Maciej: #1]}}

\newcommand{\lukas}[1]{\textcolor{orange}{[Lukas: #1]}}

\definecolor{hlL}{rgb}{0.8 0.8 0.99}

\newcounter{highlight}
\newcommand{\highlight}[1]{%
\stepcounter{highlight}\tikzmarkin{\thehighlight}(0.05,-0.1)(0.08,0.23)#1\tikzmarkend{\thehighlight}%
}

\newcounter{hlLR}
\newcommand{\hlLR}[2]{%
\stepcounter{hlLR}\hfsetfillcolor{vlgray}\tikzmarkin[set fill color=vlgray, set border color=white, above offset=0.225, right offset=#1, below offset=-0.125]{\thehighlight}#2\tikzmarkend{\thehighlight}}

\newcounter{hlLIR}

\newcounter{hlLIIR}

\newcounter{Ahighlight}

\usepackage{scalerel,stackengine}
\stackMath
\newcommand\rwh[1]{%
\savestack{\tmpbox}{\stretchto{%
  \scaleto{%
        \scalerel*[\widthof{\ensuremath{#1}}]{\kern-.6pt\bigwedge\kern-.6pt}%
                  {\rule[-\textheight/2]{1ex}{\textheight}}
                              }{\textheight}%
}{0.5ex}}%
\stackon[1pt]{#1}{\tmpbox}%
}

\usepackage[hang,flushmargin]{footmisc}

\renewcommand{\epsilon}{\ensuremath\varepsilon}

\renewcommand{\phi}{\ensuremath{\varphi}}

%% file: intro.tex
\vspaceSQ{-0.1em}
\section{Introduction}
\label{sec:intro}
\vspaceSQ{-0.35em}


\iftr

\begin{table*}[t]
\vspaceSQ{-1.5em}
\setlength{\tabcolsep}{1.4pt}
\ifsq\renewcommand{\arraystretch}{0.6}\fi
\centering
\scriptsize
\ssmall
%
\begin{tabular}{lllcclllllllllllll}
\toprule
\multirow{2}{*}{\makecell[c]{\textbf{Abstraction or}\\ \textbf{programming model}}} & 
\multirow{2}{*}{\makecell[c]{\textbf{Example}\\ \textbf{design}}} & 
\multirow{2}{*}{\makecell[c]{\textbf{Underlying}\\ \textbf{algebra?}}} & 
\multirow{2}{*}{\makecell[c]{\textbf{Key} \textbf{element}}} & 
\multirow{2}{*}{\makecell[c]{\textbf{Key} \textbf{operations}}} & 
\multicolumn{4}{c}{\textbf{Pattern M.}} & 
\multicolumn{4}{c}{\textbf{Learning}} & 
\multicolumn{4}{c}{\textbf{``Low-c.''}} &
\multirow{2}{*}{\textbf{Remarks}} \\
\cmidrule(lr){6-9} \cmidrule(lr){10-13} \cmidrule(lr){14-17}
 & & & & & 
\textbf{mc} & \textbf{kc} & \textbf{ds} & \textbf{si} & 
\textbf{vs} & \textbf{lp} & \textbf{cl} & \textbf{av} & 
\textbf{tc} & \textbf{bf} & \textbf{cc} & \textbf{pr} & \\ 
\midrule
\makecell[l]{Vertex-centric (ver-c)} & PowerGraph~\cite{gonzalez2012powergraph} & \faTimes & \makecell[c]{Vertex + its neighbors} & \makecell[c]{Vertex kernel} & \faTimes & \faTimes & \faTimes & \faTimes & \faTimes & \faTimes & \faTimes & \faTimes & \VfaBatteryHalf$^*$ & \VfaBatteryFull & \VfaBatteryHalf$^\dagger$ & \VfaBatteryFull & $^*$High comm.~costs; $^\dagger$High work and depth \\
\makecell[l]{Edge-centric (edge-c)} &  {X-Stream~\cite{roy2013x}} & \faTimes  & \makecell[c]{Edge + its endpoints} & \makecell[c]{Edge kernel} & \faTimes & \faTimes & \faTimes & \faTimes & \faTimes & \faTimes & \faTimes & \faTimes & \VfaBatteryHalf$^*$ & \VfaBatteryHalf$^*$ & \VfaBatteryHalf$^*$ & \VfaBatteryFull & $^*$High work and depth \\
\makecell[l]{Array maps} & {Ligra~\cite{shun2013ligra}} & \faTimes & \makecell[l]{Edge/vertex arrays} & \makecell[c]{Edge/vertex maps} & \faTimes & \faTimes & \VfaBatteryFull & \faTimes & \faTimes & \faTimes & \VfaBatteryHalf$^*$ & \faTimes & \VfaBatteryHalf & \VfaBatteryFull & \VfaBatteryFull & \VfaBatteryFull & $^*$Support for low-diameter decomposition only \\
GraphBLAS~\cite{kepner2016mathematical} & GraphMat~\cite{sundaram2015graphmat} & \VfaBatteryFull\ (linear) & \makecell[c]{Matrix, vector} & \makecell[c]{SpMV, SpMSpM} & \faTimes & \faTimes & \faTimes & \VfaBatteryHalf$^*$ & \faTimes & \faTimes & \faTimes & \faTimes & \VfaBatteryFull & \VfaBatteryFull & \VfaBatteryHalf$^\dagger$ & \VfaBatteryFull & $^*$Only trees as patterns; $^\dagger$High work/depth \\
\makecell[l]{GNN (graph neural\\ networks)} & {HyGCN~\cite{yan2020hygcn}} & \VfaBatteryHalf$^*$ & \makecell[c]{Various} & \makecell[c]{Various} & \faTimes & \faTimes & \faTimes & \VfaBatteryHalf$^\dagger$ & \VfaBatteryHalf & \VfaBatteryFull & \VfaBatteryFull & \VfaBatteryFull & \faTimes & \faTimes & \faTimes & \VfaBatteryHalf & \makecell[l]{$^*$Some models use linear algebra; $^\dagger$GNNs are\\ as powerful as the Weisfeiler-Lehman test~\cite{xu2018powerful}.} \\
\makecell[l]{Pattern matching} & \makecell[l]{Peregrine~\cite{jamshidi2020peregrine} and\\ others (see Section~\ref{sec:intro})} & \faTimes  & Vertex + its neighbors & Explore neighbors & \VfaBatteryHalf$^*$ & \VfaBatteryHalf$^*$ & \VfaBatteryHalf$^*$ & \VfaBatteryHalf$^*$ & \faTimes & \faTimes & \faTimes & \faTimes & \VfaBatteryHalf$^*$ & \faTimes & \faTimes & \faTimes & $^*$No bounds provided, possibly high work/depth \\
\makecell[l]{Joins~\cite{cheng2008fast}} & {RDBMS~\cite{zhao2017all}} & \VfaBatteryFull\ (relational) & Relations & Joins & \faTimes & \VfaBatteryHalf$^*$ & \VfaBatteryHalf$^*$ & \faTimes & \VfaBatteryHalf$^*$ & \VfaBatteryHalf$^*$ & \VfaBatteryHalf$^*$ & \faTimes & \VfaBatteryHalf$^*$ & \faTimes & \VfaBatteryHalf$^*$ & \VfaBatteryHalf$^*$ & $^*$No bounds provided, possibly high work/depth \\
\midrule
\makecell[l]{\textbf{Set-Centric}\\\textbf{[This work]}} & \makecell[l]{\textbf{SISA}\\\textbf{[This work]}} & \VfaBatteryFull\ (set)  & \makecell[c]{Sets of vertices/edges} & \makecell[c]{Set operations} & \VfaBatteryFull & \VfaBatteryFull & \VfaBatteryFull & \VfaBatteryFull & \VfaBatteryFull & \VfaBatteryFull & \VfaBatteryFull & \VfaBatteryFull & \VfaBatteryFull & \faTimes & \faTimes & \faTimes & \\ 
\bottomrule
\end{tabular}
\vspaceSQ{-0.5em}
\caption{
\textmd{\textbf{Comparison of the set-centric programming approach and SISA to
existing graph processing abstractions/programming models, focusing on
{support for selected graph mining problems (pattern matching, learning), and for
``low-complexity'' graph problems}}.
``\VfaBatteryFull'': Support / significant focus. ``\VfaBatteryHalf'': Partial support / some
focus. ``\faTimes'': no support / no focus.
\textbf{Pattern M.}: {selected} graph pattern matching problems,
\textbf{mc}: maximal clique listing,
\textbf{kc}: $k$-clique listing,
\textbf{ds}: dense subgraph,
\textbf{si}: subgraph isomorphism,
\textbf{Learning}: {selected} graph learning problems,
\textbf{vs}: vertex similarity,
\textbf{lp}: link prediction,
\textbf{cl}: clustering or community detection,
\textbf{av}: accuracy verification (of link prediction outcomes),
\textbf{``Low-c.''}: {selected} ``low-complexity'' problems targeted
by vast majority of existing works on graph processing.
\textbf{tc}: triangle counting,
\textbf{bf}: BFS,
\textbf{cc}: connected components,
\textbf{pr}: PageRank.
\textbf{The analysis in this table is extended in Section~\ref{sec:rw} and Table~\ref{tab:comparison_problems}} by detailing 
specific hardware accelerators for graph processing.
\all{\textbf{C?}: Does a given programming paradigm offer time complexities (of
their solutions of graph mining problems) that are competitive to those of
tuned specific graph mining algorithms?}
}}
\vspaceSQ{-2em}
\label{tab:comparison_models}
\end{table*}
\fi

\tr{Graph analytics underlies many problems in machine learning, social network
analysis, computational chemistry or biology, medicine, finances, and
others~\cite{DBLP:journals/ppl/LumsdaineGHB07}.}
\iftr
The growing importance of these
fields necessitates even more efficient large-scale graph processing. 
\fi
Research on graph analytics in computer architecture
has mostly targeted graph algorithms based on {vertex-centric}
formulations~\cite{zhuo2019graphq, ham2016graphicionado, besta2019graph, ahn2015pim,
ahn2015scalable_tes, nai2017graphpim, zhang2018graphp, kalavri2017high,
shi2018graph, batarfi2015large, mccune2015thinking}. Some works also focus on
{edge-centric} or {linear algebra} paradigms~\cite{song2018graphr,
sundaram2015graphmat, roy2013x, kepner2016mathematical}. 
\iftr
The algorithms in question are usually \emph{iterative} and have
complexities described by \emph{low-degree} polynomials~\cite{khan2016vertex},
for example $O(n+m)$ for Breadth-First Search (BFS)~\cite{cormen2009introduction},
$O(m L)$ for a power iteration scheme for PageRank (PR)~\cite{besta2017push},
and $O(mn)$ for Brandes' algorithm for Betweenness Centrality
(BC)~\cite{brandes2001faster}, where $n$ and $m$ are numbers of
vertices and edges, and $L$ is a selected iteration count.
\fi


%
Yet, there are numerous important problems and algorithms in the area of
\emph{\textbf{graph mining}}~\cite{cook2006mining, teixeira2015arabesque,
schaeffer2007graph, jiang2013survey, besta2021graphminesuite} that received little or no attention in
computer architecture. One large class is \emph{\textbf{graph pattern
matching}}~\cite{jiang2013survey}, which focuses on finding certain specific
subgraphs (also called {motifs} or {graphlets}). Examples of such
problems are $k$-clique listing~\cite{danisch2018listing, gianinazzi2021parallel}, maximal clique
listing~\cite{bron1973algorithm, cazals2008note, DBLP:conf/isaac/EppsteinLS10,
DBLP:journals/tcs/TomitaTT06}, $k$-star-clique
mining~\cite{jabbour2018pushing}, and many others~\cite{cook2006mining}.
Another class is broadly referred to as \emph{\textbf{graph
learning}}~\cite{cook2006mining}, with problems such as unsupervised learning
or clustering~\cite{jarvis1973clustering}, link prediction~\cite{liben2007link,
lu2011link, al2006link, taskar2004link}, or vertex
similarity~\cite{leicht2006vertex}. 
\iftr
All these problems are widely used in
social sciences (e.g., studying human
interactions)~\cite{DBLP:conf/isaac/EppsteinLS10}, bioinformatics (e.g.,
analyzing protein structures)~\cite{DBLP:conf/isaac/EppsteinLS10},
computational chemistry (e.g., designing chemical
compounds)~\cite{takigawa2013graph}, medicine (e.g., drug
discovery)~\cite{takigawa2013graph}, cybersecurity (e.g., identify intruder
machines)~\cite{dua2016data}, healthcare (e.g., identify groups of people who
submit fraudulent claims)~\cite{thiprungsri2011cluster}, web graph analysis
(e.g., enhance search services)~\cite{jiang2009mining}, entertainment (e.g.,
predict movie popularity)~\cite{bogdanov2013strong}, and many
others~\cite{cook2006mining, jiang2013survey, horvath2004cyclic,
chakrabarti2006graph}.
\fi
These problems often run in time at least quadratic in
the number of vertices, and many problems are NP-complete~\cite{cook2006mining,
ullmann1976algorithm, danisch2018listing, bron1973algorithm}.
Thus, they often {differ significantly in their performance properties from
``low-complexity'' problems such as BFS or PageRank}.

Importantly, the established vertex-centric model, 
originally proposed in the Pregel graph processing system~\cite{malewicz2010pregel}, does \emph{not}
effectively express graph mining problems. It 
exposes only the {local} graph structure: A thread executing a vertex
kernel for any vertex~$v$ {can only access the neighbors of~$v$}. While
this suffices for algorithms such as PageRank, graph mining often requires
{non-local} knowledge of the graph
structure~\cite{cook2006mining}. 
\iftr
Obtaining such knowledge in the
vertex-centric paradigm is hard or infeasible. As noted by multiple
researchers~\cite{salihoglu2014optimizing, yan2014pregel, liu2018empirical,
ko2018turbograph++} \emph{``(...) implementing graph algorithms efficiently on
Pregel-like systems (...) can be surprisingly difficult and require careful
optimizations.''~\cite{salihoglu2014optimizing}, ``It is challenging to design
Pregel algorithms (...)''~\cite{yan2014pregel}, ``Non-iterative graph algorithms might
be difficult to express in the vertex-centric model which heavily relies on
(...) supersteps''~\cite{kalavri2017high}, ``(...) graph algorithms, like
triangle counting, are not a good fit for the vertex-centric
model.~\cite{kalavri2017high}}.
\fi
\tr{Thus, most graph mining problems cannot be simply programmed with the
vertex-centric software (SW) frameworks~\cite{malewicz2010pregel,
low2010graphlab, gonzalez2012powergraph, kalavri2017high, mccune2015thinking}
and accelerated with vertex-centric hardware (HW)
architectures~\cite{ham2016graphicionado, song2018graphr, besta2019graph,
ahn2015scalable_tes, ahn2015pim, nai2017graphpim, zhang2018graphp}.}
Similar arguments apply to other paradigms {such as
GraphBLAS}~\cite{roy2013x, kepner2016mathematical} {and to
frameworks such as Ligra~\mbox{\cite{shun2013ligra}}}. {They do not
support many graph mining problems, and we discuss 
in Table~\mbox{\ref{tab:comparison_models}} and
Section~\mbox{\ref{sec:abstractions}}.}


Several graph mining {software} frameworks
(Peregrine~\cite{jamshidi2020peregrine} and others~\cite{teixeira2015arabesque,
dias2019fractal, mawhirter2019graphzero, yan2020prefixfpm, chen2019pangolin,
mawhirter2019automine, iyer2018asap, zhao2019kaleido, joshi2018efficient,
chen2018g, yan2017g}) were proposed. 
\iftr
Unfortunately, they focus
{exclusively} on only \emph{a few} graph pattern matching problems.
Moreover, {these frameworks usually do not have any formal guarantees
on total work~\cite{blelloch2010parallel}, and thus they do not provide time complexities and run-times
{competitive} to those of tuned parallel graph algorithms for solving
\ul{specific} mining problems}~\cite{besta2021graphminesuite}.
\fi
Overall, there is a need for a graph mining paradigm that would (1) enable
expressing many graph mining problems, and (2) offer competitive
theoretical work guarantees~\cite{blelloch2010parallel}.
\iftr
We summarize this in Table~\ref{tab:comparison_models}, which analyzes existing
programming paradigms for graph processing (a total of seven). We detail this
table in Section~\ref{sec:abstractions} -- its central message is that no
existing graph processing paradigm supports many graph mining problems or
offers competitive time complexities.
\fi

\begin{figure}[b]
\vspaceSQ{-1em}
\centering
\includegraphics[width=1.0\columnwidth]{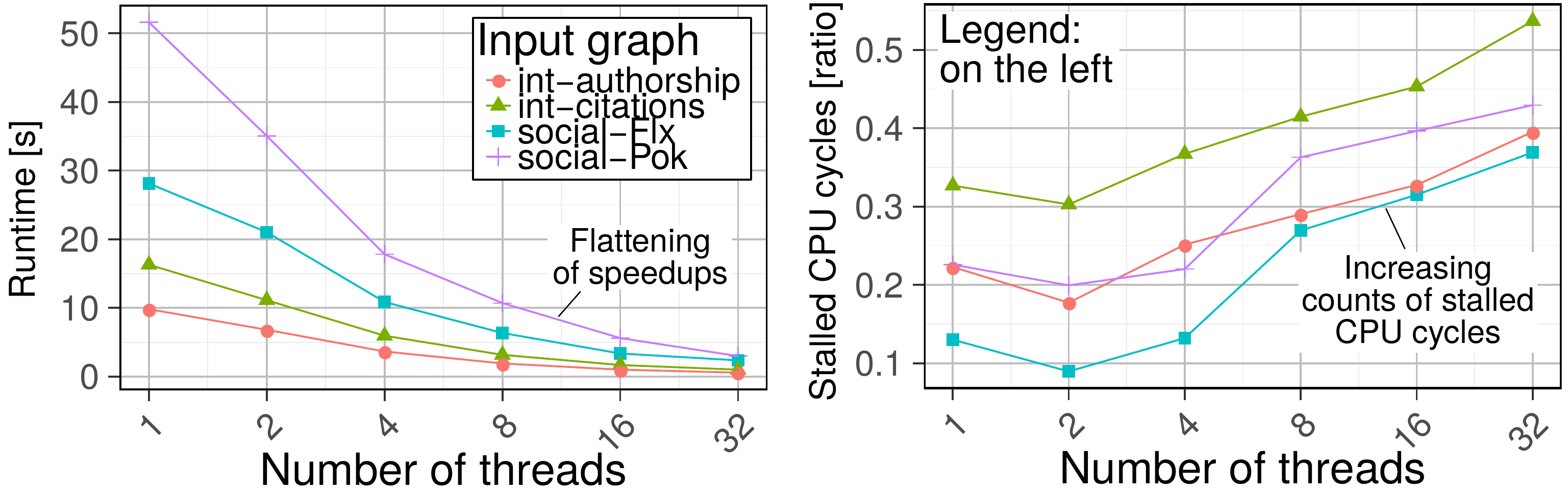}
\vspaceSQ{-1.5em}
\caption{\textbf{Runtimes and stalled CPU cycle count}, for various
numbers of parallel threads,
using the Bron-Kerbosch algorithm for listing maximal cliques
in different input graphs (Section~\ref{sec:eval}
describes our evaluation methodology).
}
\vspaceSQ{-0.75em}
\label{fig:motivation}
\end{figure}

Moreover, past works illustrated that graph mining algorithms are memory
bound~\cite{cheng2012fast, yaolocality, jamshidi2020peregrine, zhang2005genome,
eblen2012maximum}. This is because these algorithms generate and heavily
use large intermediate structures, but, similarly to algorithms
such as PageRank, they are not compute-intensive~\cite{jamshidi2020peregrine,
DBLP:conf/isaac/EppsteinLS10, yao2020locality}.
We show this in Figure~\ref{fig:motivation}: When we increase the
number of parallel threads, runtime decrease flattens out and
stalled CPU cycle count increases.
\all{For example, in the established Bron-Kerbosch algorithm for listing
maximal cliques (one of our key use-cases, details in
Section~\ref{sec:formulations}), the used storage is exponential in the worst
case~\cite{DBLP:conf/isaac/EppsteinLS10}.}
This motivates using processing-in-memory (PIM) to obtain the much needed
speedups in graph mining.
While PIM is not the only potential solution for hardware acceleration of
graph mining, we select PIM because it (1) represents one of the
most promising trends to tackle the memory
bottleneck~\mbox{\cite{ghose2019processing_pim, mutlu2019}} outperforming various other
approaches~\mbox{\cite{seshadri2017ambit}}, (2) offers well-understood
designs~\mbox{\cite{mutlu2020modern}}, and (3) 
brings very large speedups in \emph{simple} graph algorithms such as BFS or
PageRank (see more than 15 works in Table~\mbox{\ref{tab:comparison_problems}}).
Yet, graph mining algorithms are \emph{much more complex} than
PageRank, BFS, and similar: they employ deep recursion, create many
intermediate data structures with non-trivial inter-dependencies, and have high
load imbalance~\cite{DBLP:conf/isaac/EppsteinLS10, yan2017g}.
\iftr
Table~\ref{tab:comparison_problems} (shown and detailed in
Section~\ref{sec:rw}) extends Table~\ref{tab:comparison_models} by analyzing
specific hardware (HW) accelerators for graph processing, illustrating that
\fi
\emph{no existing HW design targets broad graph mining (i.e., both graph
pattern matching and graph learning), or explores PIM
techniques for accelerating broad graph mining}.

\marginparX{\Large\vspace{-16em}\colorbox{yellow}{\textbf{D}}}

%
To address all these issues, we propose a novel design that is
{high-performance} (empirically \emph{and} theoretically), applicable to
{many} graph mining problems, {and easily amenable to PIM
acceleration}. 
We first observe that {{large parts of many graph mining algorithms
can be expressed with simple set operations such as intersection $\cap$ or union
$\cup$}}, where sets contain vertices or edges. 
This drives our \textbf{set-centric programming paradigm}, in which the developer
identifies sets and set operations in a given algorithm. 
These set operations are then mapped to a small and simple yet expressive group of
instructions, offering a
rich selection of storage/performance tradeoffs. These instructions
are offloaded to PIM units. 
We call these instructions \textbf{\emph{SISA}} as they form \emph{``\textbf{S}et-centric''
\textbf{ISA}} extensions that enable a {simple interface between numerous graph
mining algorithms and PIM hardware}.
Overall, our cross-layer design consists of three key elements: a new
set-centric programming paradigm and formulations of graph algorithms
(contribution~\textbf{\#1}), set-centric ISA extensions with its
instructions, implemented set operations, and set organization
(contribution~\textbf{\#2}), and PIM acceleration
(contribution~\textbf{\#3}).
\iftr
The strength of our design comes from observing that these concepts (set
algebra/notation, set representations/algorithms, PIM) {{fit together}}
and only need minor HW extensions to provide an efficient architecture for
graph mining.
\fi


\all{This description raises the following questions about the proposed
approach: (1) is it general (applicable to many problems)? (2) is it efficient,
does it ensure strong theoretical bounds on performance? (3) }

Overall, we advocate
using set algebra as a
basis for the design of graph mining algorithms. 
Our set-centric paradigm is the first to use set operations as
fundamental general building blocks for both algorithmic formulations
\emph{and} their execution.
Using set algebra ensures that SISA set-centric algorithms are
succinct, applicable to many problems, and theoretically efficient. 
%
%
\all{When mapping set-centric
formulations to SISA code, one can use different {set representations}
(e.g., a sparse integer array or a dense bitvector), and {set operations}
such as intersection can be executed using different {set algorithms}
(e.g., merge or galloping intersection~\cite{han2018speeding}).
These choices enable flexibility as they come with different
performance/storage tradeoffs, which we analyze in detail.}
%
%
%
%
\all{Overall, distinguishing between the abstract set-algebraic algorithmic
formulations, the underlying SISA implementation, and the PIM acceleration for
SISA instructions, is key to \textbf{both programmability and high
performance}.}

For the in-memory acceleration of SISA, we investigate which types of PIM are
beneficial for which set operations. We process sets stored as
bitvectors using in-situ PIM~\cite{mutlu2020modern, ghose2019processing}, as offered in
Ambit~\cite{seshadri2017ambit, hajinazar2021simdram}, ELP2IM~\cite{xin2020elp2im},
DRISA~\cite{li2017drisa}, or ComputeDRAM~\cite{gao2019computedram}, for highest
performance and energy efficiency (``\textbf{SISA processing using memory}'' --
\textbf{SISA-PUM}). In contrast, while sets stored as sparse arrays cannot be
simply processed in situ with today's technology, they can use the high
throughput and low latency of near-memory PIM~\cite{mutlu2020modern,
ghose2019processing, loh20083d, oliveira2021damov} as offered in the 2D UPMEM
architecture~\cite{lavenier2016dna, gomez2021benchmarking} or logic layer of
3D DRAM such as Hybrid Memory
Cube (HMC)~\cite{jeddeloh2012hybrid} (``\textbf{SISA processing near memory}''
-- \textbf{SISA-PNM}).
\tr{Here, for further speedups, we also provide a small HW controller that
selects on-the-fly the best variant of a set instruction to be executed with
PIM. For example, it decides on using merge or galloping set intersection,
based on the properties of the processed graph, using our performance models.}
\cnf{For even higher performance, we provide a small HW controller that selects
the best variant of a set instruction to be executed on-the-fly.}

Overall, our results show that graph mining algorithms,  
although complex and lacking
{straightforward} parallelism, greatly benefit from PIM.
\tr{For example, Bron-Kerbosch does not offer simple vertex-level parallelism known
from algorithms such as PageRank: some vertices may belong to large cliques,
and processing such cliques results in deep recursion trees, which take a large
portion of the processing time while not offering straightforward
parallelization opportunities.} 
Our key solution is using parallelism offered by set operations and exposed
with the set-centric approach.  This solution harnesses parallelism at the
level of {bits}, DRAM subarrays, and vaults.
We show that SISA-enhanced algorithms are
theoretically efficient (contribution~\textbf{\#4}) \emph{and} empirically outperform tuned
parallel baselines (contribution~\textbf{\#5}), for example
offering more than 10$\times$ speedup for many real-world graphs over the
established Bron-Kerbosch algorithm for listing maximal
cliques~\cite{DBLP:conf/isaac/EppsteinLS10}.
Finally, for usability, we integrate SISA with the RISC-V
ISA~\cite{waterman2016design}. 
%
%
%
\all{In SISA, we ensure both programmability and high performance. On one hand, the
developer is responsible for certain \emph{simple} decisions, such as what set
representation to use. 
Here, we offer \emph{systematic guidelines} on what parts of many graph
mining algorithms benefit most from what set representations and set
operations.
SISA is also supported with theoretical analysis (which shows that SISA set-centric
algorithms are theoretically efficient) and with systematic guidelines for
developing SISA programs.}
\all{We ensure that SISA could be integrated with the
the RISC-V ISA~\cite{waterman2016design}; the small number of SISA instructions
would facilitate potential integration with other ISAs such as x86 and ARM.}
\all{We use PIM in SISA, but the generality of set notation means
that SISA could be extended to other forms of acceleration.}
\all{To the best of our knowledge, \emph{SISA offers the first hardware acceleration
for \textbf{both} graph pattern matching \textbf{and} graph learning problems.}}
\all{\maciej{finish, integrate} 
To maximize SISA's usability, we discuss a complete cross-layer design with the
set-centric ISA in its core, \emph{relying on well-established concepts
in algorithm and HW developments}.
}
\all{For manageability and to facilitate a potential real implementation, we
propose a layered design. Specifically, SISA comes with set-centric algorithmic
formulations (to facilitate algorithm design), a thin layer of software
abstractions and C-style wrappers (for programmability), high-level ISA
instructions (independent of the underlying HW), low-level ISA instructions
(directly accessing the underlying HW), and example SISA HW that manages the
execution of instructions on different accelerators, ensuring area efficiency,
high performance, and energy efficiency.}
\all{We also carefully prescribe set representations and set algorithms for
high-performance execution with in- and near-memory processing. Here, we first
illustrate that many graph mining algorithms offer different parallelization
and load balancing characteristics than low-complexity graph algorithms such as
PageRank. Simultaneously, we show that graph mining algorithms \emph{do indeed}
have potential for massive parallelization and memory acceleration, and that
\emph{the proposed set-centric formulation and execution expose this
potential}.}
%
%
%
%
%
\all{To illustrate this, we conducted an extensive analysis of related work,
considering both graph programming paradigms and hardware accelerators. The
outcomes are summarized in
Tables~\ref{tab:comparison_models}-\ref{tab:comparison_problems} and
Section~\ref{sec:abstractions}, which show that (1) SISA is the only design
that targets general graph mining and (2) unlike the majority of specific
accelerators, it comes with a detailed cross-level design, from the programming
paradigm down to the HW execution.}

\all{For this, we carefully combine processing different set representations
(used for storing graphs) on different memory acceleration designs. We show
minimal HW extensions for managing the processing of SISA sets and the
execution of SISA set instructions on these different memory accelerators.}

\iftr

\vspace{0.5em}
\noindent
To summarize, we contribute the following:

%
\begin{itemize}[leftmargin=1em,noitemsep]
\item We propose a \textbf{set-centric programming approach} for a wide
selection of graph mining problems,
in which one exposes and exploits sets and set operations in graph algorithms.
%
%
%
%
\item We develop SISA, a \textbf{set-centric ISA} interface between hardware and
software in graph mining. 
We describe the \textbf{syntax}, \textbf{semantics}, and \textbf{encoding of
SISA}.
\item We provide a careful graph \textbf{data layout} based on dense bit
vectors and sparse integer arrays, a \textbf{hardware implementation of SISA}
that harnesses {in-} and {near-memory processing}, and associated
\textbf{performance models} that enable automatic selection of fastest set
instructions.
%
%
%
%
\item We develop \textbf{programming guidelines} for SISA, covering details
such as selecting the most beneficial SISA instructions for different set
operations in graph algorithms.
%
%
\item We provide an \textbf{extensive theoretical analysis} of SISA, analyzing
the performance of graph algorithms and single set operations, for different
set representations. This analysis shows that SISA offers competitive time
complexities.
%
%
%
%
\item We use cycle-based simulations to illustrate \emph{performance advantages
of SISA over hand-tuned baselines}. 
%
%
%
\end{itemize}

\fi

\ignore{

\noindent
In this work, we contribute the following:

\begin{itemize}[leftmargin=1em] \setlength\itemsep{-0.2em}
\item We propose SISA, an Instruction Set Architecture dedicated for graph
processing that uses sets and set operations as first-class citizens to enable
a clean and expressive interface between hardware and software in the world of
graph analytics. \emph{SISA can be used to express and develop as many as at
least 33 (thirty-three) different graph algorithms.}
\item We describe in detail the syntax and semantics of SISA. To our
knowledge, SISA is the first ISA designed for graph processing using set
operators.
\item We propose a hardware implementation of SISA, orchestrated by the
dedicated Set Management Unit module. The implementation ensures high
performance by \emph{maximizing the utilization of memory-level parallelism}
and it is based on the base-and-bound organization of sets, which enables
\emph{hardware support for fully dynamic sets while minimizing changes
required to hardware.}
%
%
\item We conduct evaluation of the simulated SISA design, illustrating
\emph{performance advantages over hand-tuned baselines}. \maciej{add how much}
\item We broadly discuss the integration of SISA with a large number of
software schemes (such as the vertex-centric
approach~\cite{malewicz2010pregel}) and hardware accelerators (such as GPUs or
Tesseract~\cite{ahn2016scalable}) in the world of graph processing.
\end{itemize}

}

%% file: background.tex
\vspaceSQ{-0.5em}
\section{Notation and Background}
\label{sec:back-operations}


\iftr
We first describe background and notation, see Table~\ref{tab:symbols}.
\fi

\textbf{Graphs }
We model an undirected graph $G$ as a tuple $(V,E)$; $V$ and $E \subseteq V
\times V$ are sets of vertices and edges; $|V|=n$, $|E|=m$.  Vertices are
modeled with integers ($V = \{1, ..., n\}$).
\iftr
$\overrightarrow{G}=(V, \overrightarrow{E})$ is a directed graph;
$\overrightarrow{E}$ contains arcs.
\fi
\iftr
$N(v)$ and $N^+(v)$ denote the neighbors and the out-neighbors of $v \in V$;
$d$ and $d(v)$ denote $G$'s maximum degree and a degree of $v$.
\fi
\hl{In some cases, we consider \emph{labeled} graphs \mbox{$G = (V, E, L)$};
\mbox{$L$} is a labeling function that maps a vertex or an edge to
a label.}

\marginpar{\Large\vspace{-2em}\colorbox{yellow}{\textbf{L}}}

\textbf{Set Representations }
\iftr
A concept used heavily in SISA is the {set representation} and its
{sparsity}. 
\fi
\tr{
Figure~\ref{fig:set_reps} shows example considered
representations.}
Consider a set of $k$ vertices $S = \{v_1, ..., v_k\} \subseteq V$ 
(we focus on vertex sets, but SISA also works with edges). One can
represent $S$ as a simple contiguous \textbf{sparse array (SA)} with integers
from $S$ (``sparse'' means that only non-zero elements are explicitly
stored). SA's size is $W |S|$ [bits] where $W$ is the memory word size (we
assume that the maximum vertex ID fits in one word). One can also represent $S$
with a \textbf{dense bitvector (DB)} of size $n$ [bits]: the $i$-th set bit
indicates that a vertex $i \in S$ (``dense'' means that all zero bits are
explicitly stored).

\iftr
\textbf{Set Operations }
SISA uses all basic {set operations}: intersection $A \cap B$, union $A
\cup B$, difference $A \setminus B$, cardinality $|A|$, and membership $\in A$.
\fi
\iftr
$A$ and $B$ usually contain vertices and sometimes edges.
\fi
\iftr
We use different algorithms to implement these operations (described
later in the paper).
\fi

\iftr
\textbf{Architecture Concepts }
We outline the used architectural concepts in Section~\ref{sec:intro}; more
details are in Section~\ref{sec:sisa-implementation} and~\ref{sec:eval}.
The architecture related symbols are listed in Table~\ref{tab:symbols}
(bottom).
%
%
\fi

\iftr

\begin{table}[t]
\vspaceSQ{-1.5em}
\setlength{\tabcolsep}{1.2pt}
\ifsq
\renewcommand{\arraystretch}{0.7}
\else
\renewcommand{\arraystretch}{1.2}
\fi
%
\centering
\footnotesize
\begin{tabular}{lll@{}}
\toprule
\multirow{7}{*}{\begin{turn}{90}\textbf{Graphs}\end{turn}} & $G=(V,E)$ & An undirected graph; $V$ and $E$ are sets of vertices and edges.\\
\iftr
 & $\overrightarrow{G}=(V, \overrightarrow{E})$ & A directed graph; $V$ and $\overrightarrow{E}$ are sets of vertices and directed edges.\\
 & $G[V']$ & A subgraph (of a graph~$G$) induced on a vertex set~$V'$.\\
\fi
 & $n, m$ & The numbers of vertices and edges in $G$ $\left(|V| = n, |E| = m\right)$.\\
 & $N(v), N^+(v)$ & The neighbors and the out-neighbors of a vertex~$v$.\\
 & $d, d(v)$ & The maximum degree of $G$, the degree of $v \in V$.\\
 & $c$ & The graph degeneracy (a property used in theoretical analysis).\\
 & SA, DB & sparse array, dense bitvector \\
%
%
\ifall\m{disa}
$C_{1,A}, ..., C_{A,k}$ & A collection of $k$ chunks that constitute the representation of a set $A$ \\
\fi
\midrule
\multirow{6}{*}{\begin{turn}{90}\textbf{Architecture}\end{turn}} & $l_M$, $b_M$ & The latency and bandwidth of accessing DRAM. \\
 & $b_L$ & The bandwidth of the interconnect (e.g., QPI) between cores. \\
 & $l_I$ & The latency of one bulk bitwise operation run with in-situ PIM.\\
 & $q$ & \#rows that can be processed in parallel (e.g., in a DRAM bank).
\vspace{0.1em}\\
 & $R$ & The size [bits] of a single DRAM row. \\
 & SCU, SM  & SISA Controller Unit, Set Metadata\\
\ifall\m{disa}
SMU, SLB, ST & Set Management Unit, Set Lookaside Buffer, Set Table \\
MSHR, ASID & Miss Status Holding Registers, Address Space ID \\
\fi
\bottomrule
\end{tabular}
\vspaceSQ{-0.5em}
\caption{The most important symbols and acronyms.}
\vspaceSQ{-2.5em}
\label{tab:symbols}
\end{table}

\fi

%% file: overview.tex
\vspaceSQ{-0.5em}
\section{Overview \& Cross-Layer Design}
\label{sec:overview}

\enlargeSQ


We now overview SISA's cross-level design, see Figure~\ref{fig:overview}.
\tr{SISA's cross-layer design (see Figure~\ref{fig:overview}) consists of three
key elements: (a) set-centric formulations of graph algorithms, (b) the actual
set-centric ISA with its instructions, implemented set operations, set
organization, and a thin software layer, and (c) an example HW implementation.
We support SISA with programming guidelines and a theoretical analysis.}

\ifsq
\begin{figure}[b]
\else
\begin{figure*}[h]
\fi
\vspaceSQ{-1.5em}
\centering
\ifsq
\includegraphics[width=1.0\columnwidth]{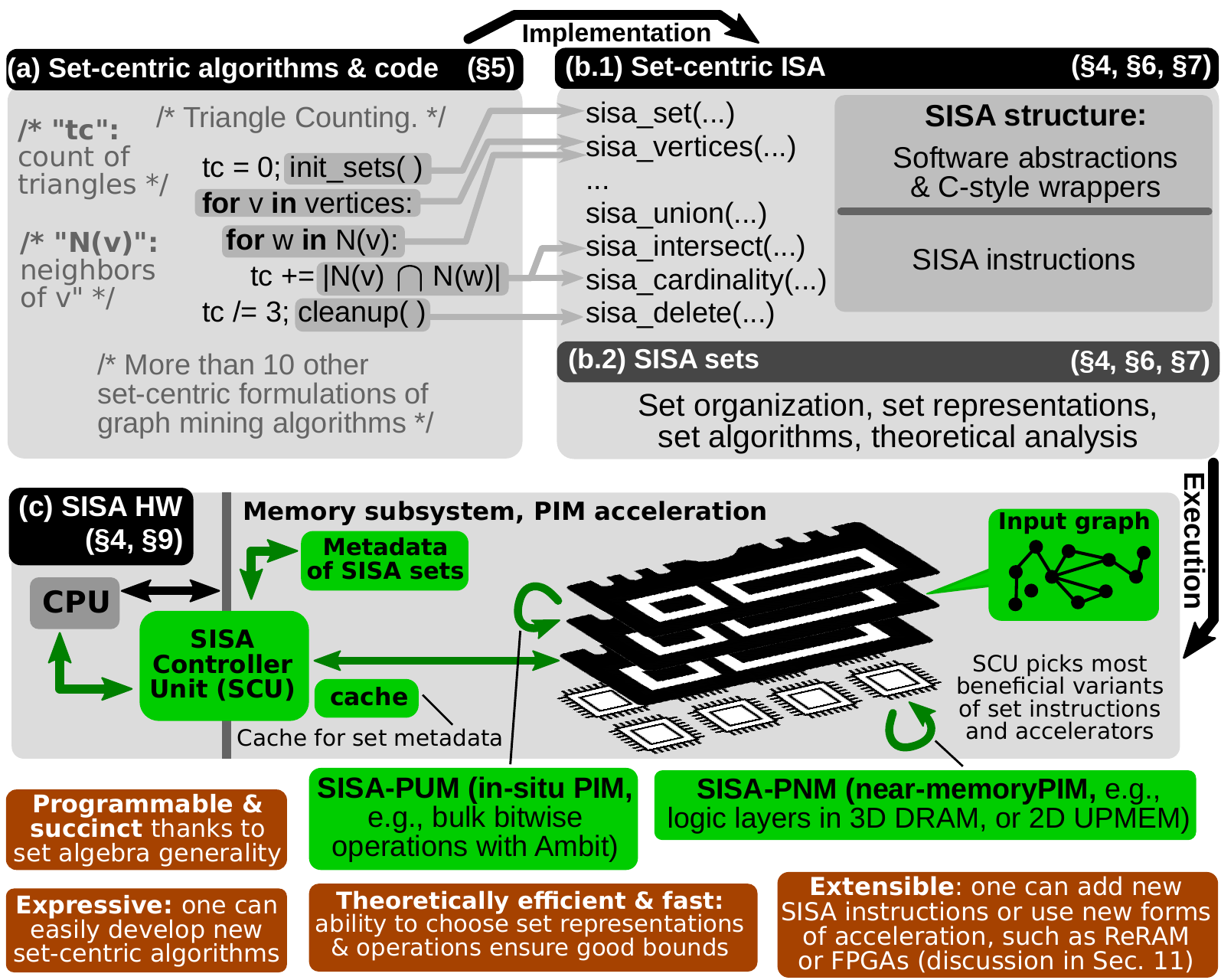}
\else
\includegraphics[width=0.65\textwidth]{sisa-overview-details___simplified___2-1___after-Juan.pdf}
\fi
\vspaceSQ{-2em}
\caption{The overview of SISA with a summary of new introduced
architecture and graph representation elements (\ul{green}) and
advantages (\ul{brown}).}
\label{fig:overview}
\vspaceSQ{-0.5em}
\ifsq
\end{figure}
\else
\end{figure*}
\fi




\textbf{(a) Set-Centric Formulations [Section~\ref{sec:formulations} \& \ref{sec:sisa-programming}] }
SISA relies on set-centric formulations of algorithms in graph mining.
While some algorithms (e.g., Bron-Kerbosch~\cite{DBLP:conf/isaac/EppsteinLS10})
by default use rich set notation, many others, such as $k$-clique listing by
Danisch et al.~\cite{danisch2018listing}, do not. In such cases, {we develop
such formulations}. Details on deriving set-centric formulations are in
Section~\ref{sec:sisa-programming}; the key common step is to express two
nested loops, commonly used to identify connections between two sets of
vertices, with a single intersection of these sets.

A set can be represented in different ways, and a set operation can
be executed using different set algorithms.
A set-centric formulation hides these details, focusing on \emph{what} a given
graph algorithm does, and not \emph{how} it is done.


\textbf{(b.1) Set-Centric ISA (Instructions) [Section~\ref{sec:sisa-syntax-semantics}] }
Our {ISA} extension implements set operations.
\iftrLOW
\else
These instructions support all variants of operations, for example
there is an instruction for both merge
and galloping set intersection (details in
Section~\ref{sec:sisa-syntax-semantics}).
We also provide a thin {software layer}: {iterators
over sets} and C-style {wrappers} for SISA instructions. 
For programmability and performance, many SISA instructions
automatize selecting the best set operation variant on-the-fly.
\fi


\textbf{(b.2) Set-Centric ISA (Organization of Sets) [Section~\ref{sec:sisa-syntax-semantics}] }
%
%
\tr{SISA instructions process \emph{sets of vertices} and \emph{edges}. Thus, a
core SISA part is high-performance set organization.}
%
%
%
We represent sets as DBs or SAs. The
former are processed by bulk bitwise in-situ PIM, harnessing huge internal DRAM
bandwidth (SISA-PUM). The latter use
near-memory PIM, for example DRAM cores in the UPMEM architecture, or logic
layers in 3D stacked DRAM, harnessing the large through-silicon via (TSV)
bandwidth (SISA-PNM).
\tr{SISA also facilitates selecting the most beneficial set representations.}
\tr{To maximize performance, SISA stores the largest neighborhoods
as dense bitvectors (in-situ PIM outperforms 
logic layers) but staying within the user-specified storage budget.}


\enlargeSQ


\textbf{(c) HW Implementation Details [Section~\ref{sec:sisa-implementation}]}
For maximum programmability and performance, we use hardware to automatically 
decide between SISA-PUM and SISA-PNM,
or a set algorithm variant (merge vs.~ galloping). For
this, we use a dedicated unit called the {SISA Controller Unit (SCU)}.
\tr{The main task of SCU is to appropriately \emph{schedule the
execution of SISA set instructions on different memory accelerators}
such that, depending on how two given sets are represented, 
the most beneficial variant of a given set operation is used.}
\iftr
The SCU can be an additional unit, or it can also be emulated by a process
occupying a dedicated core in the logic layer, to avoid any HW modifications.
The SCU receives SISA instructions from the CPU, and it appropriately schedules
their execution on SISA-PNM and SISA-PUM.  Two bitvectors are always processed
with SISA-PUM, while in other scenarios SCU uses SISA-PNM. The SCU
can also select the most advantageous set algorithm. For example, whenever two
sets have similar sizes, it is better to intersect them using a merge-based
intersection, in which input sets are streamed and they can harness high
sequential bandwidth.
\fi
\tr{For concreteness, in
Section~\ref{sec:sisa-implementation}, we pick Ambit~\cite{seshadri2017ambit}
as the implementation of SISA-PUM within a DRAM die, and logic layers for SISA-PNM, but
other designs can also be used.}

\iftr
The SCU maintains {set metadata} (SM) using a dedicated in-memory SM
structure. SM contains mappings between logical set IDs and
set addresses, and the type of the representation as well as the cardinality of a given set.
This information is used to guide SCU decisions.
Finally, the SCU has a small scratchpad, the {Set Metadata Buffer (SMB)},
to cache metadata.
\fi


%% file: abstractions.tex
\vspaceSQ{-0.5em}
\iftr
\section{\hspace{-0.2em}Sets for Simple \& Provably Fast Graph Mining}
\fi
\label{sec:abstractions}

\all{\emph{Graph mining} can be seen as a subset of a broader computational
domain of \emph{graph analytics}.}

\marginparX{\Large\vspace{3em}\colorbox{yellow}{\textbf{D}}}

The set-centric approach is
superior to other graph programming paradigms in that (1) it
supports {many} graph mining problems {and} (2) it enables algorithms 
with {competitive theoretical bounds on performance} (we discuss~(2)
in Section~\mbox{\ref{sec:theory-s}};
this is often a key to 
low runtimes~\cite{khan2016vertex, dhulipala2018theoretically}).
The analysis results for~(1) are in Table~\ref{tab:comparison_models}.


To illustrate the above points, we first extensively examined the related
literature to identify representative \textbf{graph mining problems} and
important \textbf{graph processing paradigms}~\cite{chakrabarti2006graph, washio2003state, lee2010survey, rehman2012graph,
gallagher2006matching, ramraj2015frequent, jiang2013survey,
aggarwal2010managing, tang2010graph, leicht2006vertex, liben2007link,
ribeiro2019survey, lu2011link, al2011survey}.
For the former, we pick four problems from both graph pattern matching and
graph learning areas (maximal clique listing~\cite{bron1973algorithm},
$k$-clique listing~\cite{chiba1985arboricity}, dense subgraph
discovery~\cite{lee2010survey, gibson2005discovering}, subgraph
isomorphism~\cite{ullmann1976algorithm}, vertex
similarity~\cite{leicht2006vertex, robinson2013graph}, link
prediction~\cite{liben2007link, lu2011link, al2006link, taskar2004link}, graph
clustering~\cite{schaeffer2007graph, jarvis1973clustering}, verification of
prediction accuracy~\cite{wang2014robustness}). For fairness, we also consider
four popular ``low-complexity'' problems, targeted by many past works
(triangle counting, BFS, connected components, and PageRank).
For the latter, we first select \emph{vertex-centric}~\cite{malewicz2010pregel}
and \emph{edge-centric}~\cite{roy2013x}, two established graph processing
paradigms implemented in the Pregel and X-Stream systems.  Second, we pick
\emph{vertex/edge array maps} from Ligra~\cite{shun2013ligra}, an approach for
developing graph algorithms based on transforming arrays of vertices or edges
according to a specified map.  Third, we consider \emph{GraphBLAS} and its
linear algebraic approach~\cite{kepner2016mathematical}, where graph algorithms
are expressed with linear algebra building blocks such as matrix-vector
products. Moreover, we consider \emph{pattern matching
frameworks}~\cite{gallagher2006matching} that usually employ some form of
exploring neighbors of each vertex, combined with user-specified filtering, to
search for specified graph patterns. For completeness, we also
consider recent attempts at solving graph problems with novel deep learning~\cite{ben2019modular}
paradigms such as \emph{graph neural
networks (GNN)}~\cite{wu2020comprehensive, besta2021motif} and others~\cite{gianinazzi2021learning}, 
as well as
\emph{joins} and principles from relational databases and the associated
algebra~\cite{zhao2017all}.


The analysis results are in Table~\ref{tab:comparison_models}. Overall, no
single paradigm, except for the set-centric approach, enables
efficient graph mining algorithms for the considered problems. Some paradigms,
such as the vertex-centric or the edge-centric model, do not focus on such
problems at all. Other paradigms, for example array maps or GNNs,
address only certain problems. Finally, graph pattern matching or RDBMS can
solve different graph mining problems, but they do not offer formal guarantees,
as indicated by past work.

\all{As an example, consider the vertex-centric model. Here, by agreeing to be
able to only access the neighbors of each vertex (from each vertex
perspective), one loses the global view of the graph. This works well for
simple algorithms such as PageRank, but when trying 

it does \emph{not} start with a limiting abstraction that could easily impose
theoretical overheads. Instead, it takes as input a tuned specific algorithm
(that by its design has good performance bounds), and only then it identifies
set operations that will be executed with memory hardware.}

%% file: formulations.tex
\section{Set-Centric Graph Algorithms}
\label{sec:formulations}



We now present set-centric formulations of graph mining algorithms. 
The used set operations are in Table~\ref{tab:set-forms}. 
%
%
%
\tr{We loosely categorize the considered problems and algorithms into graph
pattern matching (\cref{sec:formulations-mining}) and graph learning
(\cref{sec:formulations-learning}), based on an analysis of graph related
surveys~\cite{pingali2011tao, doekemeijer2014survey, mccune2015thinking,
quinn1984parallel, shi2018graph, beamer2015gap, cormen2009introduction,
teixeira2015arabesque, schaeffer2007graph, jiang2013survey,
besta2021graphminesuite}.}

\ifconf
\begin{table}[t]
\else
\begin{table*}[t]
\fi
%
\ifsq\renewcommand{\arraystretch}{0.6}\fi
\centering
\ifsq\scriptsize
\else
\footnotesize
\fi
\begin{tabular}{lll}
\toprule
\textbf{Problem} &
\textbf{Algorithm} &
\textbf{Used set operations} \\
\midrule
\iftr
Maximal clique list. & Bron-Kerbosch~\cite{DBLP:conf/isaac/EppsteinLS10} & $A \cup B$, $A \cap B$, $A \setminus B$, $A \cup \{b\}$, $A \setminus \{b\}$ \\
\fi
$k$-clique listing & \makecell[l]{Danisch et al.~\cite{danisch2018listing} \textbf{+ [This work]}} & $A \cap B$ \\
4-clique counting & \textbf{[This work]} & $A \cap B$, $|A \cap B|$ \\
Triangle counting & [well-known] & $|A \cap B|$ \\
$k$-clique-star listing & Jabbour et al.~\cite{jabbour2018pushing} & $A \cap B$, $A \cup B$ \\
$k$-clique-star listing & \textbf{[This work]} & $A \cap B$ \\
\iftr
Subgraph isomorphism & \textbf{[This work]} & $A \cap B$, $|A \cap B|$, $A \cup B$, $A \setminus B$, $A \cup \{b\}$, $A \setminus \{b\}$ \\ 
\fi
\midrule
Vertex similarity & Jaccard coeff., others~\cite{besta2020communication, robinson2013graph} & $|A \cap B|$, $|A \cup B|$ \\
\iftr
Clustering & Jarvis-Patrick~\cite{jarvis1973clustering} & $|A \cap B|$, $|A \cup B|$, $A \cup \{b\}$ \\
\fi
Link prediction (LP) & Jaccard coeff., others~\cite{robinson2013graph} & $|A \cap B|$, $|A \cup B|$ \\
LP accuracy testing & Wang et al.~\cite{wang2014robustness} & $A \setminus B$, $|A \cap B|$ \\
Approx.~degeneracy & Besta et al.~\cite{besta2020high} & $A \setminus B$ \\
\bottomrule
\end{tabular}
%
%
\caption{
\textbf{Overview of set-centric graph algorithms}.
In maximal clique listing, subgraph isomorphism, and clustering, one also uses
variants of union and difference where one set is always a single-element set
(i.e., $A \cup \{b\}$, $A \setminus \{b\}$).
{Bolded text indicates algorithms with set-centric formulations derived in this work.}
}
\vspaceSQ{-1em}
\label{tab:set-forms}
\ifconf
\end{table}
\else
\end{table*}
\fi

\textbf{Notes on Listings}
%
%
Set operations accelerated by SISA
are marked with the ~\tikzmarkin[set fill
color=vlgray, set border color=white, above offset=0.27, right offset=2.2em, below
offset=-0.1]{mot1}\textcolor{black}{gray}\tikzmarkend{mot1}~ color.
``\texttt{[in par]}'' indicates that in a given loop one can issue set
operations in parallel. We ensure that the parallelization does not involve
conflicting memory accesses.
\tr{We use ``$\cup$='', ``$\cap$='', ``$\setminus$='' to indicate that a set
operation mutates its first set argument.}
We now focus on {formulations} and we discuss set
representations, instructions, and parallelization later.
For clarity, we exclude unrelated optimizations from the listings. 


\iftr
\textbf{Does SISA Execute All Set Operations?}
SISA is used for executing set operations \emph{that benefit from
hardware acceleration}, but one can find certain counter-examples, i.e., it may
be more beneficial to use standard implementations of set operations, or to
exclude set notation and set-focused data structures completely.
For example, appending a vertex $v$ to a list $L$, which can be expressed as $L
\cup \{v\}$, does not necessarily benefit from memory acceleration, if $L$ is
implemented as a linked list. We provide examples of algorithms not necessarily benefiting from
the memory acceleration (as offered in SISA) later in this section.
%
%
We discovered that this is often the case with ``low-complexity''
algorithms such as Boruvka's algorithm for solving the Minimum Spanning Tree
problem.
\fi

\tr{
\textbf{Time Complexity of SISA Algorithms}
Complexities of set-centric algorithms heavily depend on many factors, such as
the used set representations.
SISA enables manipulating these
factors to ensure advantageous time complexities (Section~\ref{sec:sisa-syntax-semantics}).
We provide a theoretical analysis in Section~\ref{sec:theory-s}.
}

\iftr

\subsection{Graph Pattern Matching}
\label{sec:formulations-mining}

We first consider graph pattern matching, an important class of problems
where one searches for specific subgraphs.

\fi




\iftr
\subsubsection{Triangle Counting}

%
\ignore{ \maciej{???: We omit well-known optimizations such as ordering by
degrees~\cite{beamer2015gap} to improve listing's clarity; they do not change
the algorithmic insights related to SISA (we use them in our SISA
implementation).  }}
In the extensively researched triangle counting (TC)
problem~\cite{schank2007algorithmic, shun2015multicore, beamer2015gap,
al2018triangle}, one counts the total number of 
3-cycles $tc$. TC is used to compute clustering
coefficients~\cite{al2018triangle}.
In the set-centric formulation in Algorithm~\ref{lst:tc}, for each vertex~$v$,
one computes the cardinalities of the intersections of $N(v)$, the set of
neighbors of $v$, with the sets of the neighbors of each neighbor of~$v$
(Lines~\ref{ln:tc-main-1}-\ref{ln:tc-main-2}).  Set intersection constitutes up
to $\approx$94\% of the TC runtime~\cite{han2018speeding}.

\ignore{
Consider the well-known Node Iterator~\cite{schank2007algorithmic} algorithm
for Triangle Counting. Here, one counts the number of triangles that each
  vertex $v$ belongs to. A triangle occurs if there exist edges $\{v,w\},
  \{w,u\}, \{u,v\}$.  Node Iterator is a simple algorithm that first orders the
  vertices according to the total order $\succ$ based on vertex degrees and
  breaking ties arbitrarily; $\succ$ prevents counting the same triangle
  multiple times.
The comparison of the set-based and the traditional variant is in
Table~\ref{tab:tc-ni}. We used~ \tikzmarkin[set fill color=vlgray, set border
color=white, above offset=0.27, below offset=-0.1]{mot}\textcolor{black}{gray
background}\tikzmarkend{mot} ~and encircled numbers \encircle{1} --
\encircle{2} to indicate the corresponding parts of the algorithm in different
variants.
Two \textbf{\texttt{if}} statements marked with \encircle{1} -- \encircle{2}
become two predicates and set intersections in the set-based variant.
Specifically, to enable using the $\succ$ order in the set-based variant, one
intersects sets of neighbors with an additional set $H(v) \equiv \{w \in N_v:\
w \succ v\}$ that filters out neighbors that do not satisfy the $\succ$ order.
It is implementation-specific whether $H(v)$ is precomputed or constructed
on-the-fly.
%
%
Using such sets abstracts away any conditional statements, providing more
flexibility in compilation and implementation. It also enables new
opportunities for speedups by shifting the focus in performance engineering
towards developing fast and power-efficient set operations.
}

\ignore{

\begin{lstlisting}[float=h!,belowskip=-1em,aboveskip=-0.5em,abovecaptionskip=-0.5em,label=lst:tc,caption=Triangle Counting (Node Iterator)~\cite{schank2007algorithmic}.]
|\vspace{0.5em}|/* |\textbf{Input:}| A graph $G$. |\textbf{Output:}| Triangle count $tc \in \mathbb{N}$. */
//Derive an array $r$ so that $\forall_{v,w \in V}$ $r[v] < r[w] \Rightarrow d(v) \le d(w)$:
$r$ = /* Sort vertices by degrees, |break| ties. */
//Sets $M(v) \subset V$ prevent counting a triangle three times: 
|\vspace{0.5em}|for $v \in V$: $M(v)$ = $\{w \in N(v)\ |\ r[v] < r[w]\}$
|\label{ln:tc-main-1}|$tc$ = $0$; //Init $tc$; for all neighbor pairs, increase $tc$:
for $v \in V$ do: 
|\vspace{0.25em}|  for $w \in M(v)$ do:
|\label{ln:tc:s}||\label{ln:tc-main-2}|    $tc$ += |\hspace{-0.5em}||\highlight{ |$\mid M(v) \cap M(w)\mid$| }|
\end{lstlisting}

\begin{lstlisting}[float=h!,belowskip=-1em,aboveskip=-0.5em,abovecaptionskip=-0.5em,label=lst:tc,caption=Triangle Counting (Node Iterator)~\cite{schank2007algorithmic}.]
|\vspace{0.5em}|/* |\textbf{Input:}| A graph $G$. |\textbf{Output:}| Triangle count $tc \in \mathbb{N}$. */
//Derive an array $r$ so that $\forall_{v,w \in V}$ $r[v] < r[w] \Rightarrow d(v) \le d(w)$:
$r$ = /* Sort vertices by degrees, |break| ties. */
//Sets $M(v) \subset V$ prevent counting a triangle three times: 
|\vspace{0.5em}|for |\hspace{-0.5em}||\hfsetfillcolor{vlgray}\highlight{ |$v \in V$| }| [in par]: $M(v)$ = $\{w \in N(v)\ |\ r[v] < r[w]\}$
|\label{ln:tc-main-1}|$tc$ = $0$; //Init $tc$; for all neighbor pairs, increase $tc$:
|\vspace{0.5em}||\label{ln:tc:s}||\label{ln:tc-main-2}|for |\hspace{-0.5em}||\hfsetfillcolor{vlgray}\highlight{ |$v \in V$| }| [in par] do: 
  for |\hspace{-0.5em}||\hfsetfillcolor{vlgray}\highlight{ |$w \in M(v)$| }| [in par] do: $tc$ += |\hspace{-0.5em}||\hfsetfillcolor{vlgray}\highlight{ |$\mid M(v) \cap M(w)\mid$| }|
\end{lstlisting}

\begin{lstlisting}[float=h!,belowskip=-1em,aboveskip=-0.5em,abovecaptionskip=-0.5em,label=lst:tc,caption=Triangle Counting (Node Iterator)~\cite{schank2007algorithmic}.]
|\vspace{0.5em}|/* |\textbf{Input:}| A graph $G$. |\textbf{Output:}| Triangle count $tc \in \mathbb{N}$. */
//Derive an array $r$ so that $\forall_{v,w \in V}$ $r[v] < r[w] \Rightarrow d(v) \le d(w)$:
$r$ = /* Sort vertices by degrees, |break| ties. */
//Sets $M(v) \subset V$ prevent counting a triangle three times: 
|\vspace{0.5em}||\hspace{-0.5em}||\hfsetfillcolor{vlgray}\highlight{\textbf{for} |$v \in V$ [in par]| }| do: $M(v)$ = $\{w \in N(v)\ |\ r[v] < r[w]\}$
|\label{ln:tc-main-1}|$tc$ = $0$; //Init $tc$; for all neighbor pairs, increase $tc$:
|\vspace{0.5em}||\label{ln:tc:s}||\label{ln:tc-main-2}||\hspace{-0.5em}||\hfsetfillcolor{vlgray}\highlight{\textbf{for} |$v \in V$ [in par]| }| do: 
  |\hspace{-0.5em}||\hfsetfillcolor{vlgray}\highlight{\textbf{for} |$w \in M(v)$ [in par]| }| do: $tc$ += |\hspace{-0.5em}||\hfsetfillcolor{vlgray}\highlight{ |$\mid M(v) \cap M(w)\mid$| }|
\end{lstlisting}

}

\begin{lstlisting}[float=h, aboveskip=-0.3em,belowskip=-1em,abovecaptionskip=0em,label=lst:tc,caption=Triangle Counting (Node Iterator).]
|\vspace{0.5em}|/* |\textbf{Input:}| A graph $G$. |\textbf{Output:}| Triangle count $tc \in \mathbb{N}$. */
// Different optimizations are excluded for clarity (they
// minimize the number of times that each triangle is counted.
|\label{ln:tc-main-1}|$tc$ = $0$; //Init $tc$; for all neighbor pairs, increase $tc$:
|\label{ln:tc:s}|$v \in V$ [in par] do:
|\vspace{0.25em}|  |\label{ln:tc-main-2}|for $w \in N(v)$ [in par] do: $tc$ += |\hlLR{9em}{ $\mid N(v) \cap N(w)\mid$ }|
$tc$ /= 6 //With optimizations excluded from the listing
//for more clarity, this division is not needed.
\end{lstlisting}


\fi

\marginparX{\Large\vspace{2em}\colorbox{yellow}{\textbf{D}}}

\iftr
\subsubsection{Maximal Cliques Listing}

\fi
\iftr
A clique is a fully-connected subgraph of an input graph; a maximal clique is a
clique not contained in a larger clique. Finding all maximal cliques, an
NP-Hard problem, has many applications in social network
analysis~\cite{wasserman1994social}, bioinformatics~\cite{day1986computational,
spirin2003protein}, and computational chemistry~\cite{rhodes2003clip}.
Algorithm~\ref{lst:bk} contains the recursive backtracking Bron-Kerbosch
algorithm~\cite{bron1973algorithm} with pivoting and degeneracy
optimizations~\cite{cazals2008note, DBLP:conf/isaac/EppsteinLS10,
DBLP:journals/tcs/TomitaTT06}, an established and commonly used scheme for
finding maximal cliques (deriving the degeneracy ordering is itself an
important graph problem and SISA also provides a dedicated set-centric
formulation). 
\fi
\iftr
BK heavily uses different set operations. 
\fi
The main recursive function \texttt{BKPivot} (Line~4) has three arguments that
are dynamic sets containing vertices. $R$ is a partially constructed,
non-maximal clique~$c$, $P$ are candidate vertices that {may} belong to $c$ but
are yet to be tried, and $X$ are vertices that definitely do {not} belong to
$c$. The algorithm recursively calls \texttt{BKPivot} for each new candidate
vertex, checks if this gives a clique, and updates accordingly $P$ and $X$.
Some optimizations need more set operations, but they reduce the search space
of potential cliques~\cite{DBLP:journals/tcs/TomitaTT06}. For example, the set
of candidates (for extending a clique~$c$) is $P \setminus N(u)$ instead of
$P$, where $u \in P \cup X$. 
\tr{Second, the outermost loop iterates over $V$ using the degeneracy order and
uses it to prune $P$ and $X$, involving two additional set intersections in
each iteration.}
Overall, BK is non-trivial, with many different set
operations, \emph{including non anti-monotonic ones such as union}. Thus, it
shows SISA's ability to accelerate complex algorithms.

\marginparX{\Large\vspace{-2em}\colorbox{yellow}{\textbf{B}}}

\iftr
\begin{lstlisting}[float=h,aboveskip=0em,belowskip=0em,abovecaptionskip=0em,label=lst:bk,caption=Maximal Clique Listing (Bron-Kerbosch)~\cite{bron1973algorithm, cazals2008note}.]
/* |\textbf{Input:} A graph $G$. \textbf{Output:} Maximal clique $R$ ($R \subseteq V$).|*/
$P$ = $V$; $R$ = $\emptyset$; $X$ = $\emptyset$; //Init sets appropriately.
for $v \in V$ [in par] do: 
  BKPivot($\{v\}$, $P$, $X$)
|\vspace{0.2em}|function BKPivot($R$, $P$, $X$):
|\vspace{0.2em}|  if |\hlLR{3em}{ $\vert P\vert$ }| == $0$ and |\hlLR{3em}{ $\vert X\vert$ }| == $0$:
    return $R$ //We found a maximal clique
|\vspace{0.25em}|  $u$ = /* Choose a pivot vertex from |\hlLR{3.8em}{ $P \cup X$ }| */ 
|\vspace{0.5em}|  for $v \in $ |\hlLR{5.5em}{ $P \setminus N(u)$ }| do: 
|\vspace{0.5em}|    BKPivot( |\hlLR{5em}{ $R \cup \{v\}$ }|, |\hlLR{5.5em}{ $P \cap N(v)$ }|, |\hlLR{5.5em}{ $X \cap N(v)$ }|) 
|\hspace{-0.5em}|    $P$ = |\hlLR{4.7em}{ $P \setminus \{v\}$ }|; $X$ = |\hlLR{4.8em}{ $X \cup \{v\}$ }|
\end{lstlisting}
\fi

\iftr
\subsubsection{$k$-Clique Listing}
\label{sec:kcls}

We consider listing all $k$-cliques,
a problem important for dense subgraph discovery~\cite{danisch2018listing}.
Listing~\ref{lst:kcls} contains a set-centric variant of a recent $k$-clique
listing algorithm~\cite{danisch2018listing}. We reformulated the original
scheme (without changing its time complexity) to expose the implicitly used set
operations. The algorithm is somewhat similar to Bron-Kerbosch in that it
is also recursive backtracking. One starts with iterating
over edges (2-cliques) in Lines~8-9.
In each backtracking search step, the algorithm augments the
considered cliques by one vertex $v$ and restricts the search to neighbors of
$v$ that come after $v$ in the degeneracy order (Lines~14-15).
Set operations executed by SISA are intersection and cardinality.

\begin{lstlisting}[float=h,belowskip=0em,aboveskip=0em,abovecaptionskip=0em,label=lst:kcls,caption=$k$-Clique Counting~\cite{danisch2018listing}.]
|\vspace{0.5em}|/* |\textbf{Input:}| A graph $G$. |\textbf{Output:}| Number of k-cliques $ck \in \mathbb{N}$. */
//First, derive a degeneracy vertex order $\eta$. This is an
//optimization that we also separately accelerate with SISA
$\eta$ = deg_order($v_1, v_2, ..., v_n$) //Details in our report and in |\cite{cazals2008note}|
//Construct a directed version of $G$ using $\eta$. This is an
//additional optimization to reduce the search space:
|\vspace{0.5em}|$\overrightarrow{G}(V,\overrightarrow{E})$ = dir($G$); //Edge goes from $v$ to $u\ \text{iff}\ \eta(v) < \eta(u)$ 
|\vspace{0.1em}|$ck$ = $0$ //We start with zero counted cliques.
|\vspace{0.25em}| for $u \in V$ [in par] do: //Count $u$'s neighboring $k$-cliques
  $C_2$ = $N^+(u)$; ck += count(2, $\overrightarrow{G}$, $C_2$)
|\vspace{0.25em}|function count($i$, $\overrightarrow{G}$, $C_{i}$):
|\vspace{0.25em}|  if ($i$ == $k$): return |\hlLR{3.3em}{ $\vert C_{k} \vert$ }| //Count $k$-cliques
|\vspace{0.25em}|  else:
		ci = 0
	 |\vspace{0.25em}| for $v \in C_{i}$ do: //search within neighborhood of $v$
|\vspace{0.25em}|      $C_{i+1}$ = |\hlLR{6.3em}{ $N^+(v) \cap C_{i}$ }| // $C_i$ counts $i$-cliques.
      $ci$ += count(i+1, $\overrightarrow{G}$, $C_{i+1})$
    return ci
\end{lstlisting}
\fi

\iftr
Table~\ref{tab:uc-4}
contains the specialized version for $k=4$, where no recursion is necessary.
%

%
\fi





\marginparX{\Large\vspace{3em}\colorbox{yellow}{\textbf{D}}}


\iftr
\subsubsection{$k$-Clique-Star Listing}

\fi
\iftr
$k$-\emph{clique-stars} combine the features of
cliques \emph{and stars}. 
\fi
A $k$-clique-star is a $k$-clique with additional
adjacent vertices that are connected to all the vertices in the clique.
$k$-clique-stars relax the restrictive
nature of cliques~\cite{jabbour2018pushing}.
Algorithm~\ref{lst:kscl} shows the scheme.
We first find $k$-cliques. Then, for each $k$-clique,
one finds additional vertices that form stars with intersections
and a union.

\iftr
We also observe that those extra vertices that are connected to the $k$-clique
actually form a $(k+1)$-clique (together with this $k$-clique). For this, we
provide another \textbf{variant of $k$-clique-star} listing, see
Algorithm~\ref{lst:kscl-other}. Specifically, to find $k$-clique-stars, we
first mine $(k+1)$-cliques. Then, we find $k$-clique-stars within each
$(k+1)$-clique using set union, membership, and difference.
\fi

\tr{Large cliques are expected to be rare because \emph{every} vertex in a
clique, regardless of the clique size, must be connected to all other vertices
in this clique.}


\iftr

\begin{lstlisting}[float=h!,aboveskip=0em, belowskip=0em, abovecaptionskip=0em,label=lst:kscl,caption=$k$-clique-star listing~\cite{jabbour2018pushing}.]
|\vspace{0.5em}|/* |\textbf{Input:} A graph $G$. \textbf{Output:} All $k$-clique-stars, $S$.|*/
$C$ = /* First, find $k$-cliques (e.g., with Table |\ref{tab:uc-4}|)*/
|\vspace{0.5em}|$S$ = $\emptyset$ //$S$ is a set with identified $k$-clique-stars.
|\vspace{0.5em}|foreach $c = (V_c, E_c) \in C$ do: //For each $k$-clique...
|\vspace{0.5em}|  $X$ = |\hlLR{6.7em}{ $\bigcap_{u \in V_c} N(u)$ }| //Intersect all $N(u)$ such that $u \in V_c$
|\vspace{0.5em}|  $G_s$ = |\hlLR{4.5em}{ $X \cup V_c$ }| //Derive the actual $k$-clique-star
  $S$ = $S \cup \{G_s\}$ //Add an identified $k$-clique-star to $S$
//At the end, remove duplicates from $S$
\end{lstlisting}

\fi


\iftr

\begin{lstlisting}[float=h!,aboveskip=0em,belowskip=0em,abovecaptionskip=0em,label=lst:kscl-other,caption=$k$-clique-star listing (our variant).]
/* |\textbf{Input:} A graph $G$. \textbf{Output:} $S$ contains the maximal $k$-clique-stars of $G$.|*/|\vspace{0.5em}|
$C$ = /* First, find $(k+1)$-cliques (use Listing |\ref{lst:kcls}|)*/
$S$ = /* Empty map where the keys are $k$-cliques and the values are $k$-clique-stars */|\vspace{0.5em}|
|\vspace{0.5em}|for $c \in C$ [in par] do: //For each $(k+1)$-clique $c$...
|\vspace{0.5em}|  for $v \in c$ do:  //for each vertex in clique $c$...
     |\hlLR{8.5em}{ $S$[$c \setminus \{v\}$] $\cup$= $c$ }| //Add $c$ to a $k$-clique-star.
\end{lstlisting}

\fi

\iftr

\subsubsection{Degeneracy Order and $k$-Core}
\label{sec:degeneracy}

Several graph pattern matching algorithms use the \emph{degeneracy ordering} of
the vertices. This ordering produces an orientation of the edge of the graph
with low out-degree of the vertices. Algorithm~\ref{lst:apxcore} shows a set-centric
(and easily parallelizable) algorithm to compute an \emph{approximate}
degeneracy order (the algorithm has $O(\log n)$ iterations for any constant
$\epsilon>0$ and has an approximation ratio of $2+\epsilon$). The algorithm is
based on a streaming scheme for large
graphs~\cite{DBLP:conf/latin/Farach-ColtonT14} and uses set cardinality and
difference.
The derived degeneracy order can be directly used to compute the $k$-core of
$G$ (a maximal connected subgraph of $G$ whose all vertices have degree at
least $k$). This is done by iterating over vertices in the degeneracy order and
removing all vertices with out-degree less than $k$ (in the oriented graph).





\begin{lstlisting}[float=h!,belowskip=0em,aboveskip=0em,abovecaptionskip=0em,label=lst:apxcore,caption=Approximate Degeneracy Order~\cite{DBLP:conf/latin/Farach-ColtonT14}]
|\vspace{0.5em}|/* |\textbf{Input:}| A graph $G$. |\textbf{Output:}| Degeneracy Order $\eta$.*/
i = 0
do:
   |\vspace{0.5em}|$X$ = $\{v \in V \ : \vert N(v) \vert \leq (1+ \epsilon) (\sum_{v \in V} \vert N(v)\vert ) / \vert V \vert \}$
  |\vspace{0.5em}|for $v \in X$ [in par] do: $\eta(v)$ = i //degen. order for $v$
  |\vspace{0.5em}||\hlLR{4.8em}{ $V$ $\setminus$= $X$ }| //remove assigned vertices
  |\vspace{0.25em}|for $v \in V$ [in par] do: |\hlLR{6.5em}{ $N(v)$ $\setminus$= $X$ }| //update edges
   i = i+1
while $V \neq \emptyset$
\end{lstlisting}

\fi

\iftr
\begin{figure*}[t]
\vspaceSQ{-1.5em}
\centering
\includegraphics[width=1.0\textwidth]{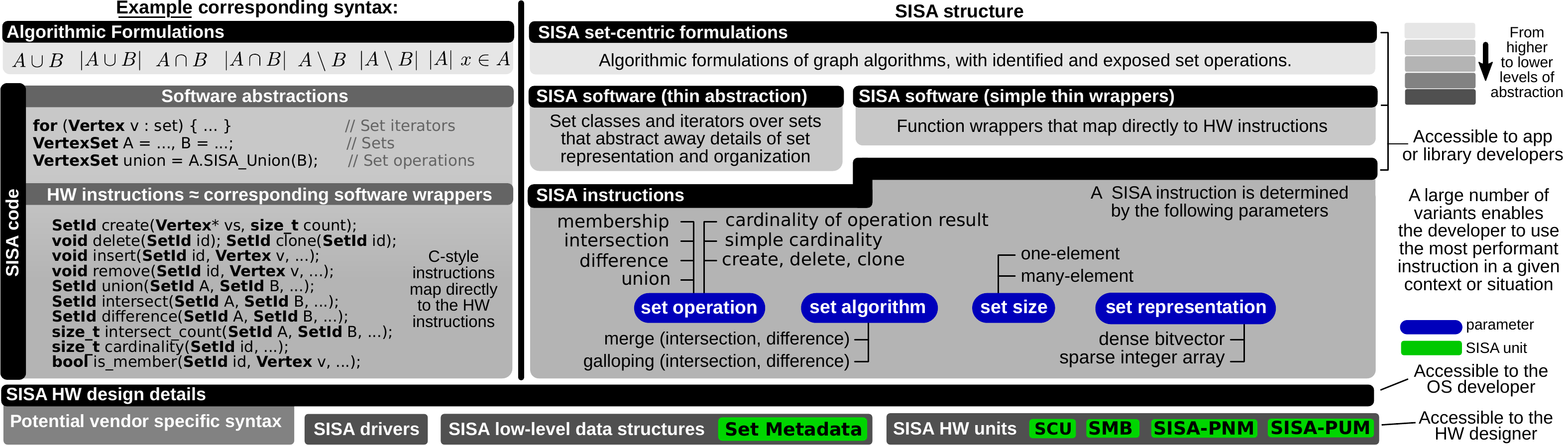}
\vspaceSQ{-1.5em}
\caption{Overview of SISA instructions and syntax at different levels of abstraction.}
\vspaceSQ{-1.5em}
\label{fig:sisa-full}
\end{figure*}
\fi

\marginparX{\Large\vspace{2em}\colorbox{yellow}{\textbf{D}}}

\iftr
\subsubsection{Subgraph Isomorphism}

\fi
Subgraph isomorphism (SI) is a key graph problem where one checks whether a given
(usually small) graph $G_2$ is a subgraph of a graph \mbox{$G_1$}. Here, we consider
an established VF2 algorithm~\cite{cordella2004sub}.
\ifconf
\else
In Algorithm~\ref{lst:si}, we first provide the most
important part that recursively constructs a candidate set of vertices from
$G_1$, and verifies if it matches the pattern $G_2$.
\fi
\iftr
\fi

\marginpar{\Large\vspace{5em}\colorbox{yellow}{\textbf{L}}}

\hl{We use SI as an example of how SISA supports \textbf{labeled graphs}.  In
VF2~\mbox{\cite{cordella2004sub}}, for each transition between states, one
first verifies if the structure of \mbox{$G_2$} matches that of \mbox{$G_1$}
(Line~\mbox{\ref{ln:si-str-match}}).  Then, label matching is verified
independently
(Lines~\mbox{\ref{ln:si-lab-match-1}}-\mbox{\ref{ln:si-lab-match-2}}).
Checking if vertex labels match, i.e., if \mbox{$L(v_1)$} equals
\mbox{$L(v_2)$}, is trivial.  Yet, a graph may also contain edge labels that
need to be matched.  This could be done with a standard approach without set
operations~\mbox{\cite{cordella2004sub}}. 
However, the generality of set notation also enables supporting label
verification. For this, we first identify all edges in~\mbox{$G_1$} where one
endpoint is the newly matched vertex $v_1$ and the other endpoint~\mbox{$v'_1$}
is already matched (i.e., \mbox{$v'_1 \in M_1(s)$}). This is done with an
intersection \mbox{$N_1(v_1) \cap M_1(s)$}. Then, we find the vertex with which
\mbox{$v'_1$} is matched, see the second loop in
Line~\mbox{\ref{ln:si-lab-loop2}}.  Finally, we verify that the respective
labels match (Line~\mbox{\ref{ln:si-lab-check}}).}


\begin{lstlisting}[float=t, aboveskip=0em, belowskip=-1em,
abovecaptionskip=0em,label=lst:si,caption=Subgraph
isomorphism~\cite{cordella2004sub}. $M_1$ and $M_2$ denote the current partial
mappings associated with $G_1$ and $G_2${,} respectively{.} $T_1$ and $T_2$
denote sets of vertices adjacent to the ones in $M_1$ and $M_2${,}
respectively. \hl{$N_1$ and $N_2$ denote neighborhoods within $G_1$ and $G_2$,
respectively. \texttt{verify\_labels} is used if graphs are labeled.} ]
/* |\textbf{Input:} target graph $G_1$, pattern $G_2$. \textbf{Output:} mapping between graphs.|*/
$s_0$ = {}; $M(s_0)$ = $\emptyset$; // Initial state
Match($s_0$); // Algorithm start
|\vspace{0.2em}|function Match($s$):
|\vspace{0.2em}| if $M(s)$ covers all nodes in pattern graph: output $M(s)$; return;
|\vspace{0.2em}| $P(s)$ = /* compute set of candidate pairs to be added to $M(s)$ */
|\vspace{0.2em}| for ($v_1$, $v_2$) $\in$ $P(s)$ do: 
|\vspace{0.2em}|    $checkCore$ = /* original $R_{core}$ rule */
|\vspace{0.5em}|    $checkTerm$ = |\hlLR{9em}{ $\vert N_1(v_1) \cap T_1(s) \vert$ }| $\ge$ |\hlLR{9em}{ $\vert N_2(v_2) \cap T_2(s) \vert$ }|
|\vspace{0.5em}|    $checkNew$ = |\hlLR{13.5em}{ $\vert N_1(v_1) \setminus (M_1(s) \cup T_1(s)) \vert$ }| $\ge$ |\hlLR{13.5em}{ $\vert N_2(v_2) \setminus (M_2(s) \cup T_2(s)) \vert$ }|
|\vspace{0.2em}|    $checkFeasibility$ = $checkCore$ $\wedge$ $checkTerm$ $\wedge$ $checkNew$ |\label{ln:si-str-match}|
    $checkSemantic$ = verify_labels($v_1$, $v_2$, $s$) //If we use labels.|\label{ln:si-lab-match-1}|
|\vspace{0.2em}|    $checkFeasibility$ = $checkFeasibility$ $\wedge$ $checkSemantic$ //If we use labels.|\label{ln:si-lab-match-2}|
|\vspace{0.2em}|    if $checkFeasibility$ : $s^\prime$ = NewState($s$, $v_1$, $v_2$); Match($s^\prime$)
//Check if labeling of $v_1$ and $v_2$ and their neighborhoods matches:
|\vspace{0.2em}|bool verify_labels($v_1$, $v_2$, $s$): 
|\vspace{0.2em}|  forall $v'_1 \in $ |\hlLR{8.5em}{ $N_1(v_1) \cap M_1(s)$ }|:  forall $(v'_1, v'_2) \in M(s)$:|\label{ln:si-lab-loop2}|
|\vspace{0.2em}|      if ($L(v_1)$ != $L(v_2)$) or ($L(v_1, v'_1)$ != $L(v_2, v'_2)$): return false|\label{ln:si-lab-check}|
|\vspace{0.2em}|  return true
\end{lstlisting}


\marginparX{\Large\vspace{1em}\colorbox{yellow}{\textbf{B}}}

\enlargeSQ

\marginpar{\Large\vspace{5em}\colorbox{yellow}{\textbf{F}}}

\iftr
\subsubsection{Frequent Subgraph Mining}

Here,
\fi
\hl{we use an established \emph{Apriori-based} scheme~\mbox{\cite{agrawal1994fast}},\mbox{\cite[Algorithm~3.1]{jiang2013survey}}. 
We show it in Algorithm~\mbox{\ref{lst:fsm}}. It first generates candidate subgraphs
\mbox{$C_k$} (Line~\mbox{\ref{ln:fsm-cand}}) and then checks their counts
\texttt{cnt} in the input graph (Line~\mbox{\ref{ln:fsm-si}}) using subgraph
isomorphism (SI) as a fundamental kernel~\mbox{\cite{jiang2013survey}}
(combining candidate generation and occurrence verification is a very popular
FSM approach~\mbox{\cite{agrawal1994fast, han2006data,
kuramochi2001frequent, kuramochi2004efficient}}, also see other references
in~\mbox{\cite{jiang2013survey}}). If the count is above a
certain user selected threshold (\mbox{$\sigma \cdot n$}), a candidate is added
as a found frequent subgraph (Line~\mbox{\ref{ln:fsm-found}}). VF2, an SI
algorithm covered in this section, was found to be an efficient kernel for FSM;
all SISA operations in SI are reused.}
\hl{Generation of candidate subgraphs (\mbox{\texttt{candidate\_gen}}) is less
time-consuming than SI~\mbox{\cite{jiang2013survey}}.} \hl{Still, it also
benefits from set operations; for example, joining trees that represent
candidates, a key operation in a kernel by Hido and
Kawano~\mbox{\cite{hido2005amiot}}, is done using set
union~\mbox{\cite{jiang2013survey}}. These trees can be implemented with either
$n$-bit dense bitvectors or sparse arrays, benefiting from SISA-PUM or PNM
(user's choice).}
%

\marginpar{\Large\vspace{-10em}\colorbox{yellow}{\textbf{L}}}

\marginpar{\Large\vspace{3em}\colorbox{yellow}{\textbf{F}}}

\begin{lstlisting}[float=h, aboveskip=0em, belowskip=-1em, abovecaptionskip=0em,label=lst:fsm,caption=\hl{Frequent subgraph mining~\mbox{\cite{jiang2013survey}}.}]
/* |\textcolor{gray}{\textbf{Input:} target graph ($G$), minimum support / count of a found pattern ($\sigma$).}| 
 * |\textcolor{gray}{\textbf{Output:} sets of frequent subgraphs of sizes $1, 2, ..., k$ ($F_1, F_2, ..., F_k$).}|*/
$F_1$ = $V$; $k$ = 2 //$k = 2$ means we start recursion from edges.
//Use all subgraphs in $F_{k-1}$ to generate candidates of size $k$:
while $F_{k-1} \neq \emptyset$ do: //$C_k$ (below) are candidate subgraphs of size $k$
  $F_k$ = $\emptyset$; $C_k$ = candidate_gen($F_{k-1}$) //Use any selected kernel|\cite{jiang2013survey}||\label{ln:fsm-cand}|
  foreach $g \in C_k$ do:  
    cnt = SI($g$, $G$) //For |\hlLR{7.5em}{set operations }| in SI, see Algorithm |\ref{lst:si}| |\label{ln:fsm-si}|
    if cnt $\ge \sigma n$ and $g \not\in F_k$: $F_k$ $\cup$= $g$|\label{ln:fsm-found}|
  k++
\end{lstlisting}

\iftr

\subsection{Graph Learning}
\label{sec:formulations-learning}

We also consider various problems related to learning.

\fi

\marginparX{\Large\vspace{1em}\colorbox{yellow}{\textbf{D}}}

\iftr
\subsubsection{Vertex Similarity} 
\label{sec:sets-similarity}

\fi
%
%
%
Various measures assess how similar two vertices $v$ and
$u$ are, see Algorithm~\ref{lst:sim}. They can be used on their own, or as a
main building block of more complex algorithms such as clustering. 
\iftr
They are used in
multiple fields, for example in graph databases~\cite{robinson2013graph}. 
\fi
\tr{These measures heavily rely on the
cardinalities of set intersection and set union.}

\ignore{
\maciej{Integrate:
Neo4j: ``We can use the Jaccard Similarity algorithm to work out the similarity
between two things. We might then use the computed similarity as part of a
recommendation query. For example, you can use the Jaccard Similarity algorithm
to show the products that were purchased by similar customers, in terms of
previous products purchased.''
Neo4j: ``We can use the Overlap Similarity algorithm to work out which things
are subsets of others. We might then use these computed subsets to learn a
taxonomy from tagged data.''
https://jbarrasa.com/2017/03/31/quickgraph5-learning-a-taxonomy-from-your-tagged-data/
``Adamic Adar is a measure used to compute the closeness of nodes based on
their shared neighbors.  The Adamic Adar algorithm was introduced in 2003 by
Lada Adamic and Eytan Adar to predict links in a social network.''
``Resource Allocation is a measure used to compute the closeness of nodes based
on their shared neighbors.  The Resource Allocation algorithm was introduced in
2009 by Tao Zhou, Linyuan Lü, and Yi-Cheng Zhang as part of a study to predict
links in various networks. ''
``Common neighbors captures the idea that two strangers who have a friend in
common are more likely to be introduced than those who don’t have any friends
in common.''
``Preferential Attachment is a measure used to compute the closeness of nodes,
based on their shared neighbors.  Preferential attachment means that the more
connected a node is, the more likely it is to receive new links. This algorithm
was popularised by Albert-László Barabási and Réka Albert through their work on
scale-free networks.''
``Total Neighbors computes the closeness of nodes, based on the number of
unique neighbors that they have. It is based on the idea that the more
connected a node is, the more likely it is to receive new links.  ''
Adamic Adar measure: "Friends and neighbors on the Web"
}
}

\iftr

\begin{lstlisting}[float=h!, aboveskip=0em,belowskip=0em,abovecaptionskip=0em,label=lst:sim,caption=Example vertex similarity measures~\cite{leicht2006vertex}.]
/* |\textbf{Input:}| A graph $G$. |\textbf{Output:}| Similarity $S \in \mathbb{R}$ of sets $A,B$.
 * Most often, $A$ and $B$ are neighborhoods $N(u)$ and $N(v)$ 
|\vspace{0.5em}| * of vertices $u$ and $v$. */
|\vspace{0.5em}|//Jaccard similarity:
|\vspace{0.5em}|$S(A,B)$ = |\hlLR{5em}{ $\vert A \cap B\vert$ }| / |\hlLR{5em}{ $\vert A \cup B\vert$ }| = |\hlLR{5em}{ $\vert A \cap B\vert$ }| / ($\vert A \vert$ + $\vert B \vert$ - |\hlLR{4.7em}{ $\vert A \cap B \vert$ }|)
|\vspace{0.5em}|//Overlap similarity:
|\vspace{0.5em}|$S(A,B)$ = |\hlLR{4.8em}{ $\vert A \cap B\vert$ }| / min($\vert A\vert$, $\vert B\vert$)
|\vspace{0.5em}|//Certain measures are only defined for neighborhoods:
|\vspace{0.5em}|$S(v,u)$ = $\sum_{w} (1 / \log|N(w)|)$ //where $w \in $ |\hlLR{7.3em}{ $N(v) \cap N(u)$ }|; Adamic Adar
|\vspace{0.5em}|$S(v,u)$ = $\sum_{w} (1 / |N(w)|)$ //where $w \in $ |\hlLR{7.3em}{ $N(v) \cap N(u)$ }|; Resource Alloc.
|\vspace{0.5em}|$S(v,u)$ = |\hlLR{8.4em}{ $\vert N(v) \cap N(u) \vert$ }| //Common Neighbors
$S(v,u)$ = |\hlLR{8.4em}{ $\vert N(v) \cup N(u) \vert$ }| //Total Neighbors
\end{lstlisting}

\fi

\ignore{

\begin{lstlisting}[float=h!, aboveskip=-0.5em,belowskip=-1em,abovecaptionskip=-0.5em,label=lst:sim,caption=Vertex similarity measures~\cite{leicht2006vertex}.]
/* |\textbf{Input:}| A graph $G$. |\textbf{Output:}| Similarity $S \in \mathbb{R}$ of sets $A,B$.
 * Most often, $A$ and $B$ are neighborhoods $N(u)$ and $N(v)$ 
|\vspace{0.5em}| * of vertices $u$ and $v$. */
|\vspace{0.25em}|//Jaccard similarity:
$S(A,B)$ = |\hspace{-0.5em}||\highlight{|$\vert A \cap B\vert$|}| / |\hspace{-0.5em}||\highlight{|$\vert A \cup B\vert$|}| = |\hspace{-0.5em}||\highlight{|$\vert A \cap B\vert$|}| / (|\hspace{-0.5em}||\highlight{|$\vert A \vert$|}| + |\hspace{-0.5em}||\highlight{|$\vert B \vert$|}| - |\hspace{-0.5em}||\highlight{|$\vert A \cap B \vert$|}|)
|\vspace{0.25em}|//Overlap similarity:
$S(A,B)$ = |\hspace{-0.5em}||\highlight{|$\vert A \cap B\vert$|}| / min(|\hspace{-0.5em}||\highlight{|$\vert A\vert$|}|, |\hspace{-0.5em}||\highlight{|$\vert B\vert$|}|)
|\vspace{0.25em}|//Certain measures are only defined for neighborhoods:
$S(v,u)$ = $\sum_{w \in N(v) \cap N(u)} (1 / \log|N(w)|)$ //Adamic Adar
$S(v,u)$ = $\sum_{w \in N(v) \cap N(u)} (1 / |N(w)|)$ //Resource Allocation
$S(v,u)$ = $|N(v) \cap N(u)|$ //Common Neighbors
$S(v,u)$ = $|N(v) \cup N(u)|$ //Total Neighbors
$S(v,u)$ = $|N(v)| \cdot |N(u)|$ //Preferential Attachment
$S(v,u)$ = $\sum_{x \in N(v)} \sum_{y \in N(u)} \delta(x,y)$ //Friends Measure, where $\delta(x,y) = 1$ 
//$\text{iff}\ (x=y \lor (x,y) \in E \lor (y,x) \in E),\ \text{and}\ \delta(x,y) = 0\ \text{otherwise}$
\end{lstlisting}

}

\iftr

\subsubsection{Link Prediction}
\label{sec:lp}

Here, one is interested in developing schemes for predicting
whether two non-adjacent vertices can become connected in the
future. There exist many schemes for such prediction~\cite{liben2007link,
lu2011link, al2006link, taskar2004link}.  Assessing the accuracy of a specific
link prediction scheme~$S$ is done with a simple set-centric
algorithm~\cite{wang2014robustness}
shown in Listing~\ref{lst:lp}.  
We start with some graph
with \emph{known} links (edges). We derive $E_{sparse} \subseteq E$, which is $E$ with
random links removed; $E_{sparse} = E \setminus E_{rndm}$.  $E_{rndm} \subseteq
E$ are randomly selected \emph{missing} links from $E$ (\emph{links to be
predicted}).  We have $E_{sparse} \cup E_{rndm} = E$ and $E_{sparse} \cap
E_{rndm} = \emptyset$.
Now, we apply the link prediction scheme~$S$ (that we want to test) to each
edge $e \in (V \times V) \setminus E_{sparse}$. The higher a value $S(e)$, the
more probable $e$ is to appear in the future (according to $S$).  Now, the
effectiveness $eff$ of $S$ is computed by verifying how many of the edges with
highest prediction scores ($E_{predict}$) actually are present in the original dataset~$E$:
$eff = |E_{predict} \cap E_{rndm}|$.

\begin{lstlisting}[float=h!, aboveskip=0em, belowskip=0em,abovecaptionskip=0em,label=lst:lp,caption=Link prediction testing.]
/* |\textbf{Input:}| A graph $G = (V,E)$. |\textbf{Output:}| Effectiveness $eff$
|\vspace{0.5em}| * of a given prediction scheme. */
$E_{rndm}$ = /* Random subset of $E$ */
|\vspace{0.5em}|$E_{sparse}$ = |\hlLR{6.3em}{ $E \setminus E_{rndm}$ }| /* Edges in $E$ after removing $E_{rndm}$ */
//For each |\hlLR{11em}{ $e \in (V \times V) \setminus E_{sparse}$ }|, derive score $S(e)$ that
//determines the chance that $e$ appears in future. Here, 
|\vspace{0.25em}|//one can use any vertex similarity scheme $S$ from |\cref{sec:sets-similarity}|.
|\vspace{0.25em}|for $e = (v,u) \in $ |\hlLR{9.5em}{ $(V \times V) \setminus E_{sparse}$ }| [in par] do: compute $S(v,u)$
|\vspace{0.25em}|$E_{predict}$ = /* Pick selected top edges with the highest $S$ scores.*/
$eff$ = |\hlLR{10.5em}{ $\vert E_{predict} \cap E_{rndm}\vert$ }| //Derive the effectiveness.
\end{lstlisting}

\fi



\iftr

\subsubsection{Clustering} 

We consider graph clustering, a widely studied problem used in a plethora of
areas. Listing~\ref{lst:cl} shows Jarvis-Patrick
clustering~\cite{jarvis1973clustering}, a scheme that uses similarity of
neighbors of two vertices to determine whether these two vertices are in the
same cluster. The set-centric formulation relies heavily on set intersection.

\begin{lstlisting}[float=h, aboveskip=0em, abovecaptionskip=0em, belowskip=0em,label=lst:cl,caption=Jarvis-Patrick clustering (based on the Common Neighbors similarity of vertices; other measures can also be used -- see Listing~\ref{lst:sim}).]
/* |\textbf{Input:}| A graph $G = (V,E)$. |\textbf{Output:}| Clustering $C \subseteq E$ */
|\vspace{0.5em}|for $e = (v,u) \in E$ [in par] do: //$\tau$ is a user-defined threshold
  if |\hlLR{8.5em}{ $\vert N(v) \cap N(u)\vert$ }| $> \tau$: $C = C \cup \{e\}$
\end{lstlisting}

\fi


\all{\maciej{fix}

\begin{lstlisting}[float=h, aboveskip=-0.75em, abovecaptionskip=-0.5em, belowskip=-1.5em,label=lst:cl,caption=Jarvis-Patrick clustering (based on the $S_C(v,u)$
similarity of vertices; other measures can also be used, see Listing~\ref{lst:sim}).]
/* |\textbf{Input:}| A graph $G = (V,E)$. |\textbf{Output:}| Clustering $C \subseteq E$ */
|\vspace{0.5em}|//Use example similarity $S_C(v,u)$ = |\hlLR{9em}{ $\vert N(v) \cap N(u)\vert$ }| (see Listing |\ref{lst:sim}|).
|\vspace{0.5em}|for $e = (v,u) \in E$ [in par] do: //$\tau$ is a user-defined threshold
  if |\hlLR{9em}{ $\vert N(v) \cap N(u)\vert$ }| $> \tau$: $C = C \cup \{e\}$
//Other clustering schemes use other similarity measures.
\end{lstlisting}

}

\ignore{
https://www.csc2.ncsu.edu/faculty/nfsamato/practical-graph-mining-with-R/slides/pdf/Graph\_Cluster\_Analysis.pdf
}


\marginparX{\Large\vspace{2em}\colorbox{green}{\textbf{F}}}

\iftr
\subsection{``Low-Complexity'' Algorithms}

SISA does not target the ``low-complexity'' algorithms, 
\fi
as they offer few opportunities for set-centric
acceleration~\cite{skiena1990dijkstra, meyer2003delta,
cormen2009introduction, solomonik2017scaling, besta2017slimsell,
gianinazzi2018communication, shiloach1982logn, yan2014pregel, suttonoptimizing,
miller2015improved, boruuvka1926jistem}.  For example, in PageRank, one
updates vertex ranks in two nested loops, which is not easily
expressible with set operations. 
\iftr 
We
analyzed many other such algorithms.
This includes Dijkstra's SSSP~\mbox{\cite{skiena1990dijkstra}},
\mbox{$\Delta$}--Stepping~\mbox{\cite{meyer2003delta}},
Bellman-Ford~\mbox{\cite{cormen2009introduction}},
Betweenness Centrality schemes~\cite{solomonik2017scaling},
traversals~\cite{besta2017slimsell},
Connected Components algorithms~\mbox{\cite{gianinazzi2018communication,
shiloach1982logn}}, Low-Diameter
Decomposition~\mbox{\cite{miller2015improved}}, or Boruvka's Minimum Spanning
Tree~\mbox{\cite{boruuvka1926jistem}}. 
\fi
Our work is already more general
than other pattern matching accelerators / frameworks, as it supports many more
\ifconf
\else
problems beyond simple pattern matching (e.g., vertex similarity, clustering,
link prediction, complex algorithms such as Bron-Kerbosch). 
\fi

\marginparX{\Large\vspace{-2em}\colorbox{yellow}{\textbf{F}}}

\sethlcolor{yellow}

\iftr

As a single example, we illustrate and discuss a set-centric formulation of
\textbf{Breadth-First Search (BFS)}~\cite{cormen2009introduction}, the basic
graph traversal algorithm, see Algorithm~\ref{lst:bfs-ps-s}.  BFS is a basis of
the established Graph500 benchmark~\cite{murphy2010introducing} and a subject
of extensive research in a past decade~\cite{murphy2010introducing,
beamer2013direction, buluc2017distributed, beamer2013distributed,
leiserson2010work, besta2017push,
yoo2005scalable, bulucc2011parallel}.
In the bottom-up part of BFS, the key element of the set-based variant is an
additional set $\Pi$ with unvisited vertices. $\Pi$ is represented as a dense bit vector, requiring
only $n$ bits of storage.
Using $\Pi$ and other sets enables abstracting away some branches.

\begin{lstlisting}[float=h,aboveskip=0em, belowskip=0em, label=lst:bfs-ps-s,caption=Set-centric BFS.]
/* |\textbf{Input:}| A graph $G$, a root vertex $r \in V$. |\textbf{Output:}| A map $p$
 * of parents of each vertex, on the way to $r$. */
$F$ = $\emptyset$ //$F$ is the frontier.
$\forall_{v \in V}\ p(v)$ = $\perp$; $p(r)$ = $r$ //First, no $v$ has a parent, except for $r$
$\Pi$ = $V$ //Initially, all vertices are unvisited.
|\vspace{0.5em}|$F$ = $\{r\}$ //Initialize frontier $F$ with the root.
while $F$ != $\emptyset$ do:
  $F_{new}$ = $\emptyset$ //Initialize the new frontier $F_{new}$.
#if TOP_DOWN_BFS
|\vspace{0.5em}|  for $u \in F\ $ [in par] do: 
|\vspace{0.5em}|    $F$ = |\hlLR{4.7em}{ $F \setminus \{u\}$ }| //Remove the current element $u$ from $F$
|\vspace{0.5em}|    for $w \in $ |\hlLR{6.5em}{ $N(u)\ \cap\ \Pi$ }| [in par] do: 
       $p(w)$ = $u$; $F_{new}$ = |\hlLR{6.5em}{ $F_{new} \cup \{w\}$ }|; $\Pi$ = |\hlLR{4.7em}{ $\Pi \setminus \{w\}$ }|
#elif BOTTOM_UP_BFS
|\vspace{0.5em}|  for $w \in \Pi$ [in par] do: 
|\vspace{0.5em}|    for $u \in $  |\hlLR{7em}{ $N(w)\ \cap\ F$ }| do: 
|\vspace{0.25em}|       $p(w)$ = $u$; $F_{new}$ = |\hlLR{6.5em}{ $F_{new} \cup \{w\}$ }|; $\Pi$ = |\hlLR{5em}{ $\Pi \setminus \{w\}$ }|; break
#endif
   $F = F_{new}$
\end{lstlisting}

\fi

\ignore{
\begin{lstlisting}[belowskip=-0.5em, label=lst:bfs-ps,caption=Traditional variant~\cite{besta2017slimsell}.]
/* |\textbf{Input:}| A graph $G$, a root vertex $r \in V$. |\textbf{Output:}| An array
 * $p$ of parents of each vertex, on the way
 * to the root $r$. */

int[] $p$ //$p$ is an array of parents.
queue<int> $F$ //$F$ is the frontier.
$p[1, ..., n] \leftarrow [\perp, ... \perp]$ //At first, no $v$ has a parent.
$p[r] \leftarrow r$ //|E|xcept for $r$ that is itself a parent.
$F \leftarrow \{r\}$ //Initialize frontier $F$ with the root.

|\tikzmarkin{col}(0.7,-0.1)(-0.05,0.25)|while $F \neq \emptyset$ do |\text{\encircle{2}}|
  for $u \in F$|\tikzmarkend{col}| do
    $F \leftarrow F \setminus \{u\}$ //Remove the current element $u$ from $F$.
    for $w \in N_u$ do //Iterate over all neighbors of $u$.
     |\highlight{ \textbf{if}| $p[w] =\ \perp$ |\textbf{then} }| |\text{\encircle{1}}| //If a neighbor is unvisited.
        $p[w] \leftarrow u$ //Update $w$'s parent.
        $F \leftarrow F \cup \{w\}$ //|E|n|q|ueue $w$ to the frontier.
\end{lstlisting}
   
\begin{lstlisting}[belowskip=-0.5em, label=lst:bfs-ps-s,caption=Set based variant.]
/* |\textbf{Input:}| A graph $G$, a root vertex $r \in V$. |\textbf{Output:}| A map $p$
 * of parents of each vertex, on the way
 * to the root $r$. */
mutable map $p: V \to V \cup \{\perp\}$ //"$\perp$" means "undefined".
mutable set $F \subseteq V$ //$F$ is the frontier.
$\forall_{v \in V}\ p(v) \leftarrow \perp$ //First, no $v$ has a parent.
$p(r) \leftarrow r$ //|E|xcept for $r$ that is its own parent.
$F \leftarrow \{r\}$ //Initialize frontier $F$ with the root.

|\highlight{\textbf{for}|$\downarrow$ $u \in F\ $|}| |\text{\encircle{2}}| //"$\downarrow$": Iterate over $F$ until it's empty.
  $F \leftarrow F \setminus \{u\}$ //Remove the current element $u$ from $F$.
  //$\Pi \equiv \{v \in V\ :\ p(v) = \perp\}$ contains unvisited vertices.
  //|V|isit neighbors of $w$. |E|quivalenty, one could use
  //$\Pi(u) \equiv \{v \in N_u\ :\ p(v) = \perp\}$.
  for $w \in N_u $|\highlight{|$\ \cap\ \{v \in V\ :\ p(v) = \perp\}\ $|}|  |\text{\encircle{1}}| 
    $p(w) \leftarrow u$ //Update $w$'s parent.
    $F \leftarrow F \cup \{w\}$ //Update the frontier $F$.
\end{lstlisting}
}

\enlargeSQ

\subsection{Deriving a Set-Centric Formulation}
\label{sec:sisa-programming}


\marginparX{\Large\vspace{2em}\colorbox{yellow}{\textbf{D}}}

\tr{One either picks a set-centric formulation of a given algorithm, or
designs one. For the former, we offer more than 10 set-centric formulations.
For the latter, one starts with a selected algorithm specification to be
``made'' set-centric.}
Often, algorithms use set notation, and
one may simply pick operations for memory acceleration. This is the
case with, for example, Jarvis-Patrick clustering.
Still, one may need to apply more
complex changes to ``expose'' set instructions.
The general rule is to associate used data structures with sets, and then
identify respective set operations. As an example, we compare a traditional
snippet for deriving the count of all 4-cliques \texttt{cnt},  a derived set-centric algorithmic
formulation, and the corresponding SISA snippet in Table~\ref{tab:uc-4}.
The key algorithmic change is using set intersections instead of explicitly
verifying if vertices are connected. For example, instead of iterating over
{all} neighbors of $v_1$-$v_3$ (Lines~4-6, the top snippet), in SISA, we intersect
neighborhoods of $v_1$-$v_3$ (Line~4 \&~6, the middle snippet) to filter 4-cliques.
\all{This exposes parallelism present within one intersection operation and across
multiple such operations.}

\input{example.tex}

\all{\maciej{ADD these listings with iterators as well}}



%% file: example.tex
\begin{table}[h]
\vspaceSQ{-1.5em}
\centering
\iftr
\begin{minipage}[t]{0.99\columnwidth}
\begin{lstlisting}[language=C++, basicstyle=\tt\ssmall,
label=lst:bk-use-case]
//Code common to all schemes below. |\textbf{Input}:| A graph $G$.
int64_t cnt = 0; //Initially, clique count is 0.
//We use directed edges indicated with ``+'' (iterate over
//out-neighbors) to avoid counting the same clique twice.
\end{lstlisting}
\end{minipage}
\\
\fi
\vspace{2em}
\begin{minipage}[t]{0.99\columnwidth}
\centering
\begin{lstlisting}[language=C++,label=lst:uc-t, basicstyle=\tt\ssmall]
//|\ul{Non set-centric code:}|
CSR_Graph g($G$); //Standard codes often use some form of CSR
#pragma omp parallel for
for (auto v1: g.V()) //For all vertices in parallel.
 for (auto v2: g.N_out(v1)) //Explore neighborhoods of v1-v4... 
  for (auto v3: g.N_out(v2)) //...searching for a 4-clique
   for (auto v4: g.N_out(v3)) //If v1-v4 are connected pairwise
     if(g.edge(v1,v3) && g.edge(v1,v4) && g.edge(v2,v4)) ++cnt;
\end{lstlisting}
%
\end{minipage}
\\
\vspace{2em}
\begin{minipage}[t]{0.99\columnwidth}
\centering
\iftr
\begin{lstlisting}[label=lst:uc-s, basicstyle=\tt\ssmall]
//|\ul{A set-centric algorithmic formulation:}|
for $v_1 \in V$ in parallel do: //For all vertices in parallel.
|\vspace{0.25em}|  for $v_2 \in N^+(v_1)$ do: //For each neighbor of $v_1$...
|\vspace{0.25em}|    $S_1$ = |\hlLR{9em}{ $N^+(v_1) \cap N^+(v_2)$ }| //Find common neighbors of $v_1$ and $v_2$.
|\vspace{0.25em}|    for $v_3 \in S_1$ do: //Narrow further search to $S_1$.
      cnt += |\hlLR{7.3em}{ $\vert S_1 \cap N^+(v_3) \vert$ }| //Common neighbors of $v_1$, $v_2$, and $v_3$
\end{lstlisting}
\fi
\end{minipage}
\\
\vspace{2em}
\begin{minipage}[t]{0.99\columnwidth}
\centering
\begin{lstlisting}[language=C++, basicstyle=\tt\ssmall,
label=lst:bk-use-case]
//|\ul{SISA (simplified) set-centric code:}|
SetGraph g = SetGraph($G$);
#pragma omp parallel for
for (auto v1: g.V()) for (auto v2: g.N_out(v1)) {
    auto S1 = intersect(g.N_out(v1), g.N_out(v2));
    for (auto v3: S1) cnt += intersect_card(S1, g.N_out(v3)); }
\end{lstlisting}
\end{minipage}
\vspace{1.5em}
\caption{Finding all 4-cliques: a traditional (non-set-centric)
snippet, a set-centric algorithmic formulation {derived in this work},
and a SISA set-centric snippet.
\tr{In all variants, we use directed edges (i.e., iterate over
out-neighbors) to avoid counting the same clique twice.}}
%
\label{tab:uc-4}
\end{table}

%% file: structure.tex
\vspace{-0.75em}
\section{SISA: Design, Syntax, Semantics}
\label{sec:sisa-syntax-semantics}


\cnf{We now detail SISA's design, see Figure~\ref{fig:sisa-full}.}

\iftr
We now present the details of representing and processing sets used in
set-centric formulations. This constitutes core parts of SISA's design.
%
%
We summarize SISA in Figure~\ref{fig:sisa-full} and we detail key SISA
instructions in Table~\ref{tab:set-algs}. 
We already outlined SISA's general structure in~\cref{sec:overview}.
\fi


\subsection{Representation of Sets}
\label{sec:sisa-data-rep}

\cnf{The first key question is how to represent sets: SISA's ``first-class
citizens''.}
\tr{The first key question is how to represent sets: SISA's ``first-class
citizens''\footnote{For clarity, as already stated
in~\cref{sec:back-operations}, we refer to sets of vertices, but the
discussion is also applicable to edges.}}
We observe that -- in each graph algorithm -- there are {two}
fundamentally different classes of data structures. One class are \textbf{(1)
vertex neighborhoods} $N(v)$ that maintain the structure of the input graph.
There are $n$ such sets, their total size is $O(m)$, and each single
neighborhood is {static} (we currently focus on static graphs) and
{sorted} (following the established practice in graph
processing~\cite{malicevic2017everything}).
Another class are \textbf{(2) auxiliary structures}, for example $P$ in
Bron-Kerbosch (Listing~\ref{lst:bk}). These sets are used to maintain some
algorithmic state. They are usually {dynamic}, they may be
{unsorted}, their number (in a given algorithm) is usually a (small)
constant, and their total size is $O(n)$.
While SISA enables using any set representation for any specific set, we offer
certain recommendations to maximize performance. 

SAs should be used for {small} neighborhoods and DBs for the
{large} ones (in the evaluation, we vary the threshold so that 5\%-30\% 
largest neighborhoods use DBs). {This approach is memory
efficient. For example, for \mbox{$|N(v)| = n/2$}, a DB takes only \mbox{$n$} bits while an SA
uses \mbox{$16n$} bits (for a 32-bit word size).}

\enlargeSQ

\marginparX{\Large\vspace{4em}\colorbox{green}{\textbf{B}}}

Auxiliary sets benefit from being stored as dense bitvectors. This is because
such sets are often dynamic, and updates or removals take \mbox{$O(1)$} time.
As in practice there is usually a small constant number of such
sets in considered algorithms, the needed storage is not excessive, e.g.,
less than 3\% on top of a CSR for a graph with the average degree
100 (such as orkut), assuming using 32 threads and the Bron-Kerbosch algorithm, with auxiliary
sets \mbox{$P$}, \mbox{$X$}, and \mbox{$R$}
\sethlcolor{yellow}(the space complexity is \mbox{$O(T n)$} where $T$ is \#threads).
We analyze and confirm it for other algorithms and datasets.
For example, in SI, the storage complexity is \mbox{$(T n P)$} (where \mbox{$P$} is
the size of the subgraph), which is also negligible
in practice as $P$ is usually small.
\sethlcolor{yellow}To control space usage,
the user may pre-specify that, above a certain number of DBs, SISA starts to use SAs only.

\marginparX{\Large\vspace{-3em}\colorbox{yellow}{\textbf{B}}}

\sethlcolor{yellow}

\marginparX{\Large\vspace{2em}\colorbox{yellow}{\textbf{C}}}

The user controls selecting a set representation. For programmability, SISA
offers a predefined graph structure, where small and large neighborhoods are
\textbf{automatically} created (when a SISA program starts) as sparse arrays and dense bitvectors,
respectively. {A given neighborhood~\mbox{$N(v)$} is
stored as a DB whenever \mbox{$|N(v)| \ge t \cdot n$} (\mbox{$t \in (0;1)$} is a user
parameter that controls a ``bias'' towards using DBs or SAs) {and} it does
not exceed a storage budget limit set by the user (SISA by default uses a limit
of 10\% of the additional storage on top of the graph size when stored only
with SAs).}
For example, $t = 0.5$ indicates that each vertex connected to at least
50\% of all vertices has its neighborhood stored as a DB. 

\input{table-sisa.tex}

\sethlcolor{yellow}

Figure~\ref{fig:set_reps} shows an SA and a DB built from
the same vertex set. Then, it illustrates an example SISA
graph representation where some neighborhoods are DBs and some
are SAs.  

\ifconf
\begin{figure}[t]
\else
\begin{figure*}[t]
\fi
\centering
\ifconf
\else
\includegraphics[width=0.65\textwidth]{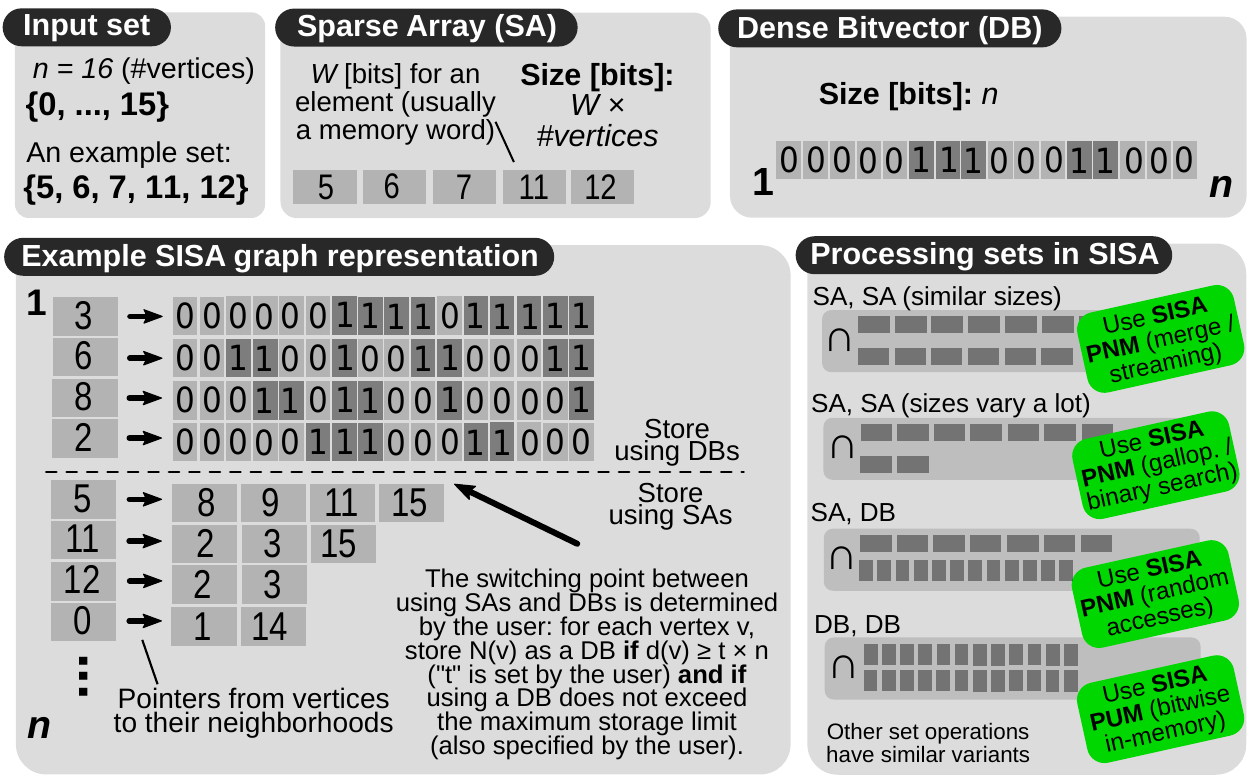}
\fi
\vspaceSQ{-1.5em}
\caption{SISA representations of sets and graphs, and processing SISA sets.}
\label{fig:set_reps}
\vspaceSQ{-1em}
\ifconf
\end{figure}
\else
\end{figure*}
\fi

\enlargeSQ


\subsection{High-Performance Set Operations}
\label{sec:set-algs-theory}

The second key challenge in SISA is how to apply set operations for 
highest performance.
For this, we detail the algorithmic aspects, a summary is in
Table~\ref{tab:set-algs}. 
HW details (used PIM 
and a performance model) are discussed in
Section~\ref{sec:sisa-implementation}.
An overview of the structure of SISA
is in Figure~\ref{fig:sisa-full}.
\tr{For each set \emph{operation} acting on sets $A$ and $B$, we
provide a number of \emph{variants} of this operation, where variants differ
based on how exactly sets $A$ and $B$ are represented.}
\all{One goal is to ensure that all these operations are \emph{fast}. The second
goal is to keep the number of such variants \emph{manageable}, i.e., avoid a
massive number of variants that would have to be provided if \emph{each}
possible combination of set representations was considered. For this, we
carefully examine set-centric formulations (\cref{sec:formulations}) and
existing related implementations, and we offer instructions that are crucial
for \emph{at least one considered graph algorithm}.}

\iftr
\subsubsection{Set Intersection $A \cap B$}

\fi
is a {key operation in SISA}, because
our analysis illustrates that it is used in {essentially all considered
graph algorithms}.
We now briefly discuss the most relevant variants of $\cap$, a summary
is in Figure~\ref{fig:set_reps}.

\begin{table*}[t]
%
\setlength{\tabcolsep}{2pt}
\renewcommand{\arraystretch}{1}
\centering
\scriptsize
\ifsq
\ssmall
\scriptsize
\else
\fi
\begin{tabular}{lcccccccc}
\toprule
& \makecell[c]{Triangle\\Counting~\cite{shun2015multicore}} & 
\makecell[c]{$k$-Clique\\Listing~\cite{danisch2018listing}} & 
\makecell[c]{$k$-Star-Clique\\Listing~\cite{jabbour2018pushing}} &
\makecell[c]{Maximal Cliques\\Listing~\cite{bron1973algorithm, DBLP:conf/isaac/EppsteinLS10}} & 
\makecell[c]{Link\\Prediction$^\text{\textdagger}$}  & 
\makecell[c]{Link\\Prediction$^\text{\textdaggerdbl}$} & 
\makecell[c]{Link\\Prediction$^\text{\textsection}$} & 
\makecell[c]{Jarvis-Patrick\\Clustering~\cite{jarvis1973clustering}} \\
\midrule
\makecell[l]{\textbf{SISA + merging} \textbf{intersection}} & $O(m c)^\text{\faStar}$ & $O(k m ({c}/{2})^{k-2})^\text{\faStar}$ & $O( k^2 m ({c}/{2})^{k-1})^\text{\faStar}$ & $O(c d n 3^{{c}/{3}})$ & $O(m d)$ & $O(n^2 + md)$ & $O(n^2)^\text{\faStar}$ & $O(md)$ \\ 
\makecell[l]{\textbf{SISA + galloping} \textbf{intersection}} & $O(mc \log c)$ & $O(k m ({c}/{2})^{k-2} \log c )$ & $O( k^2 m ({c}/{2})^{k-1} \log c ) $  & $O(c n 3^{{c}/{3}} )^{\text{\faStar}}$ & $O( m c \log c)^\text{\faStar}$ & $O(n^2 + mc \log c )^\text{\faStar}$ & $O(n^2)^\text{\faStar}$ & $O(mc\log d)^\text{\faStar}$ \\
\bottomrule
\end{tabular}
%
%
\caption{\textbf{The impact of set
intersection schemes (merging vs.~galloping) on the runtime of
graph mining algorithms.}. 
%
%
\textmd{``\faStar'' means that a given SISA variant matches asymptotically the
best known non-set-centric baseline, referenced in the top row. 
\all{``$\mathcal{B}$'' indicates that auxiliary
sets are represented as bit vectors (more details in Table~\ref{tab:set-algs}).}
$k$, $c$, and $d$ denote the size of the mined pattern, the graph degeneracy (a popular measure of graph sparsity)
and the maximum vertex degree, respectively (other symbols are described in
Section~\ref{sec:back-operations}).
Link prediction complexities are valid for the following vertex similarity
measures: 
$^\text{\textdagger}$Jaccard, Overlap, Adamic Adar, Resource Allocation, Common
Neighbors;
$^\text{\textdaggerdbl}$Total Neighbors;
$^\text{\textsection}$Preferential Attachment~\cite{leicht2006vertex,
neo4j_sim}.  }
}
\vspaceSQ{-2em}
\label{tab:theory-table}
\end{table*}

\all{We also conclude that the intersection of {almost any}
combination of SA, DB, and SB, has applications in graph mining. However, when
intersecting two SAs, at least one of them is always sorted in all the
considered algorithms. Thus, we do not consider intersecting two unsorted SAs.}

\begin{itemize}[noitemsep, leftmargin=0.5em]
\item \textbf{SA [sorted] $A$ $\cap$ SA [sorted] $B$ }
%
%
The intersection of two sorted SAs is commonly used when processing two
neighborhoods. It
comes in two ``flavors''. If $A$ and $B$ have similar
sizes ($|A| \approx |B|$), one prefers the \textbf{merge} scheme where one
simply iterates through $A$ and $B$, identifying common elements (time $O(|A| +
|B|)$). If one set is much smaller than the other ($|A| \ll |B|$), it is better
to use the \textbf{galloping} scheme~\cite{aberger2017emptyheaded}, in which
one iterates over the elements of a smaller set and uses a binary search to
check if each element is in the bigger set (time $O(|A| \log |B|)$). {SISA
offers both variants}, and a variant that automatically selects the best variant
with a performance model (described in~\cref{sec:perf_models}).
\item \textbf{SA [unsorted or sorted] $A$ $\cap$ DB $B$ }
Iterate over $A$ ($O(|A|)$) and check if each element is in $B$ ($O(1)$). 
This variant is often used to intersect a neighborhood with an
auxiliary set represented as a bitvector, for example $X \cap N(v)$ in
Listing~\ref{lst:bk}.
\item \textbf{DB $A$ $\cap$ DB $B$ }
Apply bitwise AND over both input DBs (they both have sizes of $n$
bits, giving $O(n/C)$ time, where $C$ is the maximum chunk of bits
that can be processed in $O(1)$~time using bit-level parallelism).
This variant is used for example when intersecting two dense neighborhoods.
\iftr
\item \textbf{SA [unsorted] $A$ $\cap$ SA [sorted] $B$ }
Iterate over $A$ ($O(|A|)$ time) and check if each element is in $B$ ($O(\log
|B|)$ time), for a total of $O(|A| \log |B|)$. This variant can be used to
intersect a sorted neighborhood and an auxiliary set that is implemented as an
unsorted SA (e.g., $P \cap N(v)$ in Bron-Kerbosch, see Listing~\ref{lst:bk}),
which is not uncommon in graph mining algorithms. 
\fi
\end{itemize}

\all{\noindent
Variants used less often, are described in the technical report.}


\iftr
\subsubsection{Set Union $A \cup B$, Set Difference $A \setminus B$}

\fi
$A \setminus B$ and $A \cup B$ have variants similar to those for $\cap$,
there are also corresponding merge and galloping variants.

\iftr
\subsubsection{Set Membership $x \in A$, Set Cardinality $|A|$}

\fi
Set membership takes $O(|A|)$ time for an unsorted SA (linear scan),
$O(\log|A|)$ time for a sorted SA (binary search), and $O(1)$ for a DB (a single
access to verify if $x$-th bit is set).
%
%
As for set cardinality, we keep $|A|$ for any set. This incurs
only $O(1)$ storage overhead (per set) as well $O(1)$ time overhead needed to
update the size, but it enables $O(1)$ time to resolve any set cardinality
operation.
Finally, SISA provides {dedicated instructions for computing
cardinalities of the results of set operations, for example $|A \cap B|$}. This
enables speedups as SISA avoids creating any intermediate structures needed for
keeping the results of operations such as intersection.

\enlargeSQ

\iftr
\subsubsection{Adding and Removing Elements}

\fi
Auxiliary sets often grow and shrink by one element. 
Both add and remove straightforwardly take $O(1)$ time for a DB (setting or
zeroing a corresponding bit) and $O(|A|)$ for an SA (moving data for a sorted SA).
Thus, in general, we advocate using DBs for auxiliary sets;
the size is $n$ bits.

\vspaceSQ{-0.25em}
\subsection{Additional Details of SISA Design}
\label{sec:sisa-details}

We detail several aspects of SISA's design; cf.~Figure~\ref{fig:sisa-full}.

\marginpar{\Large\vspace{2em}\colorbox{yellow}{\textbf{L}}}

\iftr
\subsubsection{Labeled Graphs}
\fi
%
\hl{As a baseline, we propose to use a sparse array to maintain labels, indexed by
vertex IDs, similarly to other works~\mbox{\cite{cordella2004sub}}. This form
benefits from SISA-PNM. The SISA user can also implement labels with a one-hot
encoding and use bit vectors. This would harness SISA-PUM.}

\iftr
\subsubsection{SISA Instructions}
\fi
SISA offers {instructions} that package the described set
operations in all the considered variants, including instructions that automatically
select merge or galloping set algorithms (cf.~\cref{sec:set-algs-theory}).
\all{Moreover, to facilitate development, we also provide separate instructions that
automatically select the best operation variant, using a pre-defined value that
we propose based on our evaluation (these values can also be influenced by the
developer). For example, a dedicated ISA
instruction for intersecting two SAs automatically decides on 
merge or galloping.}
Finally, SISA also provides instructions for creating and deleting sets.

\all{
\textbf{Low-Level SISA Instructions}
\ 
Low-level instructions manage SISA drivers (in case of the OS
support) and can directly access HW units such as SCU. We omit a detailed
specification as these instructions {strictly depend on potential 
vendor-specific HW implementations}. 
}


\iftr
\subsubsection{Programming Interface (Set Iterators \& Wrappers)}

\fi
For programmability, SISA offers a thin software layer on top of high-level
instructions that consists of {abstractions} and {wrappers}. In the
former, we provide an opaque type \texttt{Set} that is a reference to a SISA
set; this enables using C++ {iterators over sets}, see left side of
Figure~\ref{fig:sisa-full}. In the latter, SISA provides functions that
directly map to SISA set instructions. 
\tr{Function parameters determine an
instruction variant.}



\iftr
\subsubsection{Set Identification}

%
SISA {identifies sets} with unique {logical set IDs}. These IDs are mapped
by the underlying SISA HW design to any used form of \emph{physical} addresses.
\all{The mappings can be pointers, hash table indices, or offsets.}
\fi

\begin{figure}[h]
\vspaceSQ{-1em}
\centering
\includegraphics[width=1.0\columnwidth]{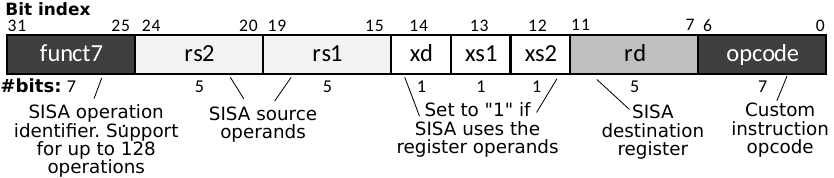}
\vspaceSQ{-2em}
\caption{Encoding of SISA instructions.}
\vspaceSQ{-1em}
\label{fig:sisa-code}
\end{figure}

\iftr
\subsubsection{RISC-V Compliant Encoding}

\fi
SISA can be integrated with the RISC-V ISA~\cite{waterman2016design}.  To
enable modularity and flexibility, SISA's new instructions are encoded using
the {custom opcode set}~\cite{waterman2011risc}. We encode the opcode and
functionality of custom RISC-V instructions using bits [31..25] and [6..0], see
Figure~\ref{fig:sisa-code}. The former represent the different SISA
instructions (up to 128). The latter are set to 0x16 to represent the custom
characteristic of the instruction. Fields \texttt{rs1}, \texttt{rs2}, and
\texttt{rd} indicate registers with IDs of input sets and the output set,
respectively.
In Table~\ref{tab:set-algs}, we assign ISA codes (bits [31..25]) to respective
instructions. The number of SISA instructions is less than 20, leaving space
for potential new variants.

%% file: table-sisa.tex
\all{\begin{table}[b]
%
\setlength{\tabcolsep}{2pt}
\renewcommand{\arraystretch}{0.6}
\centering
\ssmall
\sf
\begin{tabular}{llllcllll}
\toprule
\textbf{Code} & 
\textbf{Set op.} &
\makecell[c]{\textbf{$A$ and $B$}\\ \textbf{represent.}} & 
\makecell[c]{\textbf{Set algorithm}} &
\textbf{S?} &
\makecell[c]{\textbf{Time}} &
\makecell[c]{\textbf{Input size}\\\textbf{[bits]}} &
\makecell[l]{\textbf{Executed by}\\ \textbf{(cf.~Section~\ref{sec:sisa-implementation})}} \\ 
\midrule
\texttt{0x0} & $A \cap B$ & SA $ \cap $ SA & Merge & \faThumbsOUp, \faThumbsOUp  & $O(|A| + |B|)$ & $W|A| + W|B|$  & LL (HBM, HMC)  \\ 
\texttt{0x1} & $A \cap B$ & SA $ \cap $ SA & Galloping & \faThumbsOUp, \faThumbsOUp  & $O(|A| \log|B|)$ & $W|A| + W|B|$  & LL (HBM, HMC)  \\ 
\texttt{0x2} & $A \cap B$ & SA $ \cap $ SA & \makecell[l]{Merge/gallop.} & \faThumbsOUp, \faThumbsOUp & \makecell[l]{cf.~\texttt{0x0} \& \texttt{0x1}} & $W|A| + W|B|$  & LL (HBM, HMC) \\ 
\texttt{0x3} & $A \cap B$ & SA $ \cap $ DB & Galloping & \faThumbsOUp, na & $O(|A|)$ & $W|A| + n$  & LL (HBM, HMC)  \\ 
\texttt{0x4} & $A \cap B$ & DB $ \cap $ DB & Bitwise AND & na, na & $O(n/\sum_i S_i)$ & $n+n$  & In-situ PIM (Ambit)  \\ 
\midrule
\texttt{0x5} & $A\cup\{x\}$ & DB $\cup$ $\{x\}$ & Set bit & na, na & $O(1)$ & $n + W$ & LL (HBM, HMC)  \\
\texttt{0x6} & $A\setminus\{x\}$ & DB $\setminus$ $\{x\}$ & Clear bit & na, na & $O(1)$ & $n + W$ & LL (HBM, HMC) \\
\bottomrule
\end{tabular}
\vspace{-0.5em}
\caption{
Overview of \ul{selected} SISA instructions, \textmd{\ul{one row describes one SISA instruction (one specific set operation variant)}.
\ul{Set elements are vertices} ($A,B \subseteq V, x \in V$).
``\faThumbsOUp'' means ``yes''. ``\faThumbsDown'' means ``no''.
\textbf{``na''} means ``not applicable''.
\textbf{``Code''} is a proposed instruction opcode.
\textbf{``S (Sorted)''} indicates if an instruction assumes set representations of $A$ and $B$ to be sorted.
\all{\textbf{``I?'' (immutable)} indicates if an instruction does not modify input sets.}
\all{\textbf{``Time''} is a (sequential) time complexity oblivious of the set organization.}
}
}
\vspace{-1.5em}
\label{tab:set-algs}
\end{table}
}

\all{
\begin{table}[b]
%
\setlength{\tabcolsep}{1pt}
\renewcommand{\arraystretch}{0.6}
\centering
\ssmall
\sf
\begin{tabular}{llllclllll}
\toprule
\textbf{Code} & 
\textbf{Set op.} &
\makecell[c]{\textbf{$A$ and $B$}\\ \textbf{represent.}} & 
\makecell[c]{\textbf{Set algorithm}} &
\textbf{S?} &
\makecell[c]{\textbf{Time}} &
\makecell[c]{\textbf{Input size}\\\textbf{[bits]}} &
\makecell[l]{\textbf{Executed by}\\ \textbf{(cf.~Section~\ref{sec:sisa-implementation})}} &
\makecell[l]{\textbf{Dominating form}\\ \textbf{of data transfer}} \\
\midrule
\texttt{0x0} & $A \cap B$ & SA $ \cap $ SA & Merge & \faThumbsOUp, \faThumbsOUp  & $O(|A| + |B|)$ & $W|A| + W|B|$  & LL (HBM, HMC) & Streaming \\ 
\texttt{0x1} & $A \cap B$ & SA $ \cap $ SA & Galloping & \faThumbsOUp, \faThumbsOUp  & $O(|A| \log|B|)$ & $W|A| + W|B|$  & LL (HBM, HMC) & Random accesses \\ 
\texttt{0x2} & $A \cap B$ & SA $ \cap $ SA & \makecell[l]{Merge/gallop.} & \faThumbsOUp, \faThumbsOUp & \makecell[l]{cf.~\texttt{0x0} \& \texttt{0x1}} & $W|A| + W|B|$  & LL (HBM, HMC) & cf.~\texttt{0x0} \& \texttt{0x1} \\ 
\texttt{0x3} & $A \cap B$ & SA $ \cap $ DB & Galloping & \faThumbsOUp, na & $O(|A|)$ & $W|A| + n$  & LL (HBM, HMC) & Random accesses \\ 
\texttt{0x4} & $A \cap B$ & DB $ \cap $ DB & Bitwise AND & na, na & $O(n/\sum_i S_i)$ & $n+n$  & In-situ PIM (Ambit) & In-situ row copies  \\ 
\midrule
\texttt{0x5} & $A\cup\{x\}$ & DB $\cup$ $\{x\}$ & Set bit & na, na & $O(1)$ & $n + W$ & LL (HBM, HMC) & Random access  \\
\texttt{0x6} & $A\setminus\{x\}$ & DB $\setminus$ $\{x\}$ & Clear bit & na, na & $O(1)$ & $n + W$ & LL (HBM, HMC) & Random access \\
\bottomrule
\end{tabular}
\vspace{-0.5em}
\caption{
Overview of \ul{selected} SISA instructions, \textmd{\ul{one row describes one SISA instruction (one specific set operation variant)}.
\ul{Set elements are vertices} ($A,B \subseteq V, x \in V$).
``\faThumbsOUp'' means ``yes''. ``\faThumbsDown'' means ``no''.
\textbf{``na''} means ``not applicable''.
\textbf{``Code''} is a proposed instruction opcode.
\textbf{``S (Sorted)''} indicates if an instruction assumes set representations of $A$ and $B$ to be sorted.
\all{\textbf{``I?'' (immutable)} indicates if an instruction does not modify input sets.}
\all{\textbf{``Time''} is a (sequential) time complexity oblivious of the set organization.}
}
}
\vspace{-1.5em}
\label{tab:set-algs}
\end{table}
}

\all{
\begin{table*}[b]
\vspace{1.5em}
\setlength{\tabcolsep}{1.4pt}
\renewcommand{\arraystretch}{0.8}
\centering
\ssmall
\begin{tabular}{llllcllllll}
\toprule
\textbf{Code} & 
\textbf{Set op.} &
\makecell[c]{\textbf{$A$ and $B$}\\ \textbf{represent.}} & 
\makecell[c]{\textbf{Set algorithm}} &
\textbf{S?} &
\makecell[c]{\textbf{Time}} &
\makecell[c]{\textbf{Input size}\\\textbf{[bits]}} &
\makecell[l]{\textbf{Main form of data}\\ \textbf{transfer (\cref{sec:perf_models})}} &
\makecell[l]{\textbf{Executed by}\\ \textbf{(details in Section~\ref{sec:sisa-implementation})}} &
\makecell[l]{\textbf{Remarks, guidelines}} \\
\midrule
\texttt{0x0} & $A \cap B$ & SA $ \cap $ SA & Merge & \faThumbsOUp, \faThumbsOUp  & $O(|A| + |B|)$ & $W|A| + W|B|$ & Streaming & Logic layers (HBM, HMC) & The \textbf{merge} scheme, used best when $|A|\approx|B|$ \\ 
\texttt{0x1} & $A \cap B$ & SA $ \cap $ SA & Galloping & \faThumbsOUp, \faThumbsOUp  & $O(|A| \log|B|)$ & $W|A| + W|B|$ & Random accesses  & Logic layers (HBM, HMC) & The \textbf{galloping} scheme, used best when $|A| \ll |B|$\\ 
\texttt{0x2} & $A \cap B$ & SA $ \cap $ SA & \makecell[l]{Merge / gallop.} & \faThumbsOUp, \faThumbsOUp & \makecell[l]{cf.~\texttt{0x0} and \texttt{0x1}} & $W|A| + W|B|$ & cf.~\texttt{0x0} and \texttt{0x1} & Logic layers (HBM, HMC) & \makecell[l]{Automatic selection (\textbf{performance model based}) of {merge}/{galloping}} \\ 
\texttt{0x3} & $A \cap B$ & SA $ \cap $ DB & Galloping & \faThumbsOUp, na & $O(|A|)$ & $W|A| + n$ & Random accesses & Logic layers (HBM, HMC) & --- \\ 
\texttt{0x4} & $A \cap B$ & DB $ \cap $ DB & Bitwise AND & na, na & $O(n/(q S))$ & $n+n$ & In-situ row copies  & In-situ PIM (Ambit) & Using subarray- and row-level parallelism  \\ 
\midrule
\texttt{0x5} & $A\cup\{x\}$ & DB $\cup$ $\{x\}$ & Set bit & na, na & $O(1)$ & $n + W$ & Random access & Logic layers (HBM, HMC) & --- \\
\texttt{0x6} & $A\setminus\{x\}$ & DB $\setminus$ $\{x\}$ & Clear bit & na, na & $O(1)$ & $n + W$ & Random access & Logic layers (HBM, HMC) & --- \\
\iftr
\midrule
\texttt{0x5} & $A \cup B$ & SA $ \cap $ SA & na, na & \faThumbsOUp & ? & ? & ? & ? &  \\ 
\texttt{0x5} & $A \cup B$ & SA $ \cap $ SA & na, na & \faThumbsOUp & ? & ? & ? & ? &  \\ 
\texttt{0x5} & $A \cup B$ & SA $ \cap $ SA & na, na & \faThumbsOUp & ? & ? & ? & ? &  \\ 
\texttt{0x5} & $A \cup B$ & SB $ \cap $ SB & na, na & \faThumbsOUp & ? & ? & ? & ? &  \\ 
\texttt{0x1} & $A \cap B$ & Galloping & [S] $ \cap $ [S] & \faThumbsOUp & \faThumbsOUp & Integer arrays & \faThumbsOUp & \faThumbsOUp & $O(|A|\log|B|)^*$ & $O(|A| + |B|)$ & Used when $|A| \ll |B|$. $^*$Here, we assume $|A| \le |B|$  \\ 
& $A \cup B$ & & \texttt{0x4} & Merge & \faThumbsOUp\ ($A$ and $B$) & \faThumbsOUp & \faThumbsOUp & Integer arrays & \faThumbsOUp & \faThumbsOUp & $O(|A|+|B|)$ & $O(|A|+|B|)$ & Used with sets stored as sorted sparse arrays \\
& $A \cup B$ & & \texttt{0x5} & Concatenate & Irrelevant & \faThumbsDown & \faThumbsDown & Integer arrays & \faThumbsOUp & \faThumbsOUp & $O(1)$ & $O(|A|+|B|)$ & Used in Bron-Kerbosch~(\cref{sec:bk}) \\
& $x \in A$ & & \texttt{0x6} & Unstr.~search & \faThumbsDown & \faThumbsOUp & n/a & Integer array & \faThumbsOUp & \faThumbsOUp & $O(|A|)$ & $O(|A|)$ & Used if sorting is expensive\\
& $x \in A$ & & \texttt{0x7} & Binary search & \faThumbsOUp & \faThumbsOUp & n/a & Integer array & \faThumbsOUp & \faThumbsOUp & $O(\log|A|)$ & $O(|A|)$ & Used with sets stored as sorted sparse arrays \\
& $x \in A$ & & \texttt{0x8} & Lookup & n/a & \faThumbsOUp & n/a & Bit vector & \faThumbsDown & \faThumbsOUp & $O(1)$ & $O(n)$ & Used for fast lookups at the cost of $O(n)$ bits of storage \\
& $|A|$ & & \texttt{0x9} & Lookup & Irrelevant & \faThumbsOUp & n/a & Integer array & \faThumbsOUp & \faThumbsOUp & $O(1)$ & $O(|A|)$ & $O(1)$ time thanks to $A$'s base-and-bound organization \\ 
& $A\cup\{x\}$, $A\setminus\{x\}$ & & \texttt{0xA, 0xD} & Splitting & \faThumbsOUp & \faThumbsDown & \faThumbsOUp & Integer array & \faThumbsOUp & \faThumbsDown & $O(\log|A|)$ & $O(|A|)$ & Faster than shifting but may cause fragmentation \\
& $A\cup\{x\}$, $A\setminus\{x\}$ & & \texttt{0xB, 0xE} & Shifting & \faThumbsOUp & \faThumbsDown & \faThumbsOUp & Integer array & \faThumbsOUp & \faThumbsOUp & $O(|A|)$ & $O(|A|)$ & Slower than splitting but causes no fragmentation \\
& $A\cup\{x\}$, $A\setminus\{x\}$ & & \texttt{0xC, 0xF} & Set or zero bit & n/a ($A$) & \faThumbsDown & \faThumbsOUp & Bit vector ($A$) & \faThumbsDown & \faThumbsOUp & $O(1)$ & $O(n)$ & Faster than splitting or shifting; may take more storage \\
\midrule
\multicolumn{14}{l}{\makecell[l]{\textbf{Other Instructions ($\cdot$=):} \ul{$A$ $\cap$= $B$} (\texttt{0x10} -- \texttt{0x11}), \ul{$A$
$\setminus$= $B$} (\texttt{0x12} -- \texttt{0x13}), \ul{$A$ $\cup$= $B$}
(\texttt{0x14} -- \texttt{0x15}), \ul{$A$ $\cup$= $\{x\}$} (\texttt{0x16} --
\texttt{0x18}), \ul{$A$ $\setminus$= $\{x\}$} (\texttt{0x19} -- \texttt{0x1B}): 
these instructions are mutable (with $A$).\\Otherwise, they are identical
to their corresponding immutable variants above.
Successive instruction codes match the above variants, e.g., \texttt{0x16} -- \texttt{0x17} correspond to Splitting and Shifting}} \\
\midrule
\multicolumn{14}{l}{\makecell[l]{\textbf{Other Instructions ($|\cdot|$):} \ul{$|A
\cap B|$} (\texttt{0x1C} -- \texttt{0x1D}), \ul{$|A \setminus B|$} (\texttt{0x1E} --
\texttt{0x1F}), \ul{$|A \cup B|$} (\texttt{0x20} -- \texttt{0x21}): these
instructions do not allocate output sets. Otherwise, they are
identical to their corresponding\\non-cardinality-related variants above.
Successive instruction codes match the above variants, e.g., \texttt{0x20} -- \texttt{0x21} correspond to Merge and Concatenate.
}} \\
\fi
\bottomrule
\end{tabular}
\vspace{-0.5em}
\caption{
\textbf{Overview of SISA instructions}, one row describes one specific set operation variant.
{Set elements are vertices} ($A,B \subseteq V, x \in V$).
``\faThumbsOUp'' means ``yes''. ``\faThumbsDown'' means ``no''.
\textbf{``na''} means ``not applicable''.
\textbf{``Code''} is a proposed instruction opcode.
\textbf{``S (Sorted)''} indicates if an instruction assumes set representations of $A$ and $B$ to be sorted.
\all{\textbf{``I?'' (immutable)} indicates if an instruction does not modify input sets.}
\all{\textbf{``Time''} is a (sequential) time complexity oblivious of the set organization.}
\tr{Symbols: $l_M, b_M$: latency and bandwidth of accessing DRAM (from logic
layers), $b_L$: bandwidth between cores within a logic layer, $l_I$: latency of
executing a single bulk bitwise operation using in-situ PIM.  $S$: size [bits]
of a single DRAM sub-array.  $q$: number of sub-arrays that can be processed in
parallel (e.g., within a single DRAM bank).}
}
\vspace{-3.5em}
\label{tab:set-algs}
\end{table*}
}

\marginparX{\Large\vspace{3em}\colorbox{yellow}{\textbf{D}}}

\ifconf
\begin{table}[h]
\else
\begin{table*}[t]
\fi
\vspaceSQ{-1em}
\ifsq\renewcommand{\arraystretch}{0.8}\fi
\centering
\ssmall
\iftr
\scriptsize
\fi
\begin{tabular}{llllclll}
\toprule
\textbf{ins} & 
\textbf{Set op.} &
\makecell[c]{\textbf{$A$ and $B$}\\ \textbf{represent.}} & 
\makecell[c]{\textbf{Set}\\ \textbf{algorithm}} &
\textbf{S?} &
\makecell[l]{\textbf{Time}\\ \textbf{complexity}} &
\makecell[c]{\textbf{Input size}\\\textbf{[bits]}} &
\makecell[l]{\textbf{Main form of data}\\ \textbf{transfer (\cref{sec:perf_models})}} \\ 
\midrule
\texttt{0x0} & $A \cap B$ & SA $ \cap $ SA & Merge & \faThumbsOUp, \faThumbsOUp  & $O(|A| + |B|)$ & $W|A| + W|B|$ & Streaming \\ 
\texttt{0x1} & $A \cap B$ & SA $ \cap $ SA & Galloping & \faThumbsOUp, \faThumbsOUp  & $O(|A| \log|B|)$ & $W|A| + W|B|$ & Random accesses \\ 
\texttt{0x2} & $A \cap B$ & SA $ \cap $ SA & \makecell[l]{Merge vs.\\ gallop.} & \faThumbsOUp, \faThumbsOUp & \makecell[l]{cf.~\texttt{0x0} and \texttt{0x1}} & $W|A| + W|B|$ & cf.~\texttt{0x0} and \texttt{0x1} \\ 
\texttt{0x3} & $A \cap B$ & SA $ \cap $ DB & Galloping & \faThumbsOUp, na & $O(|A|)$ & $W|A| + n$ & Random accesses \\ 
\texttt{0x4} & $A \cap B$ & DB $ \cap $ DB & Bitwise AND & na, na & $O(n/(q S))$ & $n+n$ & In-situ row copies \\ 
\midrule
\texttt{0x5} & $A\cup\{x\}$ & DB $\cup$ $\{x\}$ & Set bit & na, na & $O(1)$ & $n + W$ & Random access \\
\texttt{0x6} & $A\setminus\{x\}$ & DB $\setminus$ $\{x\}$ & Clear bit & na, na & $O(1)$ & $n + W$ & Random access \\
\bottomrule
\end{tabular}
%
%
\caption{
\textbf{Overview of SISA instructions}, one row describes one specific set operation variant.
{Set elements are vertices} ($A,B \subseteq V, x \in V$).
``\faThumbsOUp'' means ``yes''.
\textbf{``na''} means ``not applicable''.
\textbf{``ins''} is a proposed instruction opcode.
\textbf{``S (Sorted)''} indicates if an instruction assumes set representations of $A$ and $B$ to be sorted (thus two columns).
}
%
%
\label{tab:set-algs}
\ifconf
\end{table}
\else
\end{table*}
\fi

%% file: theory.tex
\section{Theoretical Analysis}
\label{sec:theory-s}


We now show that SISA-enhanced 
algorithms are \emph{theoretically efficient}, i.e., their 
time complexities match those of hand-tuned graph mining algorithms.
This is enabled by SISA's ability to control used set
representations and set operations\tr{, facilitates tuning performance and storage
tradeoffs}.
\all{One of SISA key benefits is that, unlike paradigms such as the vertex-centric
model, SISA programs offer competitive time complexities. This is due to two
aspects. First, as the starting point of a SISA program is a selected
\emph{specific algorithm}, the general performance properties of this algorithm
are usually preserved. Second, the generality of set algebra, combined with
SISA's ability to control used set representations and operations, facilitates
tuning performance and storage tradeoffs.}
To show this, we analyze how varying a used set intersection variant (merge
vs.~galloping) impacts the runtime of set-centric algorithms, see Table~\ref{tab:theory-table}. We focus on
intersection as it is prevalent in considered algorithms. Crucially, \emph{all set-centric variants
are able to match the competitive time complexities of considered tuned graph
mining algorithms}.

\all{
Interestingly, despite the fact that SISA does not focus on algorithms such as
BFS, we illustrate that one can still develop a work-efficient set-centric BFS.
A detailed listing of such a BFS is in the extended report due to space
constraints. Intuitively, the key part of such a BFS is treating the frontier
as a set stored as a DB, and intersect it with respective neighborhoods using
the galloping variant.}

\iftr
\subsection{Parametrization with Degeneracy}

We parametrize complexities with degeneracy~$c$, a well-known measure
of graph sparsity~\cite{matula1983smallest}. The degeneracy~$c$ of a
graph $G$ is the smallest number~$x$ such that {every} subgraph in $G$ has
a vertex of degree {at most} $x$ (i.e., every subgraph has at least
one sparsely connected vertex).
Different graphs have constant degeneracy, such as planar
graphs~\cite{lick_white_1970}, certain scale-free
graphs~\cite{barabasi1999scaleFree, DBLP:conf/soda/EdenLR18}, and graphs of
bounded treewidth~\cite{DBLP:conf/soda/EdenLR18}.
We consider degeneracy as it is used by many recent graph mining algorithms to
enhance their time complexities~\cite{DBLP:journals/jgaa/ZhouN99,
DBLP:conf/isaac/EppsteinLS10, danisch2018listing}.
This is because several of the investigated graph algorithms {orient}
the graph edges according to the {degeneracy
order}~\cite{DBLP:journals/jgaa/ZhouN99,DBLP:conf/isaac/EppsteinLS10}. The
purpose of this is to (1) make the graph acyclic (2) make the out-degree as
small as possible. The smallest out-degree in a degeneracy ordering is the
{degeneracy} $c$ of the graph. 

\fi

%
\tr{By definition, the degeneracy is always less than the maximum degree: $c\leq
d$. In \Cref{tab:theory-table}, we express the bounds parameterized by $c$ and
$d$. To get worst-case bounds that hold for all $d$ and $c$, one can replace
$d$ by $n$ and $c$ by $\sqrt{m+n}$ (as $c$ also satisfies $c \leq
\sqrt{2m+n}+1$~\cite{DBLP:journals/jgaa/ZhouN99,
DBLP:journals/siamcomp/ChibaN85}).}
\tr{Note that the difference between the maximum degree $d$ and the degeneracy $c$
can be up to $n-1$: For example, a star graph has maximum degree $n-1$, but
degeneracy $1$.}

\iftr

The following observations follow directly form the definitions or the cited
literature:

\begin{observation}[\cite{DBLP:journals/jgaa/ZhouN99, DBLP:journals/siamcomp/ChibaN85}]\label{obs:deg-min-bound}
	For a graph $G=(V, E)$ with degeneracy $c$, $\sum_{(u, v)\in E} \min(d(u), d(v)) \leq 4cm$.
\end{observation}

\begin{observation}\label{obs:basic-degree-counting}
	For every graph $G=(V, E)$, we have that $\sum_{(u, v)\in E} \left( d(u) + d(v) \right ) = \sum_{i\in V} d(i)^2 \leq m d$.
\end{observation}

\begin{observation}\label{obs:degeneracy-order}
	For a graph $G=(V, E)$ directed according to its degeneracy ordering,\\ 
  $$\sum_{(u, v)\in E} \left( |N^{+}(u)| + |N^{+}(v)| \right ) \leq mc.$$
\end{observation}
\normalsize

\subsection{Derivations of Bounds}


\fi

\iftr

\noindent
Next, we discuss proving the bounds in \Cref{tab:theory-table}.

\macb{Triangle Counting} The algorithm iterates over all edges and performs an
intersection of the out-neighbor sets of the two endpoints. If Galloping is
used, the cost is $$O\left(\sum_{(u, v)\in E} \min(|N^{+}(u)|, |N^{+}(v)|)\log c \right) =
O(mc \log c),$$ by Observation~\ref{obs:deg-min-bound}. If Merging is used, the cost is
$$\sum_{(u, v)\in E} \left( |N^{+}(u)| + |N^{+}(v)| \right ) = O(mc)$$ by
Observation~\ref{obs:degeneracy-order}.

\macb{$k$-Clique Listing} Algorithm~\ref{lst:kcls} has the same
cost as the Edge-Parallel algorithm when using Merging. With Galloping, 
intersections take an additional factor $\log c$.


\sloppy
\macb{$k$-Clique-Star Listing} The algorithm computes all $k+1$ cliques and
then performs $O(k^2)$ work per clique. There are at most $O(m(c/2)^{k-1})$
cliques of $k+1$ vertices (as testified by the algorithm that lists them in
$O(k m(c/2)^{k-1})$ time~\cite{danisch2018listing}).


\macb{Jarvis-Patrick Clustering} Jarvis Patrick Clustering (for any of the
vertex-similarity measures from Algorithm~\ref{lst:sim}) iterates over
edges and perform a set intersection on the neighbors. The difference to
triangle counting is that the graphs are not oriented according to a degeneracy
ordering. This changes the cost for the Galloping approach to $O(mc \log d)$
(where only the term in the logarithm changes as we may need to search
sets of size up to $d$ instead of only up to $c$). The Merge
approach costs $\sum_{(u, v)\in E} \left( d(u) + d(v) \right )=O(md)$, by
Observation\ref{obs:basic-degree-counting}.
Similar discussions apply to macb{Link Prediction}.

\ifall

\macb{Link Prediction} Here, we iterate over all edges of an
"almost complete" graph with edges from $V\times V \setminus E'$ and compute the
similarity of the vertices (based on the original graph with edges $E$).
Similarity measures $\#5, \#7, \#9-\#11$ in \Cref{lst:sim} are
trivially $0$ for all vertex pairs that are not neighbors in the original
graph. Hence, it suffices to iterate over those edges. Similarly as for
Jarvis-Patrick Clustering, the Merge approach costs $O(m d)$ while 
the Galloping approach costs $O(mc \log d)$. 

For measure $\#12$, where $S(u, v) = |N(u) \cup N(v)| = |N(u)| +
|N(v)| - |N(u)\cap N(v)|$, one needs to iterate over all pairs of vertices, as
all vertices could have nonzero similarity. Yet, the term $|N(u)\cap N(v)|$
still only needs to be computed for neighboring $u$ and $v$. Hence, the cost is
$O(n^2 + md)$ when using Merging and $O(n^2 + mc \log d)$ when using Galloping.

For measure $\#13$, the cost to compute the similarity is constant,
given the degrees of each vertex. Thus, the cost is $O(n^2)$. 

Similarity measure $\#14$ counts for each edge the number of $4$-cycles that
contain it. Even though there can be $\Omega(n^2)$ triangles even in a planar
graph~\cite{DBLP:journals/siamcomp/ChibaN85}, they can be represented compactly
in $O(m)$ space with a $O(mc)$ time
algorithm~\cite{DBLP:journals/siamcomp/ChibaN85}. The representation is such
that we can count the number of $4$-cycles that contain each edge in $O(m)$
time.~\lukas{The connection to our setting is not obvious} 

\fi

%
\sloppy
\macb{Maximal Cliques} Computing maximal cliques takes $O(cn 3^{c/3})$
time when both good pivoting and the degeneracy ordering are
used~\cite{DBLP:conf/isaac/EppsteinLS10}. If appropriate pivoting is used
(without the degeneracy ordering), the runtime is
$O(3^{n/3})$~\cite{DBLP:journals/tcs/TomitaTT06}. 
%
%
Using Merging for intersections causes the cost of each
iteration to depend on the maximum degree $d$ in the original graph. This
does not suffice to obtain the desired
bounds~\cite{DBLP:conf/isaac/EppsteinLS10} and introduces an overhead of a
factor $d$ compared to Eppstein's~\cite{DBLP:conf/isaac/EppsteinLS10} approach.
If the sets $X$ and $P$ are stored as dense bit vectors (and the pivot vertex is chosen efficiently enough), then the runtime matches that of the original formulation.


\ifall
\macb{BFS}
In BFS we use a dense bit vector to store the frontier $F$ and the set of unexplored vertices $\Pi$.
This results in a costs equal to those in traditional tuned BFS
implementations~\cite{besta2017push}.
\fi

\ifall
\maciej{fix}
\macb{BFS} Using Merging to implement the intersection that is used to
construct the next frontier yields a cost of $\Theta(n^2)$. For the upper
bound, note that every node is in the frontier for at most one iteration of the
outer while-loop and it will be used once to construct a new frontier, the cost
of which is $O(n)$. For the lower bound, consider a path with $n$ vertices and
a BFS that starts with one of the endpoints of the path. The frontier will
always consist of a single node, and the size of the set of not-yet-visited
vertices decreases by one in each iterations. Hence, the cost of iteration $i$
is at least $n-i$ and the overall cost is $\Omega(\sum_{i=1}^n (n-i)) =
\Omega(n^2)$.

Using Galloping to implement the intersection gives a runtime of $O(m\log n)$.
This is because the cost of constructing the new frontier using a vertex $u$
that is in the frontier is bounded by $O(d(u) \log n)$. Moreover, each vertex
is involved in one such computation overall. Therefore, the overall cost is
$O(\sum_{v \in V} d(u) \log n) = O(m \log n)$.
\fi

\all{
\begin{table}[t]
%
\setlength{\tabcolsep}{3.6pt}
\renewcommand{\arraystretch}{1.5}
\centering
 \ssmall
\begin{tabular}{lcccccc}
\toprule
 & \makecell[c]{Triangle Counting} & $2$-clique-star & \makecell[c]{Link Prediction \\ \#5, \#7, \#9-\#11}  & \makecell[c]{Jarvis-Patrick\\ Clustering} & \\
\midrule
\makecell[l]{Merging} & \highlight{\hspace{0.4em}$O(m^{3/2})$} & $O(m n)$ & $O(n^3)$ & $O(mn)$ &  \\ 

\makecell[l]{Galloping} & $O(m^{3/2} \log n)$ & \highlight{\hspace{0.4em}$O(m^{3/2} \log n)$} & \highlight{\hspace{0.4em}$O( n^2 \sqrt{m} \log n)$} & \highlight{\hspace{0.2em} $O(m^{3/2}\log n)$} &  \\

\bottomrule
\end{tabular}
%
\caption{Impact of set intersection implementations (merging vs. galloping) on
the runtime of various graph algorithms \lukas{Non-parametric version of
\Cref{tab:theory-table} }. The term which is asymptotically smaller is\ 
~~\tikzmarkin[set fill color=vlgray, set border color=white, above offset=0.27,
below offset=-0.1]{mot3}\textcolor{black}{highlighted}\tikzmarkend{mot3}~ in
grey.
}
\label{tab:theory-table-nonparam}
\end{table}

}

\fi


\tr{
\textbf{Theoretical Analysis: Key Takeaway}
\ 
No single set operation variant is best for each graph problem. However,
appropriately choosing a set operation variant \emph{enables the set-centric
approach to approach or match the runtime of fast specific algorithms} for the
considered problems.
}

%% file: isa-implementation.tex
\section{Hardware Implementation}
\label{sec:sisa-implementation}

\enlargeSQ

\iftr
We now discuss details of SISA hardware implementation.

\subsection{Processing-In-Memory for Sets}

We start with how SISA uses PIM for set operations.
\fi

\textbf{SISA-PUM}
\
First, the {intersection}, {union}, and {difference} of sets represented as DBs
are processed with SISA-PUM that relies on {in-situ DRAM bulk bitwise} schemes.
{For concreteness, we pick
Ambit~\mbox{\cite{seshadri2017ambit}}, a recent design that enables
energy-efficient bulk bitwise operations fully inside DRAM, by small extensions
to the DRAM circuitry but without any changes to the DRAM interface.
}\sethlcolor{yellow}{However, SISA is generic and}{
  other designs could also be used (e.g., ELP2IM~\mbox{\cite{xin2020elp2im}},
  DRISA~\mbox{\cite{li2017drisa}},
  ComputeDRAM~\mbox{\cite{gao2019computedram}}, PCM
  (Pinatubo)~\mbox{\cite{li2016pinatubo}}).} The key extension in Ambit (for
  in-situ processing) is to modify a decoder for {three selected} DRAM rows
  (that share the same set of sense amplifiers) in such a way that one
  amplifier connects directly to three DRAM cells. 
%
%
This enables logical AND and OR over two of such three rows,
immediately computing the result in the third row
(NOT is provided by including a single row of dual-contact DRAM cells~\cite{seshadri2017ambit}).
\emph{Importantly for SISA-PUM}, only three selected designated DRAM rows (per
single DRAM subarray) are modified this way. Whenever the running code requests
an in-situ memory operation, Ambit uses a recent RowClone
technology~\cite{seshadri2013rowclone} to copy (also in-situ) the rows that
store input sets to these two designated rows, compute the result in-situ, and
again use RowClone to copy the result to the destination (unmodified) DRAM row.
Now, SISA-PUM uses Ambit's execution model and interface without any
modifications: set intersection and union are processed with an in-situ AND and
OR, respectively. Set difference is processed using set intersection, along
with the well-known set algebra rule: $A \setminus B = A \cap
B'$~\cite{jech2013set}.
\tr{Whenever needed, the negation $B'$ can be derived with the in-situ
NOT.}

\textbf{SISA-PNM}
\ 
A set operation with no bulk bitwise processing uses SISA-PNM
that relies on high bandwidth between processing units and DRAM (as in
UPMEM~\cite{lavenier2016dna}, HMC~\cite{jeddeloh2012hybrid}, or
Tesseract~\cite{ahn2015scalable_tes}).
{Adding} or {removing an element from a set stored as a DB} ($A \cup
\{x\}, A \setminus \{x\}$) is conducted with a single DRAM access to a
specific memory cell. 
Other set operations on SAs that employ \tr{either} streaming \tr{(e.g., merge~$\cap$)}
or random accesses \tr{(e.g., galloping~$\cap$)} are also executed using small in-order cores.
\tr{Here, we rely on the high TSV enabled bandwidth for high performance of
set operations dominated both by data streaming (merge) and random
accesses (galloping).}

\all{\maciej{From HRL paper: ``However, the actual performance depends on how much
processing capability one can fit within the area and power constraints of the
logic layer. The ideal compute technology for NDP should (1) be area-efficient
in order to provide sufficient compute throughput to match the high bandwidth
available through 3D stacking; (2) be power-efficient in order to reduce total
energy consumption, and to avoid causing thermal issues in the DRAM stacks; (3)
provide high flexibility to allow for reuse across multiple applications and
application domains''}}

\marginparX{\Large\vspace{1em}\colorbox{yellow}{\textbf{C}}}

\subsection{SCU \& Automatizing SISA Decisions}

\marginparX{\Large\vspace{1em}\colorbox{yellow}{\textbf{C}}\\\colorbox{yellow}{\textbf{D}}}

\marginparX{\Large\vspace{1em}\colorbox{green}{\textbf{C}}}

\sethlcolor{yellow}We use a small SISA Control Unit (SCU),
cf.~Section~\mbox{\ref{sec:overview}}, to automatically decide on (1) selecting
the PNM or PUM execution, and (2) merge or galloping.
Once the host core decodes a SISA instruction, it passes
it to the SCU. The SCU further decodes this instruction, and picks either PNM
or PUM to execute the instruction.
\sethlcolor{green}For deployment, SCU could either be added to the CPU or to the
DRAM circuitry (see the feasibility discussion later in this section), or -- to
avoid any HW modifications -- it can also be emulated by a single designated
in-order logic layer core.
\sethlcolor{yellow}SCU does not implement any complex logic
(e.g., dynamic set modifications), it only decides on variants of schemes
to execute.

\marginparX{\Large\vspace{-2em}\colorbox{yellow}{\textbf{C}},\\\colorbox{yellow}{\textbf{D}}}

\marginparX{\Large\vspace{1em}\colorbox{green}{\textbf{C}}}

\marginparX{\Large\vspace{0.5em}\colorbox{yellow}{\textbf{C}},\\\colorbox{yellow}{\textbf{D}}}

\textbf{SISA-PUM \& SISA-PNM}
\ 
\sethlcolor{green}First, SCU decides whether to use SISA-PUM or SISA-PNM for given two sets. This
decision is simple and is based on how sets are represented (this information
is stored \sethlcolor{yellow}in a simple in-memory SM (``set metadata'')
structure \sethlcolor{green}and possibly cached in SCU's cache).

\textbf{Variants of Set Operations}
\ 
Second, SCU automatically detects if it is best to use merge or galloping, and
processes input sets using the corresponding variant.
This decision is guided by our performance models.

\enlargeSQ

\subsection{Performance Models for Set Operations}
\label{sec:perf_models}

The runtime of each SISA instruction variant is dominated by either 
streaming or random accesses.

\textbf{Streaming} takes place when two sets $A$ and $B$ stored as SAs are
processed using merging. We model the runtime as
$l_{M} + W \cdot \max\{|A|,|B|\} \cdot \min\{b_{M}, b_{L}\}$.
$l_M$ and $b_M$ are latency and bandwidth of accessing DRAM, and $b_L$ is
bandwidth between cores. The model conservatively assumes that $A$ and $B$ may
be located in memory locations attached to different cores (e.g., in different
vaults), and thus (1) the overall bandwidth is bottlenecked by $\min \{b_M, b_L
\}$, and (2) we can use \mbox{$\max\{|A|,|B|\}$} as $A$ and $B$ are
streamed in parallel.

\marginparX{\Large\vspace{-1em}\colorbox{green}{\textbf{D}}}

To model \textbf{random accesses}, we simply count the number of performed
operations and multiply it by the memory access latency. This gives $l_M \cdot
\min\{|A|,|B|\} \cdot \log( \max\{|A|,|B|\} )$ for a binary search over the
larger of input sets, used when processing two SAs with galloping.
\sethlcolor{green}Then, a specific variant is \textbf{selected automatically} to minimize the
predicted runtime. To \textbf{parametrize} these models, SISA needs (1) the
sizes of processed sets, (2) their representation types, and (3)
\mbox{$b_M, b_L, l_M$}. (1) and (2) are maintained in the metadata structure. 
(3) describe the execution environment and are thus identical for each set;
they are stored directly in the SCU. We instantiate (3) to
reflect logic layers in Tesseract~\cite{ahn2015scalable_tes}.

\marginparX{\Large\vspace{-5em}\colorbox{green}{\textbf{C}}}

\subsection{Details of SISA Hardware}

\tr{We now present various details on SISA HW.}

\marginparX{\Large\vspace{3em}\colorbox{green}{\textbf{C}}}

\textbf{Life Cycle of a Set}
\
\tr{Any SISA graph application is a series of standard instructions as well as
SISA instructions that load, store, and process sets.}
A set is allocated with a standard malloc, augmented with setting the
appropriate set information in the set metadata (SM) structure.
\sethlcolor{green}Loading, processing, and storing sets is conducted by the respective existing
elements such as logic layer cores; the SCU is only responsible for selecting
the appropriate instruction variant to be executed.\sethlcolor{yellow}
Once a set is deleted, the standard free call is used, together with removing the
respective entry from the SM structure.

\marginparX{\Large\vspace{0em}\colorbox{green}{\textbf{D}}}

\sethlcolor{green}\textbf{Set Metadata}
SM forms a simple associative structure that holds constant amount
of data per set (set representation, set size). 
The total SM size is $O(n)$ as there are $n$ neighborhoods and a
constant number of auxiliary sets. Thus, while we conservatively assume that SM
is an in-memory structure, in practice it fits completely in cache or a
small scratchpad.
This is because many datasets processed by graph mining algorithms have small
\mbox{$n$}, in the order of hundreds or thousands~\mbox{\cite{nr-sigkdd16}}.
These graphs pose computational challenges, but these challenges come from high
computational complexities (e.g., listing maximal cliques is NP-hard) or from
relatively high edge counts~\mbox{$m$} (as some vertices may have high
degrees~\mbox{\cite{nr-sigkdd16}}), but \emph{not} (or to a smaller extend)
from~\mbox{$n$}.
\sethlcolor{yellow}Whenever the given SM information is not cached, there is
a single additional memory access for one set operation.
\tr{The SM information is used by the SISA performance models when deciding
which set algorithm to execute. Yet, other information could also be stored --
for example, when using other set representations such as sparse
  bitmaps~\cite{aberger2017emptyheaded, han2018speeding}. We plan on extending
  SISA with such schemes as future work.}
{Each SM entry describing one set also contains the set location.  Now, entries
in the SM structure are indexed by set IDs. A set ID is returned by a function
creating a set, cf.~Figure~\mbox{\ref{fig:sisa-full}}.} \sethlcolor{yellow}{Set
IDs and set creation (and destruction) calls are used by a developer
analogously to pointers and malloc/free calls.}

\marginparX{\Large\vspace{-8em}\colorbox{green}{\textbf{D}}}

\marginparX{\Large\vspace{-5em}\colorbox{yellow}{\textbf{D}}}


\trNEW{The ST tracks physical locations of sets. One ST entry tracks one
set~$S$. This entry contains a tag (a unique set ID) of~$S$, its base address,
its size, and an adjacent pointer to the next physical location that stores the
next base address and the corresponding size. The adjacent pointer is set to
null if it is $S$'s end. To enable full utilization of memory-level
parallelism, a selected number of bases and sizes may also be stored in the
corresponding ST entry. The ST is stored inside the main memory. The SLB
eliminates some of the memory accesses; it uses an LRU cache replacement
policy.}

\enlargeSQ

\marginparX{\Large\vspace{1em}\colorbox{green}{\textbf{C}},\\\colorbox{green}{\textbf{D}}}

\sethlcolor{green}\textbf{Caching Set Metadata}
Depending on how SISA HW is deployed, the SM information can be cached in
either a small dedicated scratchpad or cache (if the SCU is implemented as an
additional circuitry), or in the standard cache of a logic layer core (if the
SCU is emulated by a such designated core).

\marginparX{\Large\vspace{0em}\colorbox{green}{\textbf{D}},\\\colorbox{green}{\textbf{F}}}

\marginparX{\Large\vspace{1em}\colorbox{yellow}{\textbf{D}},\\\colorbox{yellow}{\textbf{F}}}

\sethlcolor{green}\textbf{SISA-PNM and SISA-PUM Together}
Ambit fully preserves the DRAM interface: the sets are always stored in
standard DRAM rows, and moved to the designated rows \emph{only} for bulk
bitwise processing~\mbox{\cite{seshadri2017ambit}}. \sethlcolor{yellow}SISA-PNM accesses run
on unmodified DRAM banks (the modifications in PNM are only related to the high
bandwidth, and the SCU in SISA). Thus, SISA-PNM and -PUM are seamlessly used
together.

\sethlcolor{yellow}

\textbf{Harnessing Parallelism}
\
\iftr
SISA HW harnesses memory parallelism at different levels, enabling parallel
execution of both a single set operation and different set operations.
\fi
First, bit-level parallelism is enabled by using Ambit's bulk bitwise
operations: bits in a row are ANDed or ORed in parallel.
Second, pairs of bitvectors placed in different subarrays (or, e.g., DRAM banks) can be processed in
parallel. 
Third, processing pairs of sets stored as integer arrays in different vaults
can also be parallelized. Here, SISA benefits from the same effect of
{bandwidth scalability} as the Tesseract graph
accelerator~\cite{ahn2015scalable_tes}.

\marginparX{\Large\vspace{0em}\colorbox{yellow}{\textbf{D}}}

\textbf{Managing Concurrency}
\
\tr{For simplicity, }SISA relies on developers using established techniques (locks, lock-free
protocols, general parallel programming principles~\cite{herlihy2020art}
and libraries such as OpenMP~\cite{chandra2001parallel}) to concurrently
access the same set. \tr{Thus, designing a parallel graph mining algorithm
that uses SISA is analogous to non-SISA based algorithms.}

\marginparX{\Large\vspace{1em}\colorbox{yellow}{\textbf{E}}}

For \textbf{cache coherence in SISA-PUM}, we rely on mechanisms
(provided by the memory controller) that flush dirty cache lines in source
rows, and invalidate cache lines in destination rows. Existing schemes also
rely on it, including Ambit~\mbox{\cite{seshadri2017ambit}}, DMA
accesses~\mbox{\cite{corbet2005linux}} and
others~\mbox{\cite{hsieh2016accelerating, seshadri2013rowclone}}. As in Ambit,
SISA-PUM accesses are always row-wise, and thus we can also rely on Dirty-Block
Index~\mbox{\cite{seshadri2014dirty}} and similar schemes for fast data
flushing. Invalidations run in parallel with Ambit operations and thus do not
incur overheads.

\textbf{Memory Layout and Storage of Sets} 
\ 
\tr{Advanced schemes for the layout of vertices and edges in
different sets (e.g., spreading large sets across different vaults) are beyond
the scope of this work. } {We ensure that storing SISA sets is feasible
(i.e., a maximum-size neighborhood, represented as SA or DB, fits into a
single vault)}.

\sethlcolor{yellow}

\all{``Algorithms to
achieve a balanced distribution of vertices and edges to vaults
are beyond the scope of this paper''}

\all{
\subsection{Feasibility Analysis}

We briefly discuss the feasibility of various design choices.

In SISA, we use two different types of memory acceleration,
in-situ PIM and logic layers. We rely on Ambit's

\maciej{Related! Also hybrid HW design: `` ReRAM alone is not sufficient for an
energy-saving graph processing system without performance degradation due to
the data access patterns of poor locality and huge amount of write traffic.
Thus, proper memory hierarchy design and data scheduling become the key to
exploit advantages of ReRAMs in graph processing accelerators.  In this paper,
we propose a Hybrid Vertex-Edge memory hierarchy, HyVE, for efficient graph
processing. In HyVE, different types of graph data are stored in either ReRAM
or conventional memories according to different access patterns.  A memory
controller is responsible for handling various requests from high-level
accelerator logic.''}

\maciej{GraphH is also important because it is also "hybrid" in a sense that they
use HMC but also add SRAM on chip}

\maciej{GraphIA: ``For each GraphIA chip, we modify the DRAM chip by designing heterogeneous
banks with peripheral circuits in the DRAM to fully exploit the huge
parallelism lying in memory banks''}

\subsubsection{Feasibility of Storage Capacities}

\maciej{Discuss: max size of a vault vs max vertex neighborhood size, 
max size of a bitvector vs max Ambit size}

\maciej{Partitioning across chips - was discussed in many schemes, maybe double
check; for example: GraphIA}
}

\all{\maciej{From GraphIA: `` Previous accelerator designs for graph processing only
focused on integrating more computing units inside memories or using more
memory layers, rather than exploiting the huge parallelism lying in memory
banks.''
``The main idea of in-situ accelerators is to modify circuits in the memory to
perform computation functions, rather than simply putting computation units
“closer” to the whole memory part.''
``Graph processing is notoriously known for the intensive irregular memory
access.'' ---> not always true for set centric!!
``For each GraphIA chip, we modify the DRAM chip by designing heterogeneous
banks with peripheral circuits in the DRAM to fully exploit the huge
parallelism lying in memory banks'' }}

\all{We overview a proposed example SISA HW implementation, {which is
independent of the set-centric formulations and the ISA syntax \& semantics}.
We focus on integration with the CPU setting and with in-memory and near-memory
accelerators; see Figure~\ref{fig:sisa-mem-design} \maciej{TODO: a new figure}. 
In~\cref{sec:integrate}, we broadly discuss how SISA could harness other units
such as GPUS, FPGAs, or accelerators such as Tesseract~\cite{ahn2016scalable}.}


\subsection{SISA Hardware Cost and Feasibility}

We also briefly discuss the hardware cost.
First, the needed \textbf{DRAM chip modifications} are minimal and identical to those
already discussed in Ambit.
Second, as the \textbf{logic to be implemented in SCU} is straightforward
decision making on what instruction variant to use, its costs are
not prohibitive, as shown by many designs proposed in the past, for example in
HyVE~\cite{huang2018hyve} (a hybrid vertex-edge memory hierarchy that uses
ReRAM and DRAM) or in GraphH~\cite{dai2018graphh} (an accelerator that combines
HMC with SRAM).
Third, the code of all SISA instructions is also straightforward: a simple
binary search (galloping), merging of two arrays (merge), or setting/clearing a
DRAM cell (set element add/remove). Thus, they can be trivially deployed in
{in-order cores in the logic layer} of 3D stacked DRAM, as shown by
other designs~\cite{dai2018graphh}. 

\all{\maciej{Related! Also hybrid HW design: `` ReRAM alone is not sufficient
for an energy-saving graph processing system without performance degradation
  due to the data access patterns of poor locality and huge amount of write
  traffic.  Thus, proper memory hierarchy design and data scheduling become the
  key to exploit advantages of ReRAMs in graph processing accelerators.  In
  this paper, we propose a Hybrid Vertex-Edge memory hierarchy, HyVE, for
  efficient graph processing. In HyVE, different types of graph data are stored
  in either ReRAM or conventional memories according to different access
  patterns.  A memory controller is responsible for handling various requests
  from high-level accelerator logic.''}

\maciej{GraphH is also important because it is also "hybrid" in a sense that
they use HMC but also add SRAM on chip}

\maciej{GraphIA: ``For each GraphIA chip, we modify the DRAM chip by designing
heterogeneous banks with peripheral circuits in the DRAM to fully exploit the
huge parallelism lying in memory banks''}}

\all{\maciej{From GraphH: ``According to the HMC 2.1 specification, the
bandwidth between memory layers and the logic layer of each vault can be up to
10 GB/s.  We implement a simple in-order core, two specific OVBs, the data
controller, and the network logic in the logic layer.''}}

\trNEW{
%
\subsection{Low-Level Details of Set Management}

We discuss various low-level aspects of set management.

\textbf{Set Allocation}
To allocate physical memory for a set, the SMU first creates a new entry in the
ST, storing a new set ID as a tag. An allocation resembles \texttt{malloc}
and is done either through the MMU or the SMU; the latter happens if
SISA keeps a private memory pool for set storage.
The SMU gets a pointer (to the starting set address) and stores
this (physical) address and the set size as another ST entry. In rare cases,
when the allocation size is greater than the available contiguous memory, the
SMU repeatedly attempts to allocate smaller chunks until successful; in such
cases a set is created in a fragmented state, with the corresponding
adjacent pointers, base addresses, and sizes properly initialized.  The SMU
ensures allocation atomicity with respect to other set operations.
%


\textbf{Loading and Storing Sets}
SISA allows both a {single load} and a {block load}. A single
load is done in two steps. First, SISA allocates the memory at the end of the
current set element (or allocate a separate base and update the ST with
the new set base). Second, SISA copies the data into the newly allocated
memory. A {block load} operation is done by appending the base and bound
list within the ST with the new block of data without copying. Unlike
the load, the {store} operation iteratively copies the entire
set to the target location\footnote{While there can be multiple ways
to implement these loads and stores in HW, our evaluation
models them similarly to x86 loads
and stores. \sf\scriptsize}.


\trNEW{\textbf{Set Reorganization} 
In the current SISA release, we focus on batch analytics, and thus in most
cases sets are contiguous.  In certain cases (e.g., for auxiliary
containers in Bron-Kerbosch), if a set is fragmented, the SMU can reorganize
the set data to improve contiguity. While the SMU could perform this task
automatically based on memory fragmentation, SISA also provides an instruction
to perform this task manually. In the former case, a dedicated
process triggered by the OS periodically checks the fragmentation
information that it can access through low-level SISA instructions. To
reorganize sets, the SMU waits for all other SISA instructions to be finished. }

\textbf{Handling Ongoing Memory Accesses} When the memory access is generated to
read the contents of a set, SISA uses core caches and Miss Status Holding
Registers (MSHR) to (1) reduce the latency of the accesses to a given set
element and (2) prevent repeating accesses to the same physical
address.  

%



\textbf{Accessing Sets} Because the ST contains a unique set ID as a tag,
accessing data within each set is done through a lookup operation.  A given set
instruction uses the set ID to identify the correct set inside the ST. When
found, the SMU iterates the list of base addresses and sizes in order to
generate parallel memory accesses to the elements of a set.


\all{\textbf{Parallel Set Operations} The ST knowledge of set sizes and
addresses enables parallel operations. For example, to answer $x \in A$, the
SMU can search different parts of $A$ in parallel.}

%

%
%
% ...
 

\textbf{Parallel Graph Algorithms}
In SISA set-centric formulations, we use the ``[in par]'' annotation to
indicate which parts of set-centric algorithms can be processed by multiple
threads without the risk of memory conflicts (cf.~Section~\ref{sec:formulations}).  For more complex
parallelizations, similarly to the MMU, the SMU relies on the SISA user to
provide a correct implementation.

\textbf{Memory Protection}
For memory protection, when multiple SISA processes are running, one could tag
(upon set allocation) respective SM entries with process IDs, similarly to the
Address Space ID (ASID) bits added to TLB entries.
}

\tr{\textbf{Integration with RocketChip}
To facilitate a potential real SISA implementation, we outline the integration
of SISA and RocketChip~\cite{asanovic2016rocket}.  Custom SISA instructions as
specified in our encoding (\cref{sec:sisa-details}) are forwarded to the SMU
and SLB tandem, which replaces the RoCC accelerator
component~\cite{asanovic2016rocket} of a Rocket/Boom tile.  The SLB is
connected to the cache network and the SMU is connected directly to the memory
bus. When the Rocket/Boom core receives an instruction whose opcode and
functionality bits match those of SISA, the core forwards the instruction,
using the existing RoCC interface~\cite{asanovic2016rocket}, to the SMU. When
the SMU completes its operation, it signals the core to continue executing the
application.}

%% file: eval.tex
\section{Evaluation}
\label{sec:eval}

\enlargeSQ

We illustrate example performance advantages from SISA. \tr{Due to a very large
evaluation space, we provide summaries; all results follow similar patterns.}

\subsection{Methodology, Setup, Parameters}

\tr{We first present our simulation setup.}

\textbf{Simulation Infrastructure}
\ 
We use Sniper~\cite{heirman2012sniper} with the Pin frontend~\cite{luk2005pin}.
Sniper is a popular cycle-level simulator used in many works proposing various
architectural extensions for both CPUs and memory
subsystem~\cite{van2013fairness, mittal2014improving}.

\textbf{SISA Implementation}
\ 
We simulate the SISA HW design and the ISA, instrumenting
the code so that the simulation toolchain can distinguish between SISA and
non-SISA instructions. 
%
%
To model each component of SISA, we add the respective set instructions and
simulate the SCU (a small fixed delay), the cache in SCU (with the LRU policy),
the SM structure (random memory accesses whenever the SCU cache is not hit),
and the execution of all used set operations by appropriate delays in the
simulation execution. For operations based on streaming and random memory
accesses, we use the performance models described in~\cref{sec:perf_models}.
To simulate SISA-PUM, we model a run-time of in-situ operations with a delay $l_M
+ l_I \cdot \lceil n / (qS) \rceil$; $l_M$ is the latency to access 
DRAM (to initiate the operation) and $l_I$ is the latency execute one 
in-situ instruction. $\lceil n / (q R) \rceil$ models a
scenario when the bitvector size $n$ exceeds the size of all DRAM rows that can
be processed in parallel.
$q$ is the count of rows within a bank that can be
used in parallel and $R$ is the size of one row.

\all{
$l_M \cdot |A|$ (checking the
existence of the elements of~$A$ in a bitvector, used when processing an SA and
a DB), and $l_M$ (clearing or setting a single bit, used when modifying a
bitvector cell).

\textbf{In-situ operations } are used when two bitvectors are processed with a
bulk bitwise operation. The model is

}

\textbf{SISA Platform \&
Parameters }
For concreteness, we set the platform {for executing SISA instructions} to
match Tesseract~\cite{ahn2015scalable_tes} (for SISA-PNM)
\sethlcolor{yellow}{and Ambit~\mbox{\cite{seshadri2017ambit}} (for
SISA-PUM)}. The former has simple in-order cores {(1
core/vault in its logic layer) with 32 KB L1 instruction/data caches, no L2, 16
8GB HMCs (128 GB in total), 32 vaults/cube, 16 banks/vault. Each vault offers
16 GB/s of memory bandwidth to its core. Thus, we assume scalable bandwidth as
proposed by Tesseract: using more vaults increases the total memory bandwidth}. 
\sethlcolor{yellow}{We set the DRAM row rank size to 8~KB,
following Ambit~\mbox{\cite{seshadri2017ambit}}.}
Next, we set the parameter $t \in [0; 1]$ (that controls the bias towards using
DBs or SAs to store neighborhoods) to 0.4 (i.e., 40\% of neighborhoods are
stored as DBs); we also analyze other values. We ensure that
the total storage used for neighborhoods does not exceed the size of the simple
CSR graph storage by more than 10\%.
{Finally, we set the size of SISA SCU's cache to be 32 KB
(matching Tesseract's L1).}

\sethlcolor{yellow}

{\textbf{Platform for non-SISA Instructions \& Baselines}
For any non-SISA instructions and baselines, we use a
high-performance Out-of-Order manycore CPU. 
Each core has a 128-entry
instruction window, a branch predictor, 32 KB L1 instruction/data caches, a 256 KB L2 cache.
All cores share an 8 MB L3 cache.
There is also a four-way associative 64-entry D-TLB, a 128-entry I-TLB, and a 512-entry S-TLB.
%
%
%
For fair comparison, \emph{we also use bandwidth scalability in
this configuration, i.e., we increase the memory bandwidth with the number of
cores, matching it with that of SISA-PNM.}}

\enlargeSQ


\textbf{Considered Mining Problems}
The graph mining problems we consider are clustering with the Jaccard
(\texttt{cl-jac}), overlap (\texttt{cl-ovr}), and total neighbors
(\texttt{cl-tot}) coefficients, listing $k$-cliques (\texttt{kcc-$k$}, $k \in
\{4, 5, 6\}$), $k$-clique-stars (\texttt{ksc-$k$}, $k \in \{4, 5, 6\}$), maximal cliques
(\texttt{mc}), triangles (\texttt{tc}), and subgraph isomorphism (\texttt{si-$k$s} for $k$-stars).

\textbf{Comparison Targets: Hand-Tuned Algorithms}
Our most important (the most challenging to outperform) baselines are
hand-optimized parallel algorithms for each graph mining problem.
Specifically, we use a tuned version from the GAP Benchmark
Suite~\cite{beamer2015gap} for \texttt{tc}, Eppstein's version of BK for \texttt{mc}~\cite{DBLP:conf/isaac/EppsteinLS10}, Danisch' scheme for
\texttt{kcc-$k$}~\cite{danisch2018listing}, enhanced Jabbour's scheme for
\texttt{ksc-$k$}~\cite{jabbour2018pushing}, parallel VF2 for \texttt{si-$k$s}~\cite{cordella2004sub}, and \texttt{cl-jac} based on counting triangles
in the GAP suite~\cite{beamer2015gap}. All used baselines {have
competitive work and depth complexities}, cf.~Table~\ref{tab:theory-table}.
\sethlcolor{green}For fair comparison, all baselines benefit from the high
bandwidth of PIM.
%
%
We consider \tr{two classes of baselines:} algorithms that do not explicitly use set
algebra (denoted with \texttt{\_non-set}) and their set-centric variants 
(denoted with \texttt{\_set-based}). \tr{SISA variants are indicated with
\texttt{\_sisa}.}

\marginparX{\Large\vspace{-3em}\colorbox{green}{\textbf{A}},\\\colorbox{green}{\textbf{F}}}

\marginparX{\Large\vspace{1em}\colorbox{yellow}{\textbf{A}},\\\colorbox{yellow}{\textbf{F}}}

\textbf{Comparison Targets: Pattern Matching Frameworks}
%
%
\sethlcolor{yellow}SISA and its underlying paradigm do not aim to outperform specific
accelerators but complement or reinforce them, by offering a novel set-centric
paradigm and building blocks. Thus, we focus on comparing to the fundamental
paradigms / algebras that underlie these accelerators: neighborhood expansion
for pattern matching implemented in
  Peregrine~\mbox{\cite{jamshidi2020peregrine}} (which represents
  GRAMER~\mbox{\cite{yao2020locality}}) and relational algebra implemented in
  RStream~\mbox{\cite{wang2018rstream}} (which represents
  TrieJax~\mbox{\cite{kalinsky2020triejax}}).
%
%
%
\sethlcolor{green}We stress that, while we consider these 
baselines for completeness, we focus on comparing to (much faster) hand-tuned
parallel algorithms for solving specific problems.

\marginparX{\Large\vspace{-1em}\colorbox{green}{\textbf{A}},\\\colorbox{green}{\textbf{F}}}

\iftr

\textbf{Comparison Baselines \& PIM}
For fairness, all considered comparison targets take advantage from the high
bandwidth of PIM setting (parametrized identically to that of SISA).

\fi

\sloppy
\textbf{Graphs }
We select different input datasets (Table~\ref{tab:graphs}) from Network
Repository~\cite{rossi2016interactive}, considering biological (\texttt{bio-}),
interaction (\texttt{int-}), brain (\texttt{bn-}), economics (\texttt{econ-}),
social (\texttt{soc-}), scientific-computing (\texttt{sc-}), discrete-math (\texttt{dimacs-}),
and wiktionary (\texttt{edit-}) networks.
We pick graphs with different structural properties (low/high density,
small/large maximum degree, low/high degree distribution skew, etc.). 
%


\begin{table}[h]
\centering
\footnotesize
%
\setlength{\tabcolsep}{2pt}
\renewcommand{\arraystretch}{1}
\begin{tabular}{l}
\toprule
\makecell[l]{
\textbf{\ul{Biological}.} 
 Gene functional associations: 
   ({\emph{bio-SC-GT}}, 1.7K, 34K), 
   ({\emph{bio-CE-PG}},\\1.8K, 48K), 
   ({\emph{bio-DM-CX}}, 4K, 77K), 
   ({\emph{bio-DR-CX}}, 3.2K, 85K), 
   ({\emph{bio-HS-LC}},\\4.2K, 39K), 
   ({\emph{bio-SC-HT}}, 2K, 63K), 
   ({\emph{bio-WormNetB3}}, 2.4K, 79K).
 Human gene \\regulatory network: 
   ({\emph{bio-humanGene}}, 14K, 9M) \textbf{\ul{(large)}},\\
   ({\emph{bio-mouseGene}}, 45K, 14.5M) \textbf{\ul{(large)}}.
%
}\\
%
%
\makecell[l]{\textbf{\ul{Interaction}.}
 Animal networks:  
    ({\emph{int-antCol3-d1}}, 161, 11.1K), 
    ({\emph{int-antCol5-d1}},\\153, 9K),
    ({\emph{int-antCol6-d2}}, 165, 10.2K),
    ({\emph{intD-antCol4}}, 134, 5K). 
 Human contact\\network:
 	({\emph{int-HosWardProx}}, 1.8k, 1.4k).
 Users-rate-users:
    ({\emph{int-dating}}, 169K, 17.3M)\\ \textbf{\ul{(large)}},
({\emph{edit-enwiktionary}}, 2.1M, 5.5M) \textbf{\ul{(large)}}.
}\\
%
%
\makecell[l]{\textbf{\ul{Brain}.}
  ({\emph{bn-flyMedulla}}, 1.8K, 8.9K),
  ({\emph{bn-mouse}}, 1.1K, 90.8K).
}\\
%
\makecell[l]{\textbf{\ul{Economic}.}
  ({\emph{econ-beacxc}}, 498, 42K),
  ({\emph{econ-beaflw}}, 508, 44.9K),\\
  ({\emph{econ-mbeacxc}}, 493, 41.6K),
  ({\emph{econ-orani678}}, 2.5K, 86.8K).
}\\
%
%
\makecell[l]{\textbf{\ul{Social}.}
  Facebook: ({\emph{soc-fbMsg}}, 1.9k, 13.8k).
  Orkut: (3.1M, 117M) \textbf{\ul{(large)}},
}\\
%
%
\makecell[l]{\textbf{\ul{Scientific computing}.}
  ({\emph{sc-pwtk}}, 217.9K, 5.6M) \textbf{\ul{(large)}},
}\\
%
%
\makecell[l]{\textbf{\ul{Discrete math}.}
	({\emph{dimacs-c500-9}}, 501, 112K),
}\\
%
%
%
\bottomrule
\end{tabular}
%
%
\caption{\textmd{Considered
graphs\cite{rossi2016interactive}}.
For each graph, we show its ``(\#vertices, \#edges)''.
\iftr
If there are multiple related graphs, for
example different snapshots of the .eu domain, we select the largest one. One
exception is the additional Italian Wikipedia snapshot selected due to its
interestingly high density.\fi
}
%
\label{tab:graphs}
\vspaceSQ{-2em}
\end{table}


\textbf{Tackling Long Simulation Runtimes }
Most benchmarks use relatively small graphs because (1) we run
cycle accurate simulations, tracing all memory accesses, which is very
time-consuming, and (2) the considered algorithms are computationally hard and
even software codes use graphs much smaller than those used with
algorithms such as PageRank~\cite{DBLP:conf/isaac/EppsteinLS10,
danisch2018listing}.
Yet, even this is often not enough to enable finishing simulations of
algorithms such as Bron-Kerbosch. Thus, we usually also pre-specify a number
of graph patterns to be found\tr{ (to eliminate unrelated
performance effects due to thread scheduling, we set a cutoff number of
patterns {per thread} and we use deterministic scheduling)}.
Past work analogously handled long simulations
graph algorithms~\cite{ahn2015scalable_tes} such as PageRank (limiting
\#iteration)\tr{, and we use this idea for graph mining}.

\textbf{Performance Measures \& Summaries:}
We focus on plain runtimes as recommended for parallel
codes~\mbox{\cite{hoefler2015scientific}} as \tr{it gives the absolute measure
of the baseline performance while} speedup may be misleading because it is
higher on unoptimized baselines.
However, for overview, we also \emph{summarize speedups}
(following~\mbox{\cite{hoefler2015scientific}}), \sethlcolor{green}i.e., we
provide (1) speedups of \emph{average runtimes}
\sethlcolor{yellow}(``speedup-of-avgs''), \sethlcolor{green}and (2)
\emph{geometric means of speedups} of all data points
\sethlcolor{yellow}(``avg-of-speedups'').

\marginparX{\Large\vspace{-2em}\colorbox{green}{\textbf{E}}}

\marginparX{\Large\vspace{-1em}\colorbox{yellow}{\textbf{E}}}

\subsection{Discussion of Results}

\tr{We now proceed to analyze the results.}

\enlargeSQ

\begin{figure}[t]
\centering
\includegraphics[width=1.0\columnwidth]{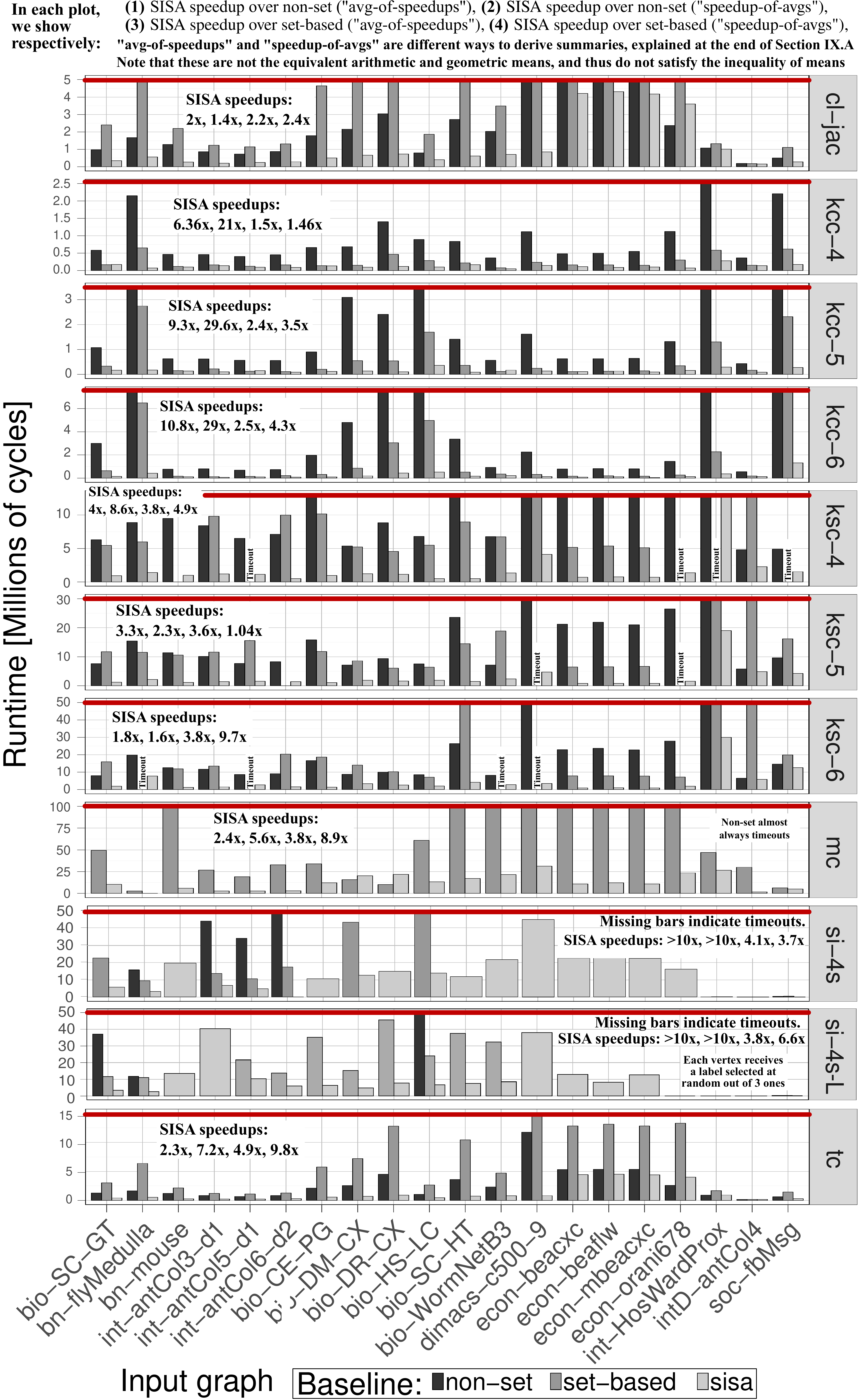}
%
\vspaceSQ{-1.5em}
\caption{\textbf{Run-times with full parallelism}.
The red line cuts off of long simulation runtimes for
readability (the bars reaching the line have much larger runtimes). No bar
indicates the timeout of the respective baseline ($>$24h). The
results for \texttt{cl-jac} (clustering based on the Jaccard coefficient) are
very similar to those that use other coefficients and for link prediction
as well as vertex similarity. 
All 32 cores are used.
\textbf{Acronyms are stated in ``Comparison Targets: Hand-Tuned Algorithms''.}
\tr{\ul{Graph classes}: biological (bio), brain (bn), interaction (int),
informative (inf), social (soc).}}
\vspaceSQ{-1.5em}
\label{fig:runtimes}
\end{figure}

\textbf{\ul{Comparison to Hand-Tuned Algorithms} }
We first analyze run-times with all available cores, comparing SISA set-centric
variants to non-set-based and set-based hand-tuned parallel baselines that all
benefit from high-bandwidth storage. The results are in
Figure~\ref{fig:runtimes}.
{SISA is almost always the fastest by a large margin of
at least 2$\times$, often more than 10$\times$} (than \texttt{non-set} schemes).
The differences vary depending on the processed graphs and the considered
problem.  Gains are usually larger on graphs with large maximum degrees, such
as brain graphs, where SISA-PUM is used more often to directly process sets
inside DRAM, reducing the latency. Such graphs are prevalent in many
computational domains~\cite{rossi2016interactive}, and this is the case for the
majority of considered datasets. 

\begin{figure*}[t]
\centering
\subfloat[{Degree distribution analysis}]{
\includegraphics[width=0.75\textwidth]{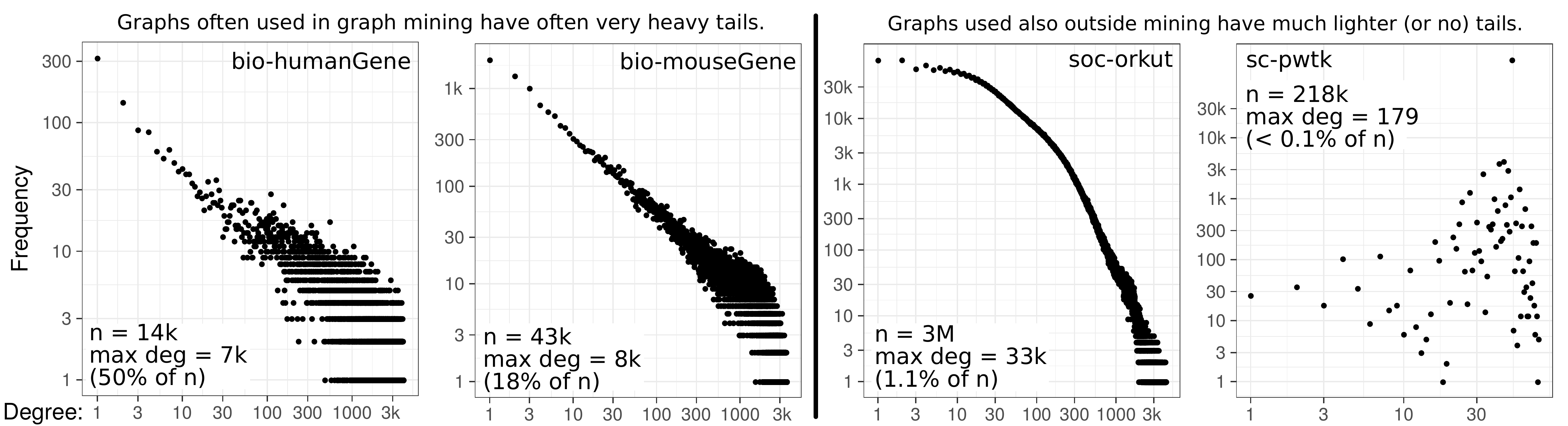}
\label{fig:eval-deg-dists}
}
\subfloat[Sensitivity analysis.]{
\includegraphics[width=0.19\textwidth]{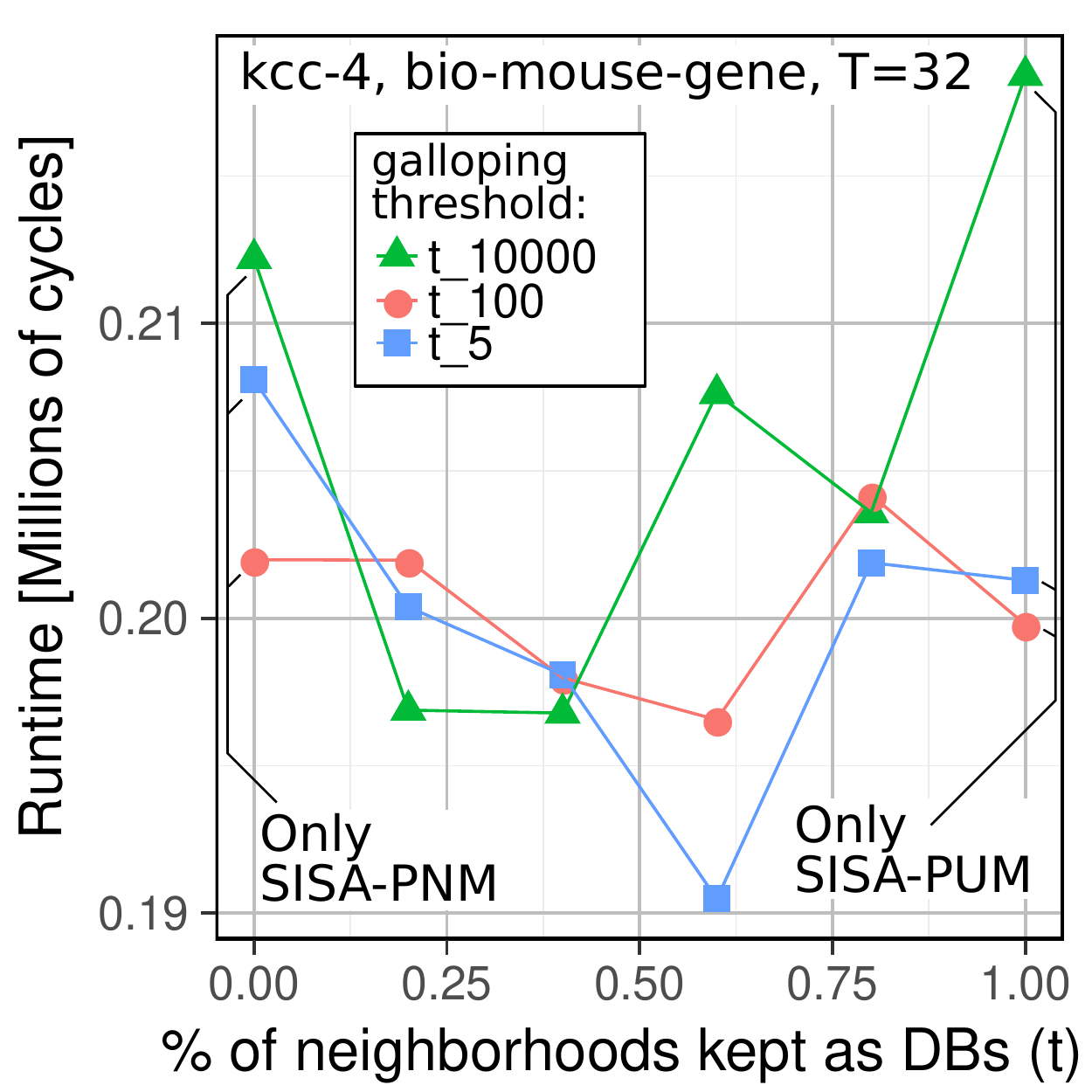}
\label{fig:eval-sens}
}
\vspaceSQ{-0.5em}
%
\vspaceSQ{-0.5em}
\caption{{Figure~\mbox{\ref{fig:eval-deg-dists}}: \textmd{Differences
between degree distributions in graphs used mostly in graph mining 
and the ones used also outside graph mining (on the right).}
Figure~\mbox{\ref{fig:eval-sens}}: \textmd{Sensitivity analysis: the
percentage of neighborhoods stored as dense bitvectors vs.~different thresholds
for using the galloping or the merging intersection.}
}}
\label{fig:eval-more-stuff}
\vspaceSQ{-1em}
\end{figure*}

\textbf{\ul{Algorithmic vs.~Architectural Speedups} }
We also observe speedups from {using only set-centric
formulations} (over non-set-based variants). 
Namely, speedups of ``\texttt{\_set-based}'' schemes over the ``\texttt{\_non-set}'' ones indicate gains from
purely \emph{algorithmic} (set-centric) changes, while speedups of ``\texttt{\_sisa}'' schemes over
the ``\texttt{\_set-based}'' indicate gains only from \emph{architectural} changes (i.e., from
using PIM).
First, the differences between \texttt{\_set-based} and \texttt{\_non-set} heavily depend on
the targeted mining algorithm. 
These speedups are particularly
visible for more complex algorithms such as \texttt{mc}, with multiple nested
loops and/or recursion. Packaging different parts of such algorithms into,
e.g., set intersections, and being able to control the used operation variant
(e.g., merging based on streaming) helps to utilize features such as high
sequential bandwidth.
Contrarily, for certain simpler schemes such as clustering,
the very tuned \texttt{\_non-set} baseline outperforms \texttt{\_set-based}
(while still falling short of \texttt{\_sisa}).
Second, the difference between \texttt{\_set-based} and \texttt{\_sisa} depend more
on the used graph. Here, in many cases, \texttt{\_sisa} is only marginally faster
than \texttt{\_set-based}, because the graph structure (e.g., sizes of neighborhoods)
favor using SAs rather than DBs, diminishing benefits from SISA-PUM (e.g., for \texttt{econ-}
graphs) and equalizing the differences because both \texttt{\_set-based} and \texttt{\_non-set}
take advantage from the high bandwidth setting. In other cases (e.g., \texttt{bio-HS-LC}), more vertices have large enough
degrees to benefit from DBs and low latencies of SISA-PUM.

\marginpar{\Large\vspace{-17em}\colorbox{yellow}{\textbf{L}}}

\hl{\mbox{\textbf{Labels }}}
\hl{We also analyze \emph{labeled} SI.  Most often, labeled graphs are 
faster to process. Despite more memory accesses, the labels form additional
constraints, which eliminates some recursive calls earlier, resulting in
performance gains.}

\marginpar{\Large\vspace{-2em}\colorbox{yellow}{\textbf{L}}}

\enlargeSQ

\textbf{\ul{Scalability} }
We also analyze how run-times change when varying numbers of threads~$T$, for a
fixed graph size (``strong scaling''), and when increasing~$T$ proportionally
to the graph size (``weak scalability''). To fix the used graph model, we use
Kronecker graphs~\cite{leskovec2010kronecker} and we vary the number of
edges/vertex.
SISA maintains its speedups, but they become less distinctive when $T$ is
small.  This is expected because fewer threads exert less pressure on the
memory subsystem, and there is overall smaller potential from using PIM in
SISA.

\marginparX{\Large\vspace{-40em}\colorbox{yellow}{\textbf{E}}}

\begin{figure}[h]
\centering
\includegraphics[width=1.0\columnwidth]{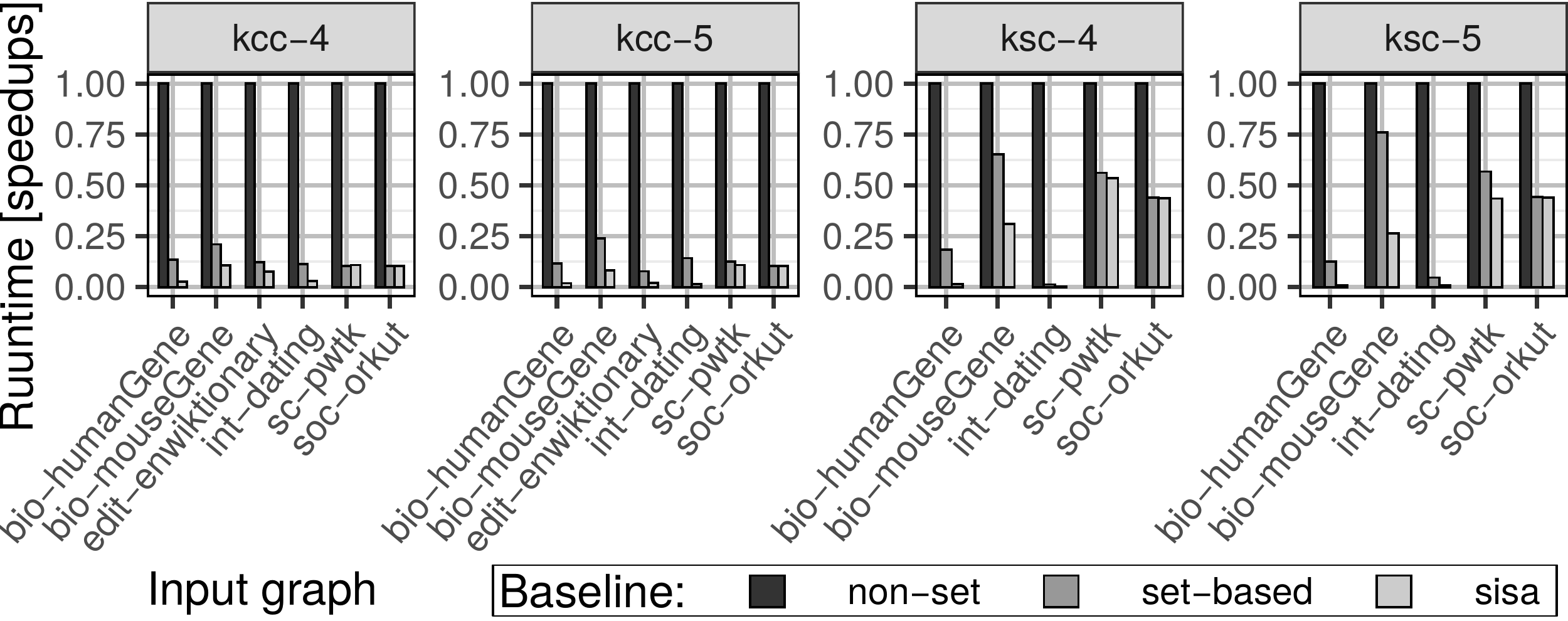}
\vspaceSQ{-1.5em}
\caption{\textbf{Run-times for large graphs}.
%
%
\tr{\ul{Graph classes}: biological (bio), brain (bn), interaction (int),
informative (inf), social (soc).}
8 cores are used.
}
\vspaceSQ{-0.5em}
\label{fig:runtimes-large}
\end{figure}

\textbf{\ul{Large Graphs}}
We execute SISA on several large graphs, 
\tr{including the Orkut social network
with 117M edges,} 
see Figure~\ref{fig:runtimes-large}.
Runtime benefits from SISA and the set-centric formulations
are similar to those in smaller graphs in Figure~\ref{fig:runtimes}.
{The only two graphs where SISA and non-SISA set baselines are comparable, are
sc-pwtk and soc-orkut. This is because these networks, due to their origin
(social and scientific) do not have large cliques or very dense clusters
(unlike, e.g., genome graphs), somewhat lowering SISA benefits.}

\enlargeSQ

\marginparX{\Large\vspace{0em}\colorbox{yellow}{\textbf{A}}\\\colorbox{yellow}{\textbf{F}}}

\sloppy
\sethlcolor{yellow}\textbf{\ul{Comparison to Other Paradigms}}
We compare SISA set-centric algorithms to neighborhood expansion and relational algebra
paradigms, representing frameworks such as Peregrine or RStream, and accelerators
such as GRAMER or TrieJax.
Peregrine is able to express only listing $k$-cliques and subgraph isomorphism,
and maximal clique listing in a limited way (i.e., it does not offer a native
scheme for MC and we implemented it by iterating over possible clique sizes and
listing maximal cliques of each size). RStream can only find
$k$-cliques. In each case, SISA baselines are \emph{much} faster:
10-100$\times$ than Peregrine (and more than 1,000$\times$ for \texttt{mc} due
to Peregrine's inability to natively support \texttt{mc}), and more than
100$\times$ for RStream.  This is because the underlying paradigms\sethlcolor{green} focus on
programmability in the first place, sacrificing performance, while in SISA we
start with tuned graph algorithms and only then restructure them with the
set-centric paradigm.\sethlcolor{yellow}

\marginparX{\Large\vspace{-9em}\colorbox{yellow}{\textbf{A}}\\\colorbox{yellow}{\textbf{F}}}

\marginparX{\Large\vspace{-1em}\colorbox{green}{\textbf{A}}\\\colorbox{green}{\textbf{F}}}

\ignore{

\tr{\textbf{Impact from \ul{Different Set Algorithms} }
}

\tr{\textbf{Impact from \ul{Automatized Execution} }
}

\tr{\textbf{Impact from \ul{Different Set Representations} }
}

\tr{\textbf{\ul{Memory Consumption} }
}

}

\textbf{\ul{Sensitivity Analysis \& Design Exploration} }
We investigate the impact from varying SISA parameters.

\textbf{SCU cache}
Not using the SCU cache lowers performance by 
$\approx$1.5$\times$ for $T=1$ and $\approx$0.05-0.1$\times$ for $T=32$. The
lower performance for high $T$ is because, with more threads
executing set operations, it becomes more difficult to ensure high
hit ratio. Overall, the behavior of the SCU cache is similar to that of
other such units such as L1, including varying cache parameters such as
size.

\marginparX{\Large\vspace{1em}\colorbox{yellow}{\textbf{C}}\\\colorbox{yellow}{\textbf{D}}}

\textbf{PNM vs.~PUM \& Sparse/Dense Neighborhoods}
\sethlcolor{yellow}PNM and PUM are synergistic in SISA.  PNM cores handle
sparse neighborhoods and SAs well, as they offer low latency and bandwidth
proportionality.
PUM is well-suited for large neighborhoods stored as DBs (common in considered
graphs due to their degree distribution skews). Yet, SISA-PUM adds overheads
when using it for sparse sets due to low utilization of very sparse rows.
\sethlcolor{green}Thus, it is relevant to not choose the DB bias parameter
to be too high. We find that 0.4 works well for most processed graphs.
We illustrate this in Figure~\mbox{\ref{fig:eval-sens}}, where we analyze how
the performance changes when varying the fraction of largest neighborhoods
stored as DBs. Smallest and largest fractions that correspond to \emph{using
only SISA-PNM or only SISA-PUM}\sethlcolor{yellow}, while technically
feasible,\sethlcolor{green} give slowest runtimes. 
We also vary the ``galloping threshold'', i.e., the relative
difference between two sets that causes the set operation to switch to the
galloping variant. For example, the value of 5 indicates that galloping is used if any of
the two sets is at least 5\mbox{$\times$} larger than the other one.
While this threshold influences performance, the general pattern stays the same.

\marginparX{\Large\vspace{-9em}\colorbox{green}{\textbf{D}}}

\marginparX{\Large\vspace{-3em}\colorbox{green}{\textbf{C}}\\\colorbox{green}{\textbf{D}}}

\enlargeSQ

We also analyze the \textbf{impact from degree distributions of datasets}, see
Figure~\mbox{\ref{fig:eval-deg-dists}}. Graphs often used in graph mining, such
as biological networks, that SISA focuses on, have often \emph{very} heavy
tails. This implies \emph{many large neighborhoods and very dense large
clusters, benefiting from SISA-PUM}.  For example, the human genome graph has
many vertices connected to more than 30\% of all other vertices.
Other graphs such as social networks have \emph{much lighter tails},
cf.~soc-orkut and sc-pwtk in Figure~\mbox{\ref{fig:eval-deg-dists}}. This is
because these networks, due to their origin (social, scientific) do not have
large cliques or very dense clusters. Such graphs benefit less from SISA-PUM.
Still, using SISA-PNM enables high performance, outperforming tuned
non-set-based baselines, cf.~Figure~\mbox{\ref{fig:runtimes-large}}.

\begin{figure}[h]
\vspaceSQ{-1em}
\centering
\subfloat[{Total amounts of stall times of different 8 parallel threads.}]{
\includegraphics[width=0.44\textwidth]{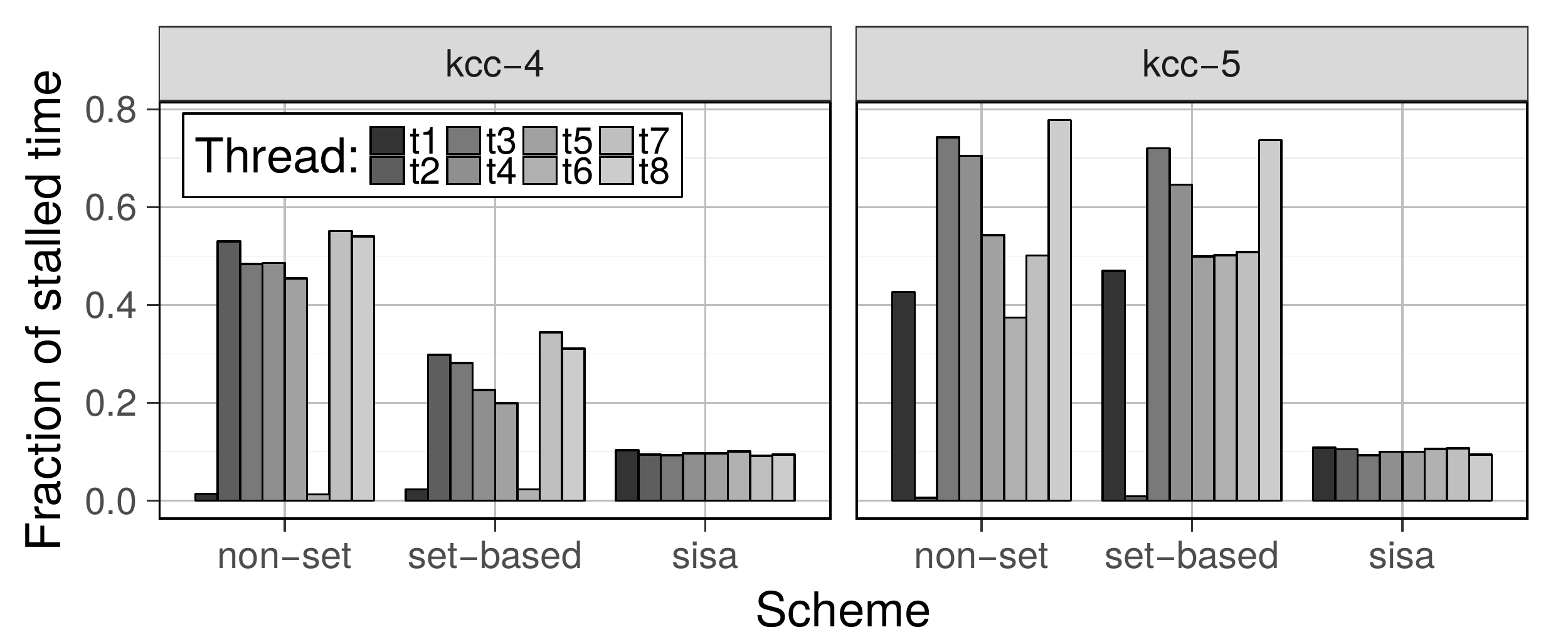}
\label{fig:eval-load-stalls}
}\\
\vspaceSQ{-0.5em}
\subfloat[{Histograms of sizes of processed sets of full vs.~partial executions, for 6 parallel threads
(the remainder of threads behave similarly). Graph: int-antCol3-d1. Problem: kcc-4.}]{
\includegraphics[width=0.48\textwidth]{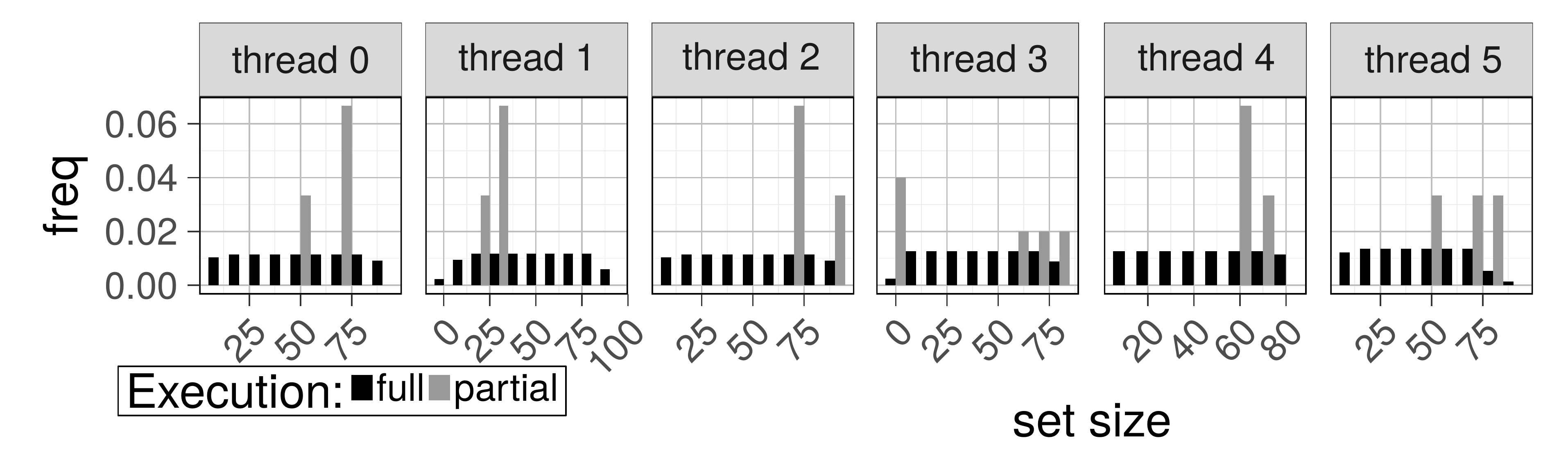}
\label{fig:eval-load-sizes}
}
%
\vspaceSQ{-0.5em}
\caption{{Load balancing analysis.
}}
\label{fig:eval-load}
\vspaceSQ{-0.5em}
\end{figure}

\textbf{Load balancing}
Figure~\mbox{\ref{fig:eval-load-stalls}} illustrates total fractions of time
during which each parallel thread is stalled when executing a given algorithm.
SISA stall times are low because its design implicitly tackles two types of
load imbalance. First, SISA's performance models enable adaptive selection of
the best variant of a set algorithm to be executed for any two sets. This
minimizes load imbalance from processing two sizes that differ a lot in sizes.
Second, load imbalance due to processing imbalanced \emph{pairs} of sets (i.e.,
two very small and two very large sets) is alleviated by the fact that very
large pairs of sets are processed with very fast SISA-PUM.

\textbf{SCU cache: shared vs. private}
We also explore sharing the SCU cache among all the cores. 
%
%
While possibly increasing the hit rate, a single
shared cache has higher access latency. This has a small ($<$1\%)
yet noticeable slowdown effect in our simulations. A potential remedy would be to
include multiple SCU cache levels. To keep the core logic simple, we do not
explore it further, and leave it for future work.

{We also show that the reduced simulation runtimes do not artificially eliminate
load imbalance. We gather traces of executed set operations
in full vs.~partial simulation executions, and we plot histograms of the sizes of processed sets,
see Figure~\mbox{\ref{fig:eval-load-sizes}}. In both types
of executions, we encounter large sets which are the primary source of load imbalance.}

\textbf{\ul{SISA Limitations}}
For some graphs with small maximum degrees (e.g., \texttt{soc-fbMsg}) in
Figure~\ref{fig:runtimes}, SISA speedups are smaller, or even (in the extreme
cases) result in slowdowns. This is because the benefits from SISA-PUM, or from
the automatic selection of the most beneficial set operation variant, are
out-weighted by having to process too many large bitvectors\tr{ (that have always
size $n$ bits)}. This effect rare, and it can be
alleviated by reducing the number of neighborhoods stored as DBs. In this case,
the performance of SISA variants gradually converges towards that of standard
CSR based set-centric algorithms. We plan on addressing it with
advanced bitvector representations.

\all{\maciej{Cool graphs (very high max degree) from NR:
ENRON-EMAIL-DYNAMIC
WIKIQUOTE-USER-EDITS
WIKIQUOTE-USER-EDITS
INTERACTION NETWORKS
}}

%% file: related.tex
\vspaceSQ{-.25em}
\section{Related Work}
\label{sec:rw}

\enlargeSQ

\iftr
In developing an ISA extension for graph mining, we follow recent footsteps of
other specialized ISAs, for example a quantum ISA~\cite{smith2016practical}, a
neural ISA~\cite{liu2016cambricon}, or even ISAs for managing cloud
resources~\cite{henry2018compiler, franke2018creating}.
\fi

Related graph processing {paradigms}
(Table~\ref{tab:comparison_models}) and {software
efforts} are described in Section~\ref{sec:intro}~\cite{besta2015accelerating,
sakr2020future, DBLP:journals/ppl/LumsdaineGHB07, besta2017push,
besta2015accelerating}.
We now briefly summarize other related areas.
First, \sethlcolor{green}we conducted an exhaustive analysis of existing
hardware accelerators as well as ISA designs for graph processing, see
Table~\mbox{\ref{tab:comparison_problems}}. The analysis indicates that SISA
offers the {only} hardware acceleration for a broad family of problems such as
maximal clique listing or clustering. The closest
designs~\cite{kalinsky2020triejax, yao2020locality, rao2021intersectx} only
focus on selected pattern matching
problems.
\sethlcolor{yellow}
\iftr
Next, some works target hardware accelerated {dynamic (time-evolving)}
graph analytics~\cite{hein2018near, bustio2015frequent, besta2019practice, besta2019demystifying, bustio2017approximate}.
Such problems are outside the scope of this work.
Moreover, several analyses illustrate how to {efficiently use existing
hardware} for graph analytics, but purely from a software development
perspective~\cite{zhang2017making, elyasi2019large, sage}.
\emph{Such works are orthogonal to SISA.}
Several works focus on {external memory} graph processing in the context
of hardware acceleration~\cite{matam2019graphssd, sage, jun2018grafboost}.
\emph{One could possibly use these designs as other SISA backends
for external memory set instructions; we leave this for
future work}.
\fi

\all{We know of only one attempt of providing a vertex-centric graph mining
algorithm~\cite{brighen2019listing}.  Its formulation is significantly more
complex than the original algorithm (29~lines for \emph{just} the user-provided
kernel compared to 10~lines for the whole standard formulation).  Moreover, the
authors only prove that \emph{the execution will finish in a finite number of
iterations}, without any tighter guarantees.
While there might exist other such formulations, their potential complexity is
prohibitive \emph{because of the inherent local approach of the vertex-centric
paradigm}.}
\all{Reliability analysis on ReRam~\cite{nien2020graphrsim}.}
\all{Using GCNs to develop better chips~\cite{wang2020gcn}.}

\input{table-hw.tex}

\iftr
While in the current SISA version we focus on implementing and executing set
operations in set-centric algorithm formulations using PIM, SISA could be
extended into different directions. This includes parallel and distributed
execution of set operations, and implementing them using high-performance
techniques such as Remote Direct Memory Access~\cite{besta2015active,
besta2014fault, gerstenberger2013enabling,
schmid2016high}.
One could also enable more efficient execution of set-centric graph mining
algorithms in the context of modern complex heterogeneous architectures that
may host massively parallel on-chip networks~\cite{besta2018slim}, NUMA and systems with locality
effects~\cite{schweizer2015evaluating, tate2014programming}, or
FPGAs~\cite{besta2020substream, besta2019graph, de2018transformations}.
One could also incorporate various forms of {graph
compression} and {summarization}~\cite{besta2019slim, besta2018survey,
liu2018graph, besta2018log}.
\fi

\all{\maciej{ADD:
In-Memory Data Parallel Processor
Batch-Aware Unified Memory Management in GPUs
for Irregular Workloads
ALL printed!
}}

\all{
\macb{ISAs}
Recently, specific ISAs were proposed, for example a quantum
ISA~\cite{smith2016practical}, a neural ISA~\cite{liu2016cambricon}, or 
ISAs for managing resources~\cite{henry2018compiler, franke2018creating}.
Moreover, efforts into open ISAs resulted in RISC-V~\cite{waterman2016design,
waterman2011risc, waterman2015risc}.
As already discussed, SISA could be implemented as a RISC-V module (custom instructions).}

\all{
To the best of our knowledge, the only collection of hardware ISA instructions used
in the context of graphs is described in a short report by
Kapre~\cite{kapre2015custom}. However, it targets only soft FPGA processors, it
only supports four operations for ``vertex-centric style'' algorithms
(\texttt{send}, \texttt{receive}, \texttt{accum}, \texttt{update}), and it does
not show or evaluate any graph algorithms.
\emph{We conclude that SISA is the first ISA for graph
mining.}
Moreover, thanks to its general set-centric nature, we predict
that it can also be used with other types of workloads.
}

\all{
\macb{Hardware-Accelerated Graph Processing}
There are many works on hardware-accelerated graph processing, both with
FPGAs~\cite{nurvitadhi2014graphgen, engelhardt2016gravf, dai2016fpgp,
zhou2017accelerating, dai2017foregraph, zhang2017boosting,
betkaoui2011framework, zhou2017tunao, weisz2013graphgen, zhou2016high,
oguntebi2016graphops, kapre2015custom} or with non-FPGA
accelerators~\cite{ham2016graphicionado, song2018graphr, zhang2018graphp,
nai2017graphpim, ahn2016scalable, ahn2015pim}.
SISA differs from all these works as it comes with a \emph{novel set-centric
approach to express graph algorithms}, and a \emph{set-centric ISA}. Thus, as
discussed in~\cref{sec:discussion-hw}, \emph{these accelerators could be used as
backends to implement 
variants of SISA instructions.}

\macb{Graph Processing Paradigms and Models}
The main difference between SISA and recent graph programming paradigms
is SISA's focus on graph mining problems, potentially difficult to express with
vertex-centric~\cite{kalavri2017high} and the edge-centric~\cite{roy2013x}
paradigms and frameworks~\cite{malewicz2010pregel, roy2013x}, or DSLs such as Green-Marl~\cite{hong2012green}.
This is because these
paradigms usually enforce certain design restrictions.  For example, the
vertex-centric approach only allows for accessing the neighbors
of each vertex (i.e., one does \emph{not have the global view of the graph}).

However, due to the generality of set algebra, SISA could also be used with
graph algorithms outside graph mining, 
cf.~Section~\ref{sec:formulations}. As discussed in~\cref{sec:sisa-soft},
SISA could also potentially accelerate graph
algorithms expressed in 
the vertex-centric paradigm (and others),
by associating certain parts of these algorithms (e.g., search)
with set operations.
}

\all{
\macb{Prefetching}
There are numerous works on hardware and software prefetching, in general
settings~\cite{oren2000survey, byna2008taxonomy, vanderwiel1996survey,
ainsworth2017software, mittal2016survey} and in the context of graph
algorithms~\cite{ainsworth2016graph, zhou2018gas, mukkara2018exploiting}.
However, the latter \emph{do not target graph mining}.
One can \emph{integrate such schemes with SISA's extensible HW
design}, for example by modifying chunk prefetching deployed in SMU.
}

\all{
\macb{Graph Mining}
Graph mining is important for large scientific domains, such as computational
biology and chemistry, medicine, machine learning, social network analysis, and
others~\cite{cook2006mining, evans2010clique, tomita2011efficient}; see
Section~\ref{sec:intro} for more examples.  While graph mining is mostly
unexplored in computer architecture, several software frameworks for graph
mining have recently been proposed~\cite{joshi2018efficient, kalnis2012mizan,
yan2019t, dias2019fractal, chen2018g, yan2017g, iyer2018asap,
teixeira2015arabesque, quamar2016nscale}.  However, they mostly focus on
techniques and optimizations for scaling mining workloads to distributed
clusters or out-of-core systems.
Contrarily, \emph{SISA is the first design that offers hardware support for
graph mining} and an \emph{ISA for graph mining}.  Moreover, we predict that
our \emph{set-centric formulations} could also be implemented and optimized for
a distributed setting.

We know of only one \emph{vertex-centric} graph mining
algorithm~\cite{brighen2019listing}.  Its formulation is significantly more
complex than the original algorithm (29~lines for \emph{just} the user-provided
kernel vs.~10~lines for the whole standard formulation).  Moreover, the authors
only prove that \emph{the execution will finish in a finite number of
iterations}, without any tighter guarantees. 
\maciej{Add: emptyheaded paper, and check the intersection paper}
}

\tr{\macb{Graph Algorithms and Set Algebra}}
Sets are used in different graph algorithms, to
simplify operations on selected data structures~\cite{pingali2011tao,
besta2017push, meyer2003delta, shiloach1980log, khaouid2015k,
schank2007algorithmic, bron1973algorithm}. For example, the BFS frontier can be
modeled as a set.
\iftr
Similarly, inserting and removing an element
from the frontier was often modeled with inserting and removing an element from
a set~\cite{beamer2013direction}. 
\fi
\iftr
The only work (that we know of) which focuses on graph processing and sets is
due to Han et al.~\cite{han2018speeding} and Aberger et
al.~\cite{aberger2017emptyheaded}, where the authors accelerate set
intersections for graph analytics.
\fi
\iftr
Contrarily, SISA is the first attempt to accelerate \emph{general graph mining}
by identifying \emph{different} set operations used in these algorithms,
formulating these operations \emph{as an ISA extension}, and supporting these
instructions with \emph{in-situ and near-memory acceleration}.
\fi
\sethlcolor{yellow}
{Here, SISA's main contribution is \emph{not} to simply use set notation.
Instead, from the algorithmic perspective, SISA is the first design that (1) 
uses set operations as the \emph{primary building blocks},
which break down complex graph mining algorithms into simple units of parallel execution,
and (2) identifies the ``appropriate'' set operations (i.e., operations
that are easily accelerated with PIM) and reformulates selected
algorithms so that they use such operations, cf.~Table~\mbox{\ref{tab:set-forms}}.}

\tr{\macb{Set Programming}}
\tr{
Some works propose to use \textbf{sets as a basis for general programming} to enhance
coding productivity, with use cases in software prototyping. These works
include SETL~\cite{schwartz2012programming, kennedy1975introduction},
ISETL~\cite{dubinsky1995isetl}, and CLAIRE~\cite{caseau2002claire}.
\emph{These efforts do not focus on graph processing or improving
performance}.

}

\marginparX{\Large\vspace{1em}\colorbox{yellow}{\textbf{B}}}

\textbf{SISA vs.~AutoMine~\mbox{\cite{mawhirter2019automine}}}
AutoMine~\mbox{\cite{mawhirter2019automine}} uses set operations to express
finding graph patterns. It focuses on automatic compilation of set schedules
into efficient code. This part is orthogonal to our work and AutoMine
\emph{could easily be combined with SISA} to, for example, generate code based
on SISA's set-centric formulations.
Note that SISA's set formulations are superior to those of AutoMine, because
SISA (1) supports \emph{all} set operations, including \emph{non anti-monotonic
ones} (not just intersection and difference), (2) it expresses whole algorithms
with the set building blocks (not just pattern generation schedules), and (3)
it targets broad graph mining (not just pattern matching).

\marginparX{\Large\vspace{-3em}\colorbox{yellow}{\textbf{B}}}

\marginparX{\Large\vspace{-10em}\colorbox{yellow}{\textbf{F}}}

SISA shows how to seamlessly integrate PUM and PNM capabilities
in a single system. They work synergistically and produce significantly better results
than working separately.

%% file: table-hw.tex
\ifsq
\begin{table}[t]
\else
\begin{table*}[hbtp]
\fi
%
%
\ifsq
\setlength{\tabcolsep}{0.1pt}
\else
\setlength{\tabcolsep}{1pt}
\fi
\ifsq\renewcommand{\arraystretch}{0.7}\fi
\centering
\scriptsize
\ifsq\ssmall\fi
%
\begin{tabular}{lllcccccccccccccc}
\toprule
\multirow{2}{*}{\makecell[c]{\textbf{Reference /}\\\textbf{Accelerator}}} & 
\multirow{2}{*}{\makecell[c]{\textbf{Prob.}}} & 
\multirow{2}{*}{\makecell[c]{\textbf{Key memory}\\\textbf{mechanism}}} & 
\multicolumn{4}{c}{\textbf{Pattern M.}} & 
\multicolumn{4}{c}{\textbf{Learning}} & 
\multicolumn{3}{c}{\textbf{``Low-c.''}} & 
\multirow{2}{*}{\makecell[c]{\textbf{is}}} & 
\multirow{2}{*}{\makecell[c]{\textbf{xl}}} & 
\multirow{2}{*}{\makecell[c]{\textbf{ab}}} \\
\cmidrule(lr){4-7} \cmidrule(lr){8-11} 
 & & & 
\textbf{mc} & \textbf{kc} & \textbf{ds} & \textbf{si} & 
\textbf{vs} & \textbf{lp} & \textbf{cl} & \textbf{av} & 
\textbf{bf} & \textbf{pr} & \textbf{cc} & & & \\ 
\midrule
\textbf{[Pi]} GaaS-X~\cite{challapallegaas} & SpMV & \textbf{[e]} CAM/MAC &  \faTimesT &  \faTimesT &  \faTimesT &  \faTimesT &  \faTimesT &  \faTimesT &  \faTimesT & \faTimesT & \faBatteryFullT & \faBatteryFullT & \faBatteryFullT & \faTimesT &  \faBatteryFullT &  \faBatteryHalfT \\
\iftr
\textbf{[Pi]} GraphSAR~\cite{dai2019graphsar} & ver-c & \textbf{[e]} ReRAM & \faTimesT &  \faTimesT &  \faTimesT &  \faTimesT &  \faTimesT &  \faTimesT &  \faTimesT & \faTimesT & \faBatteryFullT & \faBatteryFullT & \faBatteryFullT &   \faTimesT &  \faTimesT &  \faTimesT \\
\fi
\textbf{[Pi]} GraphiDe~\cite{angizi2019graphide} & low-c & \textbf{[e]} DRAM & \faTimesT & \faTimesT & \faTimesT & \faTimesT & \faBatteryFullT & \faTimesT & \faTimesT & \faTimesT & \faBatteryFullT & \faBatteryFullT & \faBatteryFullT &  \faBatteryHalfT & \faBatteryHalfT & \faTimesT \\
\iftr
\textbf{[Pi]} GraphIA~\cite{li2018graphia} & edge-c & \textbf{[e]} DRAM &  \faTimesT & \faTimesT & \faTimesT & \faTimesT & \faTimesT & \faTimesT & \faTimesT & \faTimesT & \faBatteryFullT & \faBatteryFullT & \faBatteryFullT &  \faTimesT & \faTimesT & \faTimesT \\
\fi
\iftr
\midrule
\fi
\tr{\textbf{[Pc]} GraphVine~\cite{belayneh2020graphvine} & ver-c & \textbf{[e]} 3D DRAM & \faTimesT & \faTimesT & \faTimesT & \faTimesT & \faTimesT & \faTimesT & \faTimesT & \faTimesT & \faBatteryFullT & \faBatteryFullT & \faBatteryFullT &  \faTimesT & \faTimesT & \faTimesT\\}
\tr{\textbf{[Pc]} ReGra~\cite{liu2020regra} & BFS & \textbf{[e]} ReRAM & \faTimesT & \faTimesT & \faTimesT & \faTimesT & \faTimesT & \faTimesT & \faTimesT & \faTimesT & \faBatteryFullT & \faTimesT & \faTimesT & \faTimesT & \faTimesT & \faTimesT \\}
\textbf{[Pc]} Spara~\cite{zheng2020spara} & ver-c & \textbf{[e]} ReRAM &  \faTimesT &  \faTimesT &  \faTimesT &  \faTimesT &  \faTimesT &  \faTimesT &  \faTimesT & \faTimesT & \faBatteryFullT & \faBatteryFullT & \faBatteryFullT &  \faTimesT &  \faBatteryHalfT &  \faTimesT \\
\textbf{[Pc]} GraphQ~\cite{zhuo2019graphq} & ver-c & \textbf{[e]} HMC & \faTimesT &  \faTimesT &  \faTimesT &  \faTimesT &  \faTimesT &  \faTimesT &  \faTimesT & \faTimesT & \faBatteryFullT & \faBatteryFullT & \faBatteryFullT &  \faTimesT &  \faBatteryHalfT &  \faTimesT  \\
\textbf{[Pc]} GraphS~\cite{angizi2019graphs} & low-c & \textbf{[e]} SOT-MRAM & \faTimesT & \faTimesT & \faTimesT & \faTimesT & \faBatteryFullT & \faTimesT & \faTimesT & \faTimesT & \faBatteryFullT & \faBatteryFullT & \faBatteryFullT &  \faTimesT & \faTimesT & \faTimesT \\
\textbf{[Pc]} RAGra~\cite{huang2019ragra} & ver-c & \textbf{[e]} 3D ReRAM & \faTimesT &  \faTimesT &  \faTimesT &  \faTimesT &  \faTimesT &  \faTimesT &  \faTimesT & \faTimesT & \faBatteryFullT & \faBatteryFullT & \faBatteryFullT &  \faTimesT &  \faBatteryHalfT &  \faTimesT  \\
\textbf{[Pc]} GRAM~\cite{zhou2019gram} & ver-c & \textbf{[e]} ReRAM & \faTimesT & \faTimesT & \faTimesT & \faTimesT & \faTimesT & \faTimesT & \faTimesT & \faTimesT & \faBatteryFullT & \faBatteryFullT & \faBatteryFullT &  \faTimesT & \faBatteryHalfT & \faTimesT \\
\tr{\textbf{[Pc]} Messagefusion~\cite{belayneh2019messagefusion} & ver-c & \textbf{[e]} HMC & \faTimesT & \faTimesT & \faTimesT & \faTimesT & \faTimesT & \faTimesT & \faTimesT & \faTimesT & \faBatteryFullT & \faBatteryFullT & \faBatteryFullT &  \faTimesT & \faTimesT & \faTimesT \\}
\tr{\textbf{[Pc]} Mosayebi et al.~\cite{mosayebi2019enhanced} & low-c & \textbf{[e]} HMC &  \faTimesT & \faTimesT & \faTimesT & \faTimesT & \faTimesT & \faTimesT & \faTimesT & \faTimesT & \faBatteryFullT & \faBatteryFullT & \faBatteryFullT &  \faTimesT & \faTimesT & \faTimesT \\}
\tr{\textbf{[Pc]} RPBFS~\cite{han2018novel} & BFS & \textbf{[e]} ReRAM & \faTimesT & \faTimesT & \faTimesT & \faTimesT & \faTimesT & \faTimesT & \faTimesT & \faTimesT & \faBatteryFullT & \faTimesT & \faTimesT &  \faTimesT & \faBatteryHalfT & \faTimesT   \\}
\textbf{[Pc]} GraphR~\cite{song2018graphr} & \makecell[l]{SpMV} & \textbf{[e]} ReRAM & \faTimesT & \faTimesT & \faTimesT & \faTimesT & \faTimesT & \faTimesT & \faTimesT & \faTimesT & \faBatteryFullT & \faBatteryFullT & \faBatteryFullT &  \faTimesT & \faBatteryHalfT & \faBatteryHalfT \\
\textbf{[Pc]} GraphP~\cite{zhang2018graphp} & ver-c & \textbf{[e]} HMC & \faTimesT & \faTimesT & \faTimesT & \faTimesT & \faTimesT & \faTimesT & \faTimesT & \faTimesT & \faBatteryFullT & \faBatteryFullT & \faBatteryFullT &  \faTimesT & \faTimesT & \faBatteryFullT  \\
\textbf{[Pc]} Tesseract~\cite{ahn2015scalable_tes} & low-c & \textbf{[e]} HMC & \faTimesT & \faTimesT & \faTimesT & \faTimesT & \faTimesT & \faTimesT & \faTimesT & \faTimesT & \faBatteryFullT & \faBatteryFullT & \faBatteryFullT &  \faTimesT & \faBatteryFullT & \faBatteryHalfT \\
\textbf{[Pc]} PIM-Enabled~\cite{ahn2015pim} & low-c & \textbf{[e]} HMC & \faTimesT & \faTimesT & \faTimesT & \faTimesT & \faTimesT & \faTimesT & \faTimesT & \faTimesT & \faBatteryFullT & \faBatteryFullT & \faBatteryFullT &  \faBatteryHalfT & \faBatteryFullT & \faBatteryHalfT \\
\textbf{[Pc]} Gao et al.~\cite{gao2015practical} & low-c & 3D DRAM & \faTimesT & \faTimesT & \faBatteryHalfT & \faTimesT & \faTimesT & \faTimesT & \faTimesT & \faTimesT & \faBatteryFullT & \faBatteryFullT & \faBatteryFullT &  \faTimesT & \faBatteryFullT & \faTimesT \\
\iftr
\textbf{[Pc]} LiM~\cite{zhu20133d, zhu2013accelerating} & SpMSpM & \textbf{[e]} 3D DRAM & \faTimesT & \faTimesT & \faTimesT & \faTimesT & \faTimesT & \faTimesT & \faTimesT & \faTimesT & \faBatteryFullT & \faBatteryFullT & \faBatteryFullT &  \faTimesT & \faTimesT & \faTimesT \\
\fi
\iftr
\midrule
\fi
\textbf{[A]} IntersectX~\cite{rao2021intersectx} & pattern m. & \textbf{[e]} cache & \faTimesT & \faBatteryFullT & \faBatteryFullT & \faBatteryFullT & \faTimesT & \faTimesT & \faTimesT & \faTimesT & \faTimesT & \faTimesT & \faTimesT &  \faBatteryFullT & \faBatteryFullT & \faBatteryFullT \\
\textbf{[A]} Gramer~\cite{yao2020locality} & pattern m. & DRAM, cache & \faTimesT & \faBatteryFullT & \faBatteryFullT & \faBatteryHalfT & \faTimesT & \faTimesT & \faTimesT & \faTimesT & \faTimesT & \faTimesT & \faTimesT &  \faTimesT & \faBatteryHalfT & \faTimesT \\
\textbf{[A]} TrieJax~\cite{kalinsky2020triejax} & joins & DRAM, LLC & \faTimesT & \faBatteryFullT & \faBatteryFullT & \faTimesT & \faTimesT & \faTimesT & \faTimesT & \faTimesT & \faTimesT & \faTimesT & \faTimesT &  \faTimesT & \faBatteryFullT & \faBatteryHalfT \\
\textbf{[A]} HyGCN~\cite{yan2020hygcn} & GCN & eDRAM &  \faTimesT &  \faTimesT &  \faTimesT &  \faTimesT &  \faBatteryHalfT &  \faBatteryFullT &  \faBatteryFullT & \faTimesT & \faTimesT & \faTimesT & \faBatteryHalfT &  \faTimesT &  \faBatteryHalfT &  \faBatteryFullT \\
\tr{\textbf{[A]} GCAcc~\cite{qian2018cgacc} & BFS & \textbf{[e]} HMC & \faTimesT & \faTimesT & \faTimesT & \faTimesT & \faTimesT & \faTimesT & \faTimesT & \faTimesT & \faBatteryFullT & \faTimesT & \faTimesT &  \faTimesT & \faBatteryHalfT & \faTimesT \\}
\textbf{[A]} Outerspace~\cite{pal2018outerspace} & SpMSpM & HBM & \faTimesT & \faBatteryHalfT & \faTimesT & \faTimesT & \faTimesT & \faTimesT & \faTimesT & \faTimesT & \faBatteryFullT & \faBatteryFullT & \faBatteryFullT &  \faTimesT &   \faTimesT & \faTimesT \\
\textbf{[A]} Domino~\cite{xu2018domino} & low-c & on-chip buffers &  \faTimesT & \faBatteryHalfT & \faTimesT & \faTimesT & \faTimesT & \faTimesT & \faTimesT & \faTimesT &\faBatteryFullT & \faBatteryFullT & \faBatteryFullT &  \faTimesT & \faTimesT & \faTimesT \\
\textbf{[A]} GraphPIM~\cite{nai2017graphpim} & low-c & \textbf{[e]} HMC & \faTimesT & \faBatteryHalfT & \faBatteryHalfT & \faTimesT & \faTimesT & \faTimesT & \faTimesT & \faTimesT & \faBatteryFullT & \faBatteryFullT & \faBatteryFullT &  \faBatteryFullT & \faBatteryFullT & \faTimesT \\
\textbf{[A]} Graphicionado~\cite{ham2016graphicionado} & ver-c & \textbf{[e]} eDRAM & \faTimesT & \faTimesT & \faTimesT & \faTimesT & \faTimesT & \faTimesT & \faTimesT & \faTimesT & \faBatteryFullT & \faBatteryFullT & \faBatteryFullT &  \faTimesT & \faTimesT & \faBatteryHalfT \\
\textbf{[A]} Ozdal et al.~\cite{ozdal2016energy} & ver-c & \textbf{[e]} caches & \faTimesT & \faTimesT & \faTimesT & \faTimesT & \faTimesT & \faTimesT & \faTimesT & \faTimesT & \faBatteryFullT & \faBatteryFullT & \faBatteryFullT &  \faTimesT & \faBatteryHalfT  & \faBatteryHalfT  \\
\iftr
\iftr
\midrule
\fi
\textbf{[M]} GraphSSD~\cite{matam2019graphssd} & low-c & \textbf{[e]} SSD & \faTimesT & \faTimesT & \faTimesT & \faTimesT & \faTimesT & \faTimesT & \faTimesT & \faTimesT & \faBatteryFullT & \faBatteryFullT & \faBatteryFullT &   \faTimesT & \faBatteryFullT & \faBatteryHalfT \\
\textbf{[M]} GRASP~\cite{faldu2019poster} & low-c & \textbf{[e]} LLC & \faTimesT & \faTimesT & \faTimesT & \faTimesT & \faTimesT & \faTimesT & \faTimesT & \faTimesT & \faBatteryFullT & \faBatteryFullT & \faBatteryFullT &  \faTimesT & \faBatteryHalfT & \faTimesT \\
\textbf{[M]} DROPLET~\cite{basak2019analysis} & edge-c & \textbf{[e]} DRAM pref. & \faTimesT & \faTimesT & \faTimesT & \faTimesT & \faTimesT & \faTimesT & \faTimesT & \faTimesT & \faBatteryFullT & \faBatteryFullT & \faBatteryFullT &  \faTimesT & \faBatteryFullT & \faTimesT  \\
\textbf{[M]} Ainsworth~\cite{ainsworth2018event} & low-c & \textbf{[e]} DRAM pref. & \faTimesT & \faTimesT & \faTimesT & \faTimesT & \faTimesT & \faTimesT & \faTimesT & \faTimesT &\faBatteryFullT & \faBatteryFullT & \faBatteryFullT &  \faTimesT & \faBatteryFullT & \faTimesT \\
\textbf{[M]} HyVE~\cite{huang2018hyve} & ver-c & ReRAM, SRAM & \faTimesT & \faTimesT & \faTimesT & \faTimesT & \faTimesT & \faTimesT & \faTimesT & \faTimesT & \faBatteryFullT & \faBatteryFullT & \faBatteryFullT &  \faTimesT & \faBatteryFullT &  \faTimesT \\
\textbf{[M]} HATS~\cite{mukkara2018exploiting} & low-c & \textbf{[e]} caches & \faTimesT & \faTimesT & \faTimesT & \faTimesT & \faTimesT & \faTimesT & \faTimesT & \faTimesT & \faBatteryFullT & \faBatteryFullT & \faBatteryFullT &  \faTimesT & \faBatteryFullT & \faBatteryFullT \\
\textbf{[M]} OSCAR~\cite{singapura2017oscar} & edge-c & \textbf{[e]} scratchpads & \faTimesT & \faTimesT & \faTimesT & \faTimesT & \faTimesT & \faTimesT & \faTimesT & \faTimesT & \faBatteryFullT & \faBatteryFullT & \faBatteryFullT &  \faTimesT & \faBatteryFullT & \faBatteryHalfT \\
\textbf{[M]} IMP~\cite{yu2015imp} & low-c & \textbf{[e]} caches & \faTimesT & \faBatteryHalfT & \faTimesT & \faTimesT & \faTimesT & \faTimesT & \faTimesT & \faTimesT & \faBatteryFullT & \faBatteryFullT & \faBatteryFullT &  \faTimesT & \faBatteryFullT & \faBatteryHalfT \\
\iftr
\midrule
\fi
\textbf{[F]} GraphABCD~\cite{yang2020graphabcd} & low-c & DRAM & \faTimesT & \faTimesT & \faTimesT & \faTimesT & \faBatteryHalfT & \faBatteryHalfT & \faBatteryHalfT & \faTimesT & \faBatteryFullT & \faBatteryFullT & \faBatteryFullT &  \faTimesT & \faBatteryFullT & \faBatteryFullT \\
\textbf{[F]} Wang et al.~\cite{wang2020fpga} & clustering & BRAM & \faTimesT & \faTimesT & \faTimesT & \faTimesT & \faTimesT & \faTimesT & \faBatteryHalfT & \faTimesT & \faTimesT & \faTimesT & \faTimesT &  \faBatteryFullT & \faTimesT & \faBatteryHalfT \\
\textbf{[F]} ForeGraph~\cite{dai2017foregraph, dai:foregraph} & low-c & BRAM & \faTimesT & \faTimesT & \faTimesT & \faTimesT & \faTimesT & \faTimesT & \faTimesT & \faTimesT & \faBatteryFullT & \faBatteryFullT & \faBatteryFullT &  \faTimesT & \faTimesT & \faBatteryHalfT \\
%
%
\textbf{[F]} Yang~\cite{yang2018efficient} & ver-c & DRAM & \faTimesT & \faTimesT & \faTimesT & \faTimesT & \faTimesT & \faTimesT & \faTimesT & \faTimesT & \faBatteryFullT & \faBatteryFullT & \faBatteryFullT &  \faTimesT & \faBatteryHalfT & \faBatteryHalfT \\
\textbf{[F]} Yao~\cite{yao2018efficient} & low-c & DRAM & \faTimesT & \faBatteryHalfT & \faTimesT & \faTimesT & \faTimesT & \faTimesT & \faTimesT & \faTimesT & \faBatteryFullT & \faBatteryFullT & \faBatteryFullT &  \faTimesT & \faBatteryHalfT &\faTimesT  \\
\textbf{[F]} Zhou~\cite{zhou2018framework} & edge-c & DRAM & \faTimesT & \faTimesT & \faTimesT & \faTimesT & \faTimesT & \faTimesT & \faTimesT & \faTimesT & \faBatteryFullT & \faBatteryFullT & \faBatteryFullT &  \faTimesT & \faTimesT & \faBatteryHalfT \\ 
\textbf{[F]} ExtraV~\cite{lee2017extrav} & low-c & DRAM & \faTimesT & \faTimesT & \faTimesT & \faTimesT & \faTimesT & \faTimesT & \faTimesT & \faTimesT & \faBatteryFullT & \faBatteryFullT & \faBatteryFullT &  \faTimesT & \faBatteryFullT & \faBatteryFullT  \\
\textbf{[F]} Ma~\cite{ma2017fpga} & low-c & DRAM &  \faTimesT & \faBatteryHalfT & \faTimesT & \faTimesT & \faTimesT & \faTimesT & \faTimesT & \faTimesT & \faBatteryFullT & \faBatteryFullT & \faBatteryFullT &  \faTimesT & \faBatteryHalfT & \faTimesT \\
\textbf{[F]} Zhou~\cite{zhou2017accelerating} & ver-c, edge-c & DRAM & \faTimesT & \faTimesT & \faTimesT & \faTimesT & \faTimesT & \faTimesT & \faTimesT & \faTimesT & \faBatteryFullT & \faBatteryFullT & \faBatteryFullT &  \faTimesT & \faBatteryHalfT & \faBatteryFullT  \\
\textbf{[F]} GraVF~\cite{engelhardt2016gravf} & ver-c & BRAM &  \faTimesT  &  \faTimesT  &  \faTimesT  &  \faTimesT  &  \faTimesT  &  \faTimesT  &  \faTimesT & \faTimesT & \faBatteryFullT & \faBatteryFullT & \faBatteryFullT &  \faTimesT & \faTimesT & \faBatteryFullT \\
\textbf{[F]} Zhou~\cite{zhou2016high, zhou2015pagerank} & edge-c & DRAM &  \faTimesT &  \faTimesT &  \faTimesT &  \faTimesT &  \faTimesT &  \faTimesT &  \faTimesT & \faTimesT & \faBatteryFullT & \faBatteryFullT & \faBatteryFullT &  \faTimesT & \faTimesT & \faBatteryHalfT \\
\textbf{[F]} GraphOps~\cite{oguntebi:GraphOps} & low-c & BRAM & \faTimesT &   \faTimesT &   \faTimesT &   \faTimesT &   \faTimesT &   \faTimesT &   \faTimesT & \faTimesT & \faBatteryFullT & \faBatteryFullT & \faBatteryFullT &  \faTimesT & \faBatteryHalfT & \faBatteryFullT  \\
\textbf{[F]} FPGP~\cite{dai:fpgp} & ver-c & DRAM & \faTimesT &  \faTimesT &  \faTimesT &  \faTimesT &  \faTimesT &  \faTimesT &  \faTimesT & \faTimesT & \faBatteryFullT & \faBatteryFullT & \faBatteryFullT &  \faTimesT & \faBatteryHalfT & \faTimesT \\ 
\textbf{[F]} GraphSoC~\cite{kapre:custom_graph_FPGA} & low-c, SpMV & BRAM & \faTimesT & \faTimesT & \faTimesT & \faTimesT & \faTimesT & \faTimesT & \faTimesT & \faTimesT & \faBatteryFullT & \faBatteryFullT & \faBatteryFullT &  \faBatteryFullT & \faBatteryHalfT & \faBatteryHalfT \\
\textbf{[F]} GraphGen~\cite{weisz:GraphGen} & ver-c & DRAM &  \faTimesT &  \faTimesT &  \faTimesT &  \faTimesT &  \faTimesT &  \faTimesT &  \faTimesT & \faTimesT & \faBatteryFullT & \faBatteryFullT & \faBatteryFullT &  \faTimesT & \faBatteryFullT & \faBatteryHalfT \\ 
\textbf{[F]} GraphStep~\cite{kapre2006graphstep} & low-c & BRAM & \faTimesT & \faTimesT & \faTimesT & \faTimesT & \faTimesT & \faTimesT & \faTimesT & \faTimesT & \faBatteryFullT & \faBatteryFullT & \faBatteryFullT &  \faTimesT & \faBatteryHalfT & \faBatteryFullT \\
\tr{\textbf{[F]} Besta et al.~\cite{besta2020substream, besta2019substream} & matchings & DRAM & \faTimesT & \faTimesT & \faTimesT & \faTimesT & \faTimesT & \faTimesT & \faTimesT & \faTimesT & \faTimesT & \faTimesT & \faTimesT  & \faTimesT & \faBatteryHalfT & \faBatteryFullT \\}
\textbf{[F]} Betkaoui et al.~\cite{betkaoui2011framework} & low-c & DRAM & \faTimesT & \faBatteryHalfT & \faBatteryHalfT & \faTimesT & \faTimesT & \faTimesT & \faTimesT & \faTimesT & \faBatteryFullT & \faBatteryFullT & \faBatteryFullT &  \faTimesT & \faBatteryFullT & \faBatteryHalfT \\
\tr{\multicolumn{2}{l}{\textbf{[F]} \makecell[l]{Works on SSSP~\cite{babb1996solving, dandalis1999domain, tommiska2001dijkstra}\\ \cite{mencer2002hagar, sridharan2009hardware, jagadeesh2011field, zhou2015sssp, lei2016fpga, wang2019processor}}} & \makecell[l]{Hardwired,\\ BRAM} & \faTimesT & \faTimesT & \faTimesT & \faTimesT & \faTimesT & \faTimesT & \faTimesT & \faTimesT & \faBatteryFullT & \faTimesT & \faTimesT &  \faTimesT & \faBatteryHalfT$^*$ & \faTimesT \\}
\tr{\multicolumn{2}{l}{\textbf{[F]} Works on APSP~\cite{bondhugula:APSP_FPGA, betkaoui:APSP_FPGA}} & DRAM & \faTimesT & \faTimesT & \faTimesT & \faTimesT & \faTimesT & \faTimesT & \faTimesT & \faTimesT & \faBatteryFullT & \faTimesT & \faTimesT & \faTimesT & \faBatteryHalfT$^*$ & \faTimesT \\}
\tr{\multicolumn{2}{l}{\textbf{[F]} \makecell[l]{Works on BFS\\\cite{wang2010message, betkaoui2012reconfigurable, attia2014cygraph, ni2014parallel, umuroglu:hybrid_bfs_FPGA, wang2019processor}}} & DRAM & \faTimesT & \faTimesT & \faTimesT & \faTimesT & \faTimesT & \faTimesT & \faTimesT & \faTimesT & \faBatteryFullT & \faTimesT & \faTimesT & \faTimesT & \faBatteryHalfT$^*$ & \faTimesT \\}
\fi
\iftr
\midrule
\fi
\iftr
\textbf{[M+Pc]} GraphDynS~\cite{yan2019alleviating} & low-c & HBM + others & \faTimesT & \faTimesT & \faTimesT & \faTimesT & \faTimesT & \faTimesT & \faTimesT & \faTimesT & \faBatteryFullT & \faBatteryFullT & \faBatteryFullT &  \faTimesT & \faBatteryFullT & \faBatteryFullT  \\
\fi
\iftr
\fi
\textbf{[A+Pc]} EnGN~\cite{he2019engn} & GNN & \textbf{[e]} HBM & \faTimesT & \faTimesT & \faTimesT & \faTimesT & \faBatteryHalfT & \faBatteryFullT & \faBatteryFullT & \faTimesT & \faTimesT & \faBatteryHalfT & \faTimesT &  \faTimesT & \faTimesT & \faTimesT \\
\tr{\textbf{[A+Pc]} Sadi et al.~\cite{sadi2018pagerank} & PageRank & HBM & \faTimesT & \faTimesT & \faTimesT & \faTimesT & \faTimesT & \faTimesT & \faTimesT & \faTimesT & \faTimesT & \faBatteryFullT & \faTimesT & \faTimesT & \faTimesT & \faTimesT  \\}
\textbf{[A+Pc]} OMEGA~\cite{addisie2018heterogeneous} & low-c & \textbf{[e]} Scratchpads & \faTimesT & \faBatteryHalfT & \faTimesT & \faTimesT & \faTimesT & \faTimesT & \faBatteryHalfT & \faTimesT & \faBatteryFullT & \faBatteryFullT & \faBatteryFullT &  \faTimesT & \faTimesT & \faBatteryHalfT \\
\tr{\textbf{[A+Pc]} Sadi et al.~\cite{sadi2017algorithm} & SpMV & \makecell[l]{\textbf{[e]} HBM} & \faTimesT & \faTimesT & \faTimesT & \faTimesT & \faTimesT & \faTimesT & \faTimesT & \faTimesT & \faBatteryFullT & \faBatteryFullT & \faBatteryFullT &  \faTimesT & \faTimesT & \faTimesT \\}
\iftr
\fi
\textbf{[A+Pc+M]} GraphH~\cite{dai2018graphh} & ver-c & \textbf{[e]} \makecell[l]{HMC} &  \faTimesT & \faTimesT & \faTimesT & \faTimesT & \faTimesT & \faTimesT & \faTimesT & \faTimesT & \faBatteryFullT & \faBatteryFullT & \faBatteryFullT &  \faTimesT & \faBatteryFullT &  \faTimesT \\ 
%
%
%
%
%
\iftr
\fi
\textbf{[F+Pc]} HRL~\cite{gao2016hrl} & ver-c & \makecell[l]{\textbf{[e]} 3D DRAM}  & \faTimesT & \faTimesT & \faTimesT & \faTimesT & \faTimesT & \faTimesT & \faTimesT & \faTimesT & \faBatteryFullT & \faBatteryFullT & \faBatteryFullT &  \faTimesT & \faBatteryHalfT & \faTimesT \\
\tr{\multicolumn{2}{l}{\textbf{[F+Pc]} \makecell[l]{Works on BFS \cite{zhang:graph_FPGA, zhang2018degree, khoram2018accelerating}}} & HMC & \faTimesT & \faTimesT & \faTimesT & \faTimesT & \faTimesT & \faTimesT & \faTimesT & \faTimesT & \faBatteryFullT & \faTimesT & \faTimesT & \faTimesT & \faBatteryHalfT & \faBatteryHalfT$^*$ \\}
\midrule
\makecell[l]{\textbf{[Pc+Pi]} \textbf{SISA} \textbf{[This work]}} & \makecell[l]{{Graph mining}} & PIM & \faBatteryFullT & \faBatteryFullT & \faBatteryFullT & \faBatteryFullT & \faBatteryFullT & \faBatteryFullT & \faBatteryFullT & \faBatteryFullT & \faTimesT & \faTimesT & \faTimesT &  \faBatteryFullT & \faBatteryFullT & \faBatteryFullT \\ 
\bottomrule
\end{tabular}
%
%
\caption{
\textmd{\textbf{Comparison of SISA to graph-related {accelerators}, focusing on
{supported graph mining problems} and {offered architecture elements}}.
%
%
``\faBatteryFull'': Support / significant focus. ``\faBatteryHalf'': Partial support / some
focus. ``\faTimes'': no support / no focus.
\textbf{\ul{Addressed problems}:} see~Table~\ref{tab:comparison_models}\tr{ for
details; ver-c: vertex-centric, edge-c: edge-centric, low-c: general
low-complexity problems, SpMV: sparse matrix-vector products, SpMSpM: sparse
matrix-sparse matrix products, GCN: graph convolution networks, GNN: graph
neural networks}.
\textbf{\ul{Graph problems and algorithms}: as in Table~\ref{tab:comparison_models}}.
\textbf{\ul{Considered architecture and stack elements}:}
``\textbf{is}'': an ISA, or its extensions,
``\textbf{xl}'': a cross-layer design,
``\textbf{ab}'': a programming paradigm 
\all{\maciej{fix}
(here, ``'' indicates that -- for example -- a given work introduces a new model or abstraction,
or conducts a detailed analysis of existing models, ``'' indicates that the design of
a given accelerator is based on some model or abstraction).}
\textbf{\ul{Classes of accelerators}:}
\textbf{[Pi]}: \emph{in-situ} PIM,
\textbf{[Pc]}: \emph{near memory} PIM (e.g., logic layers),
\textbf{[A]}: ASIC,
\iftr
\textbf{[M]}: focus on memory hierarchy enhancements,
\fi
\iftr
\textbf{[F]}: FPGA,
\fi
\textbf{[e]} focus on extensions and modifications to the established (already proposed) HW technology,
\tr{$^*$Applies to some works in a given group.}
Note that the generality of SISA comes from harnessing all basic set algebra operations.
}}
\vspaceSQ{-1.5em}
\label{tab:comparison_problems}
\ifsq
\end{table}
\else
\end{table*}
\fi

%% file: conclusion.tex
\vspaceSQ{-0.5em}
\section{Discussion and Conclusion}

\enlargeSQ

We develop the first hardware acceleration for 
broad graph mining.
\ifconf
\fi
First, we offer a set-centric programming paradigm, where one 
identifies and exposes set operations in graph mining algorithms\tr{,
resulting in ``set-centric'' algorithmic formulations}. This enables 
competitive time complexities and succinct formulations.
\hl{We support labeled graphs and non anti-monotonic set
operations~\mbox{\cite{rao2021intersectx, mawhirter2019automine,
kalinsky2020triejax, yao2020locality}}.}

\marginpar{\Large\vspace{-1em}\colorbox{yellow}{\textbf{L}}}

\ifconf
\fi
Second, the set-centric algorithms are mapped to SISA, a small yet expressive
\tr{family of instructions that form a} ``set-centric'' ISA extension
for graph mining. SISA could be extended \tr{into
multiple directions, for example,} with CISC-style set instructions that accept
multiple arguments (e.g., \tr{to intersect multiple sets in a single
instruction} $A_1 \cap ... \cap A_l$) to facilitate optimizations such
as vectorization with loop unrolling.
Due to the generality of set algebra, \tr{we predict that} SISA
can be used for problems beyond graph mining\tr{ and general static
graph computations, for example dynamic (time-evolving) graph processing, or
data mining beyond graphs}.
\ifconf
\fi
\iftr
Third, we pick in-situ and logic layer PIM for hardware acceleration,
and offer automatized selection of the most beneficial instruction variants,
maximizing speedups over hand-tuned baselines of parallel graph mining
algorithms. However, the interface based on set algebra could use other
hardware {backends} for SISA instructions.
For example, one could use a GPU backend for fast SIMD-based set
intersections~\cite{han2018speeding}, implement set operations on
FPGAs~\cite{besta2019graph}, execute set operations in
caches~\cite{nag2019gencache, aga2017compute}, or use
ReRAM~\cite{song2018graphr} for efficient in-memory analog matrix-vector
multiplications, which can also be used to
implement some instances of set intersection.
\fi
\ifconf
\fi

\marginparX{\Large\vspace{-2em}\colorbox{yellow}{\textbf{C}}}

Finally, our cross-layer architecture could also be extended in
other directions, for example by providing compiler support
for generating SISA programs from set-centric formulations.
Here, one could use, e.g., AutoMine's~\mbox{\cite{mawhirter2019automine}}
compiler generated schedules as input to some SISA programs.

\marginparX{\Large\vspace{-1em}\colorbox{yellow}{\textbf{B}}}

%% file: sisa_2021.bbl

\begin{thebibliography}{285}


\ifx \showCODEN    \undefined \def \showCODEN     #1{\unskip}     \fi
\ifx \showDOI      \undefined \def \showDOI       #1{#1}\fi
\ifx \showISBNx    \undefined \def \showISBNx     #1{\unskip}     \fi
\ifx \showISBNxiii \undefined \def \showISBNxiii  #1{\unskip}     \fi
\ifx \showISSN     \undefined \def \showISSN      #1{\unskip}     \fi
\ifx \showLCCN     \undefined \def \showLCCN      #1{\unskip}     \fi
\ifx \shownote     \undefined \def \shownote      #1{#1}          \fi
\ifx \showarticletitle \undefined \def \showarticletitle #1{#1}   \fi
\ifx \showURL      \undefined \def \showURL       {\relax}        \fi
\providecommand\bibfield[2]{#2}
\providecommand\bibinfo[2]{#2}
\providecommand\natexlab[1]{#1}
\providecommand\showeprint[2][]{arXiv:#2}

\bibitem[\protect\citeauthoryear{Aberger, Lamb, Tu, N{\"o}tzli, Olukotun, and
  R{\'e}}{Aberger et~al\mbox{.}}{2017}]%
        {aberger2017emptyheaded}
\bibfield{author}{\bibinfo{person}{Christopher~R Aberger},
  \bibinfo{person}{Andrew Lamb}, \bibinfo{person}{Susan Tu},
  \bibinfo{person}{Andres N{\"o}tzli}, \bibinfo{person}{Kunle Olukotun}, {and}
  \bibinfo{person}{Christopher R{\'e}}.} \bibinfo{year}{2017}\natexlab{}.
\newblock \showarticletitle{Emptyheaded: A relational engine for graph
  processing}.
\newblock \bibinfo{journal}{\emph{ACM Transactions on Database Systems (TODS)}}
  \bibinfo{volume}{42}, \bibinfo{number}{4} (\bibinfo{year}{2017}),
  \bibinfo{pages}{1--44}.
\newblock


\bibitem[\protect\citeauthoryear{Addisie, Kassa, Matthews, and
  Bertacco}{Addisie et~al\mbox{.}}{2018}]%
        {addisie2018heterogeneous}
\bibfield{author}{\bibinfo{person}{Abraham Addisie}, \bibinfo{person}{Hiwot
  Kassa}, \bibinfo{person}{Opeoluwa Matthews}, {and} \bibinfo{person}{Valeria
  Bertacco}.} \bibinfo{year}{2018}\natexlab{}.
\newblock \showarticletitle{Heterogeneous memory subsystem for natural graph
  analytics}. In \bibinfo{booktitle}{\emph{2018 IEEE International Symposium on
  Workload Characterization (IISWC)}}. IEEE, \bibinfo{pages}{134--145}.
\newblock


\bibitem[\protect\citeauthoryear{Aga, Jeloka, Subramaniyan, Narayanasamy,
  Blaauw, and Das}{Aga et~al\mbox{.}}{2017}]%
        {aga2017compute}
\bibfield{author}{\bibinfo{person}{Shaizeen Aga}, \bibinfo{person}{Supreet
  Jeloka}, \bibinfo{person}{Arun Subramaniyan}, \bibinfo{person}{Satish
  Narayanasamy}, \bibinfo{person}{David Blaauw}, {and}
  \bibinfo{person}{Reetuparna Das}.} \bibinfo{year}{2017}\natexlab{}.
\newblock \showarticletitle{Compute caches}. In \bibinfo{booktitle}{\emph{2017
  IEEE International Symposium on High Performance Computer Architecture
  (HPCA)}}. IEEE, \bibinfo{pages}{481--492}.
\newblock


\bibitem[\protect\citeauthoryear{Aggarwal and Wang}{Aggarwal and Wang}{2010}]%
        {aggarwal2010managing}
\bibfield{author}{\bibinfo{person}{Charu~C Aggarwal} {and}
  \bibinfo{person}{Haixun Wang}.} \bibinfo{year}{2010}\natexlab{}.
\newblock \bibinfo{booktitle}{\emph{Managing and mining graph data}}.
  Vol.~\bibinfo{volume}{40}.
\newblock \bibinfo{publisher}{Springer}.
\newblock


\bibitem[\protect\citeauthoryear{Agrawal, Srikant, et~al\mbox{.}}{Agrawal
  et~al\mbox{.}}{1994}]%
        {agrawal1994fast}
\bibfield{author}{\bibinfo{person}{Rakesh Agrawal},
  \bibinfo{person}{Ramakrishnan Srikant}, {et~al\mbox{.}}}
  \bibinfo{year}{1994}\natexlab{}.
\newblock \showarticletitle{Fast algorithms for mining association rules}. In
  \bibinfo{booktitle}{\emph{Proc. 20th int. conf. very large data bases,
  VLDB}}, Vol.~\bibinfo{volume}{1215}. Citeseer, \bibinfo{pages}{487--499}.
\newblock


\bibitem[\protect\citeauthoryear{Ahn, Hong, Yoo, Mutlu, and Choi}{Ahn
  et~al\mbox{.}}{2015a}]%
        {ahn2015scalable_tes}
\bibfield{author}{\bibinfo{person}{Junwhan Ahn}, \bibinfo{person}{Sungpack
  Hong}, \bibinfo{person}{Sungjoo Yoo}, \bibinfo{person}{Onur Mutlu}, {and}
  \bibinfo{person}{Kiyoung Choi}.} \bibinfo{year}{2015}\natexlab{a}.
\newblock \showarticletitle{A scalable processing-in-memory accelerator for
  parallel graph processing}. In \bibinfo{booktitle}{\emph{ISCA}}.
\newblock


\bibitem[\protect\citeauthoryear{Ahn, Yoo, Mutlu, and Choi}{Ahn
  et~al\mbox{.}}{2015b}]%
        {ahn2015pim}
\bibfield{author}{\bibinfo{person}{Junwhan Ahn}, \bibinfo{person}{Sungjoo Yoo},
  \bibinfo{person}{Onur Mutlu}, {and} \bibinfo{person}{Kiyoung Choi}.}
  \bibinfo{year}{2015}\natexlab{b}.
\newblock \showarticletitle{{PIM}-enabled instructions: a low-overhead,
  locality-aware processing-in-memory architecture}. In
  \bibinfo{booktitle}{\emph{Computer Architecture (ISCA), 2015 ACM/IEEE 42nd
  Annual International Symposium on}}. IEEE, \bibinfo{pages}{336--348}.
\newblock


\bibitem[\protect\citeauthoryear{Ainsworth and Jones}{Ainsworth and
  Jones}{2018}]%
        {ainsworth2018event}
\bibfield{author}{\bibinfo{person}{Sam Ainsworth} {and}
  \bibinfo{person}{Timothy~M Jones}.} \bibinfo{year}{2018}\natexlab{}.
\newblock \showarticletitle{An event-triggered programmable prefetcher for
  irregular workloads}.
\newblock \bibinfo{journal}{\emph{ACM SIGPLAN Notices}} \bibinfo{volume}{53},
  \bibinfo{number}{2} (\bibinfo{year}{2018}), \bibinfo{pages}{578--592}.
\newblock


\bibitem[\protect\citeauthoryear{Al~Hasan, Chaoji, Salem, and Zaki}{Al~Hasan
  et~al\mbox{.}}{2006}]%
        {al2006link}
\bibfield{author}{\bibinfo{person}{Mohammad Al~Hasan}, \bibinfo{person}{Vineet
  Chaoji}, \bibinfo{person}{Saeed Salem}, {and} \bibinfo{person}{Mohammed
  Zaki}.} \bibinfo{year}{2006}\natexlab{}.
\newblock \showarticletitle{Link prediction using supervised learning}. In
  \bibinfo{booktitle}{\emph{SDM06: workshop on link analysis, counter-terrorism
  and security}}.
\newblock


\bibitem[\protect\citeauthoryear{Al~Hasan and Dave}{Al~Hasan and Dave}{2018}]%
        {al2018triangle}
\bibfield{author}{\bibinfo{person}{Mohammad Al~Hasan} {and}
  \bibinfo{person}{Vachik~S Dave}.} \bibinfo{year}{2018}\natexlab{}.
\newblock \showarticletitle{Triangle counting in large networks: a review}.
\newblock \bibinfo{journal}{\emph{Wiley Interdisciplinary Reviews: Data Mining
  and Knowledge Discovery}} \bibinfo{volume}{8}, \bibinfo{number}{2}
  (\bibinfo{year}{2018}), \bibinfo{pages}{e1226}.
\newblock


\bibitem[\protect\citeauthoryear{Al~Hasan and Zaki}{Al~Hasan and Zaki}{2011}]%
        {al2011survey}
\bibfield{author}{\bibinfo{person}{Mohammad Al~Hasan} {and}
  \bibinfo{person}{Mohammed~J Zaki}.} \bibinfo{year}{2011}\natexlab{}.
\newblock \showarticletitle{A survey of link prediction in social networks}.
\newblock In \bibinfo{booktitle}{\emph{Social network data analytics}}.
  \bibinfo{publisher}{Springer}, \bibinfo{pages}{243--275}.
\newblock


\bibitem[\protect\citeauthoryear{Angizi and Fan}{Angizi and Fan}{2019}]%
        {angizi2019graphide}
\bibfield{author}{\bibinfo{person}{Shaahin Angizi} {and}
  \bibinfo{person}{Deliang Fan}.} \bibinfo{year}{2019}\natexlab{}.
\newblock \showarticletitle{Graphide: A graph processing accelerator leveraging
  in-dram-computing}. In \bibinfo{booktitle}{\emph{Proceedings of the 2019 on
  Great Lakes Symposium on VLSI}}. \bibinfo{pages}{45--50}.
\newblock


\bibitem[\protect\citeauthoryear{Angizi, Sun, Zhang, and Fan}{Angizi
  et~al\mbox{.}}{2019}]%
        {angizi2019graphs}
\bibfield{author}{\bibinfo{person}{Shaahin Angizi}, \bibinfo{person}{Jiao Sun},
  \bibinfo{person}{Wei Zhang}, {and} \bibinfo{person}{Deliang Fan}.}
  \bibinfo{year}{2019}\natexlab{}.
\newblock \showarticletitle{GraphS: A graph processing accelerator leveraging
  SOT-MRAM}. In \bibinfo{booktitle}{\emph{2019 Design, Automation \& Test in
  Europe Conference \& Exhibition (DATE)}}. IEEE, \bibinfo{pages}{378--383}.
\newblock


\bibitem[\protect\citeauthoryear{Asanovic, Avizienis, Bachrach, Beamer,
  Biancolin, Celio, Cook, Dabbelt, Hauser, Izraelevitz, et~al\mbox{.}}{Asanovic
  et~al\mbox{.}}{2016}]%
        {asanovic2016rocket}
\bibfield{author}{\bibinfo{person}{Krste Asanovic}, \bibinfo{person}{Rimas
  Avizienis}, \bibinfo{person}{Jonathan Bachrach}, \bibinfo{person}{Scott
  Beamer}, \bibinfo{person}{David Biancolin}, \bibinfo{person}{Christopher
  Celio}, \bibinfo{person}{Henry Cook}, \bibinfo{person}{Daniel Dabbelt},
  \bibinfo{person}{John Hauser}, \bibinfo{person}{Adam Izraelevitz},
  {et~al\mbox{.}}} \bibinfo{year}{2016}\natexlab{}.
\newblock \showarticletitle{The rocket chip generator}.
\newblock \bibinfo{journal}{\emph{EECS Department, University of California,
  Berkeley, Tech. Rep. UCB/EECS-2016-17}} (\bibinfo{year}{2016}).
\newblock


\bibitem[\protect\citeauthoryear{Attia, Johnson, Townsend, Jones, and
  Zambreno}{Attia et~al\mbox{.}}{2014}]%
        {attia2014cygraph}
\bibfield{author}{\bibinfo{person}{Osama~G Attia}, \bibinfo{person}{Tyler
  Johnson}, \bibinfo{person}{Kevin Townsend}, \bibinfo{person}{Philip Jones},
  {and} \bibinfo{person}{Joseph Zambreno}.} \bibinfo{year}{2014}\natexlab{}.
\newblock \showarticletitle{CyGraph: A Reconfigurable Architecture for Parallel
  Breadth-First Search}. In \bibinfo{booktitle}{\emph{2014 IEEE International
  Parallel \& Distributed Processing Symposium Workshops (IPDPSW)}}. IEEE,
  \bibinfo{pages}{228--235}.
\newblock


\bibitem[\protect\citeauthoryear{Babb, Frank, and Agarwal}{Babb
  et~al\mbox{.}}{1996}]%
        {babb1996solving}
\bibfield{author}{\bibinfo{person}{Jonathan~W Babb}, \bibinfo{person}{Matthew
  Frank}, {and} \bibinfo{person}{Anant Agarwal}.}
  \bibinfo{year}{1996}\natexlab{}.
\newblock \showarticletitle{Solving graph problems with dynamic computation
  structures}. In \bibinfo{booktitle}{\emph{High-Speed Computing, Digital
  Signal Processing, and Filtering Using Reconfigurable Logic}},
  Vol.~\bibinfo{volume}{2914}. International Society for Optics and Photonics,
  \bibinfo{pages}{225--237}.
\newblock


\bibitem[\protect\citeauthoryear{Barab{\'a}si and Albert}{Barab{\'a}si and
  Albert}{1999}]%
        {barabasi1999scaleFree}
\bibfield{author}{\bibinfo{person}{Albert-L{\'a}szl{\'o} Barab{\'a}si} {and}
  \bibinfo{person}{R{\'e}ka Albert}.} \bibinfo{year}{1999}\natexlab{}.
\newblock \showarticletitle{Emergence of Scaling in Random Networks}.
\newblock \bibinfo{journal}{\emph{Science}} \bibinfo{volume}{286},
  \bibinfo{number}{5439} (\bibinfo{year}{1999}), \bibinfo{pages}{509--512}.
\newblock
\showISSN{0036-8075}
\urldef\tempurl%
\url{https://doi.org/10.1126/science.286.5439.509}
\showDOI{\tempurl}
\showeprint{https://science.sciencemag.org/content/286/5439/509.full.pdf}


\bibitem[\protect\citeauthoryear{Basak, Li, Hu, Oh, Xie, Zhao, Jiang, and
  Xie}{Basak et~al\mbox{.}}{2019}]%
        {basak2019analysis}
\bibfield{author}{\bibinfo{person}{Abanti Basak}, \bibinfo{person}{Shuangchen
  Li}, \bibinfo{person}{Xing Hu}, \bibinfo{person}{Sang~Min Oh},
  \bibinfo{person}{Xinfeng Xie}, \bibinfo{person}{Li Zhao},
  \bibinfo{person}{Xiaowei Jiang}, {and} \bibinfo{person}{Yuan Xie}.}
  \bibinfo{year}{2019}\natexlab{}.
\newblock \showarticletitle{Analysis and optimization of the memory hierarchy
  for graph processing workloads}. In \bibinfo{booktitle}{\emph{2019 IEEE
  International Symposium on High Performance Computer Architecture (HPCA)}}.
  IEEE, \bibinfo{pages}{373--386}.
\newblock


\bibitem[\protect\citeauthoryear{Batarfi, El~Shawi, Fayoumi, Nouri, Barnawi,
  and Sakr}{Batarfi et~al\mbox{.}}{2015}]%
        {batarfi2015large}
\bibfield{author}{\bibinfo{person}{Omar Batarfi}, \bibinfo{person}{Radwa
  El~Shawi}, \bibinfo{person}{Ayman~G Fayoumi}, \bibinfo{person}{Reza Nouri},
  \bibinfo{person}{Ahmed Barnawi}, {and} \bibinfo{person}{Sherif Sakr}.}
  \bibinfo{year}{2015}\natexlab{}.
\newblock \showarticletitle{Large scale graph processing systems: survey and an
  experimental evaluation}.
\newblock \bibinfo{journal}{\emph{Cluster Computing}} \bibinfo{volume}{18},
  \bibinfo{number}{3} (\bibinfo{year}{2015}), \bibinfo{pages}{1189--1213}.
\newblock


\bibitem[\protect\citeauthoryear{Beamer, Asanovi{\'c}, and Patterson}{Beamer
  et~al\mbox{.}}{2013a}]%
        {beamer2013direction}
\bibfield{author}{\bibinfo{person}{Scott Beamer}, \bibinfo{person}{Krste
  Asanovi{\'c}}, {and} \bibinfo{person}{David Patterson}.}
  \bibinfo{year}{2013}\natexlab{a}.
\newblock \showarticletitle{{Direction-optimizing breadth-first search}}.
\newblock \bibinfo{journal}{\emph{Scientific Programming}}
  \bibinfo{volume}{21}, \bibinfo{number}{3-4} (\bibinfo{year}{2013}),
  \bibinfo{pages}{137--148}.
\newblock


\bibitem[\protect\citeauthoryear{Beamer, Asanovi{\'c}, and Patterson}{Beamer
  et~al\mbox{.}}{2015}]%
        {beamer2015gap}
\bibfield{author}{\bibinfo{person}{Scott Beamer}, \bibinfo{person}{Krste
  Asanovi{\'c}}, {and} \bibinfo{person}{David Patterson}.}
  \bibinfo{year}{2015}\natexlab{}.
\newblock \showarticletitle{The GAP benchmark suite}.
\newblock \bibinfo{journal}{\emph{arXiv preprint arXiv:1508.03619}}
  (\bibinfo{year}{2015}).
\newblock


\bibitem[\protect\citeauthoryear{Beamer, Buluc, Asanovic, and Patterson}{Beamer
  et~al\mbox{.}}{2013b}]%
        {beamer2013distributed}
\bibfield{author}{\bibinfo{person}{Scott Beamer}, \bibinfo{person}{Aydin
  Buluc}, \bibinfo{person}{Krste Asanovic}, {and} \bibinfo{person}{David
  Patterson}.} \bibinfo{year}{2013}\natexlab{b}.
\newblock \showarticletitle{Distributed memory breadth-first search revisited:
  Enabling bottom-up search}. In \bibinfo{booktitle}{\emph{2013 IEEE
  International Symposium on Parallel \& Distributed Processing, Workshops and
  Phd Forum}}. IEEE, \bibinfo{pages}{1618--1627}.
\newblock


\bibitem[\protect\citeauthoryear{Belayneh, Addisie, and Bertacco}{Belayneh
  et~al\mbox{.}}{2019}]%
        {belayneh2019messagefusion}
\bibfield{author}{\bibinfo{person}{Leul Belayneh}, \bibinfo{person}{Abraham
  Addisie}, {and} \bibinfo{person}{Valeria Bertacco}.}
  \bibinfo{year}{2019}\natexlab{}.
\newblock \showarticletitle{Messagefusion: On-path message coalescing for
  energy efficient and scalable graph analytics}. In
  \bibinfo{booktitle}{\emph{2019 IEEE/ACM International Symposium on Low Power
  Electronics and Design (ISLPED)}}. IEEE, \bibinfo{pages}{1--6}.
\newblock


\bibitem[\protect\citeauthoryear{Belayneh and Bertacco}{Belayneh and
  Bertacco}{2020}]%
        {belayneh2020graphvine}
\bibfield{author}{\bibinfo{person}{Leul Belayneh} {and}
  \bibinfo{person}{Valeria Bertacco}.} \bibinfo{year}{2020}\natexlab{}.
\newblock \showarticletitle{GraphVine: exploiting multicast for scalable graph
  analytics}. In \bibinfo{booktitle}{\emph{2020 Design, Automation \& Test in
  Europe Conference \& Exhibition (DATE)}}. IEEE, \bibinfo{pages}{762--767}.
\newblock


\bibitem[\protect\citeauthoryear{Ben-Nun, Besta, Huber, Ziogas, Peter, and
  Hoefler}{Ben-Nun et~al\mbox{.}}{2019}]%
        {ben2019modular}
\bibfield{author}{\bibinfo{person}{Tal Ben-Nun}, \bibinfo{person}{Maciej
  Besta}, \bibinfo{person}{Simon Huber}, \bibinfo{person}{Alexandros~Nikolaos
  Ziogas}, \bibinfo{person}{Daniel Peter}, {and} \bibinfo{person}{Torsten
  Hoefler}.} \bibinfo{year}{2019}\natexlab{}.
\newblock \showarticletitle{A modular benchmarking infrastructure for
  high-performance and reproducible deep learning}. In
  \bibinfo{booktitle}{\emph{2019 IEEE International Parallel and Distributed
  Processing Symposium (IPDPS)}}. IEEE, \bibinfo{pages}{66--77}.
\newblock


\bibitem[\protect\citeauthoryear{Besta, Carigiet, Vonarburg-Shmaria, Janda,
  Gianinazzi, and Hoefler}{Besta et~al\mbox{.}}{2020a}]%
        {besta2020high}
\bibfield{author}{\bibinfo{person}{Maciej Besta}, \bibinfo{person}{Armon
  Carigiet}, \bibinfo{person}{Zur Vonarburg-Shmaria}, \bibinfo{person}{Kacper
  Janda}, \bibinfo{person}{Lukas Gianinazzi}, {and} \bibinfo{person}{Torsten
  Hoefler}.} \bibinfo{year}{2020}\natexlab{a}.
\newblock \showarticletitle{High-performance parallel graph coloring with
  strong guarantees on work, depth, and quality}.
\newblock \bibinfo{journal}{\emph{arXiv preprint arXiv:2008.11321}}
  (\bibinfo{year}{2020}).
\newblock


\bibitem[\protect\citeauthoryear{Besta, Fischer, Ben-Nun, de~Fine~Licht, and
  Hoefler}{Besta et~al\mbox{.}}{2019a}]%
        {besta2019substream}
\bibfield{author}{\bibinfo{person}{Maciej Besta}, \bibinfo{person}{Marc
  Fischer}, \bibinfo{person}{Tal Ben-Nun}, \bibinfo{person}{Johannes de
  Fine~Licht}, {and} \bibinfo{person}{Torsten Hoefler}.}
  \bibinfo{year}{2019}\natexlab{a}.
\newblock \showarticletitle{Substream-centric maximum matchings on fpga}. In
  \bibinfo{booktitle}{\emph{Proceedings of the 2019 ACM/SIGDA International
  Symposium on Field-Programmable Gate Arrays}}. \bibinfo{pages}{152--161}.
\newblock


\bibitem[\protect\citeauthoryear{Besta, Fischer, Ben-Nun, Stanojevic, Licht,
  and Hoefler}{Besta et~al\mbox{.}}{2020b}]%
        {besta2020substream}
\bibfield{author}{\bibinfo{person}{Maciej Besta}, \bibinfo{person}{Marc
  Fischer}, \bibinfo{person}{Tal Ben-Nun}, \bibinfo{person}{Dimitri
  Stanojevic}, \bibinfo{person}{Johannes De~Fine Licht}, {and}
  \bibinfo{person}{Torsten Hoefler}.} \bibinfo{year}{2020}\natexlab{b}.
\newblock \showarticletitle{Substream-Centric Maximum Matchings on FPGA}.
\newblock \bibinfo{journal}{\emph{ACM Transactions on Reconfigurable Technology
  and Systems (TRETS)}} \bibinfo{volume}{13}, \bibinfo{number}{2}
  (\bibinfo{year}{2020}), \bibinfo{pages}{1--33}.
\newblock


\bibitem[\protect\citeauthoryear{Besta, Fischer, Kalavri, Kapralov, and
  Hoefler}{Besta et~al\mbox{.}}{2019b}]%
        {besta2019practice}
\bibfield{author}{\bibinfo{person}{Maciej Besta}, \bibinfo{person}{Marc
  Fischer}, \bibinfo{person}{Vasiliki Kalavri}, \bibinfo{person}{Michael
  Kapralov}, {and} \bibinfo{person}{Torsten Hoefler}.}
  \bibinfo{year}{2019}\natexlab{b}.
\newblock \showarticletitle{Practice of Streaming Processing of Dynamic Graphs:
  Concepts, Models, and Systems}.
\newblock \bibinfo{journal}{\emph{arXiv preprint arXiv:1912.12740}}
  (\bibinfo{year}{2019}).
\newblock


\bibitem[\protect\citeauthoryear{Besta, Grob, Miglioli, Bernold, Kwasniewski,
  Gjini, Kanakagiri, Ashkboos, Gianinazzi, Dryden, et~al\mbox{.}}{Besta
  et~al\mbox{.}}{2021a}]%
        {besta2021motif}
\bibfield{author}{\bibinfo{person}{Maciej Besta}, \bibinfo{person}{Raphael
  Grob}, \bibinfo{person}{Cesare Miglioli}, \bibinfo{person}{Nicola Bernold},
  \bibinfo{person}{Grzegorz Kwasniewski}, \bibinfo{person}{Gabriel Gjini},
  \bibinfo{person}{Raghavendra Kanakagiri}, \bibinfo{person}{Saleh Ashkboos},
  \bibinfo{person}{Lukas Gianinazzi}, \bibinfo{person}{Nikoli Dryden},
  {et~al\mbox{.}}} \bibinfo{year}{2021}\natexlab{a}.
\newblock \showarticletitle{Motif Prediction with Graph Neural Networks}.
\newblock \bibinfo{journal}{\emph{arXiv preprint arXiv:2106.00761}}
  (\bibinfo{year}{2021}).
\newblock


\bibitem[\protect\citeauthoryear{Besta, Hassan, Yalamanchili, Ausavarungnirun,
  Mutlu, and Hoefler}{Besta et~al\mbox{.}}{2018a}]%
        {besta2018slim}
\bibfield{author}{\bibinfo{person}{Maciej Besta}, \bibinfo{person}{Syed~Minhaj
  Hassan}, \bibinfo{person}{Sudhakar Yalamanchili}, \bibinfo{person}{Rachata
  Ausavarungnirun}, \bibinfo{person}{Onur Mutlu}, {and}
  \bibinfo{person}{Torsten Hoefler}.} \bibinfo{year}{2018}\natexlab{a}.
\newblock \showarticletitle{Slim noc: A low-diameter on-chip network topology
  for high energy efficiency and scalability}.
\newblock \bibinfo{journal}{\emph{ACM SIGPLAN Notices}} \bibinfo{volume}{53},
  \bibinfo{number}{2} (\bibinfo{year}{2018}), \bibinfo{pages}{43--55}.
\newblock


\bibitem[\protect\citeauthoryear{Besta and Hoefler}{Besta and Hoefler}{2014}]%
        {besta2014fault}
\bibfield{author}{\bibinfo{person}{Maciej Besta} {and} \bibinfo{person}{Torsten
  Hoefler}.} \bibinfo{year}{2014}\natexlab{}.
\newblock \showarticletitle{Fault tolerance for remote memory access
  programming models}. In \bibinfo{booktitle}{\emph{Proceedings of the 23rd
  international symposium on High-performance parallel and distributed
  computing}}. \bibinfo{pages}{37--48}.
\newblock


\bibitem[\protect\citeauthoryear{Besta and Hoefler}{Besta and Hoefler}{2015a}]%
        {besta2015accelerating}
\bibfield{author}{\bibinfo{person}{Maciej Besta} {and} \bibinfo{person}{Torsten
  Hoefler}.} \bibinfo{year}{2015}\natexlab{a}.
\newblock \showarticletitle{Accelerating irregular computations with hardware
  transactional memory and active messages}. In
  \bibinfo{booktitle}{\emph{Proceedings of the 24th International Symposium on
  High-Performance Parallel and Distributed Computing}}.
  \bibinfo{pages}{161--172}.
\newblock


\bibitem[\protect\citeauthoryear{Besta and Hoefler}{Besta and Hoefler}{2015b}]%
        {besta2015active}
\bibfield{author}{\bibinfo{person}{Maciej Besta} {and} \bibinfo{person}{Torsten
  Hoefler}.} \bibinfo{year}{2015}\natexlab{b}.
\newblock \showarticletitle{Active access: A mechanism for high-performance
  distributed data-centric computations}. In
  \bibinfo{booktitle}{\emph{Proceedings of the 29th ACM on International
  Conference on Supercomputing}}. \bibinfo{pages}{155--164}.
\newblock


\bibitem[\protect\citeauthoryear{Besta and Hoefler}{Besta and Hoefler}{2018}]%
        {besta2018survey}
\bibfield{author}{\bibinfo{person}{Maciej Besta} {and} \bibinfo{person}{Torsten
  Hoefler}.} \bibinfo{year}{2018}\natexlab{}.
\newblock \showarticletitle{Survey and Taxonomy of Lossless Graph Compression
  and Space-Efficient Graph Representations}.
\newblock \bibinfo{journal}{\emph{arXiv preprint arXiv:1806.01799}}
  (\bibinfo{year}{2018}).
\newblock


\bibitem[\protect\citeauthoryear{Besta, Kanakagiri, Mustafa, Karasikov,
  R{\"a}tsch, Hoefler, and Solomonik}{Besta et~al\mbox{.}}{2020c}]%
        {besta2020communication}
\bibfield{author}{\bibinfo{person}{Maciej Besta}, \bibinfo{person}{Raghavendra
  Kanakagiri}, \bibinfo{person}{Harun Mustafa}, \bibinfo{person}{Mikhail
  Karasikov}, \bibinfo{person}{Gunnar R{\"a}tsch}, \bibinfo{person}{Torsten
  Hoefler}, {and} \bibinfo{person}{Edgar Solomonik}.}
  \bibinfo{year}{2020}\natexlab{c}.
\newblock \showarticletitle{Communication-efficient jaccard similarity for
  high-performance distributed genome comparisons}. In
  \bibinfo{booktitle}{\emph{2020 IEEE International Parallel and Distributed
  Processing Symposium (IPDPS)}}. IEEE, \bibinfo{pages}{1122--1132}.
\newblock


\bibitem[\protect\citeauthoryear{Besta, Marending, Solomonik, and
  Hoefler}{Besta et~al\mbox{.}}{2017a}]%
        {besta2017slimsell}
\bibfield{author}{\bibinfo{person}{Maciej Besta}, \bibinfo{person}{Florian
  Marending}, \bibinfo{person}{Edgar Solomonik}, {and} \bibinfo{person}{Torsten
  Hoefler}.} \bibinfo{year}{2017}\natexlab{a}.
\newblock \showarticletitle{SlimSell: A Vectorizable Graph Representation for
  Breadth-First Search}. In \bibinfo{booktitle}{\emph{Parallel and Distributed
  Processing Symposium (IPDPS), 2017 IEEE International}}. IEEE,
  \bibinfo{pages}{32--41}.
\newblock


\bibitem[\protect\citeauthoryear{Besta, Peter, Gerstenberger, Fischer,
  Podstawski, Barthels, Alonso, and Hoefler}{Besta et~al\mbox{.}}{2019c}]%
        {besta2019demystifying}
\bibfield{author}{\bibinfo{person}{Maciej Besta}, \bibinfo{person}{Emanuel
  Peter}, \bibinfo{person}{Robert Gerstenberger}, \bibinfo{person}{Marc
  Fischer}, \bibinfo{person}{Micha{\l} Podstawski}, \bibinfo{person}{Claude
  Barthels}, \bibinfo{person}{Gustavo Alonso}, {and} \bibinfo{person}{Torsten
  Hoefler}.} \bibinfo{year}{2019}\natexlab{c}.
\newblock \showarticletitle{Demystifying graph databases: Analysis and taxonomy
  of data organization, system designs, and graph queries}.
\newblock \bibinfo{journal}{\emph{arXiv preprint arXiv:1910.09017}}
  (\bibinfo{year}{2019}).
\newblock


\bibitem[\protect\citeauthoryear{Besta, Podstawski, Groner, Solomonik, and
  Hoefler}{Besta et~al\mbox{.}}{2017b}]%
        {besta2017push}
\bibfield{author}{\bibinfo{person}{Maciej Besta}, \bibinfo{person}{Micha{\l}
  Podstawski}, \bibinfo{person}{Linus Groner}, \bibinfo{person}{Edgar
  Solomonik}, {and} \bibinfo{person}{Torsten Hoefler}.}
  \bibinfo{year}{2017}\natexlab{b}.
\newblock \showarticletitle{To Push or To Pull: On Reducing Communication and
  Synchronization in Graph Computations}. In
  \bibinfo{booktitle}{\emph{Proceedings of the 26th International Symposium on
  High-Performance Parallel and Distributed Computing}}. ACM,
  \bibinfo{pages}{93--104}.
\newblock


\bibitem[\protect\citeauthoryear{Besta, Stanojevic, Licht, Ben-Nun, and
  Hoefler}{Besta et~al\mbox{.}}{2019d}]%
        {besta2019graph}
\bibfield{author}{\bibinfo{person}{Maciej Besta}, \bibinfo{person}{Dimitri
  Stanojevic}, \bibinfo{person}{Johannes De~Fine Licht}, \bibinfo{person}{Tal
  Ben-Nun}, {and} \bibinfo{person}{Torsten Hoefler}.}
  \bibinfo{year}{2019}\natexlab{d}.
\newblock \showarticletitle{Graph Processing on {FPGAs}: Taxonomy, Survey,
  Challenges}.
\newblock \bibinfo{journal}{\emph{arXiv preprint arXiv:1903.06697}}
  (\bibinfo{year}{2019}).
\newblock


\bibitem[\protect\citeauthoryear{Besta, Stanojevic, Zivic, Singh, Hoerold, and
  Hoefler}{Besta et~al\mbox{.}}{2018b}]%
        {besta2018log}
\bibfield{author}{\bibinfo{person}{Maciej Besta}, \bibinfo{person}{Dimitri
  Stanojevic}, \bibinfo{person}{Tijana Zivic}, \bibinfo{person}{Jagpreet
  Singh}, \bibinfo{person}{Maurice Hoerold}, {and} \bibinfo{person}{Torsten
  Hoefler}.} \bibinfo{year}{2018}\natexlab{b}.
\newblock \showarticletitle{Log (graph): a near-optimal high-performance graph
  representation}. In \bibinfo{booktitle}{\emph{Proceedings of the 27th
  International Conference on Parallel Architectures and Compilation
  Techniques}}. ACM, \bibinfo{pages}{7}.
\newblock


\bibitem[\protect\citeauthoryear{Besta, Vonarburg-Shmaria, Schaffner, Schwarz,
  Kwasniewski, Gianinazzi, Beranek, Janda, Holenstein, Leisinger,
  et~al\mbox{.}}{Besta et~al\mbox{.}}{2021b}]%
        {besta2021graphminesuite}
\bibfield{author}{\bibinfo{person}{Maciej Besta}, \bibinfo{person}{Zur
  Vonarburg-Shmaria}, \bibinfo{person}{Yannick Schaffner},
  \bibinfo{person}{Leonardo Schwarz}, \bibinfo{person}{Grzegorz Kwasniewski},
  \bibinfo{person}{Lukas Gianinazzi}, \bibinfo{person}{Jakub Beranek},
  \bibinfo{person}{Kacper Janda}, \bibinfo{person}{Tobias Holenstein},
  \bibinfo{person}{Sebastian Leisinger}, {et~al\mbox{.}}}
  \bibinfo{year}{2021}\natexlab{b}.
\newblock \showarticletitle{GraphMineSuite: Enabling High-Performance and
  Programmable Graph Mining Algorithms with Set Algebra}.
\newblock \bibinfo{journal}{\emph{VLDB}} (\bibinfo{year}{2021}).
\newblock


\bibitem[\protect\citeauthoryear{Besta, Weber, Gianinazzi, Gerstenberger,
  Ivanov, Oltchik, and Hoefler}{Besta et~al\mbox{.}}{2019e}]%
        {besta2019slim}
\bibfield{author}{\bibinfo{person}{Maciej Besta}, \bibinfo{person}{Simon
  Weber}, \bibinfo{person}{Lukas Gianinazzi}, \bibinfo{person}{Robert
  Gerstenberger}, \bibinfo{person}{Andrey Ivanov}, \bibinfo{person}{Yishai
  Oltchik}, {and} \bibinfo{person}{Torsten Hoefler}.}
  \bibinfo{year}{2019}\natexlab{e}.
\newblock \showarticletitle{Slim graph: Practical lossy graph compression for
  approximate graph processing, storage, and analytics}. In
  \bibinfo{booktitle}{\emph{Proceedings of the International Conference for
  High Performance Computing, Networking, Storage and Analysis}}.
  \bibinfo{pages}{1--25}.
\newblock


\bibitem[\protect\citeauthoryear{Betkaoui, Thomas, Luk, and Przulj}{Betkaoui
  et~al\mbox{.}}{2011}]%
        {betkaoui2011framework}
\bibfield{author}{\bibinfo{person}{Brahim Betkaoui}, \bibinfo{person}{David~B
  Thomas}, \bibinfo{person}{Wayne Luk}, {and} \bibinfo{person}{Natasa Przulj}.}
  \bibinfo{year}{2011}\natexlab{}.
\newblock \showarticletitle{A framework for FPGA acceleration of large graph
  problems: Graphlet counting case study}. In
  \bibinfo{booktitle}{\emph{Field-Programmable Technology (FPT), 2011
  International Conference on}}. IEEE, \bibinfo{pages}{1--8}.
\newblock


\bibitem[\protect\citeauthoryear{Betkaoui, Wang, Thomas, and Luk}{Betkaoui
  et~al\mbox{.}}{2012a}]%
        {betkaoui:APSP_FPGA}
\bibfield{author}{\bibinfo{person}{B. Betkaoui}, \bibinfo{person}{Y. Wang},
  \bibinfo{person}{D.~B. Thomas}, {and} \bibinfo{person}{W. Luk}.}
  \bibinfo{year}{2012}\natexlab{a}.
\newblock \showarticletitle{Parallel FPGA-based all pairs shortest paths for
  sparse networks: A human brain connectome case study}. In
  \bibinfo{booktitle}{\emph{22nd International Conference on Field Programmable
  Logic and Applications (FPL)}}. \bibinfo{pages}{99--104}.
\newblock
\showISSN{1946-147X}
\urldef\tempurl%
\url{https://doi.org/10.1109/FPL.2012.6339247}
\showDOI{\tempurl}


\bibitem[\protect\citeauthoryear{Betkaoui, Wang, Thomas, and Luk}{Betkaoui
  et~al\mbox{.}}{2012b}]%
        {betkaoui2012reconfigurable}
\bibfield{author}{\bibinfo{person}{Brahim Betkaoui}, \bibinfo{person}{Yu Wang},
  \bibinfo{person}{David~B Thomas}, {and} \bibinfo{person}{Wayne Luk}.}
  \bibinfo{year}{2012}\natexlab{b}.
\newblock \showarticletitle{A reconfigurable computing approach for efficient
  and scalable parallel graph exploration}. In
  \bibinfo{booktitle}{\emph{Application-Specific Systems, Architectures and
  Processors (ASAP), 2012 IEEE 23rd International Conference on}}. IEEE,
  \bibinfo{pages}{8--15}.
\newblock


\bibitem[\protect\citeauthoryear{Blelloch and Maggs}{Blelloch and
  Maggs}{2010}]%
        {blelloch2010parallel}
\bibfield{author}{\bibinfo{person}{Guy~E. Blelloch} {and}
  \bibinfo{person}{Bruce~M. Maggs}.} \bibinfo{year}{2010}\natexlab{}.
\newblock \bibinfo{booktitle}{\emph{Parallel Algorithms} (\bibinfo{edition}{2}
  ed.)}.
\newblock \bibinfo{publisher}{Chapman \& Hall/CRC}, \bibinfo{pages}{25}.
\newblock
\showISBNx{9781584888208}


\bibitem[\protect\citeauthoryear{Bogdanov, Baumer, Basu, Bar-Noy, and
  Singh}{Bogdanov et~al\mbox{.}}{2013}]%
        {bogdanov2013strong}
\bibfield{author}{\bibinfo{person}{Petko Bogdanov}, \bibinfo{person}{Ben
  Baumer}, \bibinfo{person}{Prithwish Basu}, \bibinfo{person}{Amotz Bar-Noy},
  {and} \bibinfo{person}{Ambuj~K Singh}.} \bibinfo{year}{2013}\natexlab{}.
\newblock \showarticletitle{As strong as the weakest link: Mining diverse
  cliques in weighted graphs}. In \bibinfo{booktitle}{\emph{Joint European
  conference on machine learning and knowledge discovery in databases}}.
  Springer, \bibinfo{pages}{525--540}.
\newblock


\bibitem[\protect\citeauthoryear{Bondhugula, Devulapalli, Fernando, Wyckoff,
  and Sadayappan}{Bondhugula et~al\mbox{.}}{2006}]%
        {bondhugula:APSP_FPGA}
\bibfield{author}{\bibinfo{person}{U. Bondhugula}, \bibinfo{person}{A.
  Devulapalli}, \bibinfo{person}{J. Fernando}, \bibinfo{person}{P. Wyckoff},
  {and} \bibinfo{person}{P. Sadayappan}.} \bibinfo{year}{2006}\natexlab{}.
\newblock \showarticletitle{Parallel FPGA-based all-pairs shortest-paths in a
  directed graph}. In \bibinfo{booktitle}{\emph{Proceedings 20th IEEE
  International Parallel Distributed Processing Symposium}}. \bibinfo{pages}{10
  pp.--}.
\newblock
\showISSN{1530-2075}
\urldef\tempurl%
\url{https://doi.org/10.1109/IPDPS.2006.1639347}
\showDOI{\tempurl}


\bibitem[\protect\citeauthoryear{Boruvka}{Boruvka}{1926}]%
        {boruuvka1926jistem}
\bibfield{author}{\bibinfo{person}{Otakar Boruvka}.}
  \bibinfo{year}{1926}\natexlab{}.
\newblock \showarticletitle{O jist{\'e}m probl{\'e}mu minim{\'a}ln{\'\i}m}.
\newblock  (\bibinfo{year}{1926}).
\newblock


\bibitem[\protect\citeauthoryear{Brandes}{Brandes}{2001}]%
        {brandes2001faster}
\bibfield{author}{\bibinfo{person}{Ulrik Brandes}.}
  \bibinfo{year}{2001}\natexlab{}.
\newblock \showarticletitle{A faster algorithm for betweenness centrality}.
\newblock \bibinfo{journal}{\emph{J. of Math. Sociology}} \bibinfo{volume}{25},
  \bibinfo{number}{2} (\bibinfo{year}{2001}), \bibinfo{pages}{163--177}.
\newblock


\bibitem[\protect\citeauthoryear{Bron and Kerbosch}{Bron and Kerbosch}{1973}]%
        {bron1973algorithm}
\bibfield{author}{\bibinfo{person}{Coen Bron} {and} \bibinfo{person}{Joep
  Kerbosch}.} \bibinfo{year}{1973}\natexlab{}.
\newblock \showarticletitle{Algorithm 457: finding all cliques of an undirected
  graph}.
\newblock \bibinfo{journal}{\emph{Commun. ACM}} \bibinfo{volume}{16},
  \bibinfo{number}{9} (\bibinfo{year}{1973}), \bibinfo{pages}{575--577}.
\newblock


\bibitem[\protect\citeauthoryear{Buluc, Beamer, Madduri, Asanovic, and
  Patterson}{Buluc et~al\mbox{.}}{2017}]%
        {buluc2017distributed}
\bibfield{author}{\bibinfo{person}{Aydin Buluc}, \bibinfo{person}{Scott
  Beamer}, \bibinfo{person}{Kamesh Madduri}, \bibinfo{person}{Krste Asanovic},
  {and} \bibinfo{person}{David Patterson}.} \bibinfo{year}{2017}\natexlab{}.
\newblock \showarticletitle{Distributed-memory breadth-first search on massive
  graphs}.
\newblock \bibinfo{journal}{\emph{arXiv preprint arXiv:1705.04590}}
  (\bibinfo{year}{2017}).
\newblock


\bibitem[\protect\citeauthoryear{Bulu{\c{c}} and Madduri}{Bulu{\c{c}} and
  Madduri}{2011}]%
        {bulucc2011parallel}
\bibfield{author}{\bibinfo{person}{Aydin Bulu{\c{c}}} {and}
  \bibinfo{person}{Kamesh Madduri}.} \bibinfo{year}{2011}\natexlab{}.
\newblock \showarticletitle{Parallel breadth-first search on distributed memory
  systems}. In \bibinfo{booktitle}{\emph{Proceedings of 2011 International
  Conference for High Performance Computing, Networking, Storage and
  Analysis}}. ACM, \bibinfo{pages}{65}.
\newblock


\bibitem[\protect\citeauthoryear{Bustio, Cumplido, Hern{\'a}ndez, Bande, and
  Feregrino}{Bustio et~al\mbox{.}}{2015}]%
        {bustio2015frequent}
\bibfield{author}{\bibinfo{person}{L{\'a}zaro Bustio},
  \bibinfo{person}{Ren{\'e} Cumplido}, \bibinfo{person}{Raudel Hern{\'a}ndez},
  \bibinfo{person}{Jos{\'e}~M Bande}, {and} \bibinfo{person}{Claudia
  Feregrino}.} \bibinfo{year}{2015}\natexlab{}.
\newblock \showarticletitle{Frequent itemsets mining in data streams using
  reconfigurable hardware}. In \bibinfo{booktitle}{\emph{International Workshop
  on New Frontiers in Mining Complex Patterns}}. Springer,
  \bibinfo{pages}{32--45}.
\newblock


\bibitem[\protect\citeauthoryear{Bustio-Mart{\'\i}nez, Cumplido, Letras-Luna,
  Uribe, Hern{\'a}ndez-Le{\'o}n, and Bande-Serrano}{Bustio-Mart{\'\i}nez
  et~al\mbox{.}}{2017}]%
        {bustio2017approximate}
\bibfield{author}{\bibinfo{person}{L{\'a}zaro Bustio-Mart{\'\i}nez},
  \bibinfo{person}{Ren{\'e} Cumplido}, \bibinfo{person}{Mart{\'\i}n
  Letras-Luna}, \bibinfo{person}{Claudia~Feregrino Uribe},
  \bibinfo{person}{Raudel Hern{\'a}ndez-Le{\'o}n}, {and}
  \bibinfo{person}{Jos{\'e}~M Bande-Serrano}.} \bibinfo{year}{2017}\natexlab{}.
\newblock \showarticletitle{Approximate frequent itemsets mining on data
  streams using hashing and lexicographie order in hardware}. In
  \bibinfo{booktitle}{\emph{2017 IEEE 8th Latin American Symposium on Circuits
  \& Systems (LASCAS)}}. IEEE, \bibinfo{pages}{1--4}.
\newblock


\bibitem[\protect\citeauthoryear{Caseau, Josset, and Laburthe}{Caseau
  et~al\mbox{.}}{2002}]%
        {caseau2002claire}
\bibfield{author}{\bibinfo{person}{Yves Caseau},
  \bibinfo{person}{Fran{\c{c}}ois-Xavier Josset}, {and}
  \bibinfo{person}{Fran{\c{c}}ois Laburthe}.} \bibinfo{year}{2002}\natexlab{}.
\newblock \showarticletitle{CLAIRE: Combining sets, search and rules to better
  express algorithms}.
\newblock \bibinfo{journal}{\emph{Theory and Practice of Logic Programming}}
  \bibinfo{volume}{2}, \bibinfo{number}{6} (\bibinfo{year}{2002}),
  \bibinfo{pages}{769--805}.
\newblock


\bibitem[\protect\citeauthoryear{Cazals and Karande}{Cazals and
  Karande}{2008}]%
        {cazals2008note}
\bibfield{author}{\bibinfo{person}{Fr{\'e}d{\'e}ric Cazals} {and}
  \bibinfo{person}{Chinmay Karande}.} \bibinfo{year}{2008}\natexlab{}.
\newblock \showarticletitle{A note on the problem of reporting maximal
  cliques}.
\newblock \bibinfo{journal}{\emph{Theoretical Computer Science}}
  \bibinfo{volume}{407}, \bibinfo{number}{1-3} (\bibinfo{year}{2008}),
  \bibinfo{pages}{564--568}.
\newblock


\bibitem[\protect\citeauthoryear{Chakrabarti and Faloutsos}{Chakrabarti and
  Faloutsos}{2006}]%
        {chakrabarti2006graph}
\bibfield{author}{\bibinfo{person}{Deepayan Chakrabarti} {and}
  \bibinfo{person}{Christos Faloutsos}.} \bibinfo{year}{2006}\natexlab{}.
\newblock \showarticletitle{Graph mining: Laws, generators, and algorithms}.
\newblock \bibinfo{journal}{\emph{ACM computing surveys (CSUR)}}
  \bibinfo{volume}{38}, \bibinfo{number}{1} (\bibinfo{year}{2006}),
  \bibinfo{pages}{2}.
\newblock


\bibitem[\protect\citeauthoryear{Challapalle, Rampalli, Song, Chandramoorthy,
  Swaminathan, Sampson, Chen, and Narayanan}{Challapalle et~al\mbox{.}}{2020}]%
        {challapallegaas}
\bibfield{author}{\bibinfo{person}{Nagadastagiri Challapalle},
  \bibinfo{person}{Sahithi Rampalli}, \bibinfo{person}{Linghao Song},
  \bibinfo{person}{Nandhini Chandramoorthy}, \bibinfo{person}{Karthik
  Swaminathan}, \bibinfo{person}{John Sampson}, \bibinfo{person}{Yiran Chen},
  {and} \bibinfo{person}{Vijaykrishnan Narayanan}.}
  \bibinfo{year}{2020}\natexlab{}.
\newblock \showarticletitle{GaaS-X: Graph Analytics Accelerator Supporting
  Sparse Data Representation using Crossbar Architectures}.
\newblock \bibinfo{journal}{\emph{ISCA}} (\bibinfo{year}{2020}).
\newblock


\bibitem[\protect\citeauthoryear{Chandra, Dagum, Kohr, Menon, Maydan, and
  McDonald}{Chandra et~al\mbox{.}}{2001}]%
        {chandra2001parallel}
\bibfield{author}{\bibinfo{person}{Rohit Chandra}, \bibinfo{person}{Leo Dagum},
  \bibinfo{person}{David Kohr}, \bibinfo{person}{Ramesh Menon},
  \bibinfo{person}{Dror Maydan}, {and} \bibinfo{person}{Jeff McDonald}.}
  \bibinfo{year}{2001}\natexlab{}.
\newblock \bibinfo{booktitle}{\emph{Parallel programming in OpenMP}}.
\newblock \bibinfo{publisher}{Morgan kaufmann}.
\newblock


\bibitem[\protect\citeauthoryear{Chen, Liu, Zhao, Yan, Yan, and Cheng}{Chen
  et~al\mbox{.}}{2018}]%
        {chen2018g}
\bibfield{author}{\bibinfo{person}{Hongzhi Chen}, \bibinfo{person}{Miao Liu},
  \bibinfo{person}{Yunjian Zhao}, \bibinfo{person}{Xiao Yan},
  \bibinfo{person}{Da Yan}, {and} \bibinfo{person}{James Cheng}.}
  \bibinfo{year}{2018}\natexlab{}.
\newblock \showarticletitle{G-Miner: an efficient task-oriented graph mining
  system}. In \bibinfo{booktitle}{\emph{Proceedings of the Thirteenth EuroSys
  Conference}}. ACM, \bibinfo{pages}{32}.
\newblock


\bibitem[\protect\citeauthoryear{Chen, Dathathri, Gill, and Pingali}{Chen
  et~al\mbox{.}}{2019}]%
        {chen2019pangolin}
\bibfield{author}{\bibinfo{person}{Xuhao Chen}, \bibinfo{person}{Roshan
  Dathathri}, \bibinfo{person}{Gurbinder Gill}, {and} \bibinfo{person}{Keshav
  Pingali}.} \bibinfo{year}{2019}\natexlab{}.
\newblock \showarticletitle{Pangolin: An Efficient and Flexible Graph Mining
  System on CPU and GPU}.
\newblock \bibinfo{journal}{\emph{arXiv preprint arXiv:1911.06969}}
  (\bibinfo{year}{2019}).
\newblock


\bibitem[\protect\citeauthoryear{Cheng, Yu, Ding, Philip, and Wang}{Cheng
  et~al\mbox{.}}{2008}]%
        {cheng2008fast}
\bibfield{author}{\bibinfo{person}{Jiefeng Cheng}, \bibinfo{person}{Jeffrey~Xu
  Yu}, \bibinfo{person}{Bolin Ding}, \bibinfo{person}{S~Yu Philip}, {and}
  \bibinfo{person}{Haixun Wang}.} \bibinfo{year}{2008}\natexlab{}.
\newblock \showarticletitle{Fast graph pattern matching}. In
  \bibinfo{booktitle}{\emph{2008 IEEE 24th International Conference on Data
  Engineering}}. IEEE, \bibinfo{pages}{913--922}.
\newblock


\bibitem[\protect\citeauthoryear{Cheng, Zhu, Ke, and Chu}{Cheng
  et~al\mbox{.}}{2012}]%
        {cheng2012fast}
\bibfield{author}{\bibinfo{person}{James Cheng}, \bibinfo{person}{Linhong Zhu},
  \bibinfo{person}{Yiping Ke}, {and} \bibinfo{person}{Shumo Chu}.}
  \bibinfo{year}{2012}\natexlab{}.
\newblock \showarticletitle{Fast algorithms for maximal clique enumeration with
  limited memory}. In \bibinfo{booktitle}{\emph{Proceedings of the 18th ACM
  SIGKDD international conference on Knowledge discovery and data mining}}.
  \bibinfo{pages}{1240--1248}.
\newblock


\bibitem[\protect\citeauthoryear{Chiba and Nishizeki}{Chiba and
  Nishizeki}{1985a}]%
        {chiba1985arboricity}
\bibfield{author}{\bibinfo{person}{Norishige Chiba} {and}
  \bibinfo{person}{Takao Nishizeki}.} \bibinfo{year}{1985}\natexlab{a}.
\newblock \showarticletitle{Arboricity and subgraph listing algorithms}.
\newblock \bibinfo{journal}{\emph{SIAM Journal on computing}}
  \bibinfo{volume}{14}, \bibinfo{number}{1} (\bibinfo{year}{1985}),
  \bibinfo{pages}{210--223}.
\newblock


\bibitem[\protect\citeauthoryear{Chiba and Nishizeki}{Chiba and
  Nishizeki}{1985b}]%
        {DBLP:journals/siamcomp/ChibaN85}
\bibfield{author}{\bibinfo{person}{Norishige Chiba} {and}
  \bibinfo{person}{Takao Nishizeki}.} \bibinfo{year}{1985}\natexlab{b}.
\newblock \showarticletitle{Arboricity and Subgraph Listing Algorithms}.
\newblock \bibinfo{journal}{\emph{{SIAM} J. Comput.}} \bibinfo{volume}{14},
  \bibinfo{number}{1} (\bibinfo{year}{1985}), \bibinfo{pages}{210--223}.
\newblock
\urldef\tempurl%
\url{https://doi.org/10.1137/0214017}
\showDOI{\tempurl}


\bibitem[\protect\citeauthoryear{Cook and Holder}{Cook and Holder}{2006}]%
        {cook2006mining}
\bibfield{author}{\bibinfo{person}{Diane~J Cook} {and}
  \bibinfo{person}{Lawrence~B Holder}.} \bibinfo{year}{2006}\natexlab{}.
\newblock \bibinfo{booktitle}{\emph{Mining graph data}}.
\newblock \bibinfo{publisher}{John Wiley \& Sons}.
\newblock


\bibitem[\protect\citeauthoryear{Corbet, Rubini, and Kroah-Hartman}{Corbet
  et~al\mbox{.}}{2005}]%
        {corbet2005linux}
\bibfield{author}{\bibinfo{person}{Jonathan Corbet},
  \bibinfo{person}{Alessandro Rubini}, {and} \bibinfo{person}{Greg
  Kroah-Hartman}.} \bibinfo{year}{2005}\natexlab{}.
\newblock \bibinfo{booktitle}{\emph{Linux device drivers}}.
\newblock \bibinfo{publisher}{" O'Reilly Media, Inc."}.
\newblock


\bibitem[\protect\citeauthoryear{Cordella, Foggia, Sansone, and Vento}{Cordella
  et~al\mbox{.}}{2004}]%
        {cordella2004sub}
\bibfield{author}{\bibinfo{person}{Luigi~P Cordella}, \bibinfo{person}{Pasquale
  Foggia}, \bibinfo{person}{Carlo Sansone}, {and} \bibinfo{person}{Mario
  Vento}.} \bibinfo{year}{2004}\natexlab{}.
\newblock \showarticletitle{A (sub) graph isomorphism algorithm for matching
  large graphs}.
\newblock \bibinfo{journal}{\emph{IEEE transactions on pattern analysis and
  machine intelligence}} \bibinfo{volume}{26}, \bibinfo{number}{10}
  (\bibinfo{year}{2004}), \bibinfo{pages}{1367--1372}.
\newblock


\bibitem[\protect\citeauthoryear{Cormen, Leiserson, Rivest, and Stein}{Cormen
  et~al\mbox{.}}{2009}]%
        {cormen2009introduction}
\bibfield{author}{\bibinfo{person}{Thomas~H Cormen}, \bibinfo{person}{Charles~E
  Leiserson}, \bibinfo{person}{Ronald~L Rivest}, {and}
  \bibinfo{person}{Clifford Stein}.} \bibinfo{year}{2009}\natexlab{}.
\newblock \bibinfo{booktitle}{\emph{Introduction to algorithms}}.
\newblock \bibinfo{publisher}{MIT press}.
\newblock


\bibitem[\protect\citeauthoryear{Dai, Chi, Wang, and Yang}{Dai
  et~al\mbox{.}}{2016}]%
        {dai:fpgp}
\bibfield{author}{\bibinfo{person}{Guohao Dai}, \bibinfo{person}{Yuze Chi},
  \bibinfo{person}{Yu Wang}, {and} \bibinfo{person}{Huazhong Yang}.}
  \bibinfo{year}{2016}\natexlab{}.
\newblock \showarticletitle{FPGP: Graph Processing Framework on FPGA A Case
  Study of Breadth-First Search}. In \bibinfo{booktitle}{\emph{Proceedings of
  the 2016 ACM/SIGDA International Symposium on Field-Programmable Gate
  Arrays}} \emph{(\bibinfo{series}{FPGA '16})}. \bibinfo{publisher}{ACM},
  \bibinfo{address}{New York, NY, USA}, \bibinfo{pages}{105--110}.
\newblock
\showISBNx{978-1-4503-3856-1}
\urldef\tempurl%
\url{https://doi.org/10.1145/2847263.2847339}
\showDOI{\tempurl}


\bibitem[\protect\citeauthoryear{Dai, Huang, Chi, Xu, Wang, and Yang}{Dai
  et~al\mbox{.}}{2017a}]%
        {dai2017foregraph}
\bibfield{author}{\bibinfo{person}{Guohao Dai}, \bibinfo{person}{Tianhao
  Huang}, \bibinfo{person}{Yuze Chi}, \bibinfo{person}{Ningyi Xu},
  \bibinfo{person}{Yu Wang}, {and} \bibinfo{person}{Huazhong Yang}.}
  \bibinfo{year}{2017}\natexlab{a}.
\newblock \showarticletitle{ForeGraph: Exploring large-scale graph processing
  on multi-FPGA architecture}. In \bibinfo{booktitle}{\emph{Proceedings of the
  2017 ACM/SIGDA International Symposium on Field-Programmable Gate Arrays}}.
  ACM, \bibinfo{pages}{217--226}.
\newblock


\bibitem[\protect\citeauthoryear{Dai, Huang, Chi, Xu, Wang, and Yang}{Dai
  et~al\mbox{.}}{2017b}]%
        {dai:foregraph}
\bibfield{author}{\bibinfo{person}{Guohao Dai}, \bibinfo{person}{Tianhao
  Huang}, \bibinfo{person}{Yuze Chi}, \bibinfo{person}{Ningyi Xu},
  \bibinfo{person}{Yu Wang}, {and} \bibinfo{person}{Huazhong Yang}.}
  \bibinfo{year}{2017}\natexlab{b}.
\newblock \showarticletitle{ForeGraph: Exploring Large-scale Graph Processing
  on Multi-FPGA Architecture}. In \bibinfo{booktitle}{\emph{Proceedings of the
  2017 ACM/SIGDA International Symposium on Field-Programmable Gate Arrays}}
  \emph{(\bibinfo{series}{FPGA '17})}. \bibinfo{publisher}{ACM},
  \bibinfo{address}{New York, NY, USA}, \bibinfo{pages}{217--226}.
\newblock
\showISBNx{978-1-4503-4354-1}
\urldef\tempurl%
\url{https://doi.org/10.1145/3020078.3021739}
\showDOI{\tempurl}


\bibitem[\protect\citeauthoryear{Dai, Huang, Chi, Zhao, Sun, Liu, Wang, Xie,
  and Yang}{Dai et~al\mbox{.}}{2018}]%
        {dai2018graphh}
\bibfield{author}{\bibinfo{person}{Guohao Dai}, \bibinfo{person}{Tianhao
  Huang}, \bibinfo{person}{Yuze Chi}, \bibinfo{person}{Jishen Zhao},
  \bibinfo{person}{Guangyu Sun}, \bibinfo{person}{Yongpan Liu},
  \bibinfo{person}{Yu Wang}, \bibinfo{person}{Yuan Xie}, {and}
  \bibinfo{person}{Huazhong Yang}.} \bibinfo{year}{2018}\natexlab{}.
\newblock \showarticletitle{Graphh: A processing-in-memory architecture for
  large-scale graph processing}.
\newblock \bibinfo{journal}{\emph{IEEE Transactions on Computer-Aided Design of
  Integrated Circuits and Systems}} \bibinfo{volume}{38}, \bibinfo{number}{4}
  (\bibinfo{year}{2018}), \bibinfo{pages}{640--653}.
\newblock


\bibitem[\protect\citeauthoryear{Dai, Huang, Wang, Yang, and Wawrzynek}{Dai
  et~al\mbox{.}}{2019}]%
        {dai2019graphsar}
\bibfield{author}{\bibinfo{person}{Guohao Dai}, \bibinfo{person}{Tianhao
  Huang}, \bibinfo{person}{Yu Wang}, \bibinfo{person}{Huazhong Yang}, {and}
  \bibinfo{person}{John Wawrzynek}.} \bibinfo{year}{2019}\natexlab{}.
\newblock \showarticletitle{GraphSAR: a sparsity-aware processing-in-memory
  architecture for large-scale graph processing on ReRAMs}. In
  \bibinfo{booktitle}{\emph{Proceedings of the 24th Asia and South Pacific
  Design Automation Conference}}. \bibinfo{pages}{120--126}.
\newblock


\bibitem[\protect\citeauthoryear{Dandalis, Mei, and Prasanna}{Dandalis
  et~al\mbox{.}}{1999}]%
        {dandalis1999domain}
\bibfield{author}{\bibinfo{person}{Andreas Dandalis},
  \bibinfo{person}{Alessandro Mei}, {and} \bibinfo{person}{Viktor~K Prasanna}.}
  \bibinfo{year}{1999}\natexlab{}.
\newblock \showarticletitle{Domain specific mapping for solving graph problems
  on reconfigurable devices}. In \bibinfo{booktitle}{\emph{International
  Parallel Processing Symposium}}. Springer, \bibinfo{pages}{652--660}.
\newblock


\bibitem[\protect\citeauthoryear{Danisch, Balalau, and Sozio}{Danisch
  et~al\mbox{.}}{2018}]%
        {danisch2018listing}
\bibfield{author}{\bibinfo{person}{Maximilien Danisch}, \bibinfo{person}{Oana
  Balalau}, {and} \bibinfo{person}{Mauro Sozio}.}
  \bibinfo{year}{2018}\natexlab{}.
\newblock \showarticletitle{Listing k-cliques in sparse real-world graphs}. In
  \bibinfo{booktitle}{\emph{Proceedings of the 2018 World Wide Web Conference
  on World Wide Web}}. International World Wide Web Conferences Steering
  Committee, \bibinfo{pages}{589--598}.
\newblock


\bibitem[\protect\citeauthoryear{Day and Sankoff}{Day and Sankoff}{1986}]%
        {day1986computational}
\bibfield{author}{\bibinfo{person}{William~HE Day} {and} \bibinfo{person}{David
  Sankoff}.} \bibinfo{year}{1986}\natexlab{}.
\newblock \showarticletitle{Computational complexity of inferring phylogenies
  by compatibility}.
\newblock \bibinfo{journal}{\emph{Systematic Biology}} \bibinfo{volume}{35},
  \bibinfo{number}{2} (\bibinfo{year}{1986}), \bibinfo{pages}{224--229}.
\newblock


\bibitem[\protect\citeauthoryear{de~Fine~Licht, Besta, Meierhans, and
  Hoefler}{de~Fine~Licht et~al\mbox{.}}{2018}]%
        {de2018transformations}
\bibfield{author}{\bibinfo{person}{Johannes de Fine~Licht},
  \bibinfo{person}{Maciej Besta}, \bibinfo{person}{Simon Meierhans}, {and}
  \bibinfo{person}{Torsten Hoefler}.} \bibinfo{year}{2018}\natexlab{}.
\newblock \showarticletitle{Transformations of high-level synthesis codes for
  high-performance computing}.
\newblock \bibinfo{journal}{\emph{arXiv e-prints}} (\bibinfo{year}{2018}),
  \bibinfo{pages}{arXiv--1805}.
\newblock


\bibitem[\protect\citeauthoryear{Dhulipala, Blelloch, and Shun}{Dhulipala
  et~al\mbox{.}}{2018}]%
        {dhulipala2018theoretically}
\bibfield{author}{\bibinfo{person}{Laxman Dhulipala}, \bibinfo{person}{Guy~E
  Blelloch}, {and} \bibinfo{person}{Julian Shun}.}
  \bibinfo{year}{2018}\natexlab{}.
\newblock \showarticletitle{Theoretically efficient parallel graph algorithms
  can be fast and scalable}. In \bibinfo{booktitle}{\emph{Proceedings of the
  30th on Symposium on Parallelism in Algorithms and Architectures}}.
  \bibinfo{pages}{393--404}.
\newblock


\bibitem[\protect\citeauthoryear{Dhulipala, McGuffey, Kang, Gu, Blelloch,
  Gibbons, and Shun}{Dhulipala et~al\mbox{.}}{2020}]%
        {sage}
\bibfield{author}{\bibinfo{person}{Laxman Dhulipala}, \bibinfo{person}{Charles
  McGuffey}, \bibinfo{person}{Hongbo Kang}, \bibinfo{person}{Yan Gu},
  \bibinfo{person}{Guy Blelloch}, \bibinfo{person}{Phillip Gibbons}, {and}
  \bibinfo{person}{Julian Shun}.} \bibinfo{year}{2020}\natexlab{}.
\newblock \showarticletitle{Sage: Parallel Semi-Asymmetric Graph Algorithms for
  NVRAMs}.
\newblock \bibinfo{journal}{\emph{PVLDB}} (\bibinfo{year}{2020}).
\newblock


\bibitem[\protect\citeauthoryear{Dias, Teixeira, Guedes, Meira, and
  Parthasarathy}{Dias et~al\mbox{.}}{2019}]%
        {dias2019fractal}
\bibfield{author}{\bibinfo{person}{Vinicius Dias}, \bibinfo{person}{Carlos~HC
  Teixeira}, \bibinfo{person}{Dorgival Guedes}, \bibinfo{person}{Wagner Meira},
  {and} \bibinfo{person}{Srinivasan Parthasarathy}.}
  \bibinfo{year}{2019}\natexlab{}.
\newblock \showarticletitle{Fractal: A General-Purpose Graph Pattern Mining
  System}. In \bibinfo{booktitle}{\emph{Proceedings of the 2019 International
  Conference on Management of Data}}. ACM, \bibinfo{pages}{1357--1374}.
\newblock


\bibitem[\protect\citeauthoryear{Doekemeijer and Varbanescu}{Doekemeijer and
  Varbanescu}{2014}]%
        {doekemeijer2014survey}
\bibfield{author}{\bibinfo{person}{Niels Doekemeijer} {and}
  \bibinfo{person}{Ana~Lucia Varbanescu}.} \bibinfo{year}{2014}\natexlab{}.
\newblock \showarticletitle{A survey of parallel graph processing frameworks}.
\newblock \bibinfo{journal}{\emph{Delft University of Technology}}
  (\bibinfo{year}{2014}), \bibinfo{pages}{21}.
\newblock


\bibitem[\protect\citeauthoryear{Dua and Du}{Dua and Du}{2016}]%
        {dua2016data}
\bibfield{author}{\bibinfo{person}{Sumeet Dua} {and} \bibinfo{person}{Xian
  Du}.} \bibinfo{year}{2016}\natexlab{}.
\newblock \bibinfo{booktitle}{\emph{Data mining and machine learning in
  cybersecurity}}.
\newblock \bibinfo{publisher}{CRC press}.
\newblock


\bibitem[\protect\citeauthoryear{Dubinsky}{Dubinsky}{1995}]%
        {dubinsky1995isetl}
\bibfield{author}{\bibinfo{person}{Ed Dubinsky}.}
  \bibinfo{year}{1995}\natexlab{}.
\newblock \showarticletitle{ISETL: A programming language for learning
  mathematics}.
\newblock \bibinfo{journal}{\emph{Communications on Pure and Applied
  Mathematics}} \bibinfo{volume}{48}, \bibinfo{number}{9}
  (\bibinfo{year}{1995}), \bibinfo{pages}{1027--1051}.
\newblock


\bibitem[\protect\citeauthoryear{Eblen, Phillips, Rogers, and Langston}{Eblen
  et~al\mbox{.}}{2012}]%
        {eblen2012maximum}
\bibfield{author}{\bibinfo{person}{John~D Eblen}, \bibinfo{person}{Charles~A
  Phillips}, \bibinfo{person}{Gary~L Rogers}, {and} \bibinfo{person}{Michael~A
  Langston}.} \bibinfo{year}{2012}\natexlab{}.
\newblock \showarticletitle{The maximum clique enumeration problem: algorithms,
  applications, and implementations}. In \bibinfo{booktitle}{\emph{BMC
  bioinformatics}}, Vol.~\bibinfo{volume}{13}. Springer, \bibinfo{pages}{S5}.
\newblock


\bibitem[\protect\citeauthoryear{Eden, Levi, and Ron}{Eden
  et~al\mbox{.}}{2018}]%
        {DBLP:conf/soda/EdenLR18}
\bibfield{author}{\bibinfo{person}{Talya Eden}, \bibinfo{person}{Reut Levi},
  {and} \bibinfo{person}{Dana Ron}.} \bibinfo{year}{2018}\natexlab{}.
\newblock \showarticletitle{Testing bounded arboricity}. In
  \bibinfo{booktitle}{\emph{Proceedings of the Twenty-Ninth Annual {ACM-SIAM}
  Symposium on Discrete Algorithms, {SODA} 2018, New Orleans, LA, USA, January
  7-10, 2018}}. \bibinfo{pages}{2081--2092}.
\newblock
\urldef\tempurl%
\url{https://doi.org/10.1137/1.9781611975031.136}
\showDOI{\tempurl}


\bibitem[\protect\citeauthoryear{Elyasi, Choi, and Sivasubramaniam}{Elyasi
  et~al\mbox{.}}{2019}]%
        {elyasi2019large}
\bibfield{author}{\bibinfo{person}{Nima Elyasi}, \bibinfo{person}{Changho
  Choi}, {and} \bibinfo{person}{Anand Sivasubramaniam}.}
  \bibinfo{year}{2019}\natexlab{}.
\newblock \showarticletitle{Large-scale graph processing on emerging storage
  devices}. In \bibinfo{booktitle}{\emph{17th $\{$USENIX$\}$ Conference on File
  and Storage Technologies ($\{$FAST$\}$ 19)}}. \bibinfo{pages}{309--316}.
\newblock


\bibitem[\protect\citeauthoryear{Engelhardt and So}{Engelhardt and So}{2016}]%
        {engelhardt2016gravf}
\bibfield{author}{\bibinfo{person}{Nina Engelhardt} {and}
  \bibinfo{person}{Hayden Kwok-Hay So}.} \bibinfo{year}{2016}\natexlab{}.
\newblock \showarticletitle{Gravf: A vertex-centric distributed graph
  processing framework on {FPGAs}}. In \bibinfo{booktitle}{\emph{Field
  Programmable Logic and Applications (FPL), 2016 26th International Conference
  on}}. IEEE, \bibinfo{pages}{1--4}.
\newblock


\bibitem[\protect\citeauthoryear{Eppstein, L{\"{o}}ffler, and Strash}{Eppstein
  et~al\mbox{.}}{2010}]%
        {DBLP:conf/isaac/EppsteinLS10}
\bibfield{author}{\bibinfo{person}{David Eppstein}, \bibinfo{person}{Maarten
  L{\"{o}}ffler}, {and} \bibinfo{person}{Darren Strash}.}
  \bibinfo{year}{2010}\natexlab{}.
\newblock \showarticletitle{Listing All Maximal Cliques in Sparse Graphs in
  Near-Optimal Time}. In \bibinfo{booktitle}{\emph{Algorithms and Computation -
  21st International Symposium, {ISAAC} 2010, Jeju Island, Korea, December
  15-17, 2010, Proceedings, Part {I}}}. \bibinfo{pages}{403--414}.
\newblock
\urldef\tempurl%
\url{https://doi.org/10.1007/978-3-642-17517-6\_36}
\showDOI{\tempurl}


\bibitem[\protect\citeauthoryear{Faldu, Diamond, and Grot}{Faldu
  et~al\mbox{.}}{2019}]%
        {faldu2019poster}
\bibfield{author}{\bibinfo{person}{Priyank Faldu}, \bibinfo{person}{Jeff
  Diamond}, {and} \bibinfo{person}{Boris Grot}.}
  \bibinfo{year}{2019}\natexlab{}.
\newblock \showarticletitle{POSTER: Domain-Specialized Cache Management for
  Graph Analytics}. In \bibinfo{booktitle}{\emph{2019 28th International
  Conference on Parallel Architectures and Compilation Techniques (PACT)}}.
  IEEE, \bibinfo{pages}{473--474}.
\newblock


\bibitem[\protect\citeauthoryear{Farach{-}Colton and Tsai}{Farach{-}Colton and
  Tsai}{2014}]%
        {DBLP:conf/latin/Farach-ColtonT14}
\bibfield{author}{\bibinfo{person}{Martin Farach{-}Colton} {and}
  \bibinfo{person}{Meng{-}Tsung Tsai}.} \bibinfo{year}{2014}\natexlab{}.
\newblock \showarticletitle{Computing the Degeneracy of Large Graphs}. In
  \bibinfo{booktitle}{\emph{{LATIN} 2014: Theoretical Informatics - 11th Latin
  American Symposium, Montevideo, Uruguay, March 31 - April 4, 2014.
  Proceedings}}. \bibinfo{pages}{250--260}.
\newblock
\urldef\tempurl%
\url{https://doi.org/10.1007/978-3-642-54423-1\_22}
\showDOI{\tempurl}


\bibitem[\protect\citeauthoryear{Franke, Li, and Parris}{Franke
  et~al\mbox{.}}{2018}]%
        {franke2018creating}
\bibfield{author}{\bibinfo{person}{Hubertus Franke},
  \bibinfo{person}{Chung-Sheng Li}, {and} \bibinfo{person}{Colin~J Parris}.}
  \bibinfo{year}{2018}\natexlab{}.
\newblock \bibinfo{title}{Creating new cloud resource instruction set
  architecture}.
\newblock
\newblock
\newblock
\shownote{US Patent App. 16/041,297.}


\bibitem[\protect\citeauthoryear{Gallagher}{Gallagher}{2006}]%
        {gallagher2006matching}
\bibfield{author}{\bibinfo{person}{Brian Gallagher}.}
  \bibinfo{year}{2006}\natexlab{}.
\newblock \showarticletitle{Matching Structure and Semantics: A Survey on
  Graph-Based Pattern Matching.}. In \bibinfo{booktitle}{\emph{AAAI Fall
  Symposium: Capturing and Using Patterns for Evidence Detection}}.
  \bibinfo{pages}{45--53}.
\newblock


\bibitem[\protect\citeauthoryear{Gao, Tziantzioulis, and Wentzlaff}{Gao
  et~al\mbox{.}}{2019}]%
        {gao2019computedram}
\bibfield{author}{\bibinfo{person}{Fei Gao}, \bibinfo{person}{Georgios
  Tziantzioulis}, {and} \bibinfo{person}{David Wentzlaff}.}
  \bibinfo{year}{2019}\natexlab{}.
\newblock \showarticletitle{Computedram: In-memory compute using off-the-shelf
  drams}. In \bibinfo{booktitle}{\emph{Proceedings of the 52nd Annual IEEE/ACM
  International Symposium on Microarchitecture}}. \bibinfo{pages}{100--113}.
\newblock


\bibitem[\protect\citeauthoryear{Gao, Ayers, and Kozyrakis}{Gao
  et~al\mbox{.}}{2015}]%
        {gao2015practical}
\bibfield{author}{\bibinfo{person}{Mingyu Gao}, \bibinfo{person}{Grant Ayers},
  {and} \bibinfo{person}{Christos Kozyrakis}.} \bibinfo{year}{2015}\natexlab{}.
\newblock \showarticletitle{Practical near-data processing for in-memory
  analytics frameworks}. In \bibinfo{booktitle}{\emph{2015 International
  Conference on Parallel Architecture and Compilation (PACT)}}. IEEE,
  \bibinfo{pages}{113--124}.
\newblock


\bibitem[\protect\citeauthoryear{Gao and Kozyrakis}{Gao and Kozyrakis}{2016}]%
        {gao2016hrl}
\bibfield{author}{\bibinfo{person}{Mingyu Gao} {and} \bibinfo{person}{Christos
  Kozyrakis}.} \bibinfo{year}{2016}\natexlab{}.
\newblock \showarticletitle{HRL: Efficient and flexible reconfigurable logic
  for near-data processing}. In \bibinfo{booktitle}{\emph{2016 IEEE
  International Symposium on High Performance Computer Architecture (HPCA)}}.
  Ieee, \bibinfo{pages}{126--137}.
\newblock


\bibitem[\protect\citeauthoryear{Gerstenberger, Besta, and
  Hoefler}{Gerstenberger et~al\mbox{.}}{2013}]%
        {gerstenberger2013enabling}
\bibfield{author}{\bibinfo{person}{Robert Gerstenberger},
  \bibinfo{person}{Maciej Besta}, {and} \bibinfo{person}{Torsten Hoefler}.}
  \bibinfo{year}{2013}\natexlab{}.
\newblock \showarticletitle{Enabling highly-scalable remote memory access
  programming with MPI-3 one sided}. In \bibinfo{booktitle}{\emph{Proceedings
  of the International Conference on High Performance Computing, Networking,
  Storage and Analysis}}. \bibinfo{pages}{1--12}.
\newblock


\bibitem[\protect\citeauthoryear{Ghose, Boroumand, Kim, G{\'o}mez-Luna, and
  Mutlu}{Ghose et~al\mbox{.}}{2019a}]%
        {ghose2019processing_pim}
\bibfield{author}{\bibinfo{person}{Saugata Ghose}, \bibinfo{person}{Amirali
  Boroumand}, \bibinfo{person}{Jeremie~S Kim}, \bibinfo{person}{Juan
  G{\'o}mez-Luna}, {and} \bibinfo{person}{Onur Mutlu}.}
  \bibinfo{year}{2019}\natexlab{a}.
\newblock \showarticletitle{{Processing-in-Memory: A Workload-driven
  Perspective}}.
\newblock \bibinfo{journal}{\emph{IBM JRD}} (\bibinfo{year}{2019}).
\newblock


\bibitem[\protect\citeauthoryear{Ghose, Hsieh, Boroumand, Ausavarungnirun, and
  Mutlu}{Ghose et~al\mbox{.}}{2019b}]%
        {ghose2019processing}
\bibfield{author}{\bibinfo{person}{Saugata Ghose}, \bibinfo{person}{Kevin
  Hsieh}, \bibinfo{person}{Amirali Boroumand}, \bibinfo{person}{Rachata
  Ausavarungnirun}, {and} \bibinfo{person}{Onur Mutlu}.}
  \bibinfo{year}{2019}\natexlab{b}.
\newblock \showarticletitle{The processing-in-memory paradigm: Mechanisms to
  enable adoption}.
\newblock In \bibinfo{booktitle}{\emph{Beyond-CMOS Technologies for Next
  Generation Computer Design}}. \bibinfo{publisher}{Springer},
  \bibinfo{pages}{133--194}.
\newblock


\bibitem[\protect\citeauthoryear{Gianinazzi, Besta, Schaffner, and
  Hoefler}{Gianinazzi et~al\mbox{.}}{2021a}]%
        {gianinazzi2021parallel}
\bibfield{author}{\bibinfo{person}{Lukas Gianinazzi}, \bibinfo{person}{Maciej
  Besta}, \bibinfo{person}{Yannick Schaffner}, {and} \bibinfo{person}{Torsten
  Hoefler}.} \bibinfo{year}{2021}\natexlab{a}.
\newblock \showarticletitle{Parallel Algorithms for Finding Large Cliques in
  Sparse Graphs}. In \bibinfo{booktitle}{\emph{Proceedings of the 33rd ACM
  Symposium on Parallelism in Algorithms and Architectures}}.
  \bibinfo{pages}{243--253}.
\newblock


\bibitem[\protect\citeauthoryear{Gianinazzi, Fries, Dryden, Ben-Nun, and
  Hoefler}{Gianinazzi et~al\mbox{.}}{2021b}]%
        {gianinazzi2021learning}
\bibfield{author}{\bibinfo{person}{Lukas Gianinazzi},
  \bibinfo{person}{Maximilian Fries}, \bibinfo{person}{Nikoli Dryden},
  \bibinfo{person}{Tal Ben-Nun}, {and} \bibinfo{person}{Torsten Hoefler}.}
  \bibinfo{year}{2021}\natexlab{b}.
\newblock \showarticletitle{Learning Combinatorial Node Labeling Algorithms}.
\newblock \bibinfo{journal}{\emph{arXiv preprint arXiv:2106.03594}}
  (\bibinfo{year}{2021}).
\newblock


\bibitem[\protect\citeauthoryear{Gianinazzi, Kalvoda, De~Palma, Besta, and
  Hoefler}{Gianinazzi et~al\mbox{.}}{2018}]%
        {gianinazzi2018communication}
\bibfield{author}{\bibinfo{person}{Lukas Gianinazzi}, \bibinfo{person}{Pavel
  Kalvoda}, \bibinfo{person}{Alessandro De~Palma}, \bibinfo{person}{Maciej
  Besta}, {and} \bibinfo{person}{Torsten Hoefler}.}
  \bibinfo{year}{2018}\natexlab{}.
\newblock \showarticletitle{Communication-avoiding parallel minimum cuts and
  connected components}.
\newblock \bibinfo{journal}{\emph{ACM SIGPLAN Notices}} \bibinfo{volume}{53},
  \bibinfo{number}{1} (\bibinfo{year}{2018}), \bibinfo{pages}{219--232}.
\newblock


\bibitem[\protect\citeauthoryear{Gibson, Kumar, and Tomkins}{Gibson
  et~al\mbox{.}}{2005}]%
        {gibson2005discovering}
\bibfield{author}{\bibinfo{person}{David Gibson}, \bibinfo{person}{Ravi Kumar},
  {and} \bibinfo{person}{Andrew Tomkins}.} \bibinfo{year}{2005}\natexlab{}.
\newblock \showarticletitle{Discovering large dense subgraphs in massive
  graphs}. In \bibinfo{booktitle}{\emph{Proceedings of the 31st international
  conference on Very large data bases}}. \bibinfo{pages}{721--732}.
\newblock


\bibitem[\protect\citeauthoryear{G{\'o}mez-Luna, Hajj, Fernandez, Giannoula,
  Oliveira, and Mutlu}{G{\'o}mez-Luna et~al\mbox{.}}{2021}]%
        {gomez2021benchmarking}
\bibfield{author}{\bibinfo{person}{Juan G{\'o}mez-Luna},
  \bibinfo{person}{Izzat~El Hajj}, \bibinfo{person}{Ivan Fernandez},
  \bibinfo{person}{Christina Giannoula}, \bibinfo{person}{Geraldo~F Oliveira},
  {and} \bibinfo{person}{Onur Mutlu}.} \bibinfo{year}{2021}\natexlab{}.
\newblock \showarticletitle{Benchmarking a New Paradigm: An Experimental
  Analysis of a Real Processing-in-Memory Architecture}.
\newblock \bibinfo{journal}{\emph{arXiv preprint arXiv:2105.03814}}
  (\bibinfo{year}{2021}).
\newblock


\bibitem[\protect\citeauthoryear{Gonzalez, Low, Gu, Bickson, and
  Guestrin}{Gonzalez et~al\mbox{.}}{2012}]%
        {gonzalez2012powergraph}
\bibfield{author}{\bibinfo{person}{Joseph~E Gonzalez}, \bibinfo{person}{Yucheng
  Low}, \bibinfo{person}{Haijie Gu}, \bibinfo{person}{Danny Bickson}, {and}
  \bibinfo{person}{Carlos Guestrin}.} \bibinfo{year}{2012}\natexlab{}.
\newblock \showarticletitle{Powergraph: distributed graph-parallel computation
  on natural graphs.}. In \bibinfo{booktitle}{\emph{OSDI}},
  Vol.~\bibinfo{volume}{12}. \bibinfo{pages}{2}.
\newblock


\bibitem[\protect\citeauthoryear{Hajinazar, Oliveira, Gregorio, Ferreira,
  Ghiasi, Patel, Alser, Ghose, G{\'o}mez-Luna, and Mutlu}{Hajinazar
  et~al\mbox{.}}{2021}]%
        {hajinazar2021simdram}
\bibfield{author}{\bibinfo{person}{Nastaran Hajinazar},
  \bibinfo{person}{Geraldo~F Oliveira}, \bibinfo{person}{Sven Gregorio},
  \bibinfo{person}{Jo{\~a}o~Dinis Ferreira}, \bibinfo{person}{Nika~Mansouri
  Ghiasi}, \bibinfo{person}{Minesh Patel}, \bibinfo{person}{Mohammed Alser},
  \bibinfo{person}{Saugata Ghose}, \bibinfo{person}{Juan G{\'o}mez-Luna}, {and}
  \bibinfo{person}{Onur Mutlu}.} \bibinfo{year}{2021}\natexlab{}.
\newblock \showarticletitle{SIMDRAM: a framework for bit-serial SIMD processing
  using DRAM}. In \bibinfo{booktitle}{\emph{Proceedings of the 26th ACM
  International Conference on Architectural Support for Programming Languages
  and Operating Systems}}. \bibinfo{pages}{329--345}.
\newblock


\bibitem[\protect\citeauthoryear{Ham, Wu, Sundaram, Satish, and Martonosi}{Ham
  et~al\mbox{.}}{2016}]%
        {ham2016graphicionado}
\bibfield{author}{\bibinfo{person}{Tae~Jun Ham}, \bibinfo{person}{Lisa Wu},
  \bibinfo{person}{Narayanan Sundaram}, \bibinfo{person}{Nadathur Satish},
  {and} \bibinfo{person}{Margaret Martonosi}.} \bibinfo{year}{2016}\natexlab{}.
\newblock \showarticletitle{Graphicionado: A high-performance and
  energy-efficient accelerator for graph analytics}. In
  \bibinfo{booktitle}{\emph{Microarchitecture (MICRO), 2016 49th Annual
  IEEE/ACM International Symposium on}}. IEEE, \bibinfo{pages}{1--13}.
\newblock


\bibitem[\protect\citeauthoryear{Han and Kamber}{Han and Kamber}{2006}]%
        {han2006data}
\bibfield{author}{\bibinfo{person}{J Han} {and} \bibinfo{person}{M Kamber}.}
  \bibinfo{year}{2006}\natexlab{}.
\newblock \bibinfo{title}{Data Mining Concepts and Techniques (A. Stephan,
  Ed.), 2nd edn., vol. 40}.
\newblock
\newblock


\bibitem[\protect\citeauthoryear{Han, Shen, Liu, Shao, Huang, and Li}{Han
  et~al\mbox{.}}{2018a}]%
        {han2018novel}
\bibfield{author}{\bibinfo{person}{Lei Han}, \bibinfo{person}{Zhaoyan Shen},
  \bibinfo{person}{Duo Liu}, \bibinfo{person}{Zili Shao},
  \bibinfo{person}{H~Howie Huang}, {and} \bibinfo{person}{Tao Li}.}
  \bibinfo{year}{2018}\natexlab{a}.
\newblock \showarticletitle{A novel ReRAM-based processing-in-memory
  architecture for graph traversal}.
\newblock \bibinfo{journal}{\emph{ACM Transactions on Storage (TOS)}}
  \bibinfo{volume}{14}, \bibinfo{number}{1} (\bibinfo{year}{2018}),
  \bibinfo{pages}{1--26}.
\newblock


\bibitem[\protect\citeauthoryear{Han, Zou, and Yu}{Han et~al\mbox{.}}{2018b}]%
        {han2018speeding}
\bibfield{author}{\bibinfo{person}{Shuo Han}, \bibinfo{person}{Lei Zou}, {and}
  \bibinfo{person}{Jeffrey~Xu Yu}.} \bibinfo{year}{2018}\natexlab{b}.
\newblock \showarticletitle{Speeding Up Set Intersections in Graph Algorithms
  using SIMD Instructions}. In \bibinfo{booktitle}{\emph{Proceedings of the
  2018 International Conference on Management of Data}}. ACM,
  \bibinfo{pages}{1587--1602}.
\newblock


\bibitem[\protect\citeauthoryear{He}{He}{2019}]%
        {he2019engn}
\bibfield{author}{\bibinfo{person}{Lei He}.} \bibinfo{year}{2019}\natexlab{}.
\newblock \showarticletitle{EnGN: A High-Throughput and Energy-Efficient
  Accelerator for Large Graph Neural Networks}.
\newblock \bibinfo{journal}{\emph{arXiv preprint arXiv:1909.00155}}
  (\bibinfo{year}{2019}).
\newblock


\bibitem[\protect\citeauthoryear{Hein}{Hein}{2018}]%
        {hein2018near}
\bibfield{author}{\bibinfo{person}{Eric~Robert Hein}.}
  \bibinfo{year}{2018}\natexlab{}.
\newblock \emph{\bibinfo{title}{Near-data processing for dynamic graph
  analytics}}.
\newblock \bibinfo{thesistype}{Ph.D. Dissertation}. \bibinfo{school}{Georgia
  Institute of Technology}.
\newblock


\bibitem[\protect\citeauthoryear{Heirman, Carlson, and Eeckhout}{Heirman
  et~al\mbox{.}}{2012}]%
        {heirman2012sniper}
\bibfield{author}{\bibinfo{person}{Wim Heirman}, \bibinfo{person}{Trevor
  Carlson}, {and} \bibinfo{person}{Lieven Eeckhout}.}
  \bibinfo{year}{2012}\natexlab{}.
\newblock \showarticletitle{Sniper: Scalable and accurate parallel multi-core
  simulation}. In \bibinfo{booktitle}{\emph{8th International Summer School on
  Advanced Computer Architecture and Compilation for High-Performance and
  Embedded Systems (ACACES-2012)}}. High-Performance and Embedded Architecture
  and Compilation Network of~…, \bibinfo{pages}{91--94}.
\newblock


\bibitem[\protect\citeauthoryear{Henry, Hooker, Parks, and Reed}{Henry
  et~al\mbox{.}}{2018}]%
        {henry2018compiler}
\bibfield{author}{\bibinfo{person}{G~Glenn Henry}, \bibinfo{person}{Rodney~E
  Hooker}, \bibinfo{person}{Terry Parks}, {and} \bibinfo{person}{Douglas~R
  Reed}.} \bibinfo{year}{2018}\natexlab{}.
\newblock \bibinfo{title}{Compiler system for a processor with an expandable
  instruction set architecture for dynamically configuring execution
  resources}.
\newblock
\newblock
\newblock
\shownote{US Patent App. 10/127,041.}


\bibitem[\protect\citeauthoryear{Herlihy, Shavit, Luchangco, and Spear}{Herlihy
  et~al\mbox{.}}{2020}]%
        {herlihy2020art}
\bibfield{author}{\bibinfo{person}{Maurice Herlihy}, \bibinfo{person}{Nir
  Shavit}, \bibinfo{person}{Victor Luchangco}, {and} \bibinfo{person}{Michael
  Spear}.} \bibinfo{year}{2020}\natexlab{}.
\newblock \bibinfo{booktitle}{\emph{The art of multiprocessor programming}}.
\newblock \bibinfo{publisher}{Newnes}.
\newblock


\bibitem[\protect\citeauthoryear{Hido and Kawano}{Hido and Kawano}{2005}]%
        {hido2005amiot}
\bibfield{author}{\bibinfo{person}{Shohei Hido} {and} \bibinfo{person}{Hiroyuki
  Kawano}.} \bibinfo{year}{2005}\natexlab{}.
\newblock \showarticletitle{AMIOT: induced ordered tree mining in
  tree-structured databases}. In \bibinfo{booktitle}{\emph{Fifth IEEE
  International Conference on Data Mining (ICDM'05)}}. IEEE,
  \bibinfo{pages}{8--pp}.
\newblock


\bibitem[\protect\citeauthoryear{Hoefler and Belli}{Hoefler and Belli}{2015}]%
        {hoefler2015scientific}
\bibfield{author}{\bibinfo{person}{Torsten Hoefler} {and}
  \bibinfo{person}{Roberto Belli}.} \bibinfo{year}{2015}\natexlab{}.
\newblock \showarticletitle{Scientific benchmarking of parallel computing
  systems: twelve ways to tell the masses when reporting performance results}.
  In \bibinfo{booktitle}{\emph{Proceedings of the international conference for
  high performance computing, networking, storage and analysis}}.
  \bibinfo{pages}{1--12}.
\newblock


\bibitem[\protect\citeauthoryear{Horv{\'a}th, G{\"a}rtner, and
  Wrobel}{Horv{\'a}th et~al\mbox{.}}{2004}]%
        {horvath2004cyclic}
\bibfield{author}{\bibinfo{person}{Tam{\'a}s Horv{\'a}th},
  \bibinfo{person}{Thomas G{\"a}rtner}, {and} \bibinfo{person}{Stefan Wrobel}.}
  \bibinfo{year}{2004}\natexlab{}.
\newblock \showarticletitle{Cyclic pattern kernels for predictive graph
  mining}. In \bibinfo{booktitle}{\emph{Proceedings of the tenth ACM SIGKDD
  international conference on Knowledge discovery and data mining}}. ACM,
  \bibinfo{pages}{158--167}.
\newblock


\bibitem[\protect\citeauthoryear{Hsieh, Khan, Vijaykumar, Chang, Boroumand,
  Ghose, and Mutlu}{Hsieh et~al\mbox{.}}{2016}]%
        {hsieh2016accelerating}
\bibfield{author}{\bibinfo{person}{Kevin Hsieh}, \bibinfo{person}{Samira Khan},
  \bibinfo{person}{Nandita Vijaykumar}, \bibinfo{person}{Kevin~K Chang},
  \bibinfo{person}{Amirali Boroumand}, \bibinfo{person}{Saugata Ghose}, {and}
  \bibinfo{person}{Onur Mutlu}.} \bibinfo{year}{2016}\natexlab{}.
\newblock \showarticletitle{Accelerating pointer chasing in 3D-stacked memory:
  Challenges, mechanisms, evaluation}. In \bibinfo{booktitle}{\emph{2016 IEEE
  34th International Conference on Computer Design (ICCD)}}. IEEE,
  \bibinfo{pages}{25--32}.
\newblock


\bibitem[\protect\citeauthoryear{Huang, Dai, Wang, and Yang}{Huang
  et~al\mbox{.}}{2018}]%
        {huang2018hyve}
\bibfield{author}{\bibinfo{person}{Tianhao Huang}, \bibinfo{person}{Guohao
  Dai}, \bibinfo{person}{Yu Wang}, {and} \bibinfo{person}{Huazhong Yang}.}
  \bibinfo{year}{2018}\natexlab{}.
\newblock \showarticletitle{HyVE: Hybrid vertex-edge memory hierarchy for
  energy-efficient graph processing}. In \bibinfo{booktitle}{\emph{2018 Design,
  Automation \& Test in Europe Conference \& Exhibition (DATE)}}. IEEE,
  \bibinfo{pages}{973--978}.
\newblock


\bibitem[\protect\citeauthoryear{Huang, Zheng, Liao, Jin, Yao, and Gui}{Huang
  et~al\mbox{.}}{2019}]%
        {huang2019ragra}
\bibfield{author}{\bibinfo{person}{Yu Huang}, \bibinfo{person}{Long Zheng},
  \bibinfo{person}{Xiaofei Liao}, \bibinfo{person}{Hai Jin},
  \bibinfo{person}{Pengcheng Yao}, {and} \bibinfo{person}{Chuangyi Gui}.}
  \bibinfo{year}{2019}\natexlab{}.
\newblock \showarticletitle{RAGra: Leveraging Monolithic 3D ReRAM for
  Massively-Parallel Graph Processing}. In \bibinfo{booktitle}{\emph{2019
  Design, Automation \& Test in Europe Conference \& Exhibition (DATE)}}. IEEE,
  \bibinfo{pages}{1273--1276}.
\newblock


\bibitem[\protect\citeauthoryear{Iyer, Liu, Jin, Venkataraman, Braverman, and
  Stoica}{Iyer et~al\mbox{.}}{2018}]%
        {iyer2018asap}
\bibfield{author}{\bibinfo{person}{Anand~Padmanabha Iyer},
  \bibinfo{person}{Zaoxing Liu}, \bibinfo{person}{Xin Jin},
  \bibinfo{person}{Shivaram Venkataraman}, \bibinfo{person}{Vladimir
  Braverman}, {and} \bibinfo{person}{Ion Stoica}.}
  \bibinfo{year}{2018}\natexlab{}.
\newblock \showarticletitle{$\{$ASAP$\}$: Fast, Approximate Graph Pattern
  Mining at Scale}. In \bibinfo{booktitle}{\emph{13th $\{$USENIX$\}$ Symposium
  on Operating Systems Design and Implementation ($\{$OSDI$\}$ 18)}}.
  \bibinfo{pages}{745--761}.
\newblock


\bibitem[\protect\citeauthoryear{Jabbour, Mhadhbi, Raddaoui, and Sais}{Jabbour
  et~al\mbox{.}}{2018}]%
        {jabbour2018pushing}
\bibfield{author}{\bibinfo{person}{Said Jabbour}, \bibinfo{person}{Nizar
  Mhadhbi}, \bibinfo{person}{Badran Raddaoui}, {and} \bibinfo{person}{Lakhdar
  Sais}.} \bibinfo{year}{2018}\natexlab{}.
\newblock \showarticletitle{Pushing the Envelope in Overlapping Communities
  Detection}. In \bibinfo{booktitle}{\emph{International Symposium on
  Intelligent Data Analysis}}. Springer, \bibinfo{pages}{151--163}.
\newblock


\bibitem[\protect\citeauthoryear{Jagadeesh, Srikanthan, and Lim}{Jagadeesh
  et~al\mbox{.}}{2011}]%
        {jagadeesh2011field}
\bibfield{author}{\bibinfo{person}{George~Rosario Jagadeesh},
  \bibinfo{person}{Thambipillai Srikanthan}, {and} \bibinfo{person}{CM Lim}.}
  \bibinfo{year}{2011}\natexlab{}.
\newblock \showarticletitle{Field programmable gate array-based acceleration of
  shortest-path computation}.
\newblock \bibinfo{journal}{\emph{IET computers \& digital techniques}}
  \bibinfo{volume}{5}, \bibinfo{number}{4} (\bibinfo{year}{2011}),
  \bibinfo{pages}{231--237}.
\newblock


\bibitem[\protect\citeauthoryear{Jamshidi, Mahadasa, and Vora}{Jamshidi
  et~al\mbox{.}}{2020}]%
        {jamshidi2020peregrine}
\bibfield{author}{\bibinfo{person}{Kasra Jamshidi}, \bibinfo{person}{Rakesh
  Mahadasa}, {and} \bibinfo{person}{Keval Vora}.}
  \bibinfo{year}{2020}\natexlab{}.
\newblock \showarticletitle{Peregrine: a pattern-aware graph mining system}. In
  \bibinfo{booktitle}{\emph{Proceedings of the Fifteenth European Conference on
  Computer Systems}}. \bibinfo{pages}{1--16}.
\newblock


\bibitem[\protect\citeauthoryear{Jarvis and Patrick}{Jarvis and
  Patrick}{1973}]%
        {jarvis1973clustering}
\bibfield{author}{\bibinfo{person}{Raymond~Austin Jarvis} {and}
  \bibinfo{person}{Edward~A Patrick}.} \bibinfo{year}{1973}\natexlab{}.
\newblock \showarticletitle{Clustering using a similarity measure based on
  shared near neighbors}.
\newblock \bibinfo{journal}{\emph{IEEE Transactions on computers}}
  \bibinfo{volume}{100}, \bibinfo{number}{11} (\bibinfo{year}{1973}),
  \bibinfo{pages}{1025--1034}.
\newblock


\bibitem[\protect\citeauthoryear{Jech}{Jech}{2013}]%
        {jech2013set}
\bibfield{author}{\bibinfo{person}{Thomas Jech}.}
  \bibinfo{year}{2013}\natexlab{}.
\newblock \bibinfo{booktitle}{\emph{Set theory}}.
\newblock \bibinfo{publisher}{Springer Science \& Business Media}.
\newblock


\bibitem[\protect\citeauthoryear{Jeddeloh and Keeth}{Jeddeloh and
  Keeth}{2012}]%
        {jeddeloh2012hybrid}
\bibfield{author}{\bibinfo{person}{Joe Jeddeloh} {and} \bibinfo{person}{Brent
  Keeth}.} \bibinfo{year}{2012}\natexlab{}.
\newblock \showarticletitle{Hybrid memory cube new DRAM architecture increases
  density and performance}. In \bibinfo{booktitle}{\emph{VLSI Technology
  (VLSIT), 2012 Symposium on}}. IEEE, \bibinfo{pages}{87--88}.
\newblock


\bibitem[\protect\citeauthoryear{Jiang, Coenen, and Zito}{Jiang
  et~al\mbox{.}}{2013}]%
        {jiang2013survey}
\bibfield{author}{\bibinfo{person}{Chuntao Jiang}, \bibinfo{person}{Frans
  Coenen}, {and} \bibinfo{person}{Michele Zito}.}
  \bibinfo{year}{2013}\natexlab{}.
\newblock \showarticletitle{A survey of frequent subgraph mining algorithms}.
\newblock \bibinfo{journal}{\emph{The Knowledge Engineering Review}}
  \bibinfo{volume}{28}, \bibinfo{number}{1} (\bibinfo{year}{2013}),
  \bibinfo{pages}{75--105}.
\newblock


\bibitem[\protect\citeauthoryear{Jiang and Pei}{Jiang and Pei}{2009}]%
        {jiang2009mining}
\bibfield{author}{\bibinfo{person}{Daxin Jiang} {and} \bibinfo{person}{Jian
  Pei}.} \bibinfo{year}{2009}\natexlab{}.
\newblock \showarticletitle{Mining frequent cross-graph quasi-cliques}.
\newblock \bibinfo{journal}{\emph{ACM Transactions on Knowledge Discovery from
  Data (TKDD)}} \bibinfo{volume}{2}, \bibinfo{number}{4}
  (\bibinfo{year}{2009}), \bibinfo{pages}{1--42}.
\newblock


\bibitem[\protect\citeauthoryear{Joshi, Zhang, Bogdanov, and Hwang}{Joshi
  et~al\mbox{.}}{2018}]%
        {joshi2018efficient}
\bibfield{author}{\bibinfo{person}{Aparna Joshi}, \bibinfo{person}{Yu Zhang},
  \bibinfo{person}{Petko Bogdanov}, {and} \bibinfo{person}{Jeong-Hyon Hwang}.}
  \bibinfo{year}{2018}\natexlab{}.
\newblock \showarticletitle{An Efficient System for Subgraph Discovery}. In
  \bibinfo{booktitle}{\emph{2018 IEEE International Conference on Big Data (Big
  Data)}}. IEEE, \bibinfo{pages}{703--712}.
\newblock


\bibitem[\protect\citeauthoryear{Jun, Wright, Zhang, and Xu}{Jun
  et~al\mbox{.}}{2018}]%
        {jun2018grafboost}
\bibfield{author}{\bibinfo{person}{Sang-Woo Jun}, \bibinfo{person}{Andy
  Wright}, \bibinfo{person}{Sizhuo Zhang}, {and} \bibinfo{person}{Shuotao Xu}.}
  \bibinfo{year}{2018}\natexlab{}.
\newblock \showarticletitle{GraFBoost: Using accelerated flash storage for
  external graph analytics}. In \bibinfo{booktitle}{\emph{2018 ACM/IEEE 45th
  Annual International Symposium on Computer Architecture (ISCA)}}. IEEE,
  \bibinfo{pages}{411--424}.
\newblock


\bibitem[\protect\citeauthoryear{Kalavri, Vlassov, and Haridi}{Kalavri
  et~al\mbox{.}}{2017}]%
        {kalavri2017high}
\bibfield{author}{\bibinfo{person}{Vasiliki Kalavri}, \bibinfo{person}{Vladimir
  Vlassov}, {and} \bibinfo{person}{Seif Haridi}.}
  \bibinfo{year}{2017}\natexlab{}.
\newblock \showarticletitle{High-level programming abstractions for distributed
  graph processing}.
\newblock \bibinfo{journal}{\emph{IEEE Transactions on Knowledge and Data
  Engineering}} \bibinfo{volume}{30}, \bibinfo{number}{2}
  (\bibinfo{year}{2017}), \bibinfo{pages}{305--324}.
\newblock


\bibitem[\protect\citeauthoryear{Kalinsky, Kimelfeld, and Etsion}{Kalinsky
  et~al\mbox{.}}{2020}]%
        {kalinsky2020triejax}
\bibfield{author}{\bibinfo{person}{Oren Kalinsky}, \bibinfo{person}{Benny
  Kimelfeld}, {and} \bibinfo{person}{Yoav Etsion}.}
  \bibinfo{year}{2020}\natexlab{}.
\newblock \showarticletitle{The TrieJax Architecture: Accelerating Graph
  Operations Through Relational Joins}. In
  \bibinfo{booktitle}{\emph{Proceedings of the Twenty-Fifth International
  Conference on Architectural Support for Programming Languages and Operating
  Systems}}. \bibinfo{pages}{1217--1231}.
\newblock


\bibitem[\protect\citeauthoryear{Kapre}{Kapre}{2015}]%
        {kapre:custom_graph_FPGA}
\bibfield{author}{\bibinfo{person}{N. Kapre}.} \bibinfo{year}{2015}\natexlab{}.
\newblock \showarticletitle{Custom FPGA-based soft-processors for sparse graph
  acceleration}. In \bibinfo{booktitle}{\emph{2015 IEEE 26th International
  Conference on Application-specific Systems, Architectures and Processors
  (ASAP)}}. \bibinfo{pages}{9--16}.
\newblock
\showISSN{1063-6862}
\urldef\tempurl%
\url{https://doi.org/10.1109/ASAP.2015.7245698}
\showDOI{\tempurl}


\bibitem[\protect\citeauthoryear{Kapre, Mehta, Rizzo, Eslick, Rubin, Uribe,
  Thomas~Jr, and DeHon}{Kapre et~al\mbox{.}}{2006}]%
        {kapre2006graphstep}
\bibfield{author}{\bibinfo{person}{Nachiket Kapre}, \bibinfo{person}{Nikil
  Mehta}, \bibinfo{person}{Dominic Rizzo}, \bibinfo{person}{Ian Eslick},
  \bibinfo{person}{Raphael Rubin}, \bibinfo{person}{Tomas~E Uribe},
  \bibinfo{person}{F Thomas~Jr}, {and} \bibinfo{person}{Andre DeHon}.}
  \bibinfo{year}{2006}\natexlab{}.
\newblock \showarticletitle{GraphStep: A system architecture for sparse-graph
  algorithms}. In \bibinfo{booktitle}{\emph{Field-Programmable Custom Computing
  Machines, 2006. FCCM'06. 14th Annual IEEE Symposium on}}. IEEE,
  \bibinfo{pages}{143--151}.
\newblock


\bibitem[\protect\citeauthoryear{Kennedy and Schwartz}{Kennedy and
  Schwartz}{1975}]%
        {kennedy1975introduction}
\bibfield{author}{\bibinfo{person}{K Kennedy} {and} \bibinfo{person}{J
  Schwartz}.} \bibinfo{year}{1975}\natexlab{}.
\newblock \bibinfo{title}{An introduction to the set theoretical language
  SETL}.
\newblock
\newblock


\bibitem[\protect\citeauthoryear{Kepner, Aaltonen, Bader, Bulu{\c{c}},
  Franchetti, Gilbert, Hutchison, Kumar, Lumsdaine, and Meyerhenke}{Kepner
  et~al\mbox{.}}{2016}]%
        {kepner2016mathematical}
\bibfield{author}{\bibinfo{person}{Jeremy Kepner}, \bibinfo{person}{Peter
  Aaltonen}, \bibinfo{person}{David Bader}, \bibinfo{person}{Aydin
  Bulu{\c{c}}}, \bibinfo{person}{Franz Franchetti}, \bibinfo{person}{John
  Gilbert}, \bibinfo{person}{Dylan Hutchison}, \bibinfo{person}{Manoj Kumar},
  \bibinfo{person}{Andrew Lumsdaine}, {and} \bibinfo{person}{Henning
  Meyerhenke}.} \bibinfo{year}{2016}\natexlab{}.
\newblock \showarticletitle{Mathematical foundations of the GraphBLAS}. In
  \bibinfo{booktitle}{\emph{High Performance Extreme Computing Conference
  (HPEC), 2016 IEEE}}. IEEE, \bibinfo{pages}{1--9}.
\newblock


\bibitem[\protect\citeauthoryear{Khan}{Khan}{2016}]%
        {khan2016vertex}
\bibfield{author}{\bibinfo{person}{Arijit Khan}.}
  \bibinfo{year}{2016}\natexlab{}.
\newblock \showarticletitle{Vertex-centric graph processing: The good, the bad,
  and the ugly}.
\newblock \bibinfo{journal}{\emph{arXiv preprint arXiv:1612.07404}}
  (\bibinfo{year}{2016}).
\newblock


\bibitem[\protect\citeauthoryear{Khaouid, Barsky, Srinivasan, and
  Thomo}{Khaouid et~al\mbox{.}}{2015}]%
        {khaouid2015k}
\bibfield{author}{\bibinfo{person}{Wissam Khaouid}, \bibinfo{person}{Marina
  Barsky}, \bibinfo{person}{Venkatesh Srinivasan}, {and} \bibinfo{person}{Alex
  Thomo}.} \bibinfo{year}{2015}\natexlab{}.
\newblock \showarticletitle{K-core decomposition of large networks on a single
  PC}.
\newblock \bibinfo{journal}{\emph{Proceedings of the VLDB Endowment}}
  \bibinfo{volume}{9}, \bibinfo{number}{1} (\bibinfo{year}{2015}),
  \bibinfo{pages}{13--23}.
\newblock


\bibitem[\protect\citeauthoryear{Khoram, Zhang, Strange, and Li}{Khoram
  et~al\mbox{.}}{2018}]%
        {khoram2018accelerating}
\bibfield{author}{\bibinfo{person}{Soroosh Khoram}, \bibinfo{person}{Jialiang
  Zhang}, \bibinfo{person}{Maxwell Strange}, {and} \bibinfo{person}{Jing Li}.}
  \bibinfo{year}{2018}\natexlab{}.
\newblock \showarticletitle{Accelerating Graph Analytics by Co-Optimizing
  Storage and Access on an FPGA-HMC Platform}. In
  \bibinfo{booktitle}{\emph{Proceedings of the 2018 ACM/SIGDA International
  Symposium on Field-Programmable Gate Arrays}}. ACM,
  \bibinfo{pages}{239--248}.
\newblock


\bibitem[\protect\citeauthoryear{Ko and Han}{Ko and Han}{2018}]%
        {ko2018turbograph++}
\bibfield{author}{\bibinfo{person}{Seongyun Ko} {and}
  \bibinfo{person}{Wook-Shin Han}.} \bibinfo{year}{2018}\natexlab{}.
\newblock \showarticletitle{Turbograph++: A scalable and fast graph analytics
  system}. In \bibinfo{booktitle}{\emph{Proceedings of the 2018 International
  Conference on Management of Data}}. ACM, \bibinfo{pages}{395--410}.
\newblock


\bibitem[\protect\citeauthoryear{Kuramochi and Karypis}{Kuramochi and
  Karypis}{2001}]%
        {kuramochi2001frequent}
\bibfield{author}{\bibinfo{person}{Michihiro Kuramochi} {and}
  \bibinfo{person}{George Karypis}.} \bibinfo{year}{2001}\natexlab{}.
\newblock \showarticletitle{Frequent subgraph discovery}. In
  \bibinfo{booktitle}{\emph{Proceedings 2001 IEEE international conference on
  data mining}}. IEEE, \bibinfo{pages}{313--320}.
\newblock


\bibitem[\protect\citeauthoryear{Kuramochi and Karypis}{Kuramochi and
  Karypis}{2004}]%
        {kuramochi2004efficient}
\bibfield{author}{\bibinfo{person}{Michihiro Kuramochi} {and}
  \bibinfo{person}{George Karypis}.} \bibinfo{year}{2004}\natexlab{}.
\newblock \showarticletitle{An efficient algorithm for discovering frequent
  subgraphs}.
\newblock \bibinfo{journal}{\emph{IEEE transactions on Knowledge and Data
  Engineering}} \bibinfo{volume}{16}, \bibinfo{number}{9}
  (\bibinfo{year}{2004}), \bibinfo{pages}{1038--1051}.
\newblock


\bibitem[\protect\citeauthoryear{Lavenier, Roy, and Furodet}{Lavenier
  et~al\mbox{.}}{2016}]%
        {lavenier2016dna}
\bibfield{author}{\bibinfo{person}{Dominique Lavenier},
  \bibinfo{person}{Jean-Francois Roy}, {and} \bibinfo{person}{David Furodet}.}
  \bibinfo{year}{2016}\natexlab{}.
\newblock \showarticletitle{DNA mapping using Processor-in-Memory
  architecture}. In \bibinfo{booktitle}{\emph{2016 IEEE International
  Conference on Bioinformatics and Biomedicine (BIBM)}}. IEEE,
  \bibinfo{pages}{1429--1435}.
\newblock


\bibitem[\protect\citeauthoryear{Lee, Kim, Yoo, Choi, Hofstee, Nam, Nutter, and
  Jamsek}{Lee et~al\mbox{.}}{2017}]%
        {lee2017extrav}
\bibfield{author}{\bibinfo{person}{Jinho Lee}, \bibinfo{person}{Heesu Kim},
  \bibinfo{person}{Sungjoo Yoo}, \bibinfo{person}{Kiyoung Choi},
  \bibinfo{person}{H~Peter Hofstee}, \bibinfo{person}{Gi-Joon Nam},
  \bibinfo{person}{Mark~R Nutter}, {and} \bibinfo{person}{Damir Jamsek}.}
  \bibinfo{year}{2017}\natexlab{}.
\newblock \showarticletitle{ExtraV: boosting graph processing near storage with
  a coherent accelerator}.
\newblock \bibinfo{journal}{\emph{Proceedings of the VLDB Endowment}}
  \bibinfo{volume}{10}, \bibinfo{number}{12} (\bibinfo{year}{2017}),
  \bibinfo{pages}{1706--1717}.
\newblock


\bibitem[\protect\citeauthoryear{Lee, Ruan, Jin, and Aggarwal}{Lee
  et~al\mbox{.}}{2010}]%
        {lee2010survey}
\bibfield{author}{\bibinfo{person}{Victor~E Lee}, \bibinfo{person}{Ning Ruan},
  \bibinfo{person}{Ruoming Jin}, {and} \bibinfo{person}{Charu Aggarwal}.}
  \bibinfo{year}{2010}\natexlab{}.
\newblock \showarticletitle{A survey of algorithms for dense subgraph
  discovery}.
\newblock In \bibinfo{booktitle}{\emph{Managing and Mining Graph Data}}.
  \bibinfo{publisher}{Springer}, \bibinfo{pages}{303--336}.
\newblock


\bibitem[\protect\citeauthoryear{Lei, Dou, Li, and Xia}{Lei
  et~al\mbox{.}}{2016}]%
        {lei2016fpga}
\bibfield{author}{\bibinfo{person}{Guoqing Lei}, \bibinfo{person}{Yong Dou},
  \bibinfo{person}{Rongchun Li}, {and} \bibinfo{person}{Fei Xia}.}
  \bibinfo{year}{2016}\natexlab{}.
\newblock \showarticletitle{An fpga implementation for solving the large
  single-source-shortest-path problem}.
\newblock \bibinfo{journal}{\emph{IEEE Transactions on Circuits and Systems II:
  Express Briefs}} \bibinfo{volume}{63}, \bibinfo{number}{5}
  (\bibinfo{year}{2016}), \bibinfo{pages}{473--477}.
\newblock


\bibitem[\protect\citeauthoryear{Leicht, Holme, and Newman}{Leicht
  et~al\mbox{.}}{2006}]%
        {leicht2006vertex}
\bibfield{author}{\bibinfo{person}{Elizabeth~A Leicht}, \bibinfo{person}{Petter
  Holme}, {and} \bibinfo{person}{Mark~EJ Newman}.}
  \bibinfo{year}{2006}\natexlab{}.
\newblock \showarticletitle{Vertex similarity in networks}.
\newblock \bibinfo{journal}{\emph{Physical Review E}} \bibinfo{volume}{73},
  \bibinfo{number}{2} (\bibinfo{year}{2006}), \bibinfo{pages}{026120}.
\newblock


\bibitem[\protect\citeauthoryear{Leiserson and Schardl}{Leiserson and
  Schardl}{2010}]%
        {leiserson2010work}
\bibfield{author}{\bibinfo{person}{Charles~E Leiserson} {and}
  \bibinfo{person}{Tao~B Schardl}.} \bibinfo{year}{2010}\natexlab{}.
\newblock \showarticletitle{A work-efficient parallel breadth-first search
  algorithm (or how to cope with the nondeterminism of reducers)}. In
  \bibinfo{booktitle}{\emph{Proceedings of the twenty-second annual ACM
  symposium on Parallelism in algorithms and architectures}}. ACM,
  \bibinfo{pages}{303--314}.
\newblock


\bibitem[\protect\citeauthoryear{Leskovec, Chakrabarti, Kleinberg, Faloutsos,
  and Ghahramani}{Leskovec et~al\mbox{.}}{2010}]%
        {leskovec2010kronecker}
\bibfield{author}{\bibinfo{person}{Jure Leskovec}, \bibinfo{person}{Deepayan
  Chakrabarti}, \bibinfo{person}{Jon Kleinberg}, \bibinfo{person}{Christos
  Faloutsos}, {and} \bibinfo{person}{Zoubin Ghahramani}.}
  \bibinfo{year}{2010}\natexlab{}.
\newblock \showarticletitle{Kronecker graphs: An approach to modeling
  networks}.
\newblock \bibinfo{journal}{\emph{Journal of Machine Learning Research}}
  \bibinfo{volume}{11}, \bibinfo{number}{Feb} (\bibinfo{year}{2010}),
  \bibinfo{pages}{985--1042}.
\newblock


\bibitem[\protect\citeauthoryear{Li, Dai, Li, Wang, and Xie}{Li
  et~al\mbox{.}}{2018}]%
        {li2018graphia}
\bibfield{author}{\bibinfo{person}{Gushu Li}, \bibinfo{person}{Guohao Dai},
  \bibinfo{person}{Shuangchen Li}, \bibinfo{person}{Yu Wang}, {and}
  \bibinfo{person}{Yuan Xie}.} \bibinfo{year}{2018}\natexlab{}.
\newblock \showarticletitle{GraphIA: an in-situ accelerator for large-scale
  graph processing}. In \bibinfo{booktitle}{\emph{Proceedings of the
  International Symposium on Memory Systems}}. \bibinfo{pages}{79--84}.
\newblock


\bibitem[\protect\citeauthoryear{Li, Niu, Malladi, Zheng, Brennan, and Xie}{Li
  et~al\mbox{.}}{2017}]%
        {li2017drisa}
\bibfield{author}{\bibinfo{person}{Shuangchen Li}, \bibinfo{person}{Dimin Niu},
  \bibinfo{person}{Krishna~T Malladi}, \bibinfo{person}{Hongzhong Zheng},
  \bibinfo{person}{Bob Brennan}, {and} \bibinfo{person}{Yuan Xie}.}
  \bibinfo{year}{2017}\natexlab{}.
\newblock \showarticletitle{Drisa: A dram-based reconfigurable in-situ
  accelerator}. In \bibinfo{booktitle}{\emph{2017 50th Annual IEEE/ACM
  International Symposium on Microarchitecture (MICRO)}}. IEEE,
  \bibinfo{pages}{288--301}.
\newblock


\bibitem[\protect\citeauthoryear{Li, Xu, Zou, Zhao, Lu, and Xie}{Li
  et~al\mbox{.}}{2016}]%
        {li2016pinatubo}
\bibfield{author}{\bibinfo{person}{Shuangchen Li}, \bibinfo{person}{Cong Xu},
  \bibinfo{person}{Qiaosha Zou}, \bibinfo{person}{Jishen Zhao},
  \bibinfo{person}{Yu Lu}, {and} \bibinfo{person}{Yuan Xie}.}
  \bibinfo{year}{2016}\natexlab{}.
\newblock \showarticletitle{Pinatubo: A processing-in-memory architecture for
  bulk bitwise operations in emerging non-volatile memories}. In
  \bibinfo{booktitle}{\emph{Proceedings of the 53rd Annual Design Automation
  Conference}}. \bibinfo{pages}{1--6}.
\newblock


\bibitem[\protect\citeauthoryear{Liben-Nowell and Kleinberg}{Liben-Nowell and
  Kleinberg}{2007}]%
        {liben2007link}
\bibfield{author}{\bibinfo{person}{David Liben-Nowell} {and}
  \bibinfo{person}{Jon Kleinberg}.} \bibinfo{year}{2007}\natexlab{}.
\newblock \showarticletitle{The link-prediction problem for social networks}.
\newblock \bibinfo{journal}{\emph{Journal of the American society for
  information science and technology}} \bibinfo{volume}{58},
  \bibinfo{number}{7} (\bibinfo{year}{2007}), \bibinfo{pages}{1019--1031}.
\newblock


\bibitem[\protect\citeauthoryear{Lick and White}{Lick and White}{1970}]%
        {lick_white_1970}
\bibfield{author}{\bibinfo{person}{Don~R. Lick} {and}
  \bibinfo{person}{Arthur~T. White}.} \bibinfo{year}{1970}\natexlab{}.
\newblock \showarticletitle{k-Degenerate Graphs}.
\newblock \bibinfo{journal}{\emph{Canadian Journal of Mathematics}}
  \bibinfo{volume}{22}, \bibinfo{number}{5} (\bibinfo{year}{1970}),
  \bibinfo{pages}{1082–1096}.
\newblock
\urldef\tempurl%
\url{https://doi.org/10.4153/CJM-1970-125-1}
\showDOI{\tempurl}


\bibitem[\protect\citeauthoryear{Liu, Hua, Jin, and Zheng}{Liu
  et~al\mbox{.}}{2020}]%
        {liu2020regra}
\bibfield{author}{\bibinfo{person}{Haoqiang Liu}, \bibinfo{person}{Qiang-Sheng
  Hua}, \bibinfo{person}{Hai Jin}, {and} \bibinfo{person}{Long Zheng}.}
  \bibinfo{year}{2020}\natexlab{}.
\newblock \showarticletitle{ReGra: Accelerating Graph Traversal Applications
  Using ReRAM with Lower Communication Cost}.
\newblock \bibinfo{journal}{\emph{IEEE Access}} (\bibinfo{year}{2020}).
\newblock


\bibitem[\protect\citeauthoryear{Liu, Du, Tao, Han, Luo, Xie, Chen, and
  Chen}{Liu et~al\mbox{.}}{2016}]%
        {liu2016cambricon}
\bibfield{author}{\bibinfo{person}{Shaoli Liu}, \bibinfo{person}{Zidong Du},
  \bibinfo{person}{Jinhua Tao}, \bibinfo{person}{Dong Han},
  \bibinfo{person}{Tao Luo}, \bibinfo{person}{Yuan Xie}, \bibinfo{person}{Yunji
  Chen}, {and} \bibinfo{person}{Tianshi Chen}.}
  \bibinfo{year}{2016}\natexlab{}.
\newblock \showarticletitle{Cambricon: An instruction set architecture for
  neural networks}. In \bibinfo{booktitle}{\emph{ACM SIGARCH Computer
  Architecture News}}, Vol.~\bibinfo{volume}{44}. IEEE Press,
  \bibinfo{pages}{393--405}.
\newblock


\bibitem[\protect\citeauthoryear{Liu and Khan}{Liu and Khan}{2018}]%
        {liu2018empirical}
\bibfield{author}{\bibinfo{person}{Siyuan Liu} {and} \bibinfo{person}{Arijit
  Khan}.} \bibinfo{year}{2018}\natexlab{}.
\newblock \showarticletitle{An Empirical Analysis on Expressibility of Vertex
  Centric Graph Processing Paradigm}. In \bibinfo{booktitle}{\emph{2018 IEEE
  International Conference on Big Data (Big Data)}}. IEEE,
  \bibinfo{pages}{242--251}.
\newblock


\bibitem[\protect\citeauthoryear{Liu, Safavi, Dighe, and Koutra}{Liu
  et~al\mbox{.}}{2018}]%
        {liu2018graph}
\bibfield{author}{\bibinfo{person}{Yike Liu}, \bibinfo{person}{Tara Safavi},
  \bibinfo{person}{Abhilash Dighe}, {and} \bibinfo{person}{Danai Koutra}.}
  \bibinfo{year}{2018}\natexlab{}.
\newblock \showarticletitle{Graph summarization methods and applications: A
  survey}.
\newblock \bibinfo{journal}{\emph{ACM Computing Surveys (CSUR)}}
  \bibinfo{volume}{51}, \bibinfo{number}{3} (\bibinfo{year}{2018}),
  \bibinfo{pages}{1--34}.
\newblock


\bibitem[\protect\citeauthoryear{Loh}{Loh}{2008}]%
        {loh20083d}
\bibfield{author}{\bibinfo{person}{Gabriel~H Loh}.}
  \bibinfo{year}{2008}\natexlab{}.
\newblock \showarticletitle{3D-stacked memory architectures for multi-core
  processors}. In \bibinfo{booktitle}{\emph{ACM SIGARCH computer architecture
  news}}, Vol.~\bibinfo{volume}{36}. IEEE Computer Society,
  \bibinfo{pages}{453--464}.
\newblock


\bibitem[\protect\citeauthoryear{Low et~al\mbox{.}}{Low et~al\mbox{.}}{2010}]%
        {low2010graphlab}
\bibfield{author}{\bibinfo{person}{Yucheng Low} {et~al\mbox{.}}}
  \bibinfo{year}{2010}\natexlab{}.
\newblock \showarticletitle{{Graphlab: A new framework for parallel machine
  learning}}.
\newblock \bibinfo{journal}{\emph{preprint arXiv:1006.4990}}
  (\bibinfo{year}{2010}).
\newblock


\bibitem[\protect\citeauthoryear{L{\"u} and Zhou}{L{\"u} and Zhou}{2011}]%
        {lu2011link}
\bibfield{author}{\bibinfo{person}{Linyuan L{\"u}} {and} \bibinfo{person}{Tao
  Zhou}.} \bibinfo{year}{2011}\natexlab{}.
\newblock \showarticletitle{Link prediction in complex networks: A survey}.
\newblock \bibinfo{journal}{\emph{Physica A: statistical mechanics and its
  applications}} \bibinfo{volume}{390}, \bibinfo{number}{6}
  (\bibinfo{year}{2011}), \bibinfo{pages}{1150--1170}.
\newblock


\bibitem[\protect\citeauthoryear{Luk, Cohn, Muth, Patil, Klauser, Lowney,
  Wallace, Reddi, and Hazelwood}{Luk et~al\mbox{.}}{2005}]%
        {luk2005pin}
\bibfield{author}{\bibinfo{person}{Chi-Keung Luk}, \bibinfo{person}{Robert
  Cohn}, \bibinfo{person}{Robert Muth}, \bibinfo{person}{Harish Patil},
  \bibinfo{person}{Artur Klauser}, \bibinfo{person}{Geoff Lowney},
  \bibinfo{person}{Steven Wallace}, \bibinfo{person}{Vijay~Janapa Reddi}, {and}
  \bibinfo{person}{Kim Hazelwood}.} \bibinfo{year}{2005}\natexlab{}.
\newblock \showarticletitle{Pin: building customized program analysis tools
  with dynamic instrumentation}.
\newblock \bibinfo{journal}{\emph{Acm sigplan notices}} \bibinfo{volume}{40},
  \bibinfo{number}{6} (\bibinfo{year}{2005}), \bibinfo{pages}{190--200}.
\newblock


\bibitem[\protect\citeauthoryear{Lumsdaine, Gregor, Hendrickson, and
  Berry}{Lumsdaine et~al\mbox{.}}{2007}]%
        {DBLP:journals/ppl/LumsdaineGHB07}
\bibfield{author}{\bibinfo{person}{Andrew Lumsdaine}, \bibinfo{person}{Douglas
  Gregor}, \bibinfo{person}{Bruce Hendrickson}, {and}
  \bibinfo{person}{Jonathan~W. Berry}.} \bibinfo{year}{2007}\natexlab{}.
\newblock \showarticletitle{{Challenges in Parallel Graph Processing}}.
\newblock \bibinfo{journal}{\emph{Par. Proc. Let.}} \bibinfo{volume}{17},
  \bibinfo{number}{1} (\bibinfo{year}{2007}), \bibinfo{pages}{5--20}.
\newblock


\bibitem[\protect\citeauthoryear{Ma, Zhang, and Chiou}{Ma
  et~al\mbox{.}}{2017}]%
        {ma2017fpga}
\bibfield{author}{\bibinfo{person}{Xiaoyu Ma}, \bibinfo{person}{Dan Zhang},
  {and} \bibinfo{person}{Derek Chiou}.} \bibinfo{year}{2017}\natexlab{}.
\newblock \showarticletitle{FPGA-accelerated transactional execution of graph
  workloads}. In \bibinfo{booktitle}{\emph{Proceedings of the 2017 ACM/SIGDA
  International Symposium on Field-Programmable Gate Arrays}}. ACM,
  \bibinfo{pages}{227--236}.
\newblock


\bibitem[\protect\citeauthoryear{Malewicz, Austern, Bik, Dehnert, Horn, Leiser,
  and Czajkowski}{Malewicz et~al\mbox{.}}{2010}]%
        {malewicz2010pregel}
\bibfield{author}{\bibinfo{person}{Grzegorz Malewicz},
  \bibinfo{person}{Matthew~H Austern}, \bibinfo{person}{Aart~JC Bik},
  \bibinfo{person}{James~C Dehnert}, \bibinfo{person}{Ilan Horn},
  \bibinfo{person}{Naty Leiser}, {and} \bibinfo{person}{Grzegorz Czajkowski}.}
  \bibinfo{year}{2010}\natexlab{}.
\newblock \showarticletitle{Pregel: a system for large-scale graph processing}.
  In \bibinfo{booktitle}{\emph{Proceedings of the 2010 ACM SIGMOD International
  Conference on Management of data}}. ACM, \bibinfo{pages}{135--146}.
\newblock


\bibitem[\protect\citeauthoryear{Malicevic, Lepers, and Zwaenepoel}{Malicevic
  et~al\mbox{.}}{2017}]%
        {malicevic2017everything}
\bibfield{author}{\bibinfo{person}{Jasmina Malicevic},
  \bibinfo{person}{Baptiste Lepers}, {and} \bibinfo{person}{Willy Zwaenepoel}.}
  \bibinfo{year}{2017}\natexlab{}.
\newblock \showarticletitle{Everything you always wanted to know about
  multicore graph processing but were afraid to ask}. In
  \bibinfo{booktitle}{\emph{2017 USENIX Annual Technical Conference (USENIX
  ATC'17)}}. \bibinfo{pages}{631--643}.
\newblock


\bibitem[\protect\citeauthoryear{Matam, Koo, Zha, Tseng, and Annavaram}{Matam
  et~al\mbox{.}}{2019}]%
        {matam2019graphssd}
\bibfield{author}{\bibinfo{person}{Kiran~Kumar Matam}, \bibinfo{person}{Gunjae
  Koo}, \bibinfo{person}{Haipeng Zha}, \bibinfo{person}{Hung-Wei Tseng}, {and}
  \bibinfo{person}{Murali Annavaram}.} \bibinfo{year}{2019}\natexlab{}.
\newblock \showarticletitle{GraphSSD: graph semantics aware SSD}. In
  \bibinfo{booktitle}{\emph{Proceedings of the 46th International Symposium on
  Computer Architecture}}. \bibinfo{pages}{116--128}.
\newblock


\bibitem[\protect\citeauthoryear{Matula and Beck}{Matula and Beck}{1983}]%
        {matula1983smallest}
\bibfield{author}{\bibinfo{person}{David~W Matula} {and}
  \bibinfo{person}{Leland~L Beck}.} \bibinfo{year}{1983}\natexlab{}.
\newblock \showarticletitle{Smallest-last ordering and clustering and graph
  coloring algorithms}.
\newblock \bibinfo{journal}{\emph{JACM}} (\bibinfo{year}{1983}).
\newblock


\bibitem[\protect\citeauthoryear{Mawhirter, Reinehr, Holmes, Liu, and
  Wu}{Mawhirter et~al\mbox{.}}{2019}]%
        {mawhirter2019graphzero}
\bibfield{author}{\bibinfo{person}{Daniel Mawhirter}, \bibinfo{person}{Sam
  Reinehr}, \bibinfo{person}{Connor Holmes}, \bibinfo{person}{Tongping Liu},
  {and} \bibinfo{person}{Bo Wu}.} \bibinfo{year}{2019}\natexlab{}.
\newblock \showarticletitle{GraphZero: Breaking Symmetry for Efficient Graph
  Mining}.
\newblock \bibinfo{journal}{\emph{arXiv preprint arXiv:1911.12877}}
  (\bibinfo{year}{2019}).
\newblock


\bibitem[\protect\citeauthoryear{Mawhirter and Wu}{Mawhirter and Wu}{2019}]%
        {mawhirter2019automine}
\bibfield{author}{\bibinfo{person}{Daniel Mawhirter} {and} \bibinfo{person}{Bo
  Wu}.} \bibinfo{year}{2019}\natexlab{}.
\newblock \showarticletitle{AutoMine: harmonizing high-level abstraction and
  high performance for graph mining}. In \bibinfo{booktitle}{\emph{Proceedings
  of the 27th ACM Symposium on Operating Systems Principles}}. ACM,
  \bibinfo{pages}{509--523}.
\newblock


\bibitem[\protect\citeauthoryear{McCune, Weninger, and Madey}{McCune
  et~al\mbox{.}}{2015}]%
        {mccune2015thinking}
\bibfield{author}{\bibinfo{person}{Robert~Ryan McCune}, \bibinfo{person}{Tim
  Weninger}, {and} \bibinfo{person}{Greg Madey}.}
  \bibinfo{year}{2015}\natexlab{}.
\newblock \showarticletitle{Thinking like a vertex: a survey of vertex-centric
  frameworks for large-scale distributed graph processing}.
\newblock \bibinfo{journal}{\emph{ACM Computing Surveys (CSUR)}}
  \bibinfo{volume}{48}, \bibinfo{number}{2} (\bibinfo{year}{2015}),
  \bibinfo{pages}{25}.
\newblock


\bibitem[\protect\citeauthoryear{Mencer, Huang, and Huelsbergen}{Mencer
  et~al\mbox{.}}{2002}]%
        {mencer2002hagar}
\bibfield{author}{\bibinfo{person}{Oskar Mencer}, \bibinfo{person}{Zhining
  Huang}, {and} \bibinfo{person}{Lorenz Huelsbergen}.}
  \bibinfo{year}{2002}\natexlab{}.
\newblock \showarticletitle{HAGAR: Efficient multi-context graph processors}.
  In \bibinfo{booktitle}{\emph{International Conference on Field Programmable
  Logic and Applications}}. Springer, \bibinfo{pages}{915--924}.
\newblock


\bibitem[\protect\citeauthoryear{Meyer and Sanders}{Meyer and Sanders}{2003}]%
        {meyer2003delta}
\bibfield{author}{\bibinfo{person}{Ulrich Meyer} {and} \bibinfo{person}{Peter
  Sanders}.} \bibinfo{year}{2003}\natexlab{}.
\newblock \showarticletitle{{$\Delta$-stepping: a parallelizable shortest path
  algorithm}}.
\newblock \bibinfo{journal}{\emph{Journal of Algorithms}} \bibinfo{volume}{49},
  \bibinfo{number}{1} (\bibinfo{year}{2003}), \bibinfo{pages}{114--152}.
\newblock


\bibitem[\protect\citeauthoryear{Miller, Peng, Vladu, and Xu}{Miller
  et~al\mbox{.}}{2015}]%
        {miller2015improved}
\bibfield{author}{\bibinfo{person}{Gary~L Miller}, \bibinfo{person}{Richard
  Peng}, \bibinfo{person}{Adrian Vladu}, {and} \bibinfo{person}{Shen~Chen Xu}.}
  \bibinfo{year}{2015}\natexlab{}.
\newblock \showarticletitle{Improved parallel algorithms for spanners and
  hopsets}. In \bibinfo{booktitle}{\emph{Proceedings of the 27th ACM Symposium
  on Parallelism in Algorithms and Architectures}}. ACM,
  \bibinfo{pages}{192--201}.
\newblock


\bibitem[\protect\citeauthoryear{Mittal, Vetter, and Li}{Mittal
  et~al\mbox{.}}{2014}]%
        {mittal2014improving}
\bibfield{author}{\bibinfo{person}{Sparsh Mittal}, \bibinfo{person}{Jeffrey~S
  Vetter}, {and} \bibinfo{person}{Dong Li}.} \bibinfo{year}{2014}\natexlab{}.
\newblock \showarticletitle{Improving energy efficiency of embedded DRAM caches
  for high-end computing systems}. In \bibinfo{booktitle}{\emph{Proceedings of
  the 23rd international symposium on High-performance parallel and distributed
  computing}}. \bibinfo{pages}{99--110}.
\newblock


\bibitem[\protect\citeauthoryear{Mosayebi, Hasani, and Dehyadegari}{Mosayebi
  et~al\mbox{.}}{2019}]%
        {mosayebi2019enhanced}
\bibfield{author}{\bibinfo{person}{Mohammad~Amin Mosayebi},
  \bibinfo{person}{Arghavan~Mohammad Hasani}, {and} \bibinfo{person}{Masoud
  Dehyadegari}.} \bibinfo{year}{2019}\natexlab{}.
\newblock \showarticletitle{Enhanced graph processing in {PIM} accelerators
  with improved queue management}.
\newblock \bibinfo{journal}{\emph{Microelectronics Journal}}
  \bibinfo{volume}{94} (\bibinfo{year}{2019}), \bibinfo{pages}{104637}.
\newblock


\bibitem[\protect\citeauthoryear{Mukkara, Beckmann, Abeydeera, Ma, and
  Sanchez}{Mukkara et~al\mbox{.}}{2018}]%
        {mukkara2018exploiting}
\bibfield{author}{\bibinfo{person}{Anurag Mukkara}, \bibinfo{person}{Nathan
  Beckmann}, \bibinfo{person}{Maleen Abeydeera}, \bibinfo{person}{Xiaosong Ma},
  {and} \bibinfo{person}{Daniel Sanchez}.} \bibinfo{year}{2018}\natexlab{}.
\newblock \showarticletitle{Exploiting locality in graph analytics through
  hardware-accelerated traversal scheduling}. In \bibinfo{booktitle}{\emph{2018
  51st Annual IEEE/ACM International Symposium on Microarchitecture (MICRO)}}.
  IEEE, \bibinfo{pages}{1--14}.
\newblock


\bibitem[\protect\citeauthoryear{Murphy et~al\mbox{.}}{Murphy
  et~al\mbox{.}}{2010}]%
        {murphy2010introducing}
\bibfield{author}{\bibinfo{person}{Richard~C Murphy} {et~al\mbox{.}}}
  \bibinfo{year}{2010}\natexlab{}.
\newblock \showarticletitle{Introducing the graph 500}.
\newblock \bibinfo{journal}{\emph{Cray User’s Group (CUG)}}
  (\bibinfo{year}{2010}).
\newblock


\bibitem[\protect\citeauthoryear{Mutlu et~al\mbox{.}}{Mutlu
  et~al\mbox{.}}{2019}]%
        {mutlu2019}
\bibfield{author}{\bibinfo{person}{O. Mutlu} {et~al\mbox{.}}}
  \bibinfo{year}{2019}\natexlab{}.
\newblock \showarticletitle{{Processing Data Where It Makes Sense: {E}nabling
  In-Memory Computation}}.
\newblock \bibinfo{journal}{\emph{MicPro}} (\bibinfo{year}{2019}).
\newblock


\bibitem[\protect\citeauthoryear{Mutlu, Ghose, G{\'o}mez-Luna, and
  Ausavarungnirun}{Mutlu et~al\mbox{.}}{2020}]%
        {mutlu2020modern}
\bibfield{author}{\bibinfo{person}{Onur Mutlu}, \bibinfo{person}{Saugata
  Ghose}, \bibinfo{person}{Juan G{\'o}mez-Luna}, {and} \bibinfo{person}{Rachata
  Ausavarungnirun}.} \bibinfo{year}{2020}\natexlab{}.
\newblock \showarticletitle{A Modern Primer on Processing in Memory}.
\newblock \bibinfo{journal}{\emph{arXiv preprint arXiv:2012.03112}}
  (\bibinfo{year}{2020}).
\newblock


\bibitem[\protect\citeauthoryear{Nag, Ramachandra, Balasubramonian, Stutsman,
  Giacomin, Kambalasubramanyam, and Gaillardon}{Nag et~al\mbox{.}}{2019}]%
        {nag2019gencache}
\bibfield{author}{\bibinfo{person}{Anirban Nag}, \bibinfo{person}{CN
  Ramachandra}, \bibinfo{person}{Rajeev Balasubramonian}, \bibinfo{person}{Ryan
  Stutsman}, \bibinfo{person}{Edouard Giacomin}, \bibinfo{person}{Hari
  Kambalasubramanyam}, {and} \bibinfo{person}{Pierre-Emmanuel Gaillardon}.}
  \bibinfo{year}{2019}\natexlab{}.
\newblock \showarticletitle{Gencache: Leveraging in-cache operators for
  efficient sequence alignment}. In \bibinfo{booktitle}{\emph{Proceedings of
  the 52nd Annual IEEE/ACM International Symposium on Microarchitecture}}.
  \bibinfo{pages}{334--346}.
\newblock


\bibitem[\protect\citeauthoryear{Nai, Hadidi, Sim, Kim, Kumar, and Kim}{Nai
  et~al\mbox{.}}{2017}]%
        {nai2017graphpim}
\bibfield{author}{\bibinfo{person}{Lifeng Nai}, \bibinfo{person}{Ramyad
  Hadidi}, \bibinfo{person}{Jaewoong Sim}, \bibinfo{person}{Hyojong Kim},
  \bibinfo{person}{Pranith Kumar}, {and} \bibinfo{person}{Hyesoon Kim}.}
  \bibinfo{year}{2017}\natexlab{}.
\newblock \showarticletitle{Graphpim: Enabling instruction-level {PIM}
  offloading in graph computing frameworks}. In \bibinfo{booktitle}{\emph{High
  Performance Computer Architecture (HPCA), 2017 IEEE International Symposium
  on}}. IEEE, \bibinfo{pages}{457--468}.
\newblock


\bibitem[\protect\citeauthoryear{{Neo4j, Inc.}}{{Neo4j, Inc.}}{2019}]%
        {neo4j_sim}
\bibfield{author}{\bibinfo{person}{{Neo4j, Inc.}}}
  \bibinfo{year}{2019}\natexlab{}.
\newblock \bibinfo{title}{The Neo4j Graph Algorithms User Guide v3.5}.
\newblock
  \bibinfo{howpublished}{\url{https://neo4j.com/docs/graph-algorithms/current}}.
\newblock


\bibitem[\protect\citeauthoryear{Ni, Dou, Zou, Li, and Wang}{Ni
  et~al\mbox{.}}{2014}]%
        {ni2014parallel}
\bibfield{author}{\bibinfo{person}{Shice Ni}, \bibinfo{person}{Yong Dou},
  \bibinfo{person}{Dan Zou}, \bibinfo{person}{Rongchun Li}, {and}
  \bibinfo{person}{Qiang Wang}.} \bibinfo{year}{2014}\natexlab{}.
\newblock \showarticletitle{Parallel graph traversal for FPGA}.
\newblock \bibinfo{journal}{\emph{IEICE Electronics Express}}
  \bibinfo{volume}{11}, \bibinfo{number}{7} (\bibinfo{year}{2014}),
  \bibinfo{pages}{20130987--20130987}.
\newblock


\bibitem[\protect\citeauthoryear{Nurvitadhi, Weisz, Wang, Hurkat, Nguyen, Hoe,
  Martínez, and Guestrin}{Nurvitadhi et~al\mbox{.}}{2014}]%
        {weisz:GraphGen}
\bibfield{author}{\bibinfo{person}{E. Nurvitadhi}, \bibinfo{person}{G. Weisz},
  \bibinfo{person}{Y. Wang}, \bibinfo{person}{S. Hurkat}, \bibinfo{person}{M.
  Nguyen}, \bibinfo{person}{J.~C. Hoe}, \bibinfo{person}{J.~F. Martínez},
  {and} \bibinfo{person}{C. Guestrin}.} \bibinfo{year}{2014}\natexlab{}.
\newblock \showarticletitle{GraphGen: An FPGA Framework for Vertex-Centric
  Graph Computation}. In \bibinfo{booktitle}{\emph{2014 IEEE 22nd Annual
  International Symposium on Field-Programmable Custom Computing Machines}}.
  \bibinfo{pages}{25--28}.
\newblock
\urldef\tempurl%
\url{https://doi.org/10.1109/FCCM.2014.15}
\showDOI{\tempurl}


\bibitem[\protect\citeauthoryear{Oguntebi and Olukotun}{Oguntebi and
  Olukotun}{2016}]%
        {oguntebi:GraphOps}
\bibfield{author}{\bibinfo{person}{Tayo Oguntebi} {and} \bibinfo{person}{Kunle
  Olukotun}.} \bibinfo{year}{2016}\natexlab{}.
\newblock \showarticletitle{GraphOps: A Dataflow Library for Graph Analytics
  Acceleration}. In \bibinfo{booktitle}{\emph{Proceedings of the 2016 ACM/SIGDA
  International Symposium on Field-Programmable Gate Arrays}}
  \emph{(\bibinfo{series}{FPGA '16})}. \bibinfo{publisher}{ACM},
  \bibinfo{address}{New York, NY, USA}, \bibinfo{pages}{111--117}.
\newblock
\showISBNx{978-1-4503-3856-1}
\urldef\tempurl%
\url{https://doi.org/10.1145/2847263.2847337}
\showDOI{\tempurl}


\bibitem[\protect\citeauthoryear{Oliveira, G{\'o}mez-Luna, Orosa, Ghose,
  Vijaykumar, Fernandez, Sadrosadati, and Mutlu}{Oliveira
  et~al\mbox{.}}{2021}]%
        {oliveira2021damov}
\bibfield{author}{\bibinfo{person}{Geraldo~F Oliveira}, \bibinfo{person}{Juan
  G{\'o}mez-Luna}, \bibinfo{person}{Lois Orosa}, \bibinfo{person}{Saugata
  Ghose}, \bibinfo{person}{Nandita Vijaykumar}, \bibinfo{person}{Ivan
  Fernandez}, \bibinfo{person}{Mohammad Sadrosadati}, {and}
  \bibinfo{person}{Onur Mutlu}.} \bibinfo{year}{2021}\natexlab{}.
\newblock \showarticletitle{DAMOV: A New Methodology and Benchmark Suite for
  Evaluating Data Movement Bottlenecks}.
\newblock \bibinfo{journal}{\emph{arXiv preprint arXiv:2105.03725}}
  (\bibinfo{year}{2021}).
\newblock


\bibitem[\protect\citeauthoryear{Ozdal, Yesil, Kim, Ayupov, Greth, Burns, and
  Ozturk}{Ozdal et~al\mbox{.}}{2016}]%
        {ozdal2016energy}
\bibfield{author}{\bibinfo{person}{Muhammet~Mustafa Ozdal},
  \bibinfo{person}{Serif Yesil}, \bibinfo{person}{Taemin Kim},
  \bibinfo{person}{Andrey Ayupov}, \bibinfo{person}{John Greth},
  \bibinfo{person}{Steven Burns}, {and} \bibinfo{person}{Ozcan Ozturk}.}
  \bibinfo{year}{2016}\natexlab{}.
\newblock \showarticletitle{Energy efficient architecture for graph analytics
  accelerators}. In \bibinfo{booktitle}{\emph{Computer Architecture (ISCA),
  2016 ACM/IEEE 43rd Annual International Symposium on}}. IEEE,
  \bibinfo{pages}{166--177}.
\newblock


\bibitem[\protect\citeauthoryear{Pal, Beaumont, Park, Amarnath, Feng,
  Chakrabarti, Kim, Blaauw, Mudge, and Dreslinski}{Pal et~al\mbox{.}}{2018}]%
        {pal2018outerspace}
\bibfield{author}{\bibinfo{person}{Subhankar Pal}, \bibinfo{person}{Jonathan
  Beaumont}, \bibinfo{person}{Dong-Hyeon Park}, \bibinfo{person}{Aporva
  Amarnath}, \bibinfo{person}{Siying Feng}, \bibinfo{person}{Chaitali
  Chakrabarti}, \bibinfo{person}{Hun-Seok Kim}, \bibinfo{person}{David Blaauw},
  \bibinfo{person}{Trevor Mudge}, {and} \bibinfo{person}{Ronald Dreslinski}.}
  \bibinfo{year}{2018}\natexlab{}.
\newblock \showarticletitle{Outerspace: An outer product based sparse matrix
  multiplication accelerator}. In \bibinfo{booktitle}{\emph{2018 IEEE
  International Symposium on High Performance Computer Architecture (HPCA)}}.
  IEEE, \bibinfo{pages}{724--736}.
\newblock


\bibitem[\protect\citeauthoryear{Pingali, Nguyen, Kulkarni, Burtscher, Hassaan,
  Kaleem, Lee, Lenharth, Manevich, and M{\'e}ndez-Lojo}{Pingali
  et~al\mbox{.}}{2011}]%
        {pingali2011tao}
\bibfield{author}{\bibinfo{person}{Keshav Pingali}, \bibinfo{person}{Donald
  Nguyen}, \bibinfo{person}{Milind Kulkarni}, \bibinfo{person}{Martin
  Burtscher}, \bibinfo{person}{M~Amber Hassaan}, \bibinfo{person}{Rashid
  Kaleem}, \bibinfo{person}{Tsung-Hsien Lee}, \bibinfo{person}{Andrew
  Lenharth}, \bibinfo{person}{Roman Manevich}, {and} \bibinfo{person}{Mario
  M{\'e}ndez-Lojo}.} \bibinfo{year}{2011}\natexlab{}.
\newblock \showarticletitle{The tao of parallelism in algorithms}. In
  \bibinfo{booktitle}{\emph{ACM Sigplan Notices}}, Vol.~\bibinfo{volume}{46}.
  ACM, \bibinfo{pages}{12--25}.
\newblock


\bibitem[\protect\citeauthoryear{Qian, Childers, Huang, Guo, and Wang}{Qian
  et~al\mbox{.}}{2018}]%
        {qian2018cgacc}
\bibfield{author}{\bibinfo{person}{Cheng Qian}, \bibinfo{person}{Bruce
  Childers}, \bibinfo{person}{Libo Huang}, \bibinfo{person}{Hui Guo}, {and}
  \bibinfo{person}{Zhiying Wang}.} \bibinfo{year}{2018}\natexlab{}.
\newblock \showarticletitle{CGAcc: A Compressed Sparse Row Representation-Based
  BFS Graph Traversal Accelerator on Hybrid Memory Cube}.
\newblock \bibinfo{journal}{\emph{Electronics}} \bibinfo{volume}{7},
  \bibinfo{number}{11} (\bibinfo{year}{2018}), \bibinfo{pages}{307}.
\newblock


\bibitem[\protect\citeauthoryear{Quinn and Deo}{Quinn and Deo}{1984}]%
        {quinn1984parallel}
\bibfield{author}{\bibinfo{person}{Michael~J Quinn} {and}
  \bibinfo{person}{Narsingh Deo}.} \bibinfo{year}{1984}\natexlab{}.
\newblock \showarticletitle{Parallel graph algorithms}.
\newblock \bibinfo{journal}{\emph{ACM Computing Surveys (CSUR)}}
  \bibinfo{volume}{16}, \bibinfo{number}{3} (\bibinfo{year}{1984}),
  \bibinfo{pages}{319--348}.
\newblock


\bibitem[\protect\citeauthoryear{Ramraj and Prabhakar}{Ramraj and
  Prabhakar}{2015}]%
        {ramraj2015frequent}
\bibfield{author}{\bibinfo{person}{T Ramraj} {and} \bibinfo{person}{R
  Prabhakar}.} \bibinfo{year}{2015}\natexlab{}.
\newblock \showarticletitle{Frequent subgraph mining algorithms-a survey}.
\newblock \bibinfo{journal}{\emph{Procedia Computer Science}}
  \bibinfo{volume}{47} (\bibinfo{year}{2015}), \bibinfo{pages}{197--204}.
\newblock


\bibitem[\protect\citeauthoryear{Rao, Chen, Yik, and Qian}{Rao
  et~al\mbox{.}}{2021}]%
        {rao2021intersectx}
\bibfield{author}{\bibinfo{person}{Gengyu Rao}, \bibinfo{person}{Jingji Chen},
  \bibinfo{person}{Jason Yik}, {and} \bibinfo{person}{Xuehai Qian}.}
  \bibinfo{year}{2021}\natexlab{}.
\newblock \showarticletitle{IntersectX: An Accelerator for Graph Mining}.
\newblock \bibinfo{journal}{\emph{arXiv preprint arXiv:2012.10848}}
  (\bibinfo{year}{2021}).
\newblock


\bibitem[\protect\citeauthoryear{Rehman, Khan, and Fong}{Rehman
  et~al\mbox{.}}{2012}]%
        {rehman2012graph}
\bibfield{author}{\bibinfo{person}{Saif~Ur Rehman},
  \bibinfo{person}{Asmat~Ullah Khan}, {and} \bibinfo{person}{Simon Fong}.}
  \bibinfo{year}{2012}\natexlab{}.
\newblock \showarticletitle{Graph mining: A survey of graph mining techniques}.
  In \bibinfo{booktitle}{\emph{Seventh International Conference on Digital
  Information Management (ICDIM 2012)}}. IEEE, \bibinfo{pages}{88--92}.
\newblock


\bibitem[\protect\citeauthoryear{Rhodes, Willett, Calvet, Dunbar, and
  Humblet}{Rhodes et~al\mbox{.}}{2003}]%
        {rhodes2003clip}
\bibfield{author}{\bibinfo{person}{Nicholas Rhodes}, \bibinfo{person}{Peter
  Willett}, \bibinfo{person}{Alain Calvet}, \bibinfo{person}{James~B Dunbar},
  {and} \bibinfo{person}{Christine Humblet}.} \bibinfo{year}{2003}\natexlab{}.
\newblock \showarticletitle{CLIP: similarity searching of 3D databases using
  clique detection}.
\newblock \bibinfo{journal}{\emph{Journal of chemical information and computer
  sciences}} \bibinfo{volume}{43}, \bibinfo{number}{2} (\bibinfo{year}{2003}),
  \bibinfo{pages}{443--448}.
\newblock


\bibitem[\protect\citeauthoryear{Ribeiro, Paredes, Silva, Aparicio, and
  Silva}{Ribeiro et~al\mbox{.}}{2019}]%
        {ribeiro2019survey}
\bibfield{author}{\bibinfo{person}{Pedro Ribeiro}, \bibinfo{person}{Pedro
  Paredes}, \bibinfo{person}{Miguel~EP Silva}, \bibinfo{person}{David
  Aparicio}, {and} \bibinfo{person}{Fernando Silva}.}
  \bibinfo{year}{2019}\natexlab{}.
\newblock \showarticletitle{A Survey on Subgraph Counting: Concepts, Algorithms
  and Applications to Network Motifs and Graphlets}.
\newblock \bibinfo{journal}{\emph{arXiv preprint arXiv:1910.13011}}
  (\bibinfo{year}{2019}).
\newblock


\bibitem[\protect\citeauthoryear{Robinson, Webber, and Eifrem}{Robinson
  et~al\mbox{.}}{2013}]%
        {robinson2013graph}
\bibfield{author}{\bibinfo{person}{Ian Robinson}, \bibinfo{person}{Jim Webber},
  {and} \bibinfo{person}{Emil Eifrem}.} \bibinfo{year}{2013}\natexlab{}.
\newblock \bibinfo{booktitle}{\emph{Graph databases}}.
\newblock \bibinfo{publisher}{" O'Reilly Media, Inc."}.
\newblock


\bibitem[\protect\citeauthoryear{Rossi and Ahmed}{Rossi and Ahmed}{2016a}]%
        {nr-sigkdd16}
\bibfield{author}{\bibinfo{person}{Ryan~A. Rossi} {and}
  \bibinfo{person}{Nesreen~K. Ahmed}.} \bibinfo{year}{2016}\natexlab{a}.
\newblock \showarticletitle{An Interactive Data Repository with Visual
  Analytics}.
\newblock \bibinfo{journal}{\emph{SIGKDD Explor.}} \bibinfo{volume}{17},
  \bibinfo{number}{2} (\bibinfo{year}{2016}), \bibinfo{pages}{37--41}.
\newblock
\urldef\tempurl%
\url{http://networkrepository.com}
\showURL{%
\tempurl}


\bibitem[\protect\citeauthoryear{Rossi and Ahmed}{Rossi and Ahmed}{2016b}]%
        {rossi2016interactive}
\bibfield{author}{\bibinfo{person}{Ryan~A Rossi} {and}
  \bibinfo{person}{Nesreen~K Ahmed}.} \bibinfo{year}{2016}\natexlab{b}.
\newblock \showarticletitle{An interactive data repository with visual
  analytics}.
\newblock \bibinfo{journal}{\emph{ACM SIGKDD Explorations Newsletter}}
  \bibinfo{volume}{17}, \bibinfo{number}{2} (\bibinfo{year}{2016}),
  \bibinfo{pages}{37--41}.
\newblock


\bibitem[\protect\citeauthoryear{Roy, Mihailovic, and Zwaenepoel}{Roy
  et~al\mbox{.}}{2013}]%
        {roy2013x}
\bibfield{author}{\bibinfo{person}{Amitabha Roy}, \bibinfo{person}{Ivo
  Mihailovic}, {and} \bibinfo{person}{Willy Zwaenepoel}.}
  \bibinfo{year}{2013}\natexlab{}.
\newblock \showarticletitle{X-stream: Edge-centric graph processing using
  streaming partitions}. In \bibinfo{booktitle}{\emph{Proceedings of the
  Twenty-Fourth ACM Symposium on Operating Systems Principles}}. ACM,
  \bibinfo{pages}{472--488}.
\newblock


\bibitem[\protect\citeauthoryear{Sadi, Fileggi, and Franchetti}{Sadi
  et~al\mbox{.}}{2017}]%
        {sadi2017algorithm}
\bibfield{author}{\bibinfo{person}{Fazle Sadi}, \bibinfo{person}{Larry
  Fileggi}, {and} \bibinfo{person}{Franz Franchetti}.}
  \bibinfo{year}{2017}\natexlab{}.
\newblock \showarticletitle{Algorithm and hardware co-optimized solution for
  large SpMV problems}. In \bibinfo{booktitle}{\emph{2017 IEEE High Performance
  Extreme Computing Conference (HPEC)}}. IEEE, \bibinfo{pages}{1--7}.
\newblock


\bibitem[\protect\citeauthoryear{Sadi, Sweeney, McMillan, Low, Hoe, Pileggi,
  and Franchetti}{Sadi et~al\mbox{.}}{2018}]%
        {sadi2018pagerank}
\bibfield{author}{\bibinfo{person}{Fazle Sadi}, \bibinfo{person}{Joe Sweeney},
  \bibinfo{person}{Scott McMillan}, \bibinfo{person}{Tze~Meng Low},
  \bibinfo{person}{James~C Hoe}, \bibinfo{person}{Larry Pileggi}, {and}
  \bibinfo{person}{Franz Franchetti}.} \bibinfo{year}{2018}\natexlab{}.
\newblock \showarticletitle{PageRank Acceleration for Large Graphs with
  Scalable Hardware and Two-Step SpMV}. In \bibinfo{booktitle}{\emph{2018 IEEE
  High Performance extreme Computing Conference (HPEC)}}. IEEE,
  \bibinfo{pages}{1--7}.
\newblock


\bibitem[\protect\citeauthoryear{Sakr, Bonifati, Voigt, Iosup, Ammar, Angles,
  Aref, Arenas, Besta, Boncz, et~al\mbox{.}}{Sakr et~al\mbox{.}}{2020}]%
        {sakr2020future}
\bibfield{author}{\bibinfo{person}{Sherif Sakr}, \bibinfo{person}{Angela
  Bonifati}, \bibinfo{person}{Hannes Voigt}, \bibinfo{person}{Alexandru Iosup},
  \bibinfo{person}{Khaled Ammar}, \bibinfo{person}{Renzo Angles},
  \bibinfo{person}{Walid Aref}, \bibinfo{person}{Marcelo Arenas},
  \bibinfo{person}{Maciej Besta}, \bibinfo{person}{Peter~A Boncz},
  {et~al\mbox{.}}} \bibinfo{year}{2020}\natexlab{}.
\newblock \showarticletitle{The Future is Big Graphs! A Community View on Graph
  Processing Systems}.
\newblock \bibinfo{journal}{\emph{arXiv preprint arXiv:2012.06171}}
  (\bibinfo{year}{2020}).
\newblock


\bibitem[\protect\citeauthoryear{Salihoglu and Widom}{Salihoglu and
  Widom}{2014}]%
        {salihoglu2014optimizing}
\bibfield{author}{\bibinfo{person}{Semih Salihoglu} {and}
  \bibinfo{person}{Jennifer Widom}.} \bibinfo{year}{2014}\natexlab{}.
\newblock \showarticletitle{{Optimizing graph algorithms on Pregel-like
  systems}}.
\newblock \bibinfo{journal}{\emph{Proceedings of the VLDB Endowment}}
  \bibinfo{volume}{7}, \bibinfo{number}{7} (\bibinfo{year}{2014}),
  \bibinfo{pages}{577--588}.
\newblock


\bibitem[\protect\citeauthoryear{Schaeffer}{Schaeffer}{2007}]%
        {schaeffer2007graph}
\bibfield{author}{\bibinfo{person}{Satu~Elisa Schaeffer}.}
  \bibinfo{year}{2007}\natexlab{}.
\newblock \showarticletitle{Graph clustering}.
\newblock \bibinfo{journal}{\emph{Computer science review}}
  \bibinfo{volume}{1}, \bibinfo{number}{1} (\bibinfo{year}{2007}),
  \bibinfo{pages}{27--64}.
\newblock


\bibitem[\protect\citeauthoryear{Schank}{Schank}{2007}]%
        {schank2007algorithmic}
\bibfield{author}{\bibinfo{person}{Thomas Schank}.}
  \bibinfo{year}{2007}\natexlab{}.
\newblock \showarticletitle{Algorithmic aspects of triangle-based network
  analysis}.
\newblock \bibinfo{journal}{\emph{Phd in computer science, University
  Karlsruhe}}  \bibinfo{volume}{3} (\bibinfo{year}{2007}).
\newblock


\bibitem[\protect\citeauthoryear{Schmid, Besta, and Hoefler}{Schmid
  et~al\mbox{.}}{2016}]%
        {schmid2016high}
\bibfield{author}{\bibinfo{person}{Patrick Schmid}, \bibinfo{person}{Maciej
  Besta}, {and} \bibinfo{person}{Torsten Hoefler}.}
  \bibinfo{year}{2016}\natexlab{}.
\newblock \showarticletitle{High-performance distributed rma locks}. In
  \bibinfo{booktitle}{\emph{Proceedings of the 25th ACM International Symposium
  on High-Performance Parallel and Distributed Computing}}.
  \bibinfo{pages}{19--30}.
\newblock


\bibitem[\protect\citeauthoryear{Schwartz, Dewar, Dubinsky, and
  Schonberg}{Schwartz et~al\mbox{.}}{2012}]%
        {schwartz2012programming}
\bibfield{author}{\bibinfo{person}{Jacob~T Schwartz},
  \bibinfo{person}{Robert~BK Dewar}, \bibinfo{person}{Edward Dubinsky}, {and}
  \bibinfo{person}{Edith Schonberg}.} \bibinfo{year}{2012}\natexlab{}.
\newblock \bibinfo{booktitle}{\emph{Programming with sets: An introduction to
  SETL}}.
\newblock \bibinfo{publisher}{Springer Science \& Business Media}.
\newblock


\bibitem[\protect\citeauthoryear{Schweizer, Besta, and Hoefler}{Schweizer
  et~al\mbox{.}}{2015}]%
        {schweizer2015evaluating}
\bibfield{author}{\bibinfo{person}{Hermann Schweizer}, \bibinfo{person}{Maciej
  Besta}, {and} \bibinfo{person}{Torsten Hoefler}.}
  \bibinfo{year}{2015}\natexlab{}.
\newblock \showarticletitle{Evaluating the cost of atomic operations on modern
  architectures}. In \bibinfo{booktitle}{\emph{2015 International Conference on
  Parallel Architecture and Compilation (PACT)}}. IEEE,
  \bibinfo{pages}{445--456}.
\newblock


\bibitem[\protect\citeauthoryear{Seshadri, Bhowmick, Mutlu, Gibbons, Kozuch,
  and Mowry}{Seshadri et~al\mbox{.}}{2014}]%
        {seshadri2014dirty}
\bibfield{author}{\bibinfo{person}{Vivek Seshadri}, \bibinfo{person}{Abhishek
  Bhowmick}, \bibinfo{person}{Onur Mutlu}, \bibinfo{person}{Phillip~B Gibbons},
  \bibinfo{person}{Michael~A Kozuch}, {and} \bibinfo{person}{Todd~C Mowry}.}
  \bibinfo{year}{2014}\natexlab{}.
\newblock \showarticletitle{The dirty-block index}.
\newblock \bibinfo{journal}{\emph{ACM SIGARCH Computer Architecture News}}
  \bibinfo{volume}{42}, \bibinfo{number}{3} (\bibinfo{year}{2014}),
  \bibinfo{pages}{157--168}.
\newblock


\bibitem[\protect\citeauthoryear{Seshadri, Kim, Fallin, Lee, Ausavarungnirun,
  Pekhimenko, Luo, Mutlu, Gibbons, and Kozuch}{Seshadri et~al\mbox{.}}{2013}]%
        {seshadri2013rowclone}
\bibfield{author}{\bibinfo{person}{Vivek Seshadri}, \bibinfo{person}{Yoongu
  Kim}, \bibinfo{person}{Chris Fallin}, \bibinfo{person}{Donghyuk Lee},
  \bibinfo{person}{Rachata Ausavarungnirun}, \bibinfo{person}{Gennady
  Pekhimenko}, \bibinfo{person}{Yixin Luo}, \bibinfo{person}{Onur Mutlu},
  \bibinfo{person}{Phillip~B Gibbons}, {and} \bibinfo{person}{Michael~A
  Kozuch}.} \bibinfo{year}{2013}\natexlab{}.
\newblock \showarticletitle{RowClone: fast and energy-efficient in-DRAM bulk
  data copy and initialization}. In \bibinfo{booktitle}{\emph{Proceedings of
  the 46th Annual IEEE/ACM International Symposium on Microarchitecture}}.
  \bibinfo{pages}{185--197}.
\newblock


\bibitem[\protect\citeauthoryear{Seshadri, Lee, Mullins, Hassan, Boroumand,
  Kim, Kozuch, Mutlu, Gibbons, and Mowry}{Seshadri et~al\mbox{.}}{2017}]%
        {seshadri2017ambit}
\bibfield{author}{\bibinfo{person}{Vivek Seshadri}, \bibinfo{person}{Donghyuk
  Lee}, \bibinfo{person}{Thomas Mullins}, \bibinfo{person}{Hasan Hassan},
  \bibinfo{person}{Amirali Boroumand}, \bibinfo{person}{Jeremie Kim},
  \bibinfo{person}{Michael~A Kozuch}, \bibinfo{person}{Onur Mutlu},
  \bibinfo{person}{Phillip~B Gibbons}, {and} \bibinfo{person}{Todd~C Mowry}.}
  \bibinfo{year}{2017}\natexlab{}.
\newblock \showarticletitle{Ambit: In-memory accelerator for bulk bitwise
  operations using commodity DRAM technology}. In
  \bibinfo{booktitle}{\emph{Proceedings of the 50th Annual IEEE/ACM
  International Symposium on Microarchitecture}}. ACM,
  \bibinfo{pages}{273--287}.
\newblock


\bibitem[\protect\citeauthoryear{Shi, Zheng, Zhou, Jin, He, Liu, and Hua}{Shi
  et~al\mbox{.}}{2018}]%
        {shi2018graph}
\bibfield{author}{\bibinfo{person}{Xuanhua Shi}, \bibinfo{person}{Zhigao
  Zheng}, \bibinfo{person}{Yongluan Zhou}, \bibinfo{person}{Hai Jin},
  \bibinfo{person}{Ligang He}, \bibinfo{person}{Bo Liu}, {and}
  \bibinfo{person}{Qiang-Sheng Hua}.} \bibinfo{year}{2018}\natexlab{}.
\newblock \showarticletitle{Graph processing on GPUs: A survey}.
\newblock \bibinfo{journal}{\emph{ACM Computing Surveys (CSUR)}}
  \bibinfo{volume}{50}, \bibinfo{number}{6} (\bibinfo{year}{2018}),
  \bibinfo{pages}{81}.
\newblock


\bibitem[\protect\citeauthoryear{Shiloach and Vishkin}{Shiloach and
  Vishkin}{1980}]%
        {shiloach1980log}
\bibfield{author}{\bibinfo{person}{Yossi Shiloach} {and} \bibinfo{person}{Uzi
  Vishkin}.} \bibinfo{year}{1980}\natexlab{}.
\newblock \bibinfo{booktitle}{\emph{An O (log n) parallel connectivity
  algorithm}}.
\newblock \bibinfo{type}{{T}echnical {R}eport}. \bibinfo{institution}{Computer
  Science Department, Technion}.
\newblock


\bibitem[\protect\citeauthoryear{Shiloach and Vishkin}{Shiloach and
  Vishkin}{1982}]%
        {shiloach1982logn}
\bibfield{author}{\bibinfo{person}{Yossi Shiloach} {and} \bibinfo{person}{Uzi
  Vishkin}.} \bibinfo{year}{1982}\natexlab{}.
\newblock \showarticletitle{An O (logn) parallel connectivity algorithm}.
\newblock \bibinfo{journal}{\emph{Journal of Algorithms}} \bibinfo{volume}{3},
  \bibinfo{number}{1} (\bibinfo{year}{1982}), \bibinfo{pages}{57--67}.
\newblock


\bibitem[\protect\citeauthoryear{Shun and Blelloch}{Shun and Blelloch}{2013}]%
        {shun2013ligra}
\bibfield{author}{\bibinfo{person}{Julian Shun} {and} \bibinfo{person}{Guy~E
  Blelloch}.} \bibinfo{year}{2013}\natexlab{}.
\newblock \showarticletitle{{Ligra: a lightweight graph processing framework
  for shared memory}}. In \bibinfo{booktitle}{\emph{ACM SIGPLAN Notices}},
  Vol.~\bibinfo{volume}{48}. \bibinfo{pages}{135--146}.
\newblock


\bibitem[\protect\citeauthoryear{Shun and Tangwongsan}{Shun and
  Tangwongsan}{2015}]%
        {shun2015multicore}
\bibfield{author}{\bibinfo{person}{Julian Shun} {and} \bibinfo{person}{Kanat
  Tangwongsan}.} \bibinfo{year}{2015}\natexlab{}.
\newblock \showarticletitle{Multicore triangle computations without tuning}. In
  \bibinfo{booktitle}{\emph{Data Engineering (ICDE), 2015 IEEE 31st
  International Conference on}}. IEEE, \bibinfo{pages}{149--160}.
\newblock


\bibitem[\protect\citeauthoryear{Singapura, Srivastava, Kannan, and
  Prasanna}{Singapura et~al\mbox{.}}{2017}]%
        {singapura2017oscar}
\bibfield{author}{\bibinfo{person}{Shreyas~G Singapura},
  \bibinfo{person}{Ajitesh Srivastava}, \bibinfo{person}{Rajgopal Kannan},
  {and} \bibinfo{person}{Viktor~K Prasanna}.} \bibinfo{year}{2017}\natexlab{}.
\newblock \showarticletitle{OSCAR: Optimizing SCrAtchpad reuse for graph
  processing}. In \bibinfo{booktitle}{\emph{2017 IEEE High Performance Extreme
  Computing Conference (HPEC)}}. IEEE, \bibinfo{pages}{1--7}.
\newblock


\bibitem[\protect\citeauthoryear{Skiena}{Skiena}{1990}]%
        {skiena1990dijkstra}
\bibfield{author}{\bibinfo{person}{S Skiena}.} \bibinfo{year}{1990}\natexlab{}.
\newblock \showarticletitle{Dijkstra’s algorithm}.
\newblock \bibinfo{journal}{\emph{Implementing Discrete Mathematics:
  Combinatorics and Graph Theory with Mathematica, Reading, MA:
  Addison-Wesley}} (\bibinfo{year}{1990}), \bibinfo{pages}{225--227}.
\newblock


\bibitem[\protect\citeauthoryear{Smith, Curtis, and Zeng}{Smith
  et~al\mbox{.}}{2016}]%
        {smith2016practical}
\bibfield{author}{\bibinfo{person}{Robert~S Smith}, \bibinfo{person}{Michael~J
  Curtis}, {and} \bibinfo{person}{William~J Zeng}.}
  \bibinfo{year}{2016}\natexlab{}.
\newblock \showarticletitle{A practical quantum instruction set architecture}.
\newblock \bibinfo{journal}{\emph{arXiv preprint arXiv:1608.03355}}
  (\bibinfo{year}{2016}).
\newblock


\bibitem[\protect\citeauthoryear{Solomonik, Besta, Vella, and
  Hoefler}{Solomonik et~al\mbox{.}}{2017}]%
        {solomonik2017scaling}
\bibfield{author}{\bibinfo{person}{Edgar Solomonik}, \bibinfo{person}{Maciej
  Besta}, \bibinfo{person}{Flavio Vella}, {and} \bibinfo{person}{Torsten
  Hoefler}.} \bibinfo{year}{2017}\natexlab{}.
\newblock \showarticletitle{Scaling betweenness centrality using
  communication-efficient sparse matrix multiplication}. In
  \bibinfo{booktitle}{\emph{Proceedings of the International Conference for
  High Performance Computing, Networking, Storage and Analysis}}. ACM,
  \bibinfo{pages}{47}.
\newblock


\bibitem[\protect\citeauthoryear{Song, Zhuo, Qian, Li, and Chen}{Song
  et~al\mbox{.}}{2018}]%
        {song2018graphr}
\bibfield{author}{\bibinfo{person}{Linghao Song}, \bibinfo{person}{Youwei
  Zhuo}, \bibinfo{person}{Xuehai Qian}, \bibinfo{person}{Hai Li}, {and}
  \bibinfo{person}{Yiran Chen}.} \bibinfo{year}{2018}\natexlab{}.
\newblock \showarticletitle{{GraphR: Accelerating graph processing using
  ReRAM}}. In \bibinfo{booktitle}{\emph{High Performance Computer Architecture
  (HPCA), 2018 IEEE International Symposium on}}. IEEE,
  \bibinfo{pages}{531--543}.
\newblock


\bibitem[\protect\citeauthoryear{Spirin and Mirny}{Spirin and Mirny}{2003}]%
        {spirin2003protein}
\bibfield{author}{\bibinfo{person}{Victor Spirin} {and}
  \bibinfo{person}{Leonid~A Mirny}.} \bibinfo{year}{2003}\natexlab{}.
\newblock \showarticletitle{Protein complexes and functional modules in
  molecular networks}.
\newblock \bibinfo{journal}{\emph{Proceedings of the National Academy of
  Sciences}} \bibinfo{volume}{100}, \bibinfo{number}{21}
  (\bibinfo{year}{2003}), \bibinfo{pages}{12123--12128}.
\newblock


\bibitem[\protect\citeauthoryear{Sridharan, Priya, and Kumar}{Sridharan
  et~al\mbox{.}}{2009}]%
        {sridharan2009hardware}
\bibfield{author}{\bibinfo{person}{K Sridharan}, \bibinfo{person}{TK Priya},
  {and} \bibinfo{person}{P~Rajesh Kumar}.} \bibinfo{year}{2009}\natexlab{}.
\newblock \showarticletitle{Hardware architecture for finding shortest paths}.
  In \bibinfo{booktitle}{\emph{TENCON 2009-2009 IEEE Region 10 Conference}}.
  IEEE, \bibinfo{pages}{1--5}.
\newblock


\bibitem[\protect\citeauthoryear{Sundaram, Satish, Patwary, Dulloor, Anderson,
  Vadlamudi, Das, and Dubey}{Sundaram et~al\mbox{.}}{2015}]%
        {sundaram2015graphmat}
\bibfield{author}{\bibinfo{person}{Narayanan Sundaram},
  \bibinfo{person}{Nadathur Satish}, \bibinfo{person}{Md~Mostofa~Ali Patwary},
  \bibinfo{person}{Subramanya~R Dulloor}, \bibinfo{person}{Michael~J Anderson},
  \bibinfo{person}{Satya~Gautam Vadlamudi}, \bibinfo{person}{Dipankar Das},
  {and} \bibinfo{person}{Pradeep Dubey}.} \bibinfo{year}{2015}\natexlab{}.
\newblock \showarticletitle{Graphmat: High performance graph analytics made
  productive}.
\newblock \bibinfo{journal}{\emph{Proceedings of the VLDB Endowment}}
  \bibinfo{volume}{8}, \bibinfo{number}{11} (\bibinfo{year}{2015}),
  \bibinfo{pages}{1214--1225}.
\newblock


\bibitem[\protect\citeauthoryear{Sutton, Ben-Nun, and Barak}{Sutton
  et~al\mbox{.}}{[n. d.]}]%
        {suttonoptimizing}
\bibfield{author}{\bibinfo{person}{Michael Sutton}, \bibinfo{person}{Tal
  Ben-Nun}, {and} \bibinfo{person}{Amnon Barak}.} \bibinfo{year}{[n.
  d.]}\natexlab{}.
\newblock \showarticletitle{Optimizing Parallel Graph Connectivity Computation
  via Subgraph Sampling}.
\newblock  (\bibinfo{year}{[n. d.]}).
\newblock


\bibitem[\protect\citeauthoryear{Takigawa and Mamitsuka}{Takigawa and
  Mamitsuka}{2013}]%
        {takigawa2013graph}
\bibfield{author}{\bibinfo{person}{Ichigaku Takigawa} {and}
  \bibinfo{person}{Hiroshi Mamitsuka}.} \bibinfo{year}{2013}\natexlab{}.
\newblock \showarticletitle{Graph mining: procedure, application to drug
  discovery and recent advances}.
\newblock \bibinfo{journal}{\emph{Drug discovery today}} \bibinfo{volume}{18},
  \bibinfo{number}{1-2} (\bibinfo{year}{2013}), \bibinfo{pages}{50--57}.
\newblock


\bibitem[\protect\citeauthoryear{Tang and Liu}{Tang and Liu}{2010}]%
        {tang2010graph}
\bibfield{author}{\bibinfo{person}{Lei Tang} {and} \bibinfo{person}{Huan Liu}.}
  \bibinfo{year}{2010}\natexlab{}.
\newblock \showarticletitle{Graph mining applications to social network
  analysis}.
\newblock In \bibinfo{booktitle}{\emph{Managing and Mining Graph Data}}.
  \bibinfo{publisher}{Springer}, \bibinfo{pages}{487--513}.
\newblock


\bibitem[\protect\citeauthoryear{Taskar, Wong, Abbeel, and Koller}{Taskar
  et~al\mbox{.}}{2004}]%
        {taskar2004link}
\bibfield{author}{\bibinfo{person}{Ben Taskar}, \bibinfo{person}{Ming-Fai
  Wong}, \bibinfo{person}{Pieter Abbeel}, {and} \bibinfo{person}{Daphne
  Koller}.} \bibinfo{year}{2004}\natexlab{}.
\newblock \showarticletitle{Link prediction in relational data}. In
  \bibinfo{booktitle}{\emph{Advances in neural information processing
  systems}}. \bibinfo{pages}{659--666}.
\newblock


\bibitem[\protect\citeauthoryear{Tate, Kamil, Dubey, Gr{\"o}{\ss}linger,
  Chamberlain, Goglin, Edwards, Newburn, Padua, Unat, et~al\mbox{.}}{Tate
  et~al\mbox{.}}{2014}]%
        {tate2014programming}
\bibfield{author}{\bibinfo{person}{Adrian Tate}, \bibinfo{person}{Amir Kamil},
  \bibinfo{person}{Anshu Dubey}, \bibinfo{person}{Armin Gr{\"o}{\ss}linger},
  \bibinfo{person}{Brad Chamberlain}, \bibinfo{person}{Brice Goglin},
  \bibinfo{person}{Carter Edwards}, \bibinfo{person}{Chris~J Newburn},
  \bibinfo{person}{David Padua}, \bibinfo{person}{Didem Unat}, {et~al\mbox{.}}}
  \bibinfo{year}{2014}\natexlab{}.
\newblock \showarticletitle{Programming abstractions for data locality}. PADAL
  Workshop 2014, April 28--29, Swiss National Supercomputing Center~….
\newblock


\bibitem[\protect\citeauthoryear{Teixeira, Fonseca, Serafini, Siganos, Zaki,
  and Aboulnaga}{Teixeira et~al\mbox{.}}{2015}]%
        {teixeira2015arabesque}
\bibfield{author}{\bibinfo{person}{Carlos~HC Teixeira},
  \bibinfo{person}{Alexandre~J Fonseca}, \bibinfo{person}{Marco Serafini},
  \bibinfo{person}{Georgos Siganos}, \bibinfo{person}{Mohammed~J Zaki}, {and}
  \bibinfo{person}{Ashraf Aboulnaga}.} \bibinfo{year}{2015}\natexlab{}.
\newblock \showarticletitle{Arabesque: a system for distributed graph mining}.
  In \bibinfo{booktitle}{\emph{Proceedings of the 25th Symposium on Operating
  Systems Principles}}. ACM, \bibinfo{pages}{425--440}.
\newblock


\bibitem[\protect\citeauthoryear{Thiprungsri and Vasarhelyi}{Thiprungsri and
  Vasarhelyi}{2011}]%
        {thiprungsri2011cluster}
\bibfield{author}{\bibinfo{person}{Sutapat Thiprungsri} {and}
  \bibinfo{person}{Miklos~A Vasarhelyi}.} \bibinfo{year}{2011}\natexlab{}.
\newblock \showarticletitle{Cluster Analysis for Anomaly Detection in
  Accounting Data: An Audit Approach.}
\newblock \bibinfo{journal}{\emph{International Journal of Digital Accounting
  Research}}  \bibinfo{volume}{11} (\bibinfo{year}{2011}).
\newblock


\bibitem[\protect\citeauthoryear{Tomita, Tanaka, and Takahashi}{Tomita
  et~al\mbox{.}}{2006}]%
        {DBLP:journals/tcs/TomitaTT06}
\bibfield{author}{\bibinfo{person}{Etsuji Tomita}, \bibinfo{person}{Akira
  Tanaka}, {and} \bibinfo{person}{Haruhisa Takahashi}.}
  \bibinfo{year}{2006}\natexlab{}.
\newblock \showarticletitle{The worst-case time complexity for generating all
  maximal cliques and computational experiments}.
\newblock \bibinfo{journal}{\emph{Theor. Comput. Sci.}} \bibinfo{volume}{363},
  \bibinfo{number}{1} (\bibinfo{year}{2006}), \bibinfo{pages}{28--42}.
\newblock
\urldef\tempurl%
\url{https://doi.org/10.1016/j.tcs.2006.06.015}
\showDOI{\tempurl}


\bibitem[\protect\citeauthoryear{Tommiska and Skytt{\"a}}{Tommiska and
  Skytt{\"a}}{2001}]%
        {tommiska2001dijkstra}
\bibfield{author}{\bibinfo{person}{Matti Tommiska} {and} \bibinfo{person}{Jorma
  Skytt{\"a}}.} \bibinfo{year}{2001}\natexlab{}.
\newblock \showarticletitle{Dijkstra’s shortest path routing algorithm in
  reconfigurable hardware}. In \bibinfo{booktitle}{\emph{International
  Conference on Field Programmable Logic and Applications}}. Springer,
  \bibinfo{pages}{653--657}.
\newblock


\bibitem[\protect\citeauthoryear{Ullmann}{Ullmann}{1976}]%
        {ullmann1976algorithm}
\bibfield{author}{\bibinfo{person}{Julian~R Ullmann}.}
  \bibinfo{year}{1976}\natexlab{}.
\newblock \showarticletitle{An algorithm for subgraph isomorphism}.
\newblock \bibinfo{journal}{\emph{Journal of the ACM (JACM)}}
  \bibinfo{volume}{23}, \bibinfo{number}{1} (\bibinfo{year}{1976}),
  \bibinfo{pages}{31--42}.
\newblock


\bibitem[\protect\citeauthoryear{Umuroglu, Morrison, and Jahre}{Umuroglu
  et~al\mbox{.}}{2015}]%
        {umuroglu:hybrid_bfs_FPGA}
\bibfield{author}{\bibinfo{person}{Y. Umuroglu}, \bibinfo{person}{D. Morrison},
  {and} \bibinfo{person}{M. Jahre}.} \bibinfo{year}{2015}\natexlab{}.
\newblock \showarticletitle{Hybrid breadth-first search on a single-chip
  FPGA-CPU heterogeneous platform}. In \bibinfo{booktitle}{\emph{2015 25th
  International Conference on Field Programmable Logic and Applications
  (FPL)}}. \bibinfo{pages}{1--8}.
\newblock
\showISSN{1946-147X}
\urldef\tempurl%
\url{https://doi.org/10.1109/FPL.2015.7293939}
\showDOI{\tempurl}


\bibitem[\protect\citeauthoryear{Van~Craeynest, Akram, Heirman, Jaleel, and
  Eeckhout}{Van~Craeynest et~al\mbox{.}}{2013}]%
        {van2013fairness}
\bibfield{author}{\bibinfo{person}{Kenzo Van~Craeynest},
  \bibinfo{person}{Shoaib Akram}, \bibinfo{person}{Wim Heirman},
  \bibinfo{person}{Aamer Jaleel}, {and} \bibinfo{person}{Lieven Eeckhout}.}
  \bibinfo{year}{2013}\natexlab{}.
\newblock \showarticletitle{Fairness-aware scheduling on single-ISA
  heterogeneous multi-cores}. In \bibinfo{booktitle}{\emph{Proceedings of the
  22nd international conference on Parallel architectures and compilation
  techniques}}. IEEE, \bibinfo{pages}{177--187}.
\newblock


\bibitem[\protect\citeauthoryear{Wang, Gong, Jia, and Xuehai}{Wang
  et~al\mbox{.}}{2020}]%
        {wang2020fpga}
\bibfield{author}{\bibinfo{person}{Chao Wang}, \bibinfo{person}{Lei Gong},
  \bibinfo{person}{Fahui Jia}, {and} \bibinfo{person}{Zhou Xuehai}.}
  \bibinfo{year}{2020}\natexlab{}.
\newblock \showarticletitle{An FPGA based Accelerator for Ubiquitous Clustering
  Applications with Custom Instructions}.
\newblock \bibinfo{journal}{\emph{IEEE Trans. Comput.}} (\bibinfo{year}{2020}).
\newblock


\bibitem[\protect\citeauthoryear{Wang, Zuo, Thorpe, Nguyen, and Xu}{Wang
  et~al\mbox{.}}{2018}]%
        {wang2018rstream}
\bibfield{author}{\bibinfo{person}{Kai Wang}, \bibinfo{person}{Zhiqiang Zuo},
  \bibinfo{person}{John Thorpe}, \bibinfo{person}{Tien~Quang Nguyen}, {and}
  \bibinfo{person}{Guoqing~Harry Xu}.} \bibinfo{year}{2018}\natexlab{}.
\newblock \showarticletitle{Rstream: marrying relational algebra with streaming
  for efficient graph mining on a single machine}. In
  \bibinfo{booktitle}{\emph{13th $\{$USENIX$\}$ Symposium on Operating Systems
  Design and Implementation ($\{$OSDI$\}$ 18)}}. \bibinfo{pages}{763--782}.
\newblock


\bibitem[\protect\citeauthoryear{Wang, Hu, and Tang}{Wang
  et~al\mbox{.}}{2014}]%
        {wang2014robustness}
\bibfield{author}{\bibinfo{person}{Liang Wang}, \bibinfo{person}{Ke Hu}, {and}
  \bibinfo{person}{Yi Tang}.} \bibinfo{year}{2014}\natexlab{}.
\newblock \showarticletitle{Robustness of link-prediction algorithm based on
  similarity and application to biological networks}.
\newblock \bibinfo{journal}{\emph{Current Bioinformatics}} \bibinfo{volume}{9},
  \bibinfo{number}{3} (\bibinfo{year}{2014}), \bibinfo{pages}{246--252}.
\newblock


\bibitem[\protect\citeauthoryear{Wang, Jiang, Xia, and Prasanna}{Wang
  et~al\mbox{.}}{2010}]%
        {wang2010message}
\bibfield{author}{\bibinfo{person}{Qingbo Wang}, \bibinfo{person}{Weirong
  Jiang}, \bibinfo{person}{Yinglong Xia}, {and} \bibinfo{person}{Viktor
  Prasanna}.} \bibinfo{year}{2010}\natexlab{}.
\newblock \showarticletitle{A message-passing multi-softcore architecture on
  FPGA for breadth-first search}. In
  \bibinfo{booktitle}{\emph{Field-Programmable Technology (FPT), 2010
  International Conference on}}. IEEE, \bibinfo{pages}{70--77}.
\newblock


\bibitem[\protect\citeauthoryear{Wang, Hoe, and Nurvitadhi}{Wang
  et~al\mbox{.}}{2019}]%
        {wang2019processor}
\bibfield{author}{\bibinfo{person}{Yu Wang}, \bibinfo{person}{James~C Hoe},
  {and} \bibinfo{person}{Eriko Nurvitadhi}.} \bibinfo{year}{2019}\natexlab{}.
\newblock \showarticletitle{Processor assisted worklist scheduling for FPGA
  accelerated graph processing on a shared-memory platform}. In
  \bibinfo{booktitle}{\emph{2019 IEEE 27th Annual International Symposium on
  Field-Programmable Custom Computing Machines (FCCM)}}. IEEE,
  \bibinfo{pages}{136--144}.
\newblock


\bibitem[\protect\citeauthoryear{Washio and Motoda}{Washio and Motoda}{2003}]%
        {washio2003state}
\bibfield{author}{\bibinfo{person}{Takashi Washio} {and}
  \bibinfo{person}{Hiroshi Motoda}.} \bibinfo{year}{2003}\natexlab{}.
\newblock \showarticletitle{State of the art of graph-based data mining}.
\newblock \bibinfo{journal}{\emph{Acm Sigkdd Explorations Newsletter}}
  \bibinfo{volume}{5}, \bibinfo{number}{1} (\bibinfo{year}{2003}),
  \bibinfo{pages}{59--68}.
\newblock


\bibitem[\protect\citeauthoryear{Wasserman and Faust}{Wasserman and
  Faust}{1994}]%
        {wasserman1994social}
\bibfield{author}{\bibinfo{person}{Stanley Wasserman} {and}
  \bibinfo{person}{Katherine Faust}.} \bibinfo{year}{1994}\natexlab{}.
\newblock \bibinfo{booktitle}{\emph{Social network analysis: Methods and
  applications}}. Vol.~\bibinfo{volume}{8}.
\newblock \bibinfo{publisher}{Cambridge university press}.
\newblock


\bibitem[\protect\citeauthoryear{Waterman, Lee, Patterson, and
  Asanovic}{Waterman et~al\mbox{.}}{2011}]%
        {waterman2011risc}
\bibfield{author}{\bibinfo{person}{Andrew Waterman}, \bibinfo{person}{Yunsup
  Lee}, \bibinfo{person}{David~A Patterson}, {and} \bibinfo{person}{Krste
  Asanovic}.} \bibinfo{year}{2011}\natexlab{}.
\newblock \showarticletitle{The risc-v instruction set manual, volume i: Base
  user-level isa}.
\newblock \bibinfo{journal}{\emph{EECS Department, UC Berkeley, Tech. Rep.
  UCB/EECS-2011-62}}  \bibinfo{volume}{116} (\bibinfo{year}{2011}).
\newblock


\bibitem[\protect\citeauthoryear{Waterman}{Waterman}{2016}]%
        {waterman2016design}
\bibfield{author}{\bibinfo{person}{Andrew~Shell Waterman}.}
  \bibinfo{year}{2016}\natexlab{}.
\newblock \emph{\bibinfo{title}{Design of the RISC-V instruction set
  architecture}}.
\newblock \bibinfo{thesistype}{Ph.D. Dissertation}. \bibinfo{school}{UC
  Berkeley}.
\newblock


\bibitem[\protect\citeauthoryear{Wu, Pan, Chen, Long, Zhang, and Philip}{Wu
  et~al\mbox{.}}{2020}]%
        {wu2020comprehensive}
\bibfield{author}{\bibinfo{person}{Zonghan Wu}, \bibinfo{person}{Shirui Pan},
  \bibinfo{person}{Fengwen Chen}, \bibinfo{person}{Guodong Long},
  \bibinfo{person}{Chengqi Zhang}, {and} \bibinfo{person}{S~Yu Philip}.}
  \bibinfo{year}{2020}\natexlab{}.
\newblock \showarticletitle{A comprehensive survey on graph neural networks}.
\newblock \bibinfo{journal}{\emph{IEEE Transactions on Neural Networks and
  Learning Systems}} (\bibinfo{year}{2020}).
\newblock


\bibitem[\protect\citeauthoryear{Xin, Zhang, and Yang}{Xin
  et~al\mbox{.}}{2020}]%
        {xin2020elp2im}
\bibfield{author}{\bibinfo{person}{Xin Xin}, \bibinfo{person}{Youtao Zhang},
  {and} \bibinfo{person}{Jun Yang}.} \bibinfo{year}{2020}\natexlab{}.
\newblock \showarticletitle{ELP2IM: Efficient and Low Power Bitwise Operation
  Processing in DRAM}. In \bibinfo{booktitle}{\emph{2020 IEEE International
  Symposium on High Performance Computer Architecture (HPCA)}}. IEEE,
  \bibinfo{pages}{303--314}.
\newblock


\bibitem[\protect\citeauthoryear{Xu, Wang, Gong, Jin, Li, and Zhou}{Xu
  et~al\mbox{.}}{2018b}]%
        {xu2018domino}
\bibfield{author}{\bibinfo{person}{Chongchong Xu}, \bibinfo{person}{Chao Wang},
  \bibinfo{person}{Lei Gong}, \bibinfo{person}{Lihui Jin}, \bibinfo{person}{Xi
  Li}, {and} \bibinfo{person}{Xuehai Zhou}.} \bibinfo{year}{2018}\natexlab{b}.
\newblock \showarticletitle{Domino: Graph Processing Services on
  Energy-Efficient Hardware Accelerator}. In \bibinfo{booktitle}{\emph{2018
  IEEE International Conference on Web Services (ICWS)}}. IEEE,
  \bibinfo{pages}{274--281}.
\newblock


\bibitem[\protect\citeauthoryear{Xu, Hu, Leskovec, and Jegelka}{Xu
  et~al\mbox{.}}{2018a}]%
        {xu2018powerful}
\bibfield{author}{\bibinfo{person}{Keyulu Xu}, \bibinfo{person}{Weihua Hu},
  \bibinfo{person}{Jure Leskovec}, {and} \bibinfo{person}{Stefanie Jegelka}.}
  \bibinfo{year}{2018}\natexlab{a}.
\newblock \showarticletitle{How powerful are graph neural networks?}
\newblock \bibinfo{journal}{\emph{arXiv preprint arXiv:1810.00826}}
  (\bibinfo{year}{2018}).
\newblock


\bibitem[\protect\citeauthoryear{Yan, Chen, Cheng, {\"O}zsu, Zhang, and
  Lui}{Yan et~al\mbox{.}}{2017}]%
        {yan2017g}
\bibfield{author}{\bibinfo{person}{Da Yan}, \bibinfo{person}{Hongzhi Chen},
  \bibinfo{person}{James Cheng}, \bibinfo{person}{M~Tamer {\"O}zsu},
  \bibinfo{person}{Qizhen Zhang}, {and} \bibinfo{person}{John Lui}.}
  \bibinfo{year}{2017}\natexlab{}.
\newblock \showarticletitle{G-thinker: big graph mining made easier and
  faster}.
\newblock \bibinfo{journal}{\emph{arXiv preprint arXiv:1709.03110}}
  (\bibinfo{year}{2017}).
\newblock


\bibitem[\protect\citeauthoryear{Yan, Cheng, Xing, Lu, Ng, and Bu}{Yan
  et~al\mbox{.}}{2014}]%
        {yan2014pregel}
\bibfield{author}{\bibinfo{person}{Da Yan}, \bibinfo{person}{James Cheng},
  \bibinfo{person}{Kai Xing}, \bibinfo{person}{Yi Lu}, \bibinfo{person}{Wilfred
  Ng}, {and} \bibinfo{person}{Yingyi Bu}.} \bibinfo{year}{2014}\natexlab{}.
\newblock \showarticletitle{Pregel algorithms for graph connectivity problems
  with performance guarantees}.
\newblock \bibinfo{journal}{\emph{Proceedings of the VLDB Endowment}}
  \bibinfo{volume}{7}, \bibinfo{number}{14} (\bibinfo{year}{2014}),
  \bibinfo{pages}{1821--1832}.
\newblock


\bibitem[\protect\citeauthoryear{Yan, Qu, Guo, and Wang}{Yan
  et~al\mbox{.}}{2020b}]%
        {yan2020prefixfpm}
\bibfield{author}{\bibinfo{person}{Da Yan}, \bibinfo{person}{Wenwen Qu},
  \bibinfo{person}{Guimu Guo}, {and} \bibinfo{person}{Xiaoling Wang}.}
  \bibinfo{year}{2020}\natexlab{b}.
\newblock \showarticletitle{PrefixFPM: A Parallel Framework for General-Purpose
  Frequent Pattern Mining}. In \bibinfo{booktitle}{\emph{Proceedings of the
  36th IEEE International Conference on Data Engineering (ICDE) 2020}}.
\newblock


\bibitem[\protect\citeauthoryear{Yan, Deng, Hu, Liang, Feng, Ye, Zhang, Fan,
  and Xie}{Yan et~al\mbox{.}}{2020a}]%
        {yan2020hygcn}
\bibfield{author}{\bibinfo{person}{Mingyu Yan}, \bibinfo{person}{Lei Deng},
  \bibinfo{person}{Xing Hu}, \bibinfo{person}{Ling Liang},
  \bibinfo{person}{Yujing Feng}, \bibinfo{person}{Xiaochun Ye},
  \bibinfo{person}{Zhimin Zhang}, \bibinfo{person}{Dongrui Fan}, {and}
  \bibinfo{person}{Yuan Xie}.} \bibinfo{year}{2020}\natexlab{a}.
\newblock \showarticletitle{Hygcn: A gcn accelerator with hybrid architecture}.
  In \bibinfo{booktitle}{\emph{2020 IEEE International Symposium on High
  Performance Computer Architecture (HPCA)}}. IEEE, \bibinfo{pages}{15--29}.
\newblock


\bibitem[\protect\citeauthoryear{Yan, Hu, Li, Basak, Li, Ma, Akgun, Feng, Gu,
  and Deng}{Yan et~al\mbox{.}}{2019}]%
        {yan2019alleviating}
\bibfield{author}{\bibinfo{person}{Mingyu Yan}, \bibinfo{person}{Xing Hu},
  \bibinfo{person}{Shuangchen Li}, \bibinfo{person}{Abanti Basak},
  \bibinfo{person}{Han Li}, \bibinfo{person}{Xin Ma}, \bibinfo{person}{Itir
  Akgun}, \bibinfo{person}{Yujing Feng}, \bibinfo{person}{Peng Gu}, {and}
  \bibinfo{person}{Lei Deng}.} \bibinfo{year}{2019}\natexlab{}.
\newblock \showarticletitle{Alleviating irregularity in graph analytics
  acceleration: A hardware/software co-design approach}. In
  \bibinfo{booktitle}{\emph{Proceedings of the 52nd Annual IEEE/ACM
  International Symposium on Microarchitecture}}. \bibinfo{pages}{615--628}.
\newblock


\bibitem[\protect\citeauthoryear{Yang}{Yang}{2018}]%
        {yang2018efficient}
\bibfield{author}{\bibinfo{person}{Chengbo Yang}.}
  \bibinfo{year}{2018}\natexlab{}.
\newblock \showarticletitle{An Efficient Dispatcher for Large Scale
  GraphProcessing on OpenCL-based {FPGAs}}.
\newblock \bibinfo{journal}{\emph{arXiv preprint arXiv:1806.11509}}
  (\bibinfo{year}{2018}).
\newblock


\bibitem[\protect\citeauthoryear{Yang, Li, Deng, Liu, Yin, Wei, and Liu}{Yang
  et~al\mbox{.}}{2020}]%
        {yang2020graphabcd}
\bibfield{author}{\bibinfo{person}{Yifan Yang}, \bibinfo{person}{Zhaoshi Li},
  \bibinfo{person}{Yangdong Deng}, \bibinfo{person}{Zhiwei Liu},
  \bibinfo{person}{Shouyi Yin}, \bibinfo{person}{Shaojun Wei}, {and}
  \bibinfo{person}{Leibo Liu}.} \bibinfo{year}{2020}\natexlab{}.
\newblock \showarticletitle{GraphABCD: Scaling Out Graph Analytics with
  Asynchronous Block Coordinate Descent}. In \bibinfo{booktitle}{\emph{2020
  ACM/IEEE 47th Annual International Symposium on Computer Architecture
  (ISCA)}}. IEEE, \bibinfo{pages}{419--432}.
\newblock


\bibitem[\protect\citeauthoryear{Yao}{Yao}{2018}]%
        {yao2018efficient}
\bibfield{author}{\bibinfo{person}{Pengcheng Yao}.}
  \bibinfo{year}{2018}\natexlab{}.
\newblock \showarticletitle{An Efficient Graph Accelerator with Parallel Data
  Conflict Management}.
\newblock \bibinfo{journal}{\emph{arXiv preprint arXiv:1806.00751}}
  (\bibinfo{year}{2018}).
\newblock


\bibitem[\protect\citeauthoryear{Yao, Zheng, Zeng, Huang, Gui, Liao, Jin, and
  Xue}{Yao et~al\mbox{.}}{[n. d.]}]%
        {yaolocality}
\bibfield{author}{\bibinfo{person}{Pengcheng Yao}, \bibinfo{person}{Long
  Zheng}, \bibinfo{person}{Zhen Zeng}, \bibinfo{person}{Yu Huang},
  \bibinfo{person}{Chuangyi Gui}, \bibinfo{person}{Xiaofei Liao},
  \bibinfo{person}{Hai Jin}, {and} \bibinfo{person}{Jingling Xue}.}
  \bibinfo{year}{[n. d.]}\natexlab{}.
\newblock \showarticletitle{A Locality-Aware Energy-Efficient Accelerator for
  Graph Mining Applications}.
\newblock  (\bibinfo{year}{[n. d.]}).
\newblock


\bibitem[\protect\citeauthoryear{Yao, Zheng, Zeng, Huang, Gui, Liao, Jin, and
  Xue}{Yao et~al\mbox{.}}{2020}]%
        {yao2020locality}
\bibfield{author}{\bibinfo{person}{Pengcheng Yao}, \bibinfo{person}{Long
  Zheng}, \bibinfo{person}{Zhen Zeng}, \bibinfo{person}{Yu Huang},
  \bibinfo{person}{Chuangyi Gui}, \bibinfo{person}{Xiaofei Liao},
  \bibinfo{person}{Hai Jin}, {and} \bibinfo{person}{Jingling Xue}.}
  \bibinfo{year}{2020}\natexlab{}.
\newblock \showarticletitle{A Locality-Aware Energy-Efficient Accelerator for
  Graph Mining Applications}. In \bibinfo{booktitle}{\emph{2020 53rd Annual
  IEEE/ACM International Symposium on Microarchitecture (MICRO)}}. IEEE,
  \bibinfo{pages}{895--907}.
\newblock


\bibitem[\protect\citeauthoryear{Yoo, Chow, Henderson, McLendon, Hendrickson,
  and Catalyurek}{Yoo et~al\mbox{.}}{2005}]%
        {yoo2005scalable}
\bibfield{author}{\bibinfo{person}{Andy Yoo}, \bibinfo{person}{Edmond Chow},
  \bibinfo{person}{Keith Henderson}, \bibinfo{person}{William McLendon},
  \bibinfo{person}{Bruce Hendrickson}, {and} \bibinfo{person}{Umit
  Catalyurek}.} \bibinfo{year}{2005}\natexlab{}.
\newblock \showarticletitle{A scalable distributed parallel breadth-first
  search algorithm on BlueGene/L}. In \bibinfo{booktitle}{\emph{Proceedings of
  the 2005 ACM/IEEE conference on Supercomputing}}. IEEE Computer Society,
  \bibinfo{pages}{25}.
\newblock


\bibitem[\protect\citeauthoryear{Yu, Hughes, Satish, and Devadas}{Yu
  et~al\mbox{.}}{2015}]%
        {yu2015imp}
\bibfield{author}{\bibinfo{person}{Xiangyao Yu}, \bibinfo{person}{Christopher~J
  Hughes}, \bibinfo{person}{Nadathur Satish}, {and} \bibinfo{person}{Srinivas
  Devadas}.} \bibinfo{year}{2015}\natexlab{}.
\newblock \showarticletitle{IMP: Indirect memory prefetcher}. In
  \bibinfo{booktitle}{\emph{Proceedings of the 48th International Symposium on
  Microarchitecture}}. \bibinfo{pages}{178--190}.
\newblock


\bibitem[\protect\citeauthoryear{Zhang, Khoram, and Li}{Zhang
  et~al\mbox{.}}{2017a}]%
        {zhang:graph_FPGA}
\bibfield{author}{\bibinfo{person}{Jialiang Zhang}, \bibinfo{person}{Soroosh
  Khoram}, {and} \bibinfo{person}{Jing Li}.} \bibinfo{year}{2017}\natexlab{a}.
\newblock \showarticletitle{Boosting the Performance of FPGA-based Graph
  Processor Using Hybrid Memory Cube: A Case for Breadth First Search}. In
  \bibinfo{booktitle}{\emph{Proceedings of the 2017 ACM/SIGDA International
  Symposium on Field-Programmable Gate Arrays}} \emph{(\bibinfo{series}{FPGA
  '17})}. \bibinfo{publisher}{ACM}, \bibinfo{address}{New York, NY, USA},
  \bibinfo{pages}{207--216}.
\newblock
\showISBNx{978-1-4503-4354-1}
\urldef\tempurl%
\url{https://doi.org/10.1145/3020078.3021737}
\showDOI{\tempurl}


\bibitem[\protect\citeauthoryear{Zhang and Li}{Zhang and Li}{2018}]%
        {zhang2018degree}
\bibfield{author}{\bibinfo{person}{Jialiang Zhang} {and} \bibinfo{person}{Jing
  Li}.} \bibinfo{year}{2018}\natexlab{}.
\newblock \showarticletitle{Degree-aware Hybrid Graph Traversal on FPGA-HMC
  Platform}. In \bibinfo{booktitle}{\emph{Proceedings of the 2018 ACM/SIGDA
  International Symposium on Field-Programmable Gate Arrays}}. ACM,
  \bibinfo{pages}{229--238}.
\newblock


\bibitem[\protect\citeauthoryear{Zhang, Zhuo, Wang, Gao, Wu, Chen, Kozyrakis,
  and Qian}{Zhang et~al\mbox{.}}{2018}]%
        {zhang2018graphp}
\bibfield{author}{\bibinfo{person}{Mingxing Zhang}, \bibinfo{person}{Youwei
  Zhuo}, \bibinfo{person}{Chao Wang}, \bibinfo{person}{Mingyu Gao},
  \bibinfo{person}{Yongwei Wu}, \bibinfo{person}{Kang Chen},
  \bibinfo{person}{Christos Kozyrakis}, {and} \bibinfo{person}{Xuehai Qian}.}
  \bibinfo{year}{2018}\natexlab{}.
\newblock \showarticletitle{{GraphP: Reducing Communication for {PIM}-based
  Graph Processing with Efficient Data Partition}}. In
  \bibinfo{booktitle}{\emph{High Performance Computer Architecture (HPCA), 2018
  IEEE International Symposium on}}. IEEE, \bibinfo{pages}{544--557}.
\newblock


\bibitem[\protect\citeauthoryear{Zhang, Abu-Khzam, Baldwin, Chesler, Langston,
  and Samatova}{Zhang et~al\mbox{.}}{2005}]%
        {zhang2005genome}
\bibfield{author}{\bibinfo{person}{Yun Zhang}, \bibinfo{person}{Faisal~N
  Abu-Khzam}, \bibinfo{person}{Nicole~E Baldwin}, \bibinfo{person}{Elissa~J
  Chesler}, \bibinfo{person}{Michael~A Langston}, {and}
  \bibinfo{person}{Nagiza~F Samatova}.} \bibinfo{year}{2005}\natexlab{}.
\newblock \showarticletitle{Genome-scale computational approaches to
  memory-intensive applications in systems biology}. In
  \bibinfo{booktitle}{\emph{SC'05: Proceedings of the 2005 ACM/IEEE Conference
  on Supercomputing}}. IEEE, \bibinfo{pages}{12--12}.
\newblock


\bibitem[\protect\citeauthoryear{Zhang, Kiriansky, Mendis, Amarasinghe, and
  Zaharia}{Zhang et~al\mbox{.}}{2017b}]%
        {zhang2017making}
\bibfield{author}{\bibinfo{person}{Yunming Zhang}, \bibinfo{person}{Vladimir
  Kiriansky}, \bibinfo{person}{Charith Mendis}, \bibinfo{person}{Saman
  Amarasinghe}, {and} \bibinfo{person}{Matei Zaharia}.}
  \bibinfo{year}{2017}\natexlab{b}.
\newblock \showarticletitle{Making caches work for graph analytics}. In
  \bibinfo{booktitle}{\emph{2017 IEEE International Conference on Big Data (Big
  Data)}}. IEEE, \bibinfo{pages}{293--302}.
\newblock


\bibitem[\protect\citeauthoryear{Zhao, Zhang, Xu, Zheng, and Cheng}{Zhao
  et~al\mbox{.}}{2019}]%
        {zhao2019kaleido}
\bibfield{author}{\bibinfo{person}{Cheng Zhao}, \bibinfo{person}{Zhibin Zhang},
  \bibinfo{person}{Peng Xu}, \bibinfo{person}{Tianqi Zheng}, {and}
  \bibinfo{person}{Xueqi Cheng}.} \bibinfo{year}{2019}\natexlab{}.
\newblock \showarticletitle{Kaleido: An Efficient Out-of-core Graph Mining
  System on A Single Machine}.
\newblock \bibinfo{journal}{\emph{arXiv preprint arXiv:1905.09572}}
  (\bibinfo{year}{2019}).
\newblock


\bibitem[\protect\citeauthoryear{Zhao and Yu}{Zhao and Yu}{2017}]%
        {zhao2017all}
\bibfield{author}{\bibinfo{person}{Kangfei Zhao} {and}
  \bibinfo{person}{Jeffrey~Xu Yu}.} \bibinfo{year}{2017}\natexlab{}.
\newblock \showarticletitle{All-in-one: Graph processing in rdbmss revisited}.
  In \bibinfo{booktitle}{\emph{Proceedings of the 2017 ACM International
  Conference on Management of Data}}. \bibinfo{pages}{1165--1180}.
\newblock


\bibitem[\protect\citeauthoryear{Zheng, Zhao, Huang, Wang, Zeng, Xue, Liao, and
  Jin}{Zheng et~al\mbox{.}}{2020}]%
        {zheng2020spara}
\bibfield{author}{\bibinfo{person}{Long Zheng}, \bibinfo{person}{Jieshan Zhao},
  \bibinfo{person}{Yu Huang}, \bibinfo{person}{Qinggang Wang},
  \bibinfo{person}{Zhen Zeng}, \bibinfo{person}{Jingling Xue},
  \bibinfo{person}{Xiaofei Liao}, {and} \bibinfo{person}{Hai Jin}.}
  \bibinfo{year}{2020}\natexlab{}.
\newblock \showarticletitle{Spara: An Energy-Efficient ReRAM-Based Accelerator
  for Sparse Graph Analytics Applications}. In \bibinfo{booktitle}{\emph{2020
  IEEE International Parallel and Distributed Processing Symposium (IPDPS)}}.
  IEEE, \bibinfo{pages}{696--707}.
\newblock


\bibitem[\protect\citeauthoryear{Zhou, Imani, Gupta, Kim, and Rosing}{Zhou
  et~al\mbox{.}}{2019}]%
        {zhou2019gram}
\bibfield{author}{\bibinfo{person}{Minxuan Zhou}, \bibinfo{person}{Mohsen
  Imani}, \bibinfo{person}{Saransh Gupta}, \bibinfo{person}{Yeseong Kim}, {and}
  \bibinfo{person}{Tajana Rosing}.} \bibinfo{year}{2019}\natexlab{}.
\newblock \showarticletitle{GRAM: graph processing in a ReRAM-based
  computational memory.}. In \bibinfo{booktitle}{\emph{ASP-DAC}}.
  \bibinfo{pages}{591--596}.
\newblock


\bibitem[\protect\citeauthoryear{Zhou, Chelmis, and Prasanna}{Zhou
  et~al\mbox{.}}{2015a}]%
        {zhou2015sssp}
\bibfield{author}{\bibinfo{person}{Shijie Zhou}, \bibinfo{person}{Charalampos
  Chelmis}, {and} \bibinfo{person}{Viktor~K Prasanna}.}
  \bibinfo{year}{2015}\natexlab{a}.
\newblock \showarticletitle{Accelerating large-scale single-source shortest
  path on FPGA}. In \bibinfo{booktitle}{\emph{Parallel and Distributed
  Processing Symposium Workshop (IPDPSW), 2015 IEEE International}}. IEEE,
  \bibinfo{pages}{129--136}.
\newblock


\bibitem[\protect\citeauthoryear{Zhou, Chelmis, and Prasanna}{Zhou
  et~al\mbox{.}}{2015b}]%
        {zhou2015pagerank}
\bibfield{author}{\bibinfo{person}{Shijie Zhou}, \bibinfo{person}{Charalampos
  Chelmis}, {and} \bibinfo{person}{Viktor~K Prasanna}.}
  \bibinfo{year}{2015}\natexlab{b}.
\newblock \showarticletitle{Optimizing memory performance for FPGA
  implementation of pagerank.}. In \bibinfo{booktitle}{\emph{ReConFig}}.
  \bibinfo{pages}{1--6}.
\newblock


\bibitem[\protect\citeauthoryear{Zhou, Chelmis, and Prasanna}{Zhou
  et~al\mbox{.}}{2016}]%
        {zhou2016high}
\bibfield{author}{\bibinfo{person}{Shijie Zhou}, \bibinfo{person}{Charalampos
  Chelmis}, {and} \bibinfo{person}{Viktor~K Prasanna}.}
  \bibinfo{year}{2016}\natexlab{}.
\newblock \showarticletitle{High-throughput and energy-efficient graph
  processing on fpga}. In \bibinfo{booktitle}{\emph{Field-Programmable Custom
  Computing Machines (FCCM), 2016 IEEE 24th Annual International Symposium
  on}}. IEEE, \bibinfo{pages}{103--110}.
\newblock


\bibitem[\protect\citeauthoryear{Zhou, Kannan, Zeng, and Prasanna}{Zhou
  et~al\mbox{.}}{2018}]%
        {zhou2018framework}
\bibfield{author}{\bibinfo{person}{Shijie Zhou}, \bibinfo{person}{Rajgopal
  Kannan}, \bibinfo{person}{Hanqing Zeng}, {and} \bibinfo{person}{Viktor~K
  Prasanna}.} \bibinfo{year}{2018}\natexlab{}.
\newblock \showarticletitle{An FPGA framework for edge-centric graph
  processing}. In \bibinfo{booktitle}{\emph{Proceedings of the 15th ACM
  International Conference on Computing Frontiers}}. ACM,
  \bibinfo{pages}{69--77}.
\newblock


\bibitem[\protect\citeauthoryear{Zhou and Prasanna}{Zhou and Prasanna}{2017}]%
        {zhou2017accelerating}
\bibfield{author}{\bibinfo{person}{Shijie Zhou} {and} \bibinfo{person}{Viktor~K
  Prasanna}.} \bibinfo{year}{2017}\natexlab{}.
\newblock \showarticletitle{Accelerating Graph Analytics on CPU-FPGA
  Heterogeneous Platform}. In \bibinfo{booktitle}{\emph{2017 29th International
  Symposium on Computer Architecture and High Performance Computing
  (SBAC-PAD)}}. IEEE, \bibinfo{pages}{137--144}.
\newblock


\bibitem[\protect\citeauthoryear{Zhou and Nishizeki}{Zhou and
  Nishizeki}{1999}]%
        {DBLP:journals/jgaa/ZhouN99}
\bibfield{author}{\bibinfo{person}{Xiao Zhou} {and} \bibinfo{person}{Takao
  Nishizeki}.} \bibinfo{year}{1999}\natexlab{}.
\newblock \showarticletitle{Edge-Coloring and f-Coloring for Various Classes of
  Graphs}.
\newblock \bibinfo{journal}{\emph{J. Graph Algorithms Appl.}}
  \bibinfo{volume}{3}, \bibinfo{number}{1} (\bibinfo{year}{1999}),
  \bibinfo{pages}{1--18}.
\newblock
\urldef\tempurl%
\url{https://doi.org/10.7155/jgaa.00012}
\showDOI{\tempurl}


\bibitem[\protect\citeauthoryear{Zhu, Akin, Sumbul, Sadi, Hoe, Pileggi, and
  Franchetti}{Zhu et~al\mbox{.}}{2013a}]%
        {zhu20133d}
\bibfield{author}{\bibinfo{person}{Qiuling Zhu}, \bibinfo{person}{Berkin Akin},
  \bibinfo{person}{H~Ekin Sumbul}, \bibinfo{person}{Fazle Sadi},
  \bibinfo{person}{James~C Hoe}, \bibinfo{person}{Larry Pileggi}, {and}
  \bibinfo{person}{Franz Franchetti}.} \bibinfo{year}{2013}\natexlab{a}.
\newblock \showarticletitle{A 3D-stacked logic-in-memory accelerator for
  application-specific data intensive computing}. In
  \bibinfo{booktitle}{\emph{2013 IEEE international 3D systems integration
  conference (3DIC)}}. IEEE, \bibinfo{pages}{1--7}.
\newblock


\bibitem[\protect\citeauthoryear{Zhu, Graf, Sumbul, Pileggi, and
  Franchetti}{Zhu et~al\mbox{.}}{2013b}]%
        {zhu2013accelerating}
\bibfield{author}{\bibinfo{person}{Qiuling Zhu}, \bibinfo{person}{Tobias Graf},
  \bibinfo{person}{H~Ekin Sumbul}, \bibinfo{person}{Larry Pileggi}, {and}
  \bibinfo{person}{Franz Franchetti}.} \bibinfo{year}{2013}\natexlab{b}.
\newblock \showarticletitle{Accelerating sparse matrix-matrix multiplication
  with 3D-stacked logic-in-memory hardware}. In \bibinfo{booktitle}{\emph{2013
  IEEE High Performance Extreme Computing Conference (HPEC)}}. IEEE,
  \bibinfo{pages}{1--6}.
\newblock


\bibitem[\protect\citeauthoryear{Zhuo, Wang, Zhang, Wang, Niu, Wang, and
  Qian}{Zhuo et~al\mbox{.}}{2019}]%
        {zhuo2019graphq}
\bibfield{author}{\bibinfo{person}{Youwei Zhuo}, \bibinfo{person}{Chao Wang},
  \bibinfo{person}{Mingxing Zhang}, \bibinfo{person}{Rui Wang},
  \bibinfo{person}{Dimin Niu}, \bibinfo{person}{Yanzhi Wang}, {and}
  \bibinfo{person}{Xuehai Qian}.} \bibinfo{year}{2019}\natexlab{}.
\newblock \showarticletitle{Graphq: Scalable {PIM}-based graph processing}. In
  \bibinfo{booktitle}{\emph{Proceedings of the 52nd Annual IEEE/ACM
  International Symposium on Microarchitecture}}. \bibinfo{pages}{712--725}.
\newblock


\end{thebibliography}
